To the blessed memory of my parents
who taught me to think.

# World of Movable Objects

## *Preface*

Suppose that you are sitting at the writing table. In front of you there are books, text-books, pens, pencils, pieces of paper with some short notes, and a lot of other things. You need to do something, you have to organize your working place, and for this you will start moving the things around the table. There are no restrictions on moving all these things and there are no regulations on how they must be placed. You can put some items side by side or atop each other, you can put several books in an accurate pile, or you can move the bunch of small items to one side in a single movement and forget about them, because you do not need them at the moment. You do not need any instructions for the process of organizing your working place; you simply start putting the things in such an order, which is the best for you at this particular moment. Later, if something does not satisfy you or if you do not need some items in their places, you will move some of the items around and rearrange everything in whatever order you need without even paying attention to this process. You make your working place comfortable for your work at each moment and according to your wish.

The majority of those who are going to read this text nearly forgot everything about paper, books, hand writing… A personal computer became the only instrument and the working place for millions of people. Screens of our computers are occupied with different programs; the area of each program is filled with many different objects. Whenever you need to do something, you start the needed program and in each of them you know exactly, what you want to do and what you can do. Did it ever come to your attention that in any program you know exactly the whole set of possible movements and actions? Have you ever understood that the set of allowed actions is extremely small and you try to do your work within a strictly limited area of allowed steps? Those limits are the same in all the programs; that is why there are no questions about the fairness of such situation. You can press a button; you can select a line or several lines in a list; you can do several other typical things, but you never go out of the standard actions which you do in any other program. You and everyone else know exactly what they are allowed to do. Other things are neither tried nor discussed. They simply do not exist.

Now try to forget for a minute that for many years you were taught what you could do. Let us say that you know, how to use the mouse (press – move – release), but there are no restrictions on what you are allowed to do with a mouse. Switch off the rules of your behaviour "in programs" according to which all your work was going for years.

Try the new set of rules:

- You can press ANY object and move it to any new location.
- You can press the border of any object and change its size in the same easy way; there are no limitations on such resizing (except some obvious and natural).
- A lot of objects you can reconfigure in the same simple way by moving one side or another.

It must be obvious to you that the reaction of a program on your clicking a button does not depend on the screen position of this button, so if you change the size or location of a button or a list it is not going to change any code that is linked with clicking a button, selecting a line in the list or making a choice between several positions. All the programs are still going to work according to their purposes. Buttons and lists represent a tiny part of objects that occupy the screens. Now make one more step and imagine the programs, in which ALL the objects are under your control in exactly the same way. You can do whatever you want with the objects, while the programs continue to work according to their purposes.

What would you say about such a world? Do not be in a hurry with your answer. You never worked with such programs; you would better try before giving an answer. This book is about the screen world of movable and resizable objects. Those objects can be of very different shapes, behaviour, and origin; that is why there are more than 80 examples in the accompanying program. Some of those examples are simple; others are in reality complex and big enough applications by themselves. And there is <u>not a single unmovable object in all of them</u>.

The world of movable objects – the world of *user-driven applications*.

# Contents

# Introduction

Some time ago I wrote two articles to describe the main results of my work throughout the last four years. The first of them appeared in the late fall of 2009 and was called "*On the theory of moveable objects*" [5]. The second article appeared in spring of 2010 and was called "*User-driven applications*" [6]. These papers came in a row with my previous articles [7 - 11] and summarized the results of years of work. Both papers [5, 6] were accompanied by their own demonstration programs and the full codes for both projects. I tried to make those articles as short as possible because they were also published on the site www.codeproject.com, where the average and expected publications are usually short. Even with all my efforts, each of those two articles was around 30 pages. It was the limit below which I could not go; even such volume required the removal of many important explanations. As a result, both texts were left without some needed pieces, though they gave the main idea of my work. To reduce the volume of those articles, I often included into the texts some remarks on the code, but without putting the code itself next to the explanation. The codes can be looked through separately because all of them are available, but this process requires much more efforts from the readers. I got some complaints about it, but I knew these things before publishing the articles, so I had only to agree with those remarks.

The two articles [5, 6] are strongly related and represent the halves of one general problem and its solution. The first article was about "How to do…"; the second was about "What would happen if the previous results were used". The division in two parts was done only due to the technical (publishing) problems; in reality those two things must be always regarded as a single theory. So, this is the main idea of this book: to introduce the whole theory without any divisions and to do it in a more detailed form.

The book is divided into two big parts which correspond to those two articles, but there is some problem in such presentation. The book has a linear structure with the chapters going one after another. Any next chapter can use the explanations from the previous part of the book and introduce something new. At the same time, very few real algorithms have a linear structure; the structure of a tree is much more common in programming than anything else. Each new piece opens the way for a new branch or even several branches of ideas. It would be nice to have the book with the same tree structure, but I do not know how to write such a book, so I continue to write this one in a traditional linear way.

The book is accompanied by a program with a lot of examples. Even from the beginning, the examples are designed according to all the ideas of the user-driven applications, but the explanation of these ideas and the discussion of how they were born and why the applications must be designed in such a way appear much later, in the second half of the book. It cannot be done in another way: before talking about the ideas of user-driven applications I have to demonstrate the design of all the elements, without which such applications simply cannot exist.

I would suggest one thing. If you are not familiar with those two articles [5, 6], look through them first and at the programs that come with them. (Applications and articles are available at www.sourceforge.net.) Look at them at least in a quick way; this will give you some understanding of the whole area of discussion. But be aware that even if later you see the similar examples from the articles in the book, the later versions from the book usually have some significant additions.

This book is about two things:

- The design of movable / resizable objects.
- The development of user-driven applications.

The two theories can be looked at as the independent things because:

- The design of movable objects is not influenced in any way by the afterthoughts of where these objects are going to be used.
- The design of user-driven applications is independent of the real algorithm for constructing movable objects.

At the same time there is a very strong relation between the two things, as the user-driven applications can be designed only on the basis of movable / resizable objects and all the extraordinary features of such applications are the results of their construction exclusively and entirely of movable objects. The movable / resizable objects can be used by themselves in the existing applications of the standard type, but only invention of an algorithm turning an arbitrary object into movable / resizable allowed to design an absolutely new type of programs – user-driven applications.

Often enough, when I tell people that I have thought out an algorithm of making movable any screen object, a lot of people are a bit (or strongly) surprised: "What are you talking about? Is there anything new in it? We have seen objects moving around the screen for years and years?" Certainly, they saw; as the demonstration of moving objects has the history of several decades. Only all those objects moved according to the scenario written by their developers. Users could do nothing about that moving except watching. The thing I was working on for years and which I am going to describe here is absolutely different: it is the moving of screen objects by the users of the programs. This article is about the development of screen objects, movements of which are not predicted by the designer of an application. It is not about a film developed on a predetermined scenario. It is about an absolutely new type of programs – *user-driven applications*. In these programs all objects, from the simplest to the most complicated, consisting of many independent or related parts, are designed to be moved and resized only by USERS. The objects are designed with these special features, and then the whole control of WHAT, WHEN, and HOW is going to appear on the screen is given to the users.

I would like to mention beforehand that the overall behaviour of user-driven applications is so different from whatever you experienced before that you can feel a shock or at least a great amusement at the first try. From my point of view, such a reaction from the people who are introduced to the user-driven applications for the first time supposed to be absolutely natural. A lot of people had the same feeling of a shock when, after years of work under DOS, they tried the Windows system for the first time. By the way, the only visual difference of the Windows system from the familiar DOS was the existence of several movable / resizable windows (in comparison with a single one and unmovable) and the possibility of moving icons across and along the screen. That was all! And even that was a shock. From the users' point of view, the step from the currently used programs to the user-driven applications is much bigger than that old step from DOS to Windows. In user-driven applications EVERYTHING is movable and resizable, and this is done not according to some predefined scenario, but by users themselves.

Movability of elements is not some kind of an extra feature that is simply added to well known objects in order to improve their behaviour. Technically (from the programming point of view), it is adding a new feature, but it turned out to be not simply a small addition to the row of other features. The movability of objects which are well known for years changes the way of using these objects absolutely. The most remarkable result of this change is that the applications have to be redesigned, while movability of all their parts must be taken into consideration. Movable and unmovable elements cannot coexist on the screen in any way; they immediately begin to conflict and demand the transformation of any unmovable objects into movable. Movability of elements changes the whole system of relations between the objects of applications. Movability of objects is the main, the fundamental feature of the new design. .

Throughout the whole history of programming we have a basic rule which was never doubted since the beginning and up till now and, as a result, turned into an axiom: <u>any application is used according to the developer's understanding of the task and the scenario that was coded at the stage of the design</u>. After reading this, you will definitely announce a statement: "Certainly. How else can it be?" Well, for centuries there was a general view, which eventually turned into an axiom, that the Sun was going around the Earth. There were no doubts about it. Yet, it turned out to be not correct.

40 years ago the majority of programs were aimed at solving some scientific or engineering tasks and the overwhelming majority of those programs were written in FORTRAN. I think that not too many readers of this book can explain or even remember the origin of this name; I would remind that it stands for **FOR**mula **TRAN**slation. The main purpose of the language was to translate the equations and algorithms into the intermediate notation, which was, in its turn, translated into the inner machine codes. The main goal of the language was to deal with formulas! Computers were big calculators and nothing else. It was not strange at all that whatever commands there were for visualizing the data and results, those commands were intermixed with other commands for calculations. At that time nobody asked the questions about such development of programs. The author of a program tried to write the consecution of instructions to turn formulas into the final results; it would be nice to see some intermediate results, if the calculations were long and complicated. They were often very long and complicated, so few extra operations for showing out intermediate values were incorporated into the block of calculations just at the places, where those values were obtained.

Years later much better visualization was achieved both with the hardware improvement and the design of new languages. With this there came understanding that calculations and visualization had to be separated. They were separated from the point of programming, but the same person – developer – is still responsible for everything. A developer knows all the insides of the calculations and he decides what, when, and how to show. Eventually this developers' dictate came into conflict with the wide variety of users' requests for visualization; the adaptive interface was thought out to solve the problem. Many forms of adaptive interface were proposed (the dynamic layout is only one of its popular branches), but all those numerous solutions are only softening the problems but not solving them. The main defect of the adaptive interface is in its own base: the designer puts into the code the list of available choices for each and all situations he can think about. Users have no chances to step out of the designer's view and understanding of any situation. It is the dead end of evolution programs under those ideas which were proposed around 25 years ago.

With the movability of all the parts from the tiny elements to the most complex objects, there is another way of application design, when a program continues to be absolutely correct from the point of fulfilling its main purpose (whatsoever this purpose is), but at the same time does not work according to the predefined scenario and does not even need such a scenario. To do such a thing, an application has to be developed not as a program in which whatever can be done with it has to be thought out by the developer beforehand and hard coded; instead an application is turned into an instrument of solving problems in particular area. An instrument has no fixed list of things that can be done with it, but only an idea of how it can be used; then an instrument is developed according to this idea. A user of an instrument has full control of it; only the user decides when, how, and for what purpose it must be used. Exactly the same thing happens with programs that are turned into instruments.

I call the programs, based on movable / resizable objects, ***user-driven applications***. When you get a car, you get an instrument of transportation. Its manual contains some suggestions on maintenance, but there is no fixed list of destinations for this car. You are the driver, you decide about the place to go and the way to go. That is the meaning of the term *user-driven application*: you run the program and make all the decisions about its use; a designer only provides you with an instrument which allows you to drive.

The first half of this book is about the design of movable objects. I have already mentioned the first misunderstanding of the importance of this task, based on not realizing the difference between the moving according to the predefined scenario (it is simply an animation) and the moving of objects according to the user's wish. But when I explain the obvious difference between these two things, I often hear another statement. "Everyone can move and resize the windows at any moment and in any way he wants, so what is the novelty of your approach?" The answer is simple, but a bit longer than on the first question.

Rectangular windows are the basic screen elements of the Windows operating system.* You can easily move all these windows, resize them, overlap them or put them side by side. At any moment you can reorganize the whole screen view to whatever you really need. It was not this way at the beginning of the computer era; it became the law after Windows conquered the world. This is **axiom 1** in modern day programming design: *On the upper level, all objects are movable and resizable*. To make these features obvious and easy to use, windows have title bars by which they can be moved, and borders by which they can be resized. Being movable and resizable are standard features of all the windows, and only for special purposes these features can be eliminated.

Usually the goal of switching on the computer is not to move some rectangular windows around the screen, but to do something special in applications which are represented by those windows. You start an application you need, you step onto the inner level, and then everything changes. Here, inside the applications, you are doing the real work you are interested in, and at the same time you are stripped of all the flexibility of the upper level – you can only do what the designer of the program allows you to do. The design can be excellent or horrible, it can influence the effectiveness of your work in different ways, but still it is awkward that users are absolutely deprived of any control of the situation. Have you ever questioned the cause of this abrupt change? If you have, then you belong to the tiny percentage of those who did. And I would guess that the answer was: "Just because. These are the rules."

Unfortunately, these ARE the rules, but rules are always based on something. The huge difference between the levels is that on the upper level there is only one type of objects – windows, while on the inner level there are two different types: controls, inheriting a lot from windows, and graphical objects that have no inheritance from them and that are absolutely different. The addition of these graphical objects changes the whole inner world of the applications. (In reality, whatever you see at the screen is a graphical object, but as it was declared in the famous book [1] decades ago: "All animals are equal, but some animals are more equal than others". Controls are those "more equal" elements on the screen and their behaviour is absolutely different from the behaviour of ordinary graphical objects.)

The inheritance of controls from windows is not always obvious, as controls often do not look like windows. Controls have no title bars, so there is no indication that they can be moved; usually there are no borders that indicate the possibility of resizing. But programmers can easily use these features of all the controls and from time to time they do use them, for example, via anchoring and docking. The most important thing is not *how* controls can be moved and resized, but that for them moving and resizing *can be organized*, though I have to mention that the programmers use this moving and resizing of controls as their secret weapon and never give users direct access to it. The designer decides what will be good for users in one or another situation, and, for example, when a user changes the size of the window, then the controls inside can change their size and position, but only according to the decisions previously coded by the designer.

Graphical objects are of an absolutely different origin than controls and, by default, they are neither movable nor resizable. There are ways to make things look different than what they are in reality (programmers are even paid for their knowledge of such tricks). One technique that programmers use is to paint on top of a control: any panel is a control, so it is resizable by default; with the help of anchoring / docking features, it is fairly easy to make an impression as if you have a resizable graphics, which is changing its sizes according to the resizing of the form (dialog). Simply paint on top of a panel and make this panel a subject of anchoring / docking. By default, panels have no visible borders, and if the background color of the panel is the same as its parent form, then there is no way to distinguish between painting in the form or on the panel, which resides on it. Certainly, such "resizing" of graphics is very limited, but in some cases it is just enough; all depends on the

* There are other multi-window operating systems which also use rectangular windows as the basic element. When I write about moving of the windows, it is applied not only to the particular Windows system from Microsoft, but to the whole class of multi-window operating systems. So, in further text, I will not add "*and other similar operating systems*" every time when I mention Windows.

purpose of application. Another solution for resizing of rectangular graphical objects is the use of bitmap operations, but in most cases it cannot be used because of quality problems, especially for enlarging images. Both of these tricky solutions (painting on a panel or using bitmap operations) have one common defect – they can be used only with the rectangular objects.

If any limited area is populated with two different types of tenants (in our case – controls and graphical objects) which prefer to live under different rules, then the only way to organize their peaceful residence and avoid any mess is to force them to live under ONE law. Because graphics are neither movable nor resizable, the easiest solution is to ignore these controls' features, as if they do not exist. That is why so few applications allow users to move around any inner parts. Thus we have **axiom 2**: *On the inner level, objects are usually neither movable nor resizable*. Interestingly, the combined use of these two axioms creates this absolutely paradoxical situation:

- On the upper level, which is not so important for real work, any user has an absolute control of all the components, and any changes are done easily.
- On the inner level, which is much more important for any user because the real tasks are solved here, users have nearly no control at all. If they do have some control, then it is very limited and is always organized indirectly through some additional windows or features.

Axioms I mentioned were never declared as axioms in a strict mathematical way; at the same time I have never seen, read, or heard about even a single attempt to look at this awkward situation any other way than as an axiom and to design any kind of application on a different foundation. Programmers received these undeclared axioms from Microsoft and have worked under these rules for years without questioning them. If you project these same rules on your everyday life, it would be like this: you are free to move around the city or country, but somebody tells you where to put each piece of furniture inside your house. Would you question such a situation?

So, in the world of programs we have a situation when users have to do their most important work inside the applications while being deprived of the real control of these applications; the work can go on only according to the previously developed scenarios. (The questions of whether any form of adaptive interface can really solve the problems are discussed at the beginning of the second part of this book.) I realized years ago that the immovability of all the elements inside the applications became the main problem in design of many programs, but especially in complicated ones. My goal was to find a general solution which would allow to move and resize objects of an arbitrary shape. I had been looking for the general solution and I found it.

## About the structure of this book

There are two main parts in this book:

- Part 1. The design of movable / resizable objects.
- Part 2. The development of user-driven applications.

The first part contains 14 chapters.

Chapter 1 includes the description of the ideas which are transformed into the algorithm for turning any object into movable and resizable.

Chapter 2 describes nodes and covers – the basic level of movability; also the moving of the first real objects – lines – is demonstrated.

Next several chapters either analyse movability of widely used objects of the most popular shapes or describe movements that are often used by objects of different shapes.

Chapter 3 describes various movable rectangles which differ by the type of resizing.

Chapter 4 analyses the basic principles of rotation.

Chapter 5 describes texts in different movements.

Chapter 6 analyses the moving of different polygons (regular, convex, etc.). Also the first example of not the abstract figure, but the real object in moving and rotation is demonstrated here.

Chapter 7 describes the design of the N-node covers which are used in resizing of objects with curved borders.

Chapter 8 analyses movability of objects with the unusual shapes; transparent nodes can be very useful for rings, crescents, sectors, and objects with the holes.

Chapter 9 is about the inner movements – the movements of sliding partitions in already familiar objects - rectangles, circles, and rings. An example of moving an unlimited number of objects of many different shapes is also included into this chapter.

Chapter 10 discusses complex objects, parts of which are involved in individual, related, and synchronous movements. Here questions of identification of objects are discussed. Interesting elements, like track bars, which are widely used in more complex applications are also discussed in this chapter. One example demonstrates a simplified version of scientific applications.

Chapter 11 is about different types of movement restrictions and overlapping prevention. The involvement of many parts and many different objects begins to turn examples into real applications.

All the previous chapters of the first part were devoted to graphical objects; the last three chapters of this part are mostly about controls.

Chapter 12 is about moving / resizing of the solitary controls.

Chapter 13 analyses the widely used pair of elements "control + text" in which the text is represented not as another control (`Label`), but as a painted object.

Chapter 14 is about the movable / resizable groups. Such groups can be organized on absolutely different principles with different levels of fixation for relative positions of elements. The elements of groups can be either pure controls or some combinations of controls and graphical objects, which can influence the design of different groups and produce some unusual solutions.

The discussion of the groups completes the analysis of the movable / resizable elements which can be used in all types of applications. When it is known how to turn into movable / resizable an object of any shape and origin; when it is demonstrated how to organize not only individual, but also the synchronous and related movements of the objects, and after the detailed analysis of movable groups, it is time to turn to the development of applications based on all these elements.

The second part contains five chapters. There are no more small examples to demonstrate one or another detail; this is all about the design of new type of programs, so all the examples are the real applications; some of them are really big.

Chapter 15 postulates the basic rules of user-driven applications and demonstrates their use in the first relatively small application. Another example demonstrates the design on the basis of that class of groups, which I think is the best for development of programs for many different areas. At least, this class of groups is used in all the demonstrated applications beginning from the main form and up to the auxiliary forms at all the levels.

Chapter 16 is about the scientific and engineering programs. The request for the movability of elements was born in this area; this is the area in which I try all my ideas on movability and design; this is the area of my greatest interest throughout my entire professional life. Not surprisingly that the design of user-driven applications in this area is discussed at the most detailed level.

Chapter 17 discusses another attractive area for applying user-driven applications – design of programs for financial and economical analysis. Three types of plots are used in the examples (bar charts, pie charts, and ring sets), but other similar and not very similar plots can be used in the same way. The variety of movable / resizable plots is not the crucial thing; the main discussion is about HOW such applications can be developed on the basis of movable elements.

Chapter 18 is devoted to the transformation of the applications already in use into user-driven ones. How much efforts can be needed? What changes have to be done on the programmers' side? Is it difficult for users to switch from the old style programs to the new one? The example program of this chapter is the one with which everybody is familiar – Calculator.

Chapter 19 demonstrates that the principles of user-driven applications can be very well used even in the areas which are far away from the most serious scientific applications, where they were born.

The book includes the summary of rules for design of movable objects and development of user-driven applications. There are also references and links to other programs and documents, which were developed for better explanation of these items. In addition, there are three appendices.

Appendix A discusses the visualization of covers. Such visualization is never used in the real applications, but it is used in the first part of this book for better explanation of the design of covers.

Appendix B discusses the opportunity of using several movers. There are no examples with more than one mover throughout the entire book (except this appendix), but there are situations when more than one mover is needed.

Appendix C contains the list of all the examples of this book. There are more than 80 different examples; it would be difficult to navigate through and quickly find the needed example without such a reminder. The table includes a small picture of each form and the information about the chapter and the page where its explanation and discussion can be found.

## Preliminary remarks

When any well known object gets new features, the users should be aware of these new things. The goal of my work is to develop an easy to use algorithm for turning ANY screen object into movable and resizable. All those innumerable objects were developed at the best level by their designers, and I do not want in any way to interfere in their design. The appearance of the objects must be exactly the same as it was thought out by their designers. But if there is no visual indication of the new features, then how the users will know that the objects on the screen are movable and resizable? They need to know this fact! Try not to exclaim any astonishment or indignation on reading the previous sentence. I only want to remind you that there is no indication that windows on the upper level can be moved and resized; everyone should know this fact, and this knowledge is enough to navigate throughout the Windows or similar systems. There were no such things before the Windows era, so everyone, who was going to use it for the first time, had to be told beforehand that all those windows could be moved and resized and how he was going to do it. Some people are old enough to remember what was before Windows and at the same time are not old enough to forget it; the younger generation may think that they are already born with the knowledge of movability of windows, but I have a feeling that even they have to be informed about this fact at one moment or another.

After this reminder, I think that the fact that users of the new user-driven applications have to be informed at least once about the movability of each and all objects is not an outrageous one.

Now, if you are aware that all the objects on the screen are movable, then how are you going to move them? The mouse seems to be an obvious and the best instrument, as there are programs in which one or another element is moved, and this is always done with a mouse. Again, the moving of windows is the first example which comes to one's mind. But there is a big difference between the moving of windows on the upper level and my algorithm for all kinds of elements. Each window has a title bar by which it is moved. This decision seems to be good enough when the fact that all the windows have the standard rectangular shape is taken into consideration. I propose the algorithm which works for the objects of an arbitrary shape, so my vision is that objects must be moved by any inner point. And if an object is moved by any point, then there is no need at all in any indication of its movability. You simply should know this fact and that is enough.

If there is no indication of the movability of an object, then how are we going to decide whether an object is movable or not? We do not need to do it: **all the objects must be movable**. I hope that this time you contained your indignation. Have you ever had doubts about the movability of any particular window? So all the objects in the Demo application **WorldOfMoveableObjects** are movable; if any object in this application is not movable, then it was done purposely for better comparison and explanation.

The movability of all the elements is the basis of *user-driven applications*. I have heard the complaints about the underlined statement, but only from the people who never tried the applications on movable / resizable elements. (The famous outcry that "driving the car faster than 15 miles per hour is dangerous for animals and people and thus must be forbidden", that outcry came only from those who never tried the car, but not from those who had ever made a single car trip.)

There are a lot of different objects in our applications; these objects are supposed to be involved in different types of movements; if there are no indications of any movements at all, how are the users going to understand which movements can be applied to each object? This is one of the problems that are mentioned in a lot of examples which I am going to demonstrate and discuss in further chapters.

**Figure I.1** demonstrates the first view that you see on starting the **WorldOfMoveableObjects** application. This is, I think, also the first of my Demo applications in which I did not put on the screen the reminder that ALL the objects in this program (not only in this form, but in all the forms of this application!) are movable. The objects that you see in this figure are discussed in detail in one or another chapter of the book. I think that a quick look across this picture from one object to another may give you a better understanding of what you can find further on. Like a scheme at the entrance of a big museum. Only instead of the "Do not touch the exhibits", you have "Any object can be moved by any inner point", so I do not need to repeat it for each of them.

Polygon in the middle. Configuration of this polygon can be changed by moving any end point of any segment of the perimeter or the central point. By changing the configuration you can literally turn the figure inside out. All other points of the perimeter, except the apices, can be used for scaling the polygon. The polygon can be rotated (right mouse press) by any inner point. Class `ChatoyantPolygon` is discussed in the subsection of the chapter *Polygons*.

Group in the top right corner.

The group consists of controls paired with their comments. Any control is moved by the frame; its comment moves synchronously. Comments can be moved independently and placed anywhere. Class `CommentedControl` is discussed in the chapter *Control + text*. A set of elements constitutes the group. The frame of the group adjusts to the positions of all the inner elements and is always shown around them. The title of the group can be moved left and right between the two sides of the group. Class `ElasticGroup` is discussed in the chapter *Groups of elements* and is widely used in all the complex applications demonstrated in the second part of the book. Class `ElasticGroup` has a special tuning form which is used in all other places, but not in this form.

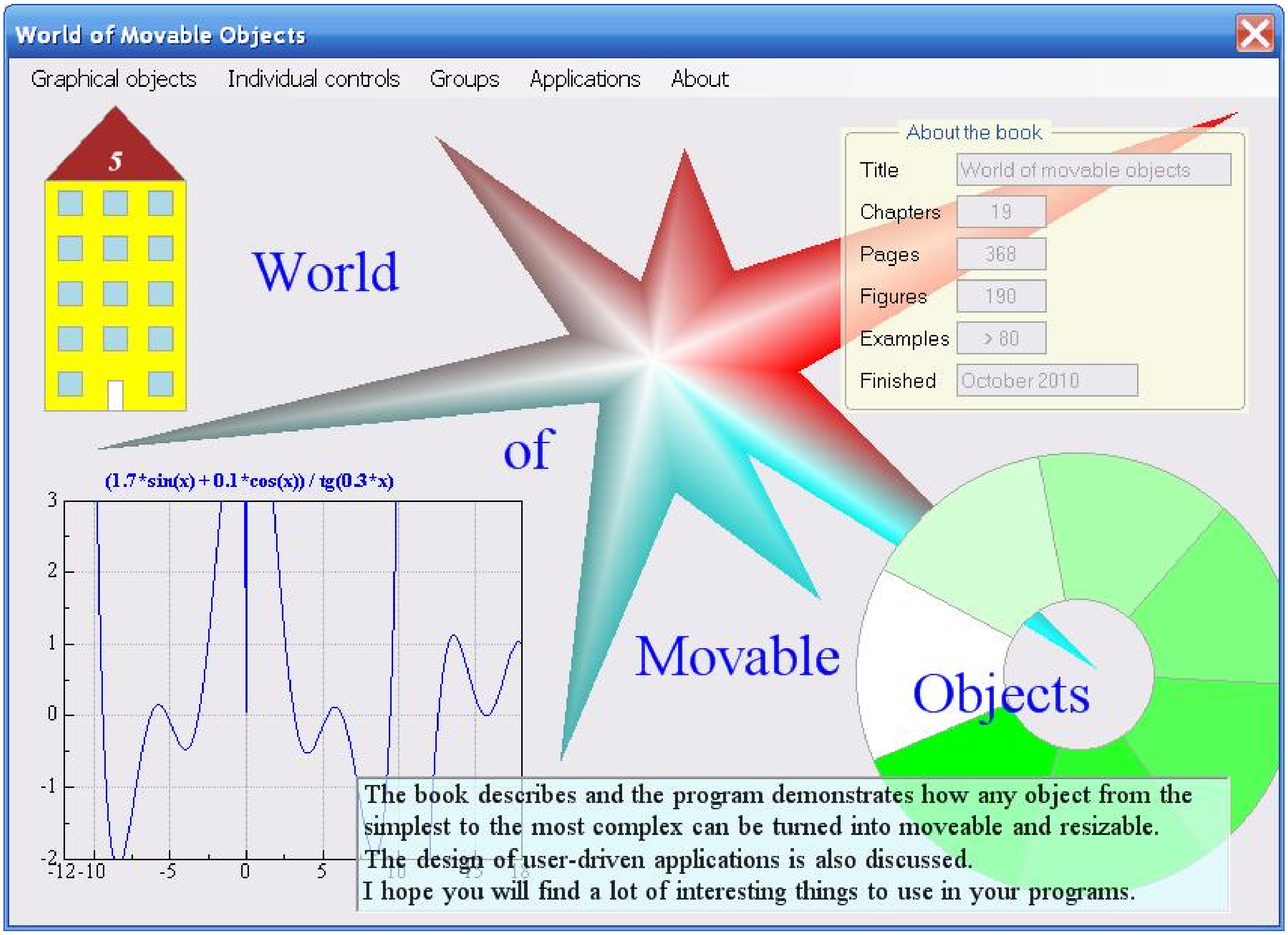

**Fig.I.1** The first view that you will see on starting the application

Plot in the bottom left corner.

The main plotting area is resized by borders and corners. The scales can be moved individually and positioned anywhere (with some restrictions, as there must be the conformity between the scales and the main plotting area). Comment belongs to the `CommentToRect` class; the identical copy of this class is discussed in the chapter *Complex objects*. Classes `Scale` and `Plot` are discussed in the chapter *Applications for science and engineering*. Both classes have special tuning forms which are used in all other places (forms), but not in this one.

Ring in the bottom right corner.

Ring can be resized by any point of the outer or inner circle; resizing of rings is discussed in the chapter *Curved borders. N-node covers*. The cover of a ring uses a special technique which is discussed in the chapter *Transparent nodes*. This ring has the sliding partitions. Class `PrimitiveRing` is discussed in the chapter *Data visualization*.

Information at the bottom.

This text can be moved but not rotated. Class `TextM` is discussed in the chapter *Texts*.

| | |
|---|---|
| Words across the form. | Can be moved and rotated by any point. Class `TextMR` is discussed in the chapter *Texts*. |
| House in the top left corner. | House can be resized by all four sides and all four corners of the rectangular part. The roof top can be moved not only up or down, but also to the sides (no requirement for the symmetry of the roof). Class `SimpleHouse` appears only in the *Appendix B*, but there are similar classes of houses in the chapters *Polygons* and *An exercise in painting*. |

There are two big differences between using objects of the same classes in the **Form_Main.cs** and further on. You can move the objects in the **Form_Main.cs**, you can resize them, but you cannot tune them, and the results are not saved for later use. Design of programs without these two things is against the laws of the user-driven applications, but I decided not to introduce them here. Everything will come at a proper time. Let us start.

# Requirements, ideas, algorithm

This chapter describes the basic requirements for the movable objects to be used. It also explains the algorithm and the basic steps to turn any object into movable and resizable.

## *Basic requirements*

"In science, finding the right formulation of a problem is often the key to solving it…" [2]. As a designer of very complicated systems, I know exactly, what features the system of movable / resizable elements must have to be useful for my work.

- I need an easy way to declare the objects in the form (dialog) movable and resizable; just touch them and the objects become movable ([3]).
- Easy does not mean that it can be applied only to primitive objects. The configuration of objects can be at any level of complexity; changing of configuration might be influenced by a lot of different things. For example, some objects may allow any changes, others may need fixing of several parameters (sizes), and parts of the objects may generate restrictions on changes of other parts. But in any case the whole variety of possible reconfigurations must be easy to understand and implement.
- These features – movable and resizable – must be added like the extra "invisible" features. They will not destroy any image, but it will be obvious that objects are movable and resizable. The best thing would be simply to know about these features without any additional indication. In some cases, I may need some instrument for indication of these new possibilities, but usually I would prefer to go on without any extra lines or marks.
- These features should not be an "all or nothing" case. Objects of the same classes can be used with or without these new features; so that the decision of declaring an object movable / unmovable, or changeable / fixed can be even done by a user, while the application is running.
- The whole technique of dealing with the movable / resizable elements must be extremely simple: press and move, press and reconfigure, and even press and rotate, which is not organized for windows, but can be very useful for many graphical objects. And if it is useful, it must be organized without any limitations.

Some of these "nice to have" features look like they conflict (simple but with all the possibilities you can imagine) and even provide alternatives (not visible and obvious), but I am not writing the article about the future of programming in 2025. The biggest secret is that all these things are already designed and work now; a programmer can apply these features to any object.

I constantly work with the C# language, so all the programs, I am going to mention here, are written in this language; the code examples are in C#, and I am going to use some terms from this language. But the algorithm and designed classes are not linked only to C# – they can be easily developed with other instruments; I simply prefer to use C#. The examples are from the **WorldOfMoveableObjects** application; all the codes from this project are available (see information in the section *Programs and Documents*).

The easiest way to move any object around the screen is to drag it, so only three mouse events are used for the whole process: `MouseDown`, `MouseMove`, and `MouseUp`. No additional keyboard pressing is used for moving / resizing anywhere throughout my applications. If an element is involved both in moving and resizing, then starting of one of these processes is determined only by the place, where an object was pressed by a mouse. The distinction between the forward movement and rotation, and a lot of elements are involved in both, is made on the pressed button: I use left button for forward movement and right button for rotation.

Applications are populated both with graphical objects and controls. Graphical objects can be of an arbitrary shape; the proposed technique is especially designed for turning into movable / resizable the objects of any shape. Standard controls have rectangular shape, which simplify some problems. Very important things for turning objects into movable / resizable can be automated in the case of controls. I will write about controls and groups of controls in the special chapters, but throughout the whole explanation, beginning from the simplest cases, you will see the movable controls and groups of controls in the same forms, where some features of moving / resizing the graphical objects are demonstrated. Now let us turn to the algorithm.

## Algorithm

My idea of making graphical objects movable and resizable is based on covering an object by a set of sensitive areas, which are used to start one or another type of movement. These sensitive areas are called ***nodes*** and belong to the `CoverNode` class. The nodes are usually invisible, so there is no direct visual indication about this new, but very important feature of the screen objects – their movability. The nodes can be visualized, but if such visualization is used in the real applications, then only on the initial stage of introducing the new ideas. It is enough for the users to know that all the objects are movable and resizable; after understanding and accepting this fact users do not need any more reminders. Movement of any node is transferred into the real movement of the screen object, with which it is associated. Four different types of movement are considered.

- If the size of an object is not changed and an object is moved without any change of relative positions of its parts and without rotation, then it is a forward movement.
- If all the parts of an object are increased or decreased with the same coefficient, but the general shape is not changed, then it is a resizing.
- If a movement of any part changes the relative positions of the parts and the general shape, then it is a reconfiguring.
- If the general shape is not changed, but the whole object is turned from the original position, then it is a rotation.

The first three movements (forward movement of the whole object, resizing, and reconfiguring) are started with the left button press; the choice between these movements is decided by the starting point and the node, to which this point belongs. There are special areas to start reconfiguring (it depends on the object) and resizing (usually it is the border); at any inner point of an object the forward movement of the whole object is usually started. The start of rotation is distinguished from other three movements by using the different button (the right one). The rotation of an object usually starts at any point.

There are no limits on the size of the nodes, but their possible shapes are limited to three variants and described by this enumeration.

```
enum NodeShape { Circle, Polygon, Strip };
```

A circular node is defined by its central point and radius.

A polygonal node is described by an array of points, representing its vertices. Polygon must be convex!

A strip node is defined by two points and radius. The strip node can be also looked at as a rectangle with two additional semicircles on the opposite sides. Those two points are the middles of those sides and the centers of the semicircles; the diameter of the semicircles is equal to the width of the strip.

In addition to the location, size, and shape, each node has a parameter, which determines the shape of the mouse cursor, when the mouse is moved across this node. Usually the change of the cursor, while the mouse is moved from one object to another and across the different parts of the objects, is the only prompt for the users that they can grab an object under the mouse and move it or change it one way or another, so it is better to make the shape of the cursor informative and not confusing. Throughout all the further programming examples you will find out that, for example, if the rectangular object can be resized in the horizontal or vertical direction, then the `Cursors.SizeWE` and `Cursors.SizeNS` are used. The resizing possibilities for objects of an arbitrary form are usually signaled by the `Cursors.Hand;` if the whole object can be moved in any direction, then is often signalled by the `Cursors.SizeAll.`

Each node has one more very important parameter, which describes its overall behaviour. The possible values of this parameter are described by the `Behavior` enumeration; I will return to this parameter a bit later.

An array of nodes, covering an object, is called a ***cover*** (class `Cover`). A cover must include at least one node. There are no other restrictions on the number of nodes, their types or sizes. The relative positions of the nodes are not regulated at all. If you want an object to be movable by any point, then the whole area of object must be covered with the nodes; any gaps in such situation will result in the appearance of the places, by which an object cannot be moved. The overlapping of the nodes is not a problem at all; in many cases such overlapping is used for better design of covers. If the overlapping nodes are used for different movements, then their order in the cover is very important, as the decision about the possible movement is made according to this order. Usually the number of nodes is small even for very complicated objects, but there are cases, when a significant number of nodes is needed. This often happens for the objects with the curved borders; covers of such type are called the ***N-node covers***.

Technically I saw two ways to add movable / resizable features to the objects: either to use an interface or an abstract class; after trying both ways I decided upon an abstract class. To turn any graphical object into being movable and resizable, it must be derived from the abstract class `GraphicalObject` and three crucial methods of this class must be overridden.

```
public abstract class GraphicalObject
{
    public abstract void DefineCover ();
    public abstract void Move (int dx, int dy);
    public abstract bool MoveNode (int i, int dx, int dy, Point ptMouse,
                                   MouseButtons catcher);
```

For graphical objects, the cover is organized in the **`DefineCover`**() method. Controls are automatically wrapped by a graphical object of the `SolitaryControl` class; for controls none of these three methods is needed, as everything is automated.

Each node in the cover has its personal identification number (from the range between 0 and the number of nodes in the cover); this number is often used for the node identification, when the decision about the movement is made in the `MoveNode`() method. It will be shown in many examples of this book that the number of the node is not the only way to make such a decision; the nodes of different shapes are often used to organize different kind of movements, so the shape of a node can be as helpful, as its number. But the number is the only way of an absolutely correct identification of the touched node in the case, when all or some of the nodes have similar shape.

**Note**. Though the whole technique of moving and resizing of objects is based on the idea of cover, the `DefineCover()` method is usually the only place, where you have to think about the cover! The rule is: organize the cover and forget about it. There is one exception of this rule; which is mentioned several lines below in the explanation of the `MoveNode()` method.

**`Move`** `(dx, dy)` is the method for forward moving of the whole object for a number of pixels, passed as the parameters.

The drawing of any graphical object with any level of complexity is usually based on one or few very simple elements (`Point` and `Rectangle`) and some additional parameters (sizes). While moving the whole object, the sizes are not changed, so only the positions of these basic elements have to be changed in this method.

**`MoveNode`** `(i, dx, dy, ptMouse, catcher)` is the method for individual moving of the nodes. The method returns a Boolean value, indicating whether the required movement is allowed. In the case of forward movement, the `true` value must be returned if any of the proposed movements along the X or Y axes is allowed. If the movement of one node results in relocation of other nodes, it is natural to call the `DefineCover`() method from inside the `MoveNode`() method, and then it does not matter, what value is returned by the `MoveNode`() method. The call for the `DefineCover`() from inside the `MoveNode()` method may happen even for the forward movement, when movement of one node affects the relocation of other nodes, and it usually happens with rotation, when all the nodes must be relocated. This is the exception of the previously mentioned rule that cover is not even thought about anywhere outside the method of its definition.

Though the call of the `DefineCover`() method from inside the `MoveNode`() method looks like a very good idea, it has a limitation of its own. In some cases, the movement of a node starts the resizing of an object, which in turn demands the change of the number of nodes in the cover (this depends on the design of cover); in such situation the call to the `DefineCover`() method from inside the `MoveNode`() method may cause a problem, because in the new cover the node with the same number (the one, which was originally pressed and caught) may belong to the node with different rules for moving. To avoid this problem, in such cases the cover is not changed throughout the resizing, but only when an object is released; a detailed explanation of such a case is given in the chapter *Curved borders. N-node covers*.

`MoveNode()` method has five parameters:

| | |
|---|---|
| `int i` | Identification number of the node – the same number that was used in the `DefineCover()` method on design of this node. |
| `int dx` | Movement (in pixels) along the horizontal scale; positive number means the moving from left to right. Use this parameter if you write the code for forward movement. |
| `int dy` | Movement (in pixels) along the vertical scale; positive number means the moving from top to bottom. Use this parameter if you write the code for forward movement. |
| `Point ptMouse` | The mouse cursor position. For calculations of forward movement I simply ignore this parameter and use the previous pair; for calculations of rotation I ignore the previous pair and use only this mouse position; I found it much more reliable for organizing any rotations. |

`MouseButtons catcher` Informs you, which mouse button was used to grab the object. If by the logic of an application the object can be grabbed by any mouse button, then ignore this parameter; if the move is allowed by one button only, then use this parameter; in case the node can be involved in both types of movement (forward movement and rotation), this is a very useful parameter to distinguish between them.

The `MoveNode()` method can be the longest of all three methods, because it must include the code for moving each node, but it is always uncomplicated, as more often than not the code for different nodes is partly the same. There are interesting situations, when the `MoveNode()` method is very short, though the number of nodes is big. For example, the *N-node* covers often consist of a significant number of nodes, but for such covers the `MoveNode()` method is usually very short, because the behaviour of all those nodes is identical.

When you work with the unmovable graphical objects, which is the standard situation for today, you deal with all of them as placed on the same surface, though even in this case there is an order of drawing such objects. This order determines the way you see these objects, as those that are drawn later appear atop of others that were drawn before. When all the objects are movable, they can be easily moved across the screen occasionally either closing other objects or going underneath. The system of graphical objects on the screen becomes multilevel with each object occupying its personal level. This is like a move from the old model of the world with all the starts positioned at the same sphere to the understanding of the Universe's infinity with each object having its personal distance from the Earth and moving in its own way.

An object, derived from the `GraphicalObject`, gets an ability to be moved and resized, but it can be really moved and resized only if it is registered with the ***mover*** object (of the `Mover` class). Mover supervises the whole moving / resizing process and in addition can provide a lot of information, associated with it. Regardless of the number of different objects, involved in moving / resizing, usually there is a single mover per form (dialog). However, there are situations when it is better (easier and more reliable) to have several movers in the form; for example, when you have movable objects on different panels.

Only three mouse events - **`MouseDown`**, **`MouseMove`**, and **`MouseUp`** - are used for the whole moving / resizing process, and these are the methods, in which mover works and must be mentioned. To organize the moving / resizing of the objects in the form, several steps must be made.

Declare and initialize a `Mover` object.

```
Mover mover;
mover = new Mover (this);
```

Register with the mover all the objects that you want to move and / or resize.

```
mover .Add (…);
mover .Insert (…);
```

Add the code for three mouse events.

```
private void OnMouseDown (object sender, MouseEventArgs e)
{
    mover .Catch (e .Location, e .Button);
}
private void OnMouseUp (object sender, MouseEventArgs e)
{
    mover .Release ();
}
private void OnMouseMove (object sender, MouseEventArgs e)
{
    if (mover .Move (e .Location))
    {
        Invalidate ();
    }
}
```

This is not a simplification; this is the code, which you can see in nearly every form of the Demo applications, regardless of the number or complexity of the objects, involved in moving / resizing. The calls to three mover methods (one call per each mouse event) are the only needed lines of code! Further on you will see some additional lines of code inside these methods, but they are used only for some auxiliary things like changing the order of objects on the screen or calling different context menus on different objects.

Those three mouse events – MouseDown, MouseMove, and MouseUp – are the standard and often the only places, where a mover is used. There are two other events – **MouseDoubleClick** and **Paint**, where mover can be mentioned and used, but this happens only on special occasions.

I often use the **MouseDoubleClick** event for opening the tuning forms of the complex objects, for example, scales and different plotting areas. Without mover at hand (before implementing the moving / resizing of objects), I had to decide about the clicked object by comparison of the mouse location and the boundaries of the objects. Mover can do this job much better, as it informs not only about the occurrence of any catch, but also about the class of the caught object, its order in the queue, and other useful things. As any object, derived from the GraphicalObject, gets a unique identification number, then with this I get an easy access to the object itself.

```
private void OnMouseDoubleClick (object sender, MouseEventArgs e)
{
    if (mover .Catch (e .Location, MouseButtons .Left)) {
        int iInMover = mover .CaughtObject;
        if (mover [iInMover] .Source is Plot)
        {
```

Inside the **Paint** event, mover can be mentioned on those rare occasions, when covers have to be visualized, but it is really a rare thing, as good examples of design are those, which do not require such visualization.

```
private void OnPaint (object sender, PaintEventArgs e)
{
    mover .DrawCovers (e .Graphics);
```

There can be a lot of different situations for moving / resizing of the stand alone screen objects and there are even more variants, when you have a set of objects on the screen, which can be independent of each other or be in some kind of relations. Those objects can be placed in an arbitrary way to each other; they can stay apart from each other or overlap. Objects can be involved exclusively in individual movements, or they can move synchronously, or influence each others movements in different ways. Objects can be solid, or they might have some holes, through which the underlying objects can be seen. If the underlying object is seen through the hole in the upper object, then you would expect that that underlying object can be grabbed for moving ("caught") through the hole. By allowing such things, the objects on different levels can be moved around and resized without changing their levels. Any type of these numerous movement possibilities originates from the movement of some node; for this purpose any node has one more very important parameter, which describes the main feature of its behaviour. The possible values of this parameter are defined by the members of the Behaviour enumeration*.

```
enum Behaviour { Nonmoveable, Moveable, Transparent = 4, Frozen };
```

While making the decision about the possibility of catching any object, the mover checks the covers according to their order in the queue; the cover of each object is analysed node after node according to their order in the cover. When the first node, containing the mouse point, is found, then there are several possible reactions, depending on the Behaviour parameter of this node:

- If it is Behaviour.Nonmoveable, then the object is unmovable by this point. At the same time, such node do not allow mover to look anywhere further; all other nodes and objects at this spot are blocked from mover. The analysis is over, try another point.
- If it is Behaviour.Frozen, then the object under the mouse cannot be moved by this point, but it is recognized by the mover as any other object, so, for example, the context menu can be easily called for it.

---

* This parameter was used beginning from the very first version of my movable graphics, but throughout the years its main idea and the use in different situations have changed. At the beginning, it was used only for describing the possibility of movement, so the enumeration was called MovementFreedom. From the very first version there were four members in that enumeration, but two of them turned out to be redundant, as the difference in the node's behaviour was defined not by this parameter, but by the code in the MoveNode() method of an object. At the same time the new and very interesting possibilities of the nodes' behaviour were added to the algorithm. This new possibilities were described by the new members, added to the same enumeration. After the exclusion of the redundant members of the enumeration and adding new their total number is still four. But with this change of the enumeration's members, I felt more and more that the enumeration's name became very confusing. Beginning from version 6.08, I decided to rename this enumeration, so that its new name – **Behaviour** - better describes its purpose. The change of the enumeration's name required to change the names of some methods in the Cover and CoverNode classes; otherwise the names of those methods would have no sense at all. These changes are mentioned in the **MoveGraphLibrary_Classes.doc** in the section *Important changes*.

- If it is `Behaviour.Moveable`, then the possibility of movement is decided by the `MoveNode()` method of this object, according to the number or shape of the node and the movements restrictions, if there are any.
- If it is `Behaviour.Transparent`, then mover skips this and all other nodes of the same object and continues the analysis of the situation from the next object in its queue.

## Safe and unsafe moving and resizing

The design of applications on the basis of movable and resizable elements opens an opportunity for the purposeful or accidental disappearance of the elements from view. This never happens in the ordinary applications with the unmovable elements, but now users can simply move any element out of view across the border, release it there, and what then? If in the resizable form an object is moved across the right or bottom border, then it is not a problem, as the form can be enlarged and the object returned back into play. Such temporary relocation of the currently unneeded objects is often used in the user-driven applications. But if an object is moved across the upper or left border of the form and dropped there, then there is no way to return it back into view by resizing the form. The mover can take care of this situation and prevent such disappearance of the objects, but only if you want it to overlook this process. For this purpose, the mover has to be initialized with an additional parameter – the form itself. (If the mover works on a panel, then this panel is used as a parameter.)

```
mover = new Mover (this);
```

You can find throughout the code of the **WorldOfMoveableObjects** application that such type of initialization for mover is used in all the forms. In such a case, when mover grabs any element for moving or resizing, the clipping is organized inside the borders of the client area. The level of clipping can be changed with one of the properties of the `Mover` class; three levels of clipping are implemented

```
public enum Clipping { Visual, Safe, Unsafe };
```

- `Visual` – elements can be moved only inside the client area.
- `Safe` – elements can be moved from view only across the right and bottom borders of the form.
- `Unsafe` – elements can be moved from view across any border.

By default, the `Visual` level is used, but can be changed at any moment. No clipping is used, if the mover is initialized without that additional parameter. When the `Clipping.Visual` level is used, it does not mean that any object is always fully in view. The clipping is organized not for the objects, but for the mouse, when it has grabbed an object. Under such clipping the mouse cannot go outside the visible part of the form, so this guarantees that an object cannot be entirely moved beyond the borders. A part of an object can cross the border, but the mouse, which is stuck to some point inside an object, cannot, so, regardless of where the movement ends, some part of an object is still in view. By pressing this part, though it can be very small, an object can be returned back into full view.

The unlimited shrinking of an object can be another cause of its disappearance. To avoid this, the minimum sizes must be declared for any class of resizable objects; these restrictions are used in the `MoveNode()` method. The restriction on minimum size of objects is used in the majority of classes, which can be seen in the Demo application. However, there are two other possibilities, which are required from time to time; both of them are also demonstrated in Demo application.

The first approach is to allow the shrinking of an object to a tiny size or even to total disappearance. The shrinking of an object to the size less than the predetermined minimum is considered as the user's command to delete this object from view; so the object is removed.

In some very rare situations an object can be squeezed to a tiny size or even disappear from view, but continue to exist; there is still a possibility to grab this invisible object and increase it, thus making it visible again. This approach is used only on those rare occasions, when there are several objects side by side and still visible neighbours inform users that the empty space between objects is not absolutely empty, but there exists an invisible object, which can be found and caught by the mouse.

**Technical note.** To avoid the screen flicker, do not forget to switch ON the double-buffering in any form, where you use the moving / resizing algorithm. It has nothing to do with the described technique, but it is a nice feature from Visual Studio.

## From algorithm to working programs

Beginning from the very first example of the accompanying program I will demonstrate the moving / resizing of different objects and write about the associated code, but even before commenting on some piece of code I want to mention several common things.

In every form you will have different objects involved in different types of movement. In addition to movements, there is often the reordering of objects to change their appearance on the screen; there is also calling of different context menus on different classes of objects or even their parts. All these things are started by the mouse, and it is obvious that it would be not enough to take into consideration only the mouse button, which starts one or another action. For example, the forward movement of objects (plus their resizing and reconfiguring) and putting an object on top of others are all started with the left button. Rotation of objects and calling of context menus are started with the right button, so something else must be considered in addition to mouse button to distinguish all these possibilities.

To distinguish one command from another, I often pay attention to the distance between the points of `MouseDown` and `MouseUp` events. In the ideal case the decision must be based on having at all any distance between two points or not, but the request of not moving a mouse even for one pixel between pressing it and releasing can be too strong for a lot of users, so I consider the move of less than 3 pixels as “not moved”. (Three pixels for “not moved” decision is my own suggestion; this number can be easily changed in either way.) Implementation of this rule in my programs results in such system of decisions, depending on the pressed button:

**Left** button. If the mouse was not moved between two events, then I recognize it as a command to popup an object; otherwise it is simply a move of an object.

**Right** button. If the mouse was not moved between two events, then I recognize it as a menu call; otherwise it is a rotation.

When an object is caught, moved, and released, then a sequence of events in which it is involved is this one: **MouseDown** – **MouseMove** – **MouseUp**. Throughout all three events the mover plays the main role as the organizer and conductor of the whole process, but I will start some explanations of the role which mover plays from the last stage.

A lot of actions are decided inside the `OnMouseUp()` method after the `mover.Release()` call. First, this method informs by its returned value whether any object was released or not. If any object was released, then it would be very helpful to have more information on this object. The `Release()` method has variants which inform about the order of released object in the queue `(iWasObject)`, the number of node by which this object was caught `(iWasNode)`, and the shape of this node `(shape)`.

```
bool Release ()
bool Release (out int iWasObject)
bool Release (out int iWasObject, out int iWasNode)
bool Release (out int iWasObject, out int iWasNode, out NodeShape shape)
```

Even if the first of these variants is used, the order of the released object in the queue can be obtained with the `Mover.WasCaughtObject` property

```
int iInMover = mover .WasCaughtObject;
```

When the order of an object in the queue is known, it is easy to get the object itself.

```
GraphicalObject grobj = mover [mover .WasCaughtObject] .Source
```

There is an alternate way to find the same released object; it uses not the order of object in the queue, but another property.

```
GraphicalObject grobj = mover.WasCaughtSource
```

Any object, derived from the `GraphicalObject`, automatically gets the unique identification number. The ID property of the base class allows to get this number for any object at any moment and to use this number for identification of objects. This is widely used, when you need to find, which object was pressed, to what class it belongs, and so on.

Mover is the conductor of the whole moving / resizing process for all the objects in the form. To some extent mover can be associated with the mouse, as it catches the objects by the mouse, moves them with the mouse, and then releases them at the place, to which the mouse was moved. When the mouse is moved around the screen, then a lot of information can be obtained inside the `OnMouseMove()` method from the `mover.Move()` call and several of properties of the `Mover` class. This call itself informs (by the return value) if any object is moved at the moment or not. The link between the mover and the mouse is especially strong during the moving process, because the only parameter of the `mover.Move()` method is the mouse position.

The same set of data about the caught object can be needed throughout the process of moving: the order of object in the queue, the number of node, by which this object is moved, and the shape of this node. All the needed information about the currently caught object can be obtained with three properties of the Mover class.

```
int iObject      = mover .CaughtObject;          // Gets the index of the caught element.
int iNode        = mover .CaughtNode;            // Gets the number of the caught node.
NodeShape shape = mover .CaughtNodeShape;        // Gets the shape of the caught node.
```

The caught object itself can be received with either of these calls

```
GraphicalObject grobj = mover .CaughtSource;                           or
GraphicalObject grobj = mover [mover .CaughtObject] .Source;
```

There is also a property to inform, if any object is caught by the mover at this moment or not

```
bool bCaught = mover .Caught;
```

Even if no object was caught for moving, the mover can produce all the needed data about the objects that are under the mouse cursor. Mover "senses" the movable objects under the mouse cursor and gives the standard set of information about them.

```
bool bSensed = mover .Sensed;            Gets the indication of any movable object under the cursor.
int iObject  = mover .SensedObject;      Gets the index of the element underneath.
int iNode    = mover .SensedNode;        Gets the number of the node underneath.
```

Mover is like a shark, which is not hungry at the moment and lazy to attack, but watches carefully and keeps track of everything that is going around. In this way mover can produce information about any point of the form and not especially under the cursor; the information is returned in the form of the MoverPointInfo object. In addition to the number of object in the queue, the number of node in the cover, and the shape of node there are also the behaviour of the node (of the Behaviour enumeration) and the shape of cursor above this node. It is possible to get either the information about the upper node (and object) at the particular point or the list of data for all the nodes that overlap at that place.

Every move of an object starts by pressing the object with a mouse. The associated with it method – mover.Catch() – returns the value, which indicates if any object was really caught or not. This method has three variants, which can be used in deferent scenarios, depending on what is needed. All three variations have one mandatory parameter – the point, where the mouse was pressed, but differ in using other parameters.

`mover .Catch (Point ptMouse, MouseButtons catcher)`

This is the most frequently used variant of this method. The second parameter specifies the mouse button, by which an object was caught. In nearly all the situations it is a very important parameter, as it allows to distinguish the forward movement from rotation; it also indicates the request for calling a context menu.

`mover .Catch (Point ptMouse)`

The variant without any additional parameter is used rarely enough; you can use it, if you do not care by what button an object is caught. In such case the system simply substitutes the left button as the default value and goes on.

`mover .Catch (Point ptMouse, MouseButtons catcher, bool bShowAngle)`

This version of the method is often used for the forms with the textual comments. The majority of these comments are derived from the TextMR class; the moving and rotation of all these objects are used so often that they are automated and do not need to be mentioned anywhere in the code. However, throughout the rotation of any comment, its angle can be shown nearby, and this is regulated by the third parameter of the method. Usually, this parameter is controlled by the user and can be changed at any moment, for example, via some context menu.

With all this information from mover, it is possible to organize any kind of moving and resizing. In the further examples you will see the use of these properties and methods, but some of them can be needed only for the most complicated forms. I think that it is better to have some information about them in one place, then to look for it throughout the book. Now we can start with the real examples and see how it all works from the simplest programs to the most complicated.

Just a reminder: there is not a single unmovable object in this application. You can move everything.

# The first acquaintance with nodes and covers

File: **Form_Nodes.cs**
Menu position: *Graphical objects – Node shapes*

Covers may consist of an arbitrary number of nodes. The minimum number of nodes in the cover is one, but there is no upper limit for this number. When a cover consists of several nodes, then some of them can be responsible for resizing of an object, others – for moving the whole object. If the cover of an object consists of a single node, then such object is either movable or resizable, but not both. In some cases there might be a need for a resizable, but not movable object, though it would be an extremely rare situation. The request for a movable, but non-resizable object is not so rare, so let us consider such a case. As there are three different types of nodes (circles, strips, and convex polygons), then the **Form_Nodes.cs** demonstrates three different objects with the covers consisting of a single node. This is a nearly unique situation, when the area of cover is exactly the same as the area of object.

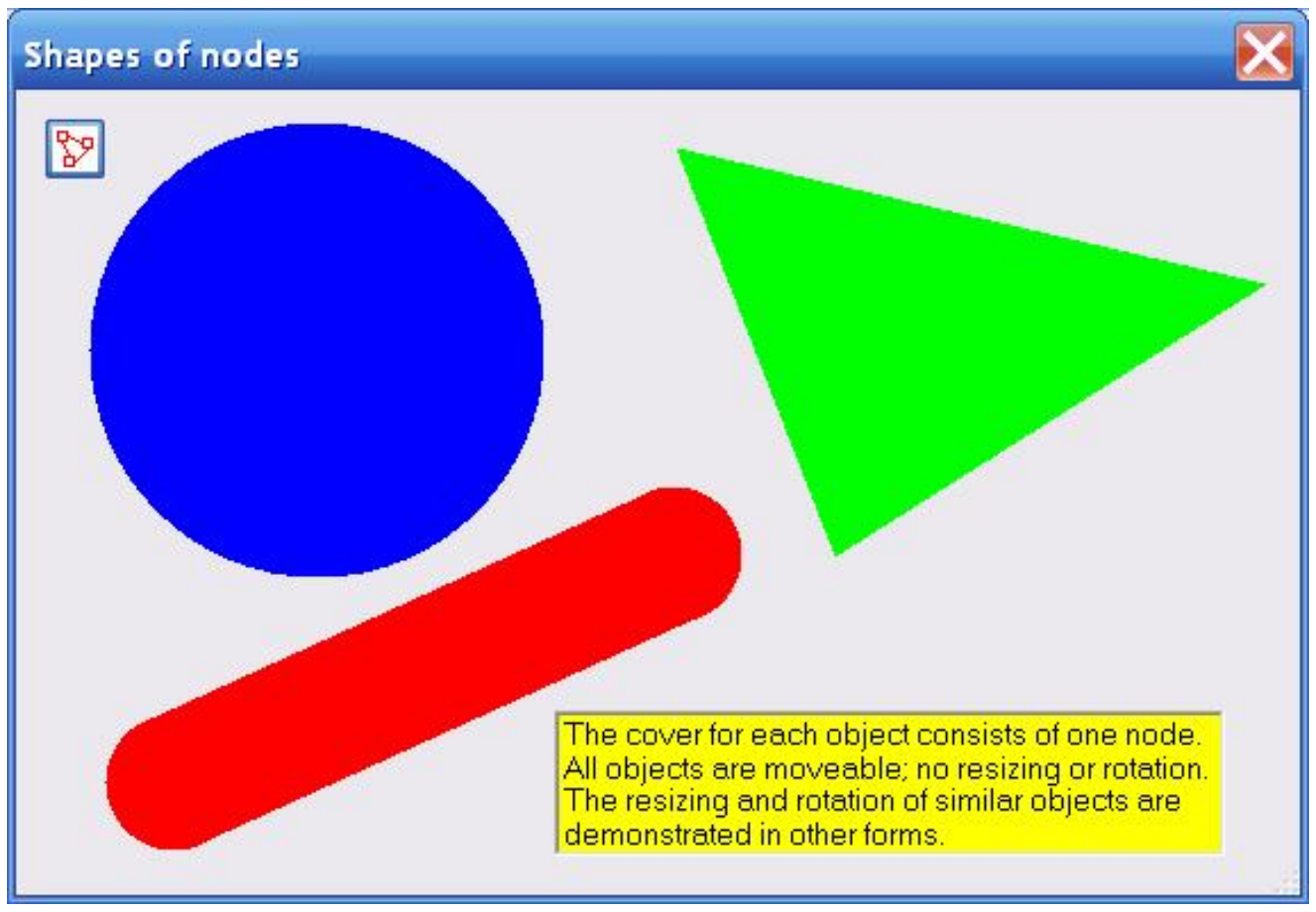

**Fig.2.1** Shapes of nodes

I write in many places that nodes can be of *three different types* or *three different shapes*, but the second statement is a bit incorrect. Circles can differ by the radius, but all of them look similar. Different ratio between the width and length of a strip can change its view, but even in such a case we can say that nearly all strips look similar. But nobody is going to agree that triangle and rectangle look similar, so declaring that two such nodes have the same shape is definitely wrong. Yet, they belong to the same type of polygonal nodes! All nodes in the form of any convex polygon belong to the same type of nodes! The correct statement is: "There are three different **types** of nodes". Whenever you see the statement "three shapes of nodes" it means "three types of nodes".

In addition to three colored figures, the **Form_Nodes.cs** includes two more objects (**figure 2.1**). The button 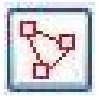 can be moved by its border; I will write about the movability of the individual controls and the groups of controls further on in the special chapters. By clicking this button, it is possible to switch ON / OFF the visualization of covers. The visualization of covers is widely used in the examples of the first part of the book, where I want to explain, how the cover can be organized. The use of this button is absolutely identical throughout all these examples. The button changes the value of the flag, which regulates the drawing of the covers.

```
private void Click_btnCovers (object sender, EventArgs e)
{
    bShowCovers = !bShowCovers;
    Invalidate ();
}
```

The drawing of the covers is provided by the mover and depends on this flag.

```
private void OnPaint (object sender, PaintEventArgs e)
{
    Graphics grfx = e .Graphics;
    … …
    if (bShowCovers)
    {
        mover .DrawCovers (grfx);
    }
}
```

The text, the short explanation on the yellow background, which you can see in this form, belongs to the `TextM` class; such objects can be moved by any inner point. The cover of any `TextM` object consists of a single rectangular node. A lot of informative texts in this application are organized in such a way. The `TextM` class is discussed in details in the *Texts* chapter further on.

Objects of the same shape may have different types of covers; the design of cover depends on the way, in which an object is supposed to be moved and resized. Later on I will show objects of the same shape (circles, strips, and polygons), which are involved in different types of movement and resizing, so the covers for similar looking objects in later examples will be more complicated and consist of more than one node. As the colored figures in the **Form_Nodes.cs** can be only moved and nothing else; I included into the names of their classes the word *primitive*.

Any class, derived from the `GraphicalObject` class, has to implement three methods, which were declared abstract in the base class. In case of the `PrimitiveCircle` class it looks like this.

There are eight different constructors for the circular nodes. The constructor, used in the `PrimitiveCircle` class, specifies the central point, radius of the node, and the mouse cursor over it.

```
public override void DefineCover ()
{
    CoverNode node = new CoverNode (0, center, radius, Cursors .SizeAll);
    cover = new Cover (new CoverNode [] { node });
}
```

The position of a circle is fully defined by its center, so the moving of such an object is entirely described by the moving of this central point.

```
public override void Move (int dx, int dy)
{
    center += new Size (dx, dy);
}
```

The moving of a node is described by the `MoveNode()` method. In all the cases of the multi-node covers there are always variants of what to do if one or another node is involved in movement; later you will see that such a decision is usually based on either the number of node (the most common case) or the shape. In such a primitive case of a single node there is no decision of such type and the `MoveNode()` method automatically calls the `Move()` method.

```
public override bool MoveNode (int i, int dx, int dy, Point ptM, MouseButtons btn)
{
    bool bRet = false;
    if (btn == MouseButtons .Left)
    {
        Move (dx, dy);
        bRet = true;
    }
    return (bRet);
}
```

The example with these primitive objects, which can be only moved, demonstrates a lot of things that are important and used for all the types of moving / resizing even for much more complicated objects. Let us look at some important details.

Default and non-default parameters

The cover for any object is organized as a set of nodes. Nodes can be of three different types, so there are three big groups of the `CoverNode` constructors. The parameters which must be always declared on construction of a node include identification number, position, and sizes. Other parameters often get the default values, but can be declared either during the construction or later. Of the three types of nodes, the polygons are often used to move the objects around the screen, so their default cursor is `Cursors.SizeAll;` nodes of two other types are usually used in a small size and mostly for resizing or reconfiguring; their default cursor is `Cursors.Hand.` Because in the **Form_Nodes.cs** all three types of nodes are used in a big size and for moving the objects around the screen, I decided to keep their cursors consistent and used for circles and strips those constructors that allow to set the cursor. Here is the `PrimitiveStrip.DefineCover()` method; the strip nodes are defined by two points and the radius of the semicircles.

```
public override void DefineCover ()
{
    cover = new Cover (new CoverNode [] { new CoverNode (0, pt0, pt1, radius,
                                                         Cursors .SizeAll) });
}
```

Visualization of covers is discussed at the end of the book in *Appendix A*, but I would like to make one remark here at the very beginning. Visualization of covers means visualization of nodes; this includes for each node possible painting of the interior area and drawing its border with a thin line. Because the nodes of different shapes are usually used in different sizes, they have different default values even for visualization: the interior of circles and strips are wiped out, the interior of polygons is not. Again, for consistency in this example, I added one line into the PrimitivePolygon.DefineCover() method so that this cover is also wiped out on request of visualization.

```
        public override void DefineCover ()
        {
            CoverNode node = new CoverNode (0, pts);
            node .Clearance = true;
            cover = new Cover (new CoverNode [] { node });
        }
```

You can comment the extra line in this method and see the results; this will show you the standard view of the polygonal nodes, when they are visualized.

<u>Deciding on the buttons to move an object</u>

The MoveNode() method allows to specify, by which button each node can be moved. For the PrimitiveCircle and PrimitivePolygon classes the possibility of moving only with the left button is declared.

```
public override bool MoveNode (int i, int dx, int dy, Point ptM, MouseButtons btn)
{
    bool bRet = false;
    if (btn == MouseButtons .Left)
    {
        Move (dx, dy);
        bRet = true;
    }
    return (bRet);
}
```

For the PrimitiveStrip class I did not include into the mentioned method the checking of the pressed button, thus allowing the moving of a strip by any button.

```
public override bool MoveNode (int i, int dx, int dy, Point ptM, MouseButtons btn)
{
    Move (dx, dy);
    return (true);
}
```

Certainly, in any real application it would be a big mistake to put on the screen the objects, which can be moved only by one button, together with the objects that can be moved by any button. This will be the most confusing thing for the users of any program; in further examples of this **WorldOfMoveableObjects** application you will not see such a mix. But to demonstrate different possibilities, I did such a thing in this **Form_Nodes.cs.** If you decide to make the moving of all three primitive figures in this form identical, you can either add the same checking of the pressed button into the PrimitiveStrip.MoveNode() method or delete those checkings from other two classes. In the first case the strip will become movable by the left button only; in the second case all three objects will become movable by any button.

<u>The order of objects</u>

In order to become movable, all objects must be registered in the queue of a mover. When the objects are moved around the screen, they can overlap. When two or more objects share the same part of the screen and somebody wants to pickup an element at the point of overlapping, then the common expectation would be that the object on top has to be caught. Mover does not know anything about the drawing of objects and their appearance at the screen; mover makes the decision about the object to catch only according to their order in the queue, so it is the developer's responsibility first to place the objects in the appropriate order in the queue and then to draw them in the correct order.

While deciding about the order of objects in the queue, first take into consideration some restrictions that are imposed by an operating system. The elements viewed on top of others must precede them in the queue, but there is the strict rule enforced by the system: all the controls are always shown atop all the graphical objects. Thus we receive the rule 1.

**Rule 1.** All the controls must precede all the graphical objects in the queue.

There are four graphical objects and one control in the **Form_Nodes.cs**; this control must be placed at the head of the queue. Here is the FillMoversQueue() method to register all the objects; it is obvious that the button takes the leading position.

```
private void FillMoversQueue ()
{
    mover .Add (info);
    mover .Add (circle);
    mover .Add (poly);
    mover .Add (strip);
    mover .Insert (0, btnCovers);
}
```

The second rule coordinates the order of graphical objects in the queue with the order of their drawing and is based on the fact that graphical objects are painted from the bottom layer to the top.

**Rule 2.** The order of drawing objects must be opposite to their order in the queue.

In other words, the drawing of the graphical objects must go from the tail of the queue to the head. In such a way the graphical object, shown on top of others, always precede them in the queue, and the top one is always caught by the mover. Exactly as expected.

```
private void OnPaint (object sender, PaintEventArgs e)
{
    Graphics grfx = e .Graphics;
    strip .Draw (grfx);
    poly .Draw (grfx);
    circle .Draw (grfx);
    info .Draw (grfx);
```

The whole process of moving any object in the **Form_Nodes.cs** is organized in such a way.

1. Declare a mover.

```
Mover mover;
```

2. Initialize a mover, construct all the objects to be moved, and register them with the mover.

```
public Form_Nodes ()
{
    InitializeComponent ();
    mover = new Mover (this);
    circle = new PrimitiveCircle (new Point (150, 120), 100, Color .Blue);
    poly = new PrimitivePolygon (new Point [] { new Point (340, 60),
                new Point (600, 120), new Point (410, 240) }, Color .Lime);
    strip = new PrimitiveStrip (new Point (100, 340), new Point (320, 240),
                                30, Color .Red);
    info = new TextM (this, new Point (300, 300), infotext);
    info .BackColor = Color .Yellow;
    FillMoversQueue ();
}
```

3. Write the code for three mouse events, through which the whole process is organized. All three methods are used in the **Form_Nodes.cs** in the simplest form, as there is nothing else except moving the objects around.

```
private void OnMouseDown (object sender, MouseEventArgs e)
{
    mover .Catch (e .Location, e .Button);
}
private void OnMouseMove (object sender, MouseEventArgs e)
{
    if (mover .Move (e .Location))
    {
        Invalidate ();
    }
}
private void OnMouseUp (object sender, MouseEventArgs e)
{
    mover .Release ();
}
```

## Solitary lines

File: **Form_Lines_Solitary.cs**
Menu position: *Graphical objects – Basic elements – Lines – Solitary lines*

This example with the straight lines includes only few objects of the `LineRsRt` class, so the diversity of objects in the **Form_Lines_Solitary.cs** (**figure 2.2**) is less than in the previous example. But the objects of the `LineRsRt` class can be moved and resized, so the covers of such objects are more interesting and have to consist of more than one node. Each line can be moved by any inner point and resized by both end points (`ptA` and `ptB`), so the cover in this class consists of three nodes.

```
public override void DefineCover ()
{
    cover = new Cover (new CoverNode [] {new CoverNode (0, ptA),
                                         new CoverNode (1, ptB),
                                new CoverNode (2, ptA, ptB, Cursors .SizeAll)});
}
```

Two circular nodes are placed at the end points of the line; the strip node covers the line itself. Both types of nodes have multiple constructors, which allow either to declare all the parameters or omit some of them, thus assigning them the default values. In the case of `LineRsRt` class, I use the simplest type of constructor for circular nodes, when only two mandatory parameters are declared: the number of the node and its point. This means that the end nodes get the default radius (3 pixels) and the default view of the mouse cursor (`Cursors.Hand`). For the strip node there are three mandatory parameters: the number of the node and two end points. I use not the default cursor for this form of nodes, but set it to `Cursors.SizeAll`; this was done on purpose to inform users that pressing the line at the ends or anywhere inside start different actions. Not declared parameters of these nodes get the default values. The default radius of the circular nodes is 3 pixels; the default radius of semicircles at the ends of a strip node is the same, so the default width of the strip node is 6 pixels.

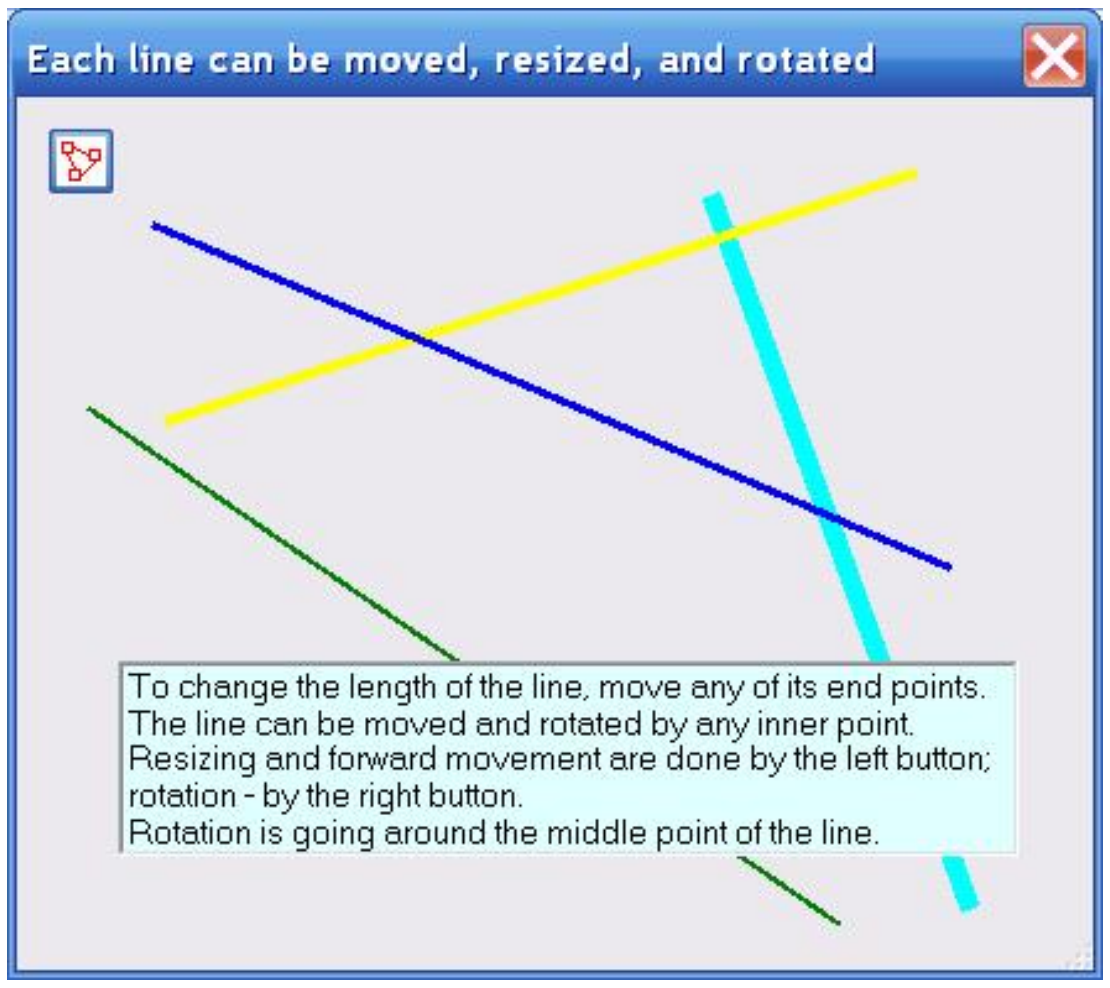

**Fig.2.2** Each LineRsRt object can be moved, resized, and rotated

Even such a simple cover of several nodes has some interesting aspects, which must be taken into consideration, while designing the cover. Circular nodes at the ends and the semicircles of the strip node have the same radius; centers of the circular nodes and of those semicircles are also the same. If the strip node would be placed in the cover ahead of the circles, then those circles would be entirely covered by the strip; there would be no chances for resizing of a line. To avoid this, the strip node must be the last in the cover.

The width of a real line can be from one pixel and up, but this parameter is not mentioned anywhere in the design of cover, so regardless of the width of a real line its cover has the same width. When a line is thin enough (less than six pixels), the difference between the width of a line and the width of the cover is not a problem; the bigger width of a strip node only makes the grabbing of this line for a movement easier. However, in those rare situations, when the line is wider than six pixels, such a cover design with the fixed width is not a good idea, as there will be the dots inside the line closer to its sides which will be not covered; it would be impossible to grab the line by those points. In such special cases of wide lines it would be much better to make the width of the cover equal to the real width of a line by using the `CoverNode` constructor with the parameter declaring its size.

To avoid the disappearance of line, the `LineRsRt` class has the minimum allowed size; the length of any line is not allowed to be less than this value. If you try to construct a shorter line, the angle is calculated from the two proposed points, but the length is increased according to this limitation.

```
int minLen = 20;

public LineRsRt (PointF pt0, PointF pt1, Pen pn)
{
    ptA = pt0;
    if (Auxi_Geometry .Distance (pt0, pt1) >= minLen)
    {
        ptB = pt1;
```

```
        }
        else
        {
            ptB = Auxi_Geometry .PointToPoint (ptA,
                                  Auxi_Geometry .Line_Angle (pt0, pt1), minLen);
        }
        pen = pn;
    }
```

Forward moving of a line has to change only the end points, on which the line is based.

```
        public override void Move (int dx, int dy)
        {
            Size size = new Size (dx, dy);
            ptA += size;
            ptB += size;
        }
```

If a line is caught by the mover (pressed by a mouse), then the decision about the possibility of movement is made inside the `MoveNode()` method. If a line is caught by the left button, then the decision on either to change the line or simply move it forward without any changes depends on the number of the caught node. If it is a strip node, covering the whole line ( `i = 2` ), then the line is moved; the `Move()` method is called from inside the `MoveNode()` method. For two other nodes the proposed new length is checked against the minimum allowed length; only if the new length would be not less than the allowed minimum, then the end point is moved to the new position.

```
public override bool MoveNode (int i, int dx, int dy, Point ptM, MouseButtons btn)
{
    bool bRet = false;
    if (btn == MouseButtons .Left)
    {
        PointF ptNew;
        switch (i)
        {
            case 0:
                ptNew = new PointF (ptA .X + dx, ptA .Y + dy);
                if (Auxi_Geometry .Distance (ptNew, ptB) >= minLen)
                {
                    A_Point = ptNew;
                    bRet = true;
                }
                break;
            case 1:
                ptNew = new PointF (ptB .X + dx, ptB .Y + dy);
                if (Auxi_Geometry .Distance (ptA, ptNew) >= minLen)
                {
                    B_Point = ptNew;
                    bRet = true;
                }
                break;
            case 2:
                Move (dx, dy);
                break;
        }
    }
```

Lines can be rotated. You press the line by the right button and start the rotation. The rotation will go exactly as it is supposed to go, but the code for rotation in this example is not correct, though visually it looks fine. I do not want to discuss the rotation here; it is too early. There will be a special chapter *Rotation*, in which I describe this process in details. After it you can return back to this example and find yourself, what is wrong with this code.*

---

* The rotation of these lines looks OK, because the line is painted by the same color, you cannot visually distinguish one end of line from another, and the rotation is always organized around the middle point. There are two simple ways to find that

This **Form_Lines_Solitary.cs** demonstrates an interesting solution for the problem of the correct order of drawing. The order of movable objects in the queue is determined in the `RenewMover()` method. Button is at the head of the queue, then goes the information, and then all the lines.

```
private void RenewMover ()
{
    mover .Clear ();
    mover .Add (btnCovers);
    mover .Add (info);
    for (int i = 0; i < lines .Count; i++)
    {
        mover .Add (lines [i]);
    }
}
```

The order of drawing must be opposite: all the lines, beginning from the last one, then information. This is the right order of drawing; **figure 2.2** shows the information over the lines, so everything works correctly. Suppose that you decide to change the order and to show the information below the lines; in this case you would have to make changes both in the `RenewMover()` method and in the `OnPaint()` method. It is easy to do, when you have such a simple form with only few objects to draw; what about the examples with a lot of objects? It would be much better, if you would have to make the changes of order only in the `RenewMover()` method and let the painting go in the correct order automatically. Is it possible to obtain such a thing? Yes, if you rely on the mover! Mover can help to check the class of any object from its queue; this solves the problem.

```
private void OnPaint (object sender, PaintEventArgs e)
{
    Graphics grfx = e .Graphics;
    GraphicalObject grobj;
    for (int i = mover .Count - 1; i >= 0; i--)
    {
        grobj = mover [i] .Source;
        if (grobj is LineRsRt)
        {
            (grobj as LineRsRt) .Draw (grfx);
        }
        else if (grobj is TextM)
        {
            (grobj as TextM) .Draw (grfx);
        }
        if (bShowCovers)
        {
            mover [i] .DrawCover (grfx);
        }
    }
}
```

The objects have to be drawn in order from the end of queue to its head. Here I have organized the loop through the objects of the queue in this needed order; mover checks the class of each of the objects and calls the appropriate method for drawing. The appropriate means the method of the identified class.

Let us return back to the problem of the length restriction. The length of a line is not allowed to become less than the predetermined minimum. This is a standard situation with many but not all of the objects that I demonstrate. Suppose that you would allow the squeezing of a line to a zero size. What are you going to show at the screen in such a situation? The first problem will be simply with the drawing, but this you can go around by adding the small checking of the size in the `Draw()` method. Much more serious problem is the total disappearance of an object from view, while it still exists somewhere in the queue. If users do not see an object, they do not know about it. It becomes a phantom; a ghost of that

---

something is incorrect with the rotation of lines. For example, you can either mark one end of line by different color (I even put one line of code into the `Draw()` method and commented it; you can turn this comment into the working code) or organize the rotation around another point, for example, around one of the ends; in both cases you will see something strange. Not always, but occasionally. Maybe this will give you some tip on what is wrong with this code. But after reading the chapter *Rotation*, you will make the needed changes of the code in a minute.

rare type, which never shows itself to anyone. Do you need such objects in the program? What will be the purpose of keeping alive such invisible elements?

Usually I do not allow such situations by imposing the minimum limits on any resizable objects. However, there can be different solutions, which are demonstrated in several examples, beginning from the next chapter.

- The first solution is to get rid of such an object, if it is released with an extremely small or zero size. Usually there is another way to get rid of the unneeded objects, for example, through the context menu. But if it is declared that users can delete such objects by squeezing them to tiny size, then it is a normal solution. In such a case you have to take out the restriction from the `MoveNode()` method, but add similar checking into the `OnMouseUp()` method. If you find out that the released object is a line of extremely short or zero length, then this line must be deleted from the `List` of lines.
- Another variant is to restore an object to some minimum size, if it is released with a size less than the allowed minimum. This variant is nearly the same, as not allowing to go below minimum.
- The third and rarely used variant is to allow such invisible but existing objects. It is done on those rare occasions, when it is not a stand alone element, but used with other objects that clearly indicate that there is an object, though it can be invisible at the moment. The presence of other objects not only informs about the invisible one, but shows where it stays now, so there is no problem in grabbing the invisible one and enlarging it to some normal size. One example of such a type is the bar chart, in which users can change the size of each bar. Any bar can be reduced to zero, but the very space in a row of other bars clearly indicates that there is a bar, which can be grabbed and enlarged. Such example is demonstrated in one program in the chapter *Data visualization*.

# Rectangles

This chapter shows the moving / resizing of one type of primitive objects: rectangles. But though they are primitive, there are different possibilities of their resizing.

Rectangle is the most common of all the shapes that can be seen at the screen. It is also considered as a simple enough shape for moving and resizing, so from the very beginning of the Windows system the only allowed form of a window was a rectangle. In the hierarchy of standard classes window or a dialog (now it is called `Form`) is related to the `Control` class. The problems of moving / resizing of different controls and their combinations are discussed in this book much later; this chapter is about the moving / resizing of only graphical rectangles. Historically any window could be independently resized by any border; I will start with the similar case, though it is not the only possibility for rectangles.

## *Independent moving of borders*

File: **Form_Rectangles_StandardCase.cs**
Menu position: *Graphical objects – Basic elements – Rectangles – Standard case*

On opening the **Form_Rectangles_StandardCase.cs** you see four colored rectangles, which demonstrate four different types of resizing. Visually the differences become obvious only on switching ON the cover visualization (**figure 3.1**); without those covers in view, the rectangles do not give you any visual tip on how they can be changed. Though there is another prompt – the changing of the mouse cursor, when the mouse crosses any area of possible resizing or moving an object.

The rectangular shape is so popular among the screen objects that a cover for such an object can be constructed by using one of the standard constructors of the `Cover` class. That was also the reason to call this class of rectangles `RectStandard`. Maybe even the bigger reason to include the word *standard* into the name of the class was the desire to declare from the very beginning the type of resizing for such objects: they can be resized by any side and by any corner. This is the standard type of resizing, which you expect from any rectangular object on the screen; certainly, when you are absolutely sure that every screen object is resizable.

Construction of any resizable rectangle of the `RectStandard` class requires the declaration of its initial position, size, the range for its allowed resizing, and the color. There are also two additional parameters, which define the sizes of the nodes in the cover, but these parameters can be omitted and get the default values.

```
public RectStandard (Rectangle rect,        // initial position and size
                     RectRange range,       // range of the allowed resizing
                     int rad,               // radius of the circular nodes in the corners (default = 6)
                     int half,              // halh width of the nodes along the borders (default = 3)
                     Color color)           // color of rectangle
```

The `RectRange` class is included into the **MoveGraphLibrary.dll**. On initializing, an object of this class gets four values for minimum and maximum sizes along both axes.

```
public RectRange (int wMin, int wMax, int hMin, int hMax)
```

By comparison of these four values the `RectRange` object can decide, which type of resizing is allowed; the type of resizing is described by the `Resizing` enumeration. This enumeration has four different members; I hope that their names speak for themselves.*

```
enum Resizing { None, NS, WE, Any };
```

The `RectStandard` class has minimum allowed size to prevent its accidental disappearance. When you construct such an object, the initial declared sizes are first checked against this minimum size; after it the sizes of the rectangle are compared with the range, which comes as another parameter. If the range is not declared, then the rectangle is set to be non-resizable; otherwise the range parameter calculates the resizing type for this particular rectangle.

* The majority of the objects, which are discussed in the samples of this book, are resizable. Many of those objects use some parameter of the `Resizing` type to set the needed type of resizing. In all these situations I write that "an object can be used with four different types of resizing". Each time it means that there are three different variations of resizing plus a non-resizable variant.

```
    public RectStandard (Rectangle rect, RectRange range, int rad, int half,
                         Color color)
    {
        rc = new Rectangle (rect .X, rect .Y, Math .Max (minsize, rect .Width),
                                              Math .Max (minsize, rect .Height));
        if (range == null)
        {
            wMin = wMax = rc .Width;
            hMin = hMax = rc .Height;
        }
        else
        {
            wMin = Math .Max (minsize, Math .Min (rc .Width, range .MinWidth));
            wMax = Math .Max (rc .Width, range .MaxWidth);
            hMin = Math .Max (minsize, Math .Min (rc .Height, range .MinHeight));
            hMax = Math .Max (rc .Height, range .MaxHeight);
        }
        RectRange realrange = new RectRange (wMin, wMax, hMin, hMax);
        resize = realrange .Resizing;
```

Four rectangles in the **Form_Rectangles_StandardCase.cs** get four different types of resizing.

```
public Form_Rectangles_StandardCase ()
{
    rects .Add (new RectStandard (new Rectangle (50, 50, 100, 70), Color .Blue));
                                                            // Resizing .None
    RectRange rr = new RectRange (70, 70, 40, 300);
    rects .Add (new RectStandard (new Rectangle (100, 160, 70, 120), rr, radius,
                halfstrip, Color .Yellow));                 // Resizing .NS
    rr = new RectRange (30, 400, 80, 80);
    rects .Add (new RectStandard (new Rectangle (220, 100, 230, 80), rr, radius,
                halfstrip, Color .LightGreen));             // Resizing .WE
    rr = new RectRange (30, 300, 30, 300);
    rects .Add (new RectStandard (new Rectangle (320, 210, 190, 120), rr, radius,
                halfstrip, Color .Cyan));                   // Resizing .Any
```

As I have already mentioned, these rectangles can use one of the standard constructors of the `Cover` class; the `resize` parameter for the cover is defined from the analysis of the parameters of initialization.

```
    public override void DefineCover ()
    {
        cover = new Cover (rc, resize, radius, halfstrip);
    }
```

This `resize` parameter totally defines the number of nodes in the cover, their types, and order.

- For a non-resizable rectangle (the Blue one at **figure 3.1**), there is a single node, covering the whole rectangle.
- If a rectangle is resizable only in one direction (Yellow and Green rectangles from the picture), then a cover consists of three nodes: the two appropriate opposite sides are covered with two narrow rectangular nodes and then comes the same big node for the whole area.
- For a fully resizable rectangle (the Cyan one), there are nine nodes in a cover: four small circular nodes in the corners, then the rectangular nodes on borders, and then the big rectangular node, covering the whole area.

The nodes overlap with each other. At the points of their overlapping, the resizing to be started is decided by the first of those nodes in the cover. To start any kind of movement, the appropriate node must be found with a mouse cursor. It is more difficult to find the small node than the big one (remember that normally the covers are not shown), so the smaller nodes are usually placed first in this array. That is why for the fully resizable rectangle I decided about such order of nodes: circular nodes from the corners, then the nodes on borders, and then the big node for the whole area.

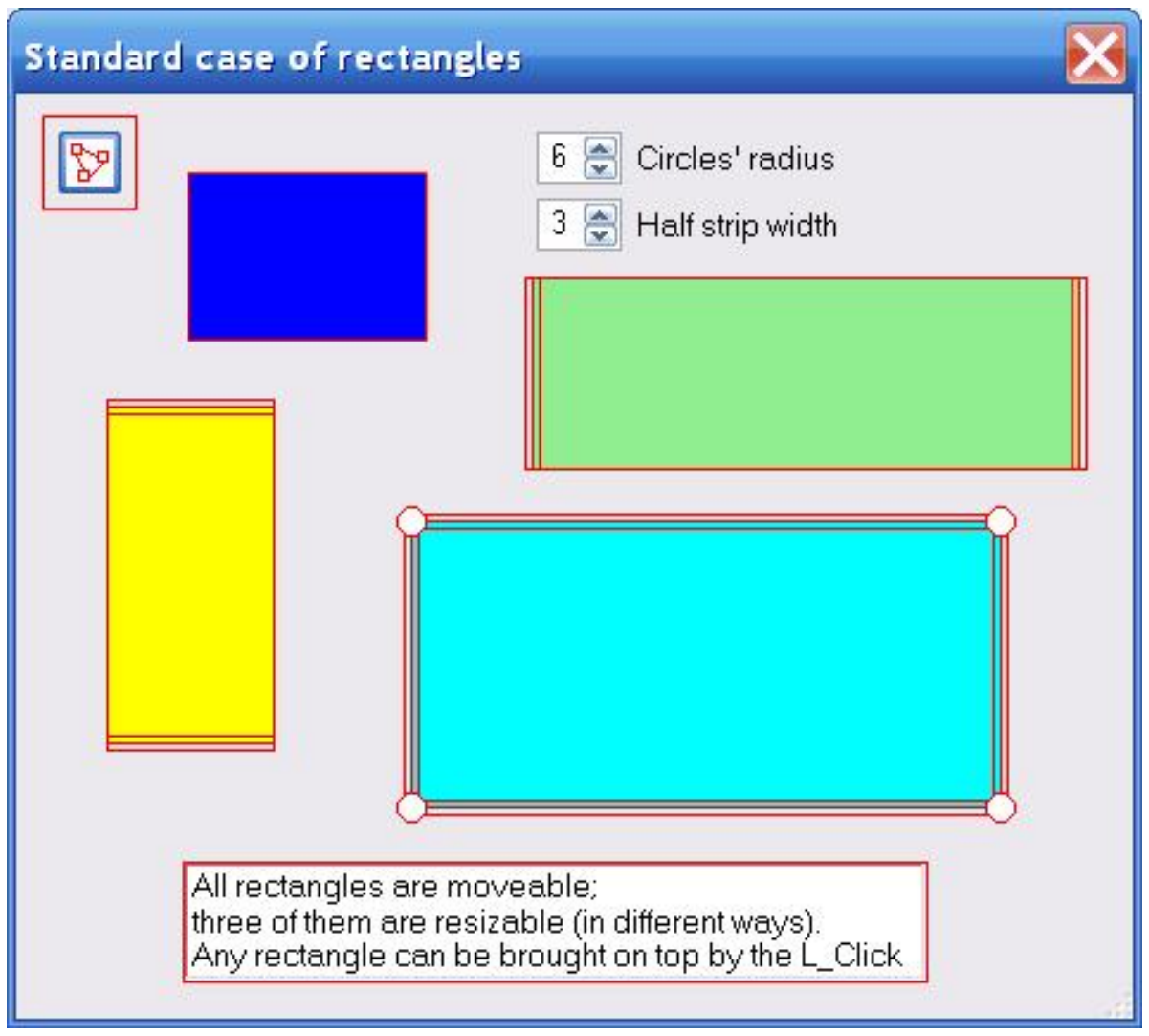

**Fig.3.1** Rectangles; standard case

The rectangle is based on a single point – its top left corner, so for the forward movement only the location of this point must be changed; the `Move()` method is simple and short.

```
public override void Move (int dx, int dy)
{
    rc .X += dx;
    rc .Y += dy;
}
```

On the contrary, the `RectStandard.MoveNode()` method is lengthy enough, but, as can be seen from the code of this method, it is simple. The method is lengthy, because:

- There are four different cases of resizing.
- Each case has its own combination of nodes.
- On moving a lot of nodes the proposed new size must be checked against the allowed minimum and maximum sizes.

Though the method becomes longer, it is still very simple, because all those checks for different nodes are either similar or even identical. Here is the part of the `MoveNode()` method for the circular node in the top left corner ( `i = 0` ) of the fully resizable rectangle.

```
public override bool MoveNode (int i, int dx, int dy, Point ptM, MouseButtons btn)
{
    bool bRet = false;
    if (btn == MouseButtons .Left)
    {
        int wNew, hNew;
        switch (resize)
        {
            case Resizing .Any:
                … …
                else if (i == 0)        //LT corner
                {
                    hNew = rc .Height - dy;
                    if (hMin <= hNew && hNew <= hMax)
                    {
                        MoveBorder_Top (dy);
                        bRet = true;
                    }
                    wNew = rc .Width - dx;
                    if (wMin <= wNew && wNew <= wMax)
                    {
                        MoveBorder_Left (dx);
                        bRet = true;
                    }
                }
```

There are two controls in the **Form_Rectangles_StandardCase.cs** (**figure 3.1**), which allow to change the sizes of the nodes and see, how the easiness of the resizing depends on those values. In this example the radius of the corner circles and the width of the sensitive strips along the movable borders are totally independent, but in many similar examples I put some restrictions and impose some correlation between them. The circles in the corners are the best and the most obvious places for resizing; for this reason I often do them bigger than the nodes on the borders. I also do not see any sense in making these circles smaller than the border nodes. But these are only my suggestions based on the years of implementing similar covers in a lot of different rectangular objects; the standard covers do not have any of these restrictions.

Though this part of the book is about the covers for the standard and popular graphical objects, but for the purpose of better demonstration this and other examples include some controls and more complex objects. Those controls or combination of controls allow to change some needed parameters. The controls and more complicated objects are the subjects of the same moving / resizing mechanism; the involvement of controls and groups of objects into the moving will be described in details further on, but all the movable / resizable objects are mentioned in the same methods of the forms, so I would like to say several words of explanation here for better understanding of the whole process.

Two controls with the comments can be easily moved around the screen by grabbing them anywhere near those controls or at any point of their comments. These two controls with their comments constitute a single object of the `LinkedRectangles` class – one of the classes, widely used for design of the user-driven applications. Overall there are seven movable objects in the **Form_Rectangles_StandardCase.cs**; initially they are registered with the mover in such an order:

1. The button to switch the visualization of covers ON / OFF.
2. The `LinkedRectangles` object to change the parameters of the nodes. Both controls with their comments can move only synchronously and never change their relative positions, so it is a single object for the mover.
3. Information (a `TextM` object).
4. Four rectangles (objects of the `RectStandard` class) in such an order: blue, yellow, green, and cyan.

I have already mentioned the rules of correct order for objects in the queue and their drawing. Regardless of the content of the form, the objects must be included into the queue in such an order.

1. Controls and complicated objects, consisting exclusively of controls.
2. Objects based on combination of controls and graphical parts.
3. Graphical objects.

If the order of objects in the form is set at its initialization and is not going to change throughout the lifetime of this form, then the mover's queue can be organized once and is fixed from this moment. The drawing of objects must be done in the opposite order to their placement in the mover's queue. If the order of objects can be changed throughout the form's lifetime, then, whenever it happens, the mover's queue is reorganized and the form needs a repainting. Changing of the mover's queue can be caused not only by reordering of the existing objects, but also by their removing from the screen or by adding the new objects. There are two ways to keep the mover's queue in order. The first one is simply to look through the mover's queue, find the needed object and move it to the new position in this queue. The second way is much more reliable: to have in the form a `RenewMover()` method, which is doing exactly the same thing of rearranging the mover's queue, but guarantees that all the objects are included into the mover's queue in correct order. Beginning from this example and further on you will find the `RenewMover()` method in nearly any form (example), which I am going to demonstrate. Here is this method for the **Form_Rectangles_StandardCase.cs.**

```
private void RenewMover ()
{
    mover .Clear ();
    mover .Insert (0, lrsView);
    mover .Insert (0, btnCovers);
    mover .Add (info);
    for (int i = 0; i < rects .Count; i++)
    {
        mover .Add (rects [i]);
    }
}
```

The order of rectangles in the **Form_Rectangles_StandardCase.cs** can be changed by a single mouse click. If a rectangle is clicked by the left button and the distance between the occurring of **MouseDown** and **MouseUp** events is not bigger than 3 pixels, then this rectangle must be moved atop of all others.

```
private void OnMouseUp (object sender, MouseEventArgs e)
{
    if (mover .Release ())
    {
        if (e .Button == MouseButtons .Left &&
            Auxi_Geometry .Distance (ptMouse_Down, e .Location) <= 3)
        {
            GraphicalObject grobj = mover [mover .WasCaughtObject] .Source;
            if (grobj is RectStandard)
```

```
                {
                    PopupRectangle (grobj .ID);
                }
            }
        }
    }
```

Any movable object has a unique identification number; this **id** of the clicked rectangle is easily obtained. By checking the **id** of each rectangle in the `List` of all of them, the pressed rectangle is easily found. Then this rectangle is moved to the head of the `List`, the mover's queue is reorganized according to the new order of elements, and the form is repainted, demonstrating the new rectangle on top.

```
    private void PopupRectangle (long id)
    {
        for (int i = rects .Count - 1; i > 0; i--)
        {
            if (id == rects [i] .ID)
            {
                RectStandard elem = rects [i];
                rects .RemoveAt (i);
                rects .Insert (0, elem);
                RenewMover ();
                Invalidate ();
                break;
            }
        }
    }
```

There is no need in rewriting the `OnPaint()` method according to the new order of rectangles. As was explained in the previous example, the `OnPaint()` method goes through the mover's queue of objects, starting from its end, and calls the appropriate `Draw()` method for each of the objects. This automatically draws all the rectangles correctly regardless of their order at any particular moment.

```
    private void OnPaint (object sender, PaintEventArgs e)
    {
        Graphics grfx = e .Graphics;
        GraphicalObject grobj;
        for (int i = mover .Count - 1; i >= 0; i--)
        {
            grobj = mover [i] .Source;
            if (grobj is RectStandard)
            {
                (grobj as RectStandard) .Draw (grfx);
            }
            else if (grobj is TextM)
            {
                (grobj as TextM) .Draw (grfx);
            }
            else if (grobj is LinkedRectangles)
            … …
```

This **Form_Rectangles_StandardCase.cs** is the first example, where you can see both the request for renewing a cover for some existing object and a request for renewing the mover's queue. These two things, though they both work with covers, are used for absolutely different purposes and in different situations. There must be a clear understanding of these differences.

When any of the parameters, regulating the size of a node or the number of nodes in the cover, are changed, then the cover of an object has to be renewed by calling the `DefineCover()` method. For example, in this form there is an easy way to change the radius of the circular nodes or the width of the nodes along the borders. Each rectangle has to be notified about the new radius of the circular nodes.

```
private void ValueChanged_numericRadius (object sender, EventArgs e)
{
    radius = Convert .ToInt32 (numericUD_Radius .Value);
    foreach (RectStandard rect in rects)
    {
        rect .Radius = radius;
    }
    Invalidate ();
}
```

Setting the new radius of the circular node via the `RectStandard.Radius` property requires the new definition of the cover for the involved rectangle. In this form it means for all the rectangles, as the changes are spread on all of them simultaneously.

```
public int Radius
{
    get { return (radius); }
    set
    {
        radius = Math .Abs (value);
        DefineCover ();
    }
}
```

This sequence of calls does not affect in any way the objects in the mover's queue. The objects are the same, they retain the same order, so there is absolutely no need in calling the `RenewMover()` method. On the other hand, if there is any change in the number of objects or in their order, then the call for the `RenewMover()` method is mandatory.

There will be some difference, when we come to the examples with the complex objects (the objects consisting of individually movable parts), but until then these are the rules for renewal of covers and queues.

## *Rectangles with a possibility of disappearance*

File: **Form_Rectangles_StandardToDisappear.cs**
Menu position: *Graphical objects – Basic elements – Rectangles – Standard case; can disappear*

All the objects, belonging to the `RectStandard` class from the previous example, can be moved around the screen. Their resizing is determined personally for each of them at the moment of construction on the basis of two parameters: the initial size and the declared range of resizing. The `RectStandard` class prevents the accidental disappearance of its representatives by imposing the minimum allowed sizes, so in need to erase such an object, it must be done, for example, via the command of context menu or in similar way. On the other side, the objects are resized by a mouse, so in some cases the squeezing of an object to a tiny size can be looked at as a command to delete this object. If users know that they can erase some objects in such a way and do it, then it is definitely not an accidental disappearance.* In many cases the squeezing of objects to a tiny size or even to zero can be more natural way of their deleting, than through the menu command. Let us look at the rectangles that can be deleted after the squeezing.

On opening the **Form_Rectangles_StandardToDisappear.cs** you see exactly the same picture as in the previous example (**figure 3.1**). Only the title of the form is slightly changed and the information has an extra line, but all the main objects look the same. The same four rectangles with different types of resizing, the same small button to switch the covers ON and OFF, the same two controls to change the sizes of the nodes.

The rectangles in this form look the same (well, all the rectangles are similar), but they are slightly different. They belong to the `RectStandardToDisappear` class, which is derived from the `RectStandard` class, so they are very close in many aspects. The main difference is that there is no minimum size as a protection against disappearance. On the contrary, there is a "disappearance size" for the objects of the new class: if at the moment of release any dimension of rectangle is less than this size, then the rectangle is taken out. The limit of disappearance is now set to 5 pixels, but can be easily changed.

* I do not share an opinion, widely spread among the developers, that users are so stupid that the fool-proof procedures must be installed in all the corners of the programs. If users know that they can delete an object by diminishing it to zero or tiny size and if they do it, then they really want to get rid of this object. In any way, you can add an additional question to confirm the request for deleting an object.

```
    public class RectStandardToDisappear : RectStandard
    {
        static int nSizeDisappearance = 5;
```

There is nothing interesting in the derived class; there is hardly anything at all. The only addition is the checking in the `Draw()` method to avoid an attempt to draw any rectangle of a zero size.

```
    new public void Draw (Graphics grfx)
    {
        if (rc .Width > 0 && rc .Height > 0)
        {
            grfx .FillRectangle (brush, rc);
        }
    }
```

The decision about possible erasing of an object and the real erasing, if it is needed, are made after releasing the mouse button and checking the sizes at this moment, so the new and interesting changes of the code can be found in the `OnMouseUp()` method of the form.

```
private void OnMouseUp (object sender, MouseEventArgs e)
{
    int iWasCaught, iNode;
    if (mover .Release (out iWasCaught, out iNode))
    {
        if (e .Button == MouseButtons .Left)
        {
            GraphicalObject grobj = mover [iWasCaught] .Source;
            if (grobj is RectStandardToDisappear)
            {
                if (iNode != grobj .NodesCount - 1)
                {
                    RectStandardToDisappear rect =grobj as RectStandardToDisappear;
                    if (Math.Min (rect.RectAround .Width, rect .RectAround.Height)
                               <= RectStandardToDisappear .SizeDisappearance)
                    {
                        long id = grobj .ID;
                        for (int i = rects .Count - 1; i >= 0; i--)
                        {
                            if (id == rects [i] .ID)
                            {
                                rects .RemoveAt (i);
                                RenewMover ();
                                Invalidate ();
                                break;
                            }
                        }
                    }
                }
```

Some of the methods and properties, which are useful in this case, were briefly mentioned in the subsection *From algorithm to working programs*; now it is time to look at their use in a real program.

1. If anything has to be done in the `OnMouseUp()` method, then only in case of releasing an object. The return value of the `Mover.Release()` method is the best indication of whether it happened or not. There are several variants of this method. For the post-release analysis I need to identify both the released object and the node, so I use the variant, which provides this information.

    ```
    if (mover .Release (out iWasCaught, out iNode))
    ```

2. The number of the released object in the mover's queue allows to check the class of the released object and to get the released rectangle itself, if it was an object of the `RectStandardToDisappear` class.

```
        GraphicalObject grobj = mover [iWasCaught] .Source;
        if (grobj is RectStandardToDisappear)
        {
            RectStandardToDisappear rect = grobj as RectStandardToDisappear;
```

3. The released rectangle may have different types of resizing, but in any case the last node in the cover is the big polygonal (rectangular) node, which is responsible not for resizing of this object, but for its moving. When an object is simply moved around, its size never changes, so there is no chance that at the end of such movement an object becomes small enough to be deleted. Thus, the situation with the last node being caught and released is not considered further on.

```
        if (iNode != grobj .NodesCount - 1)
```

4. If the released rectangle is so small that it has to be erased from the screen, then this rectangle must be found in the list of all the rectangles and deleted from there. Identification is done with the help of the `GraphicalObject.ID` property.

```
    if (Math .Min (rect .RectAround .Width, rect .RectAround .Height) <=
                                            RectStandardToDisappear .SizeDisappearance)
    {
        long id = grobj .ID;
        for (int i = rects .Count - 1; i >= 0; i--)
        {
            if (id == rects [i] .ID)
            {
                rects .RemoveAt (i);
```

5. Any change in the number of the movable objects inside the form requires the renewal of the mover's queue, which is done by the `RenewMover()` method. The change in the number of rectangles also requires the repainting of the form.

```
                RenewMover ();
                Invalidate ();
```

## *Rectangles with a single moving border*

File: **Form_Rectangles_SingleSideResizing.cs**
Menu position: *Graphical objects – Basic elements – Rectangles – Single side resizing*

Let us look at the case of a rectangle, which can be resized only by one side. It looks like a significant simplification of the previous case, where a rectangle could be resized by any of its borders, but such simplified case of rectangles is used in some situations as a part of the complex objects, so it would be nice to have this part already discussed and prepared.

**Form_Rectangles_SingleSideResizing.cs** demonstrates four different rectangles (**figure 3.2**). All of them belong to the `OneSideRectangle` class; the difference between them is only on the side, by which they can be resized. The side, by which the particular rectangle can be changed (resized), is marked by a darker line. This indication is an option; which can be switched ON and OFF via the context menu that can be called on any rectangle. Without such indication, there is no visual prompt that one side of any rectangle differs from other sides. Though there is another kind of prompt: the mouse cursor is changed, when it is in the vicinity of this "special" border.

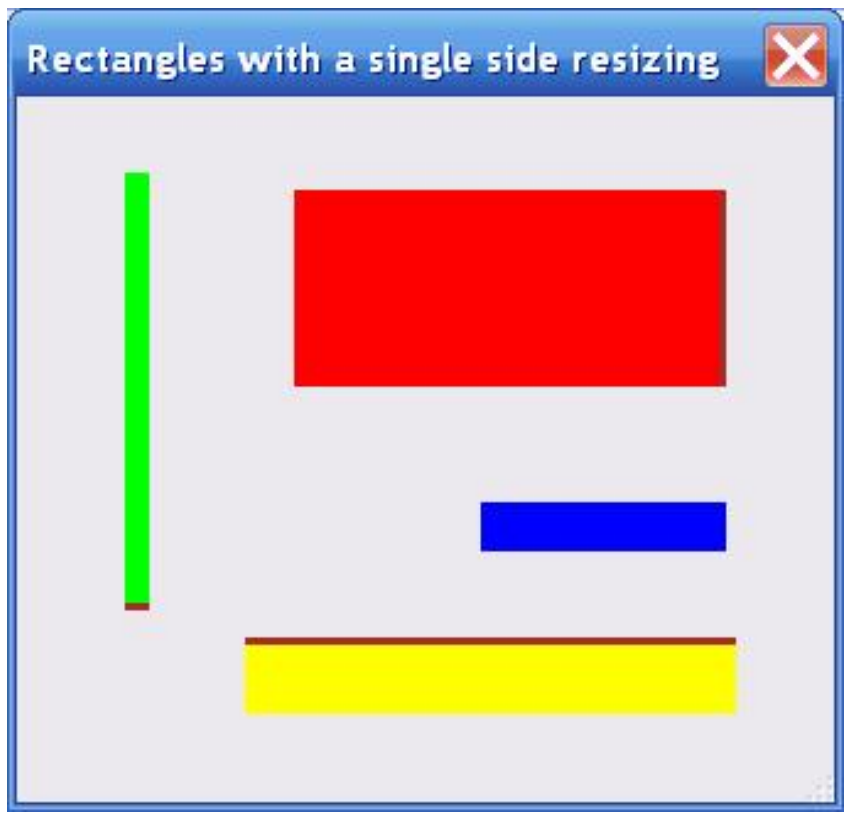

**Fig.3.2** Rectangles with a single side resizing

Objects of the `OneSideRectangle` class have minimum allowed length, which prevents their disappearance during the resizing; the maximum allowed length is passed as a parameter on initialization.

```
    public Form_Rectangles_SingleSideResizing ()
    {
        InitializeComponent ();
        mover = new Mover (this);
        stripRed = new OneSideRectangle (new Rectangle (60, 60, 120, 80),
                                         Side .E, Color .Red, 300);
        stripLime = new OneSideRectangle (new Rectangle (80, 160, 10, 130),
```

```
                                            Side .S, Color .Lime, 300);
        stripBlue = new OneSideRectangle (new Rectangle (200, 200, 100, 20),
                                            Side .W, Color .Blue, 700);
        stripYellow = new OneSideRectangle (new Rectangle (360, 360, 200, 30),
                                              Side .N, Color .Yellow, 400);
```

Any `OneSideRectangle` object can be moved and resized, so a cover for such an object must consist of two nodes. At least of two nodes and two nodes are enough in this case. Both nodes are polygons (rectangles); the first one is narrow, covers the movable border, and is used for resizing. Another one covers the whole area of an object and is used for its moving. The second node is the same for all four variants of resizing. The position of the first node and the cursor above it depend on the side, which is chosen for resizing.

```
public override void DefineCover ()
{
    CoverNode [] nodes = new CoverNode [2];
    switch (sideMove)
    {
        case Side .W:
            nodes [0] = new CoverNode (0, new Rectangle (rc .Left - halfsense,
                                 rc .Top, 2 * halfsense, rc .Height), Cursors .SizeWE);
            break;
        case Side .N:
            nodes [0] = new CoverNode (0, new RectangleF (rc .Left,
                 rc .Top - halfsense, rc .Width, 2 * halfsense), Cursors .SizeNS);
                break;
        … …
    }
    nodes [1] = new CoverNode (1, rc, Cursors .SizeAll);
    cover = new Cover (nodes);
}
```

The `MoveNode()` method in this case is much simpler, than in the previous, because the covers for all four cases are similar, consist of two nodes only, and in any case there is a check for one dimension only. To avoid disappearance, there is a limit on the minimum size of such rectangle. The limit on the maximum size is passed as a parameter during the construction, so the `MoveNode()` method has to check the proposed new size against both limits. All the checks are needed only for one node (`i = 0`); for another node there is an automatic call of the `Move()` method.

```
public override bool MoveNode (int i, int dx, int dy, Point ptM, MouseButtons btn)
{
    bool bRet = false;
    if (catcher == MouseButtons .Left)
    {
        if (i == 0)
        {
            int newW, newH;
            switch (sideMove)
            {
                case Side .W:
                    newW = rc .Width - dx;
                    if (minLen <= newW && newW <= maxLen)
                    {
                        rc .X += dx;
                        rc .Width -= dx;
                        bRet = true;
                    }
                    break;
                … …
        }
        else
        {
            Move (dx, dy);
        }
```

By clicking any rectangle with the right button you call the context menu, which allows to set the sliding side and to switch the visual prompt on the border ON or OFF (**figure 3.3**). This is the first example in this book, which uses the context menu for setting the parameters; further on I will use the menus more and more.

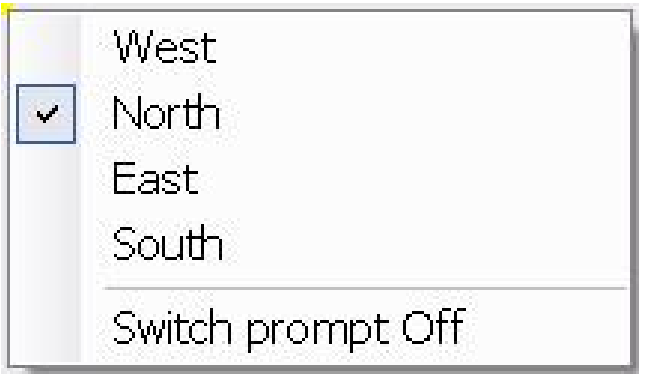

**Fig.3.3** Menu on rectangles

```
private void OnMouseUp (object sender, MouseEventArgs e)
{
    ptMouse_Up = e .Location;
    double dist = Auxi_Geometry .Distance (ptMouse_Down, e .Location);
    if (mover .Release () && e .Button == MouseButtons .Right && dist <= 3)
    {
        GraphicalObject grobj = mover .WasCaughtSource;
        if (grobj is OneSideRectangle)
        {
            stripPressed = grobj as OneSideRectangle;
            ContextMenuStrip = menuOnStrip;
        }
    }
}
```

In all the examples of this Demo application the context menus are called on the release of the right button and only if the distance between the points of the **MouseDown** and **MouseUp** events is small.

## Rectangle between two lines

File: **Form_Rectangles_LineToLine.cs**
Menu position: *Graphical objects – Basic elements – Rectangles – Turn out by moving one side*

This case is very similar to the previous one, but there is no limitation on the minimum size; the movable border can cross the unmovable opposite side and turn out the whole rectangle.

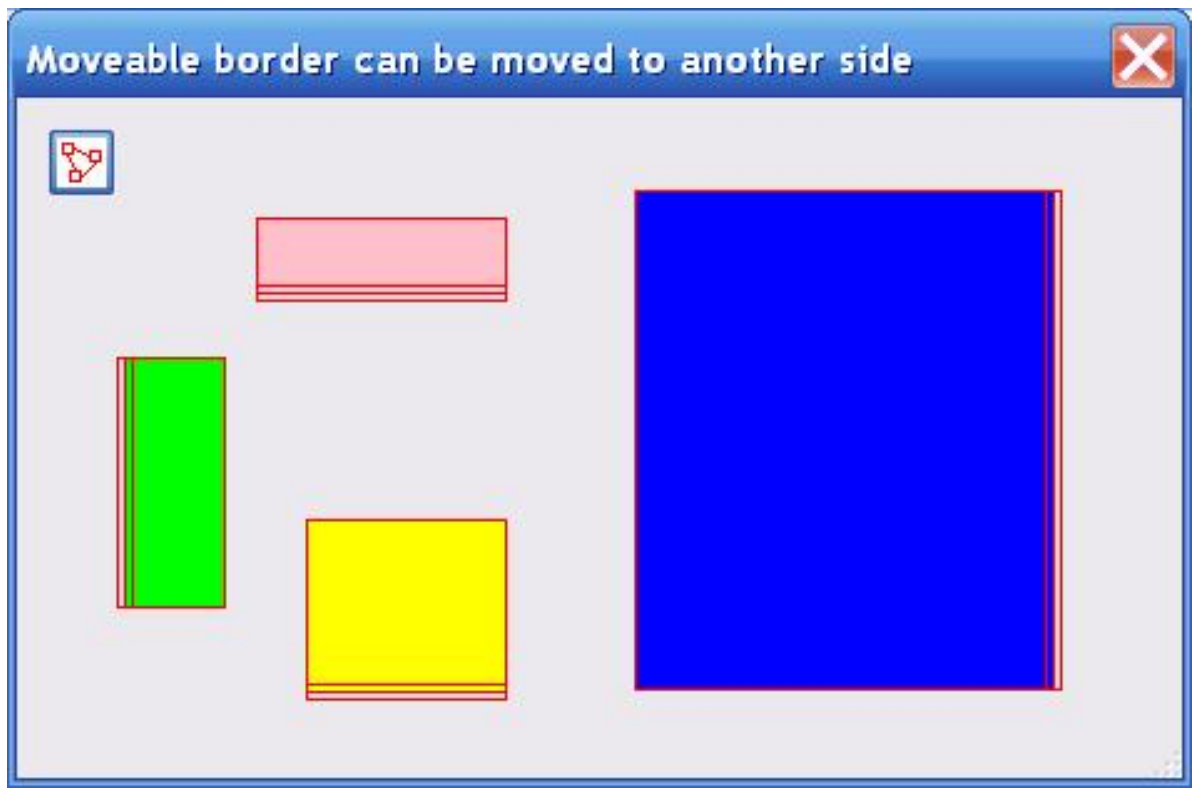

**Fig.3.4** The movable side of these rectangles can cross the opposite side

**Form_Rectangles_LineToLine.cs** demonstrates four different rectangles (**figure 3.4**). All of them belong to the `SideOfRectangle` class; the difference between them is only on the side, by which they can be resized. In the normal view, there is no indication of the side, by which each rectangle can be resized. Even more: a rectangle can be resized up to total disappearance. The movable side can be placed exactly at the same line, as the opposite side, and this rectangle disappears from view, though it still exists and sensed by the mover. While moving the mouse across the place, which looks absolutely empty, you can see the change of the cursor; this is the only indication that there is some object, which mover recognizes. If you press the mouse button at that moment and move it aside, the previously invisible rectangle reappears again. Usually, this is not the good situation and signals about the poor level of design. However, similar situations are possible and can be normal, if, for example, it is used for tuning of the bars in the bar chart, where an empty place between visible neighbouring bars signals only about the zero value of the bar in between. In the current **Form_Rectangles_LineToLine.cs** even the invisible rectangle can be visualized in special way, as the covers are shown for all the rectangles regardless of their visual size.

Here is the constructor for the `SideOfRectangle` class; the words of explanation are below.

```
public SideOfRectangle (Point pt, LineDir dirSlider, int lenSlider,
                        int dist, int mindist, int maxdist, Color clr)
```

The movable side is called a **slider**; the opposite side is fixed.

| | |
|---|---|
| `Point pt` | This is one of the corners of the initial rectangle; which corner it is exactly, depends on the next three parameters. |
| `LineDir dirSlider` | Direction of the slider. |
| `int lenSlider` | Length of the slider. The length can be even negative, which means that the point (parameter `pt`) is either on the right side (for horizontal slider) or at the bottom (for vertical slider). |
| `int dist` | Distance from the fixed side to the slider. Negative values mean that the slider is either to the left or above the fixed side. |
| `int mindist` | Negative value of this parameter means how far the slider can go from the fixed side to the left or above. |
| `int maxdist` | Positive value of this parameter means how far the slider can go from the fixed side to the right or below. |

The cover for any object of the `SideOfRectangle` class is similar to the previous case and consists of the same two polygonal nodes used in exactly the same way, but there is a small trick, which allows to avoid the possible crash in the case of disappearing rectangle.

```
public override void DefineCover ()
{
    int cxL, cxR, cyT, cyB;
    CoverNode [] nodes = new CoverNode [2];
    switch (linedir)
    {
        case LineDir .Hor:
            cxL = Math .Min (ptAnchor .X, ptAnchor .X + nFixedLen);
            cxR = Math .Max (ptAnchor .X, ptAnchor .X + nFixedLen);
            int cy = ptAnchor .Y + distance;
            nodes [0] = new CoverNode (0, new Rectangle (cxL, cy - halfsense,
                                         cxR - cxL, 2 * halfsense), Cursors .SizeNS);
            cyT = Math .Min (ptAnchor .Y, ptAnchor .Y + distance);
            cyB = Math .Max (ptAnchor .Y, ptAnchor .Y + distance);
            nodes [1] = new CoverNode (1, new Rectangle (cxL, cyT, cxR - cxL,
               Math .Max (2, cyB - cyT)), Behaviour .Moveable, Cursors .SizeAll);
            break;
```

The first node is a narrow rectangle, which covers the movable border by a strip of 6 pixels width. The second node has to cover the whole area of rectangle, but its changeable size is never set to less than 2 pixels, even when the size of an object is decreased to zero. In this case an object is invisible, but the `DefineCover()` method does not allow the size of this node to be less than two pixels, so the node is initialized in a normal way in any case. At the same time, when the changeable size of a rectangle is less than three pixels, then the second node is smaller and is totally covered by the first node, so the mover catches the first (bigger) node, which is responsible for resizing, and increases the object first.

The `MoveNode()` method is simple enough. For the calculation of the new proposed distance either one of the parameters is used or another; this depends on the direction of slider. After it the proposed distance is checked against the range of the allowed distances. All these things are only for the first node; the second node is used to move the whole object.

```
public override bool MoveNode (int i, int dx, int dy, Point ptM, MouseButtons btn)
{
    bool bRet = false;
    if (catcher == MouseButtons .Left)
    {
        if (i == 0)
        {
            int newDist = (linedir == LineDir.Hor) ? (distance+dy) : (distance+dx);
            if (minDistance <= newDist && newDist <= maxDistance)
            {
                distance = newDist;
                bRet = true;
            }
        }
        else
        {
            Move (dx, dy);
        }
    }
    return (bRet);
}
```

## *Symmetrically changing rectangles*

File: **Form_Rectangles_SymmetricalChange.cs**
Menu position: *Graphical objects – Basic elements – Rectangles – Symmetrical change*

All four previous examples of this chapter demonstrate the rectangles, in which the moving of one side does not affect other sides. But in some situations there is a need of symmetrical change of an object; this is the case of the next example. There are four different rectangles in the **Form_Rectangles_SymmetricalChange.cs**; these four objects represent the four different cases of resizing (**figure 3.5**).

The main difference of the `SymmetricallyChangedRectangle` objects from other rectangles is in their positioning: instead of the `Rectangle` field there is a central point and two half sizes, which together describe an area of an object. Such design of a class makes all the calculations much easier (in this specific case!).

```
    public class SymmetricallyChangedRectangle : GraphicalObject
    {
        Resizing resize;
        Point ptCenter;
        int wHalf, hHalf;
        SolidBrush brush;
        int minHalfSize = 10;
```

But when you need to initiate any type of rectangle in your form, it is natural to declare the area of this object as a normal rectangle, so the construction of these objects goes in a standard way.

```
    public SymmetricallyChangedRectangle (Rectangle rc, Resizing res, Color clr)
```

The design of cover is obvious. Whenever the border of rectangle must be movable, it is covered by a node in the form of a narrow strip; the number of such strips varies depending on the type of the needed resizing. There is no sense in declaring a rectangle symmetrically changing and at the same time making only one of the opposite sides movable; it would be natural to have two movable opposite sides, so the number of nodes on the borders is 0, 2 or 4. As all the rectangles are movable, there must one big rectangular node to cover (and move) the whole object; thus the total number of nodes is 1, 3 or 5. This is the part of the `DefineCover()` method for a horizontally resizable rectangle (the Green one from **figure 3.5**).

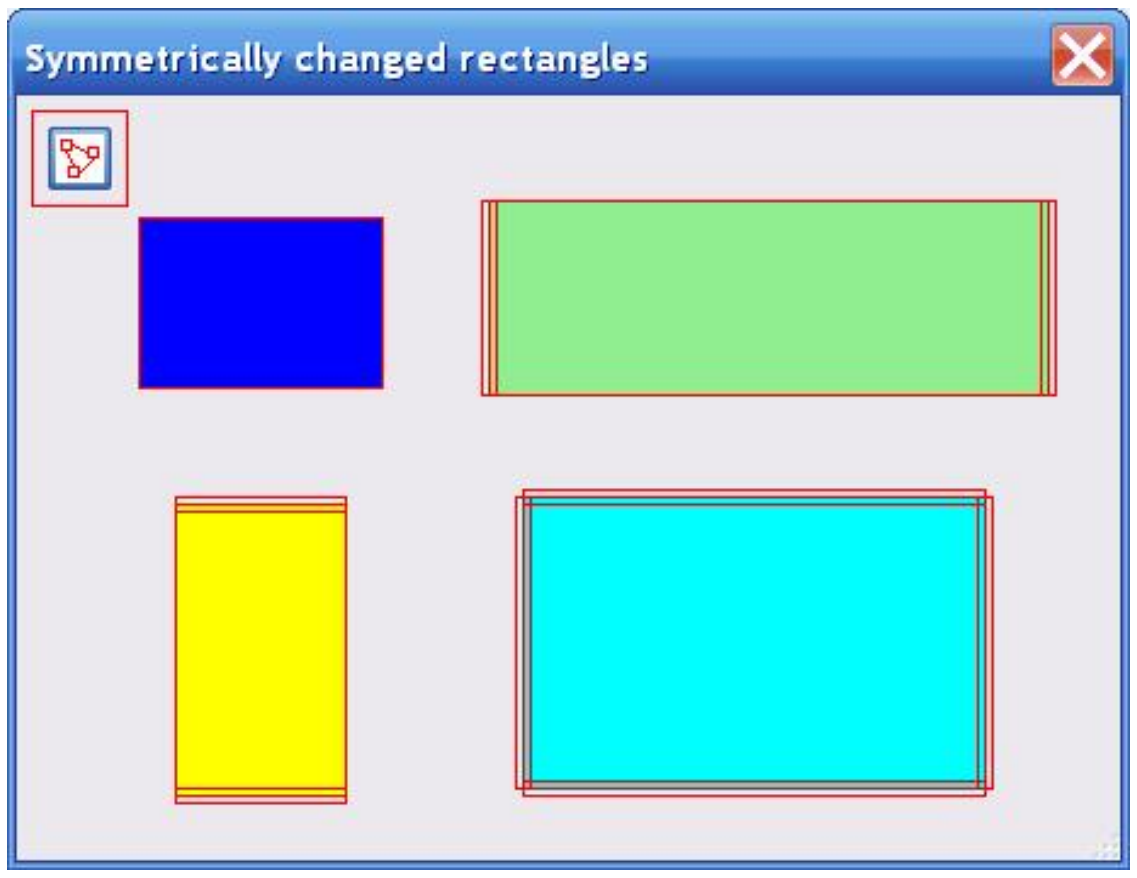

**Fig.3.5** Symmetrically changing rectangles

```
public override void DefineCover ()
{
    int cxL = ptCenter .X - wHalf;
    int cxR = ptCenter .X + wHalf;
    int cyT = ptCenter .Y - hHalf;
    int cyB = ptCenter .Y + hHalf;
    int half = 3;
    CoverNode [] nodes;
    switch (resize)
    {
        … …
        case Resizing .WE:
            nodes = new CoverNode [] {
                new CoverNode (0, new Point [] {new Point (cxL - half, cyT),
                                                new Point (cxL + half, cyT),
                                                new Point (cxL + half, cyB),
                                                new Point (cxL - half, cyB)},
                                    Cursors .SizeWE),
                new CoverNode (1, new Point [] {new Point (cxR - half, cyT),
                                                new Point (cxR + half, cyT),
                                                new Point (cxR + half, cyB),
                                                new Point (cxR - half, cyB)},
                                    Cursors .SizeWE),
                new CoverNode (2, new Point [] {new Point (cxL, cyT),
                                                new Point (cxR, cyT),
```

```
                                                   new Point (cxR, cyB),
                                                   new Point (cxL, cyB)}) };
            break;
        … …
    cover = new Cover (nodes);
```

The `SymmetricallyChangedRectangle` class has the minimum allowed size, which prevents any rectangle from disappearance. Because everywhere in the code of this class not the full sizes of rectangle are used, but half sizes, it makes sense to turn the minimum size of rectangle into the limitation on the half size; all the calculations become easier in such a way. Here is the part of the `MoveNode()` method for the same case of the horizontally resizable rectangles.

```
public override bool MoveNode (int i, int dx, int dy, Point ptM, MouseButtons btn)
{
    bool bRet = false;
    if (btn == MouseButtons .Left)
    {
        switch (resize)
        {
            … …
            case Resizing .WE:
                if (i == 0)      // on left side
                {
                    if (wHalf - dx >= minHalfSize)
                    {
                        wHalf -= dx;
                        bRet = true;
                    }
                }
                else if (i == 1)      // on right side
                {
                    if (wHalf + dx >= minHalfSize)
                    {
                        wHalf += dx;
                        bRet = true;
                    }
                }
                else
                {
                    Move (dx, dy);
                }
                break;
```

## Rectangles with the fixed ratio of the sides

File: **Form_Rectangles_FixedRatio.cs**
Menu position: *Graphical objects – Basic elements – Rectangles – Fixed ratio of the sides*

Rectangles with the fixed ratio of their sides are not used too often, but at the same time it is not an absolutely unusual task. For example, this is the case of demonstrating photos: you often want to change the size of the picture, but the ratio between its sides must be fixed, otherwise there will be image distortions.

There are four different rectangles in the **Form_Rectangles_FixedRatio.cs** (**figure 3.6**), but this is done only to bring this example in conformity with all others of the chapter. There is no such thing as different types of resizing for the `Rectangle_FixedRatio` objects; they all have the same resizing. A rectangle can be resized by any side. As seen from **figure 3.6**, the cover of each rectangle consists of four narrow rectangles on the borders and one big rectangular node to cover the whole object.

```
public override void DefineCover ()
{
    int cxL = rc .Left;
    int cxR = rc .Right;
    int cyT = rc .Top;
```

```
    int cyB = rc .Bottom;
    int nW = rc .Width;
    int nH = rc .Height;
    CoverNode [] nodes = new CoverNode [5];
    nodes [0] = new CoverNode (0, new RectangleF (cxL - halfstrip, cyT,
                                  2 * halfstrip, nH), Cursors .SizeWE);    // Left
    nodes [1] = new CoverNode (1, new RectangleF (cxR - halfstrip, cyT,
                                  2 * halfstrip, nH), Cursors .SizeWE);    // Right
    nodes [2] = new CoverNode (2, new RectangleF (cxL, cyT - halfstrip, nW,
                                  2 * halfstrip), Cursors .SizeNS);        // Top
    nodes [3] = new CoverNode (3, new RectangleF (cxL, cyB - halfstrip, nW,
                                  2 * halfstrip), Cursors .SizeNS);        // Bottom
    nodes [4] = new CoverNode (4, new RectangleF (cxL, cyT, cxR - cxL, nH),
                                  Cursors .SizeAll);
    cover = new Cover (nodes);
}
```

There is no sense to add circular nodes into the corners, as they can only become the cause of ambiguity. When you move the corner of rectangle which has the fixed ratio of sides than you cannot move both sides according to the movement of the mouse. One of the adjacent sides must be declared a dominant one and move with the mouse; position of another side must be calculated from the position of that dominant side by using the predetermined ratio of the sides.

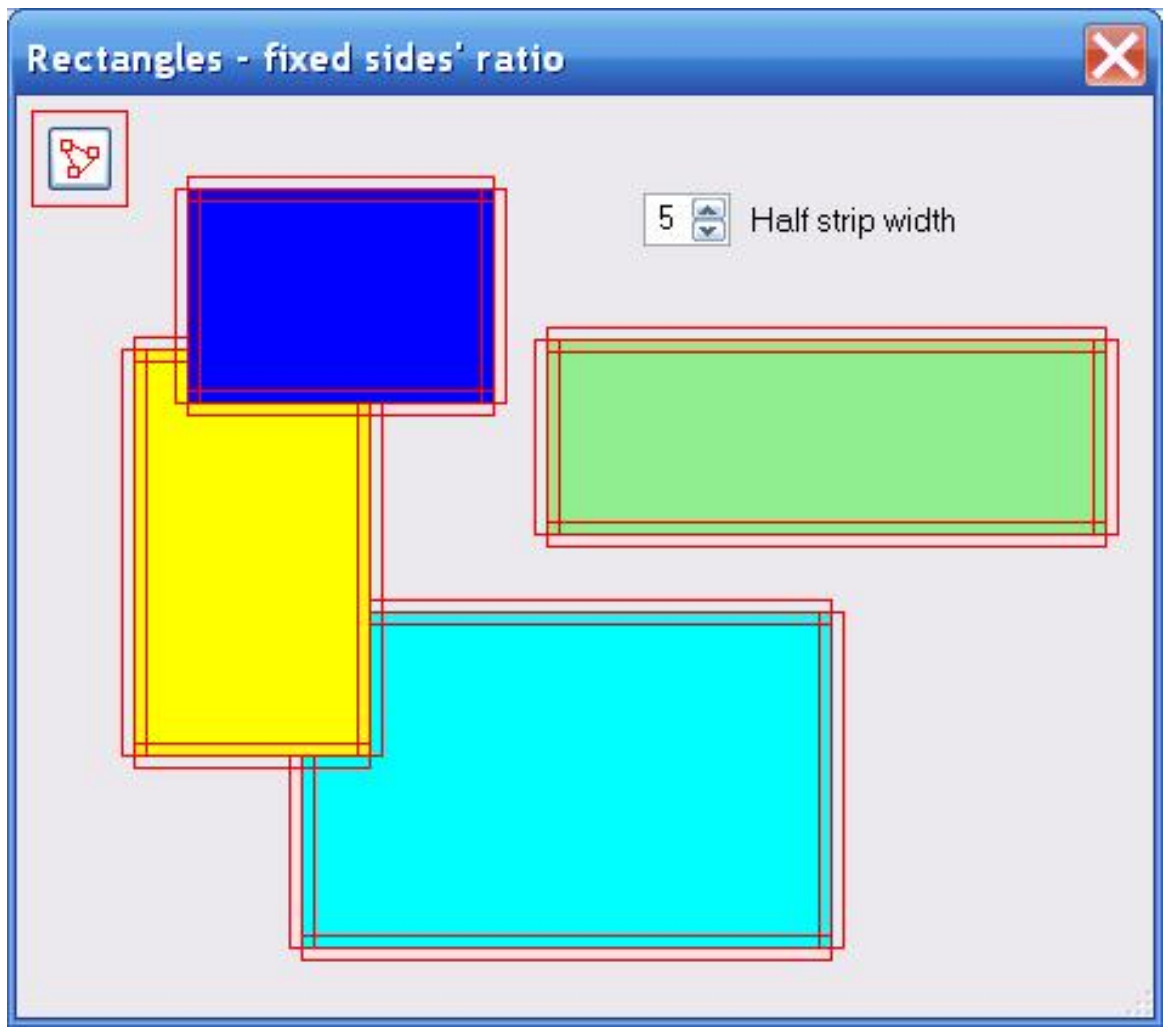

**Fig.3.6** Rectangles, which maintain the fixed sides' ratio throughout the resizing

Even without movable corners, there is a problem of the same type; maybe on a smaller scale. When any side is moved, one of its neighbouring sides must be moved also to retain the ratio of the sides. Which one of the neighbours must be moved? Such a decision was made for the `Rectangle_FixedRatio` class:

- If the left or right borders are moved, then the top of the rectangle is fixed and the bottom is moved automatically to retain the ratio.
- If the upper or lower borders are moved, then the left border is fixed, while the right border is moved automatically to retain the ratio.

Here is the part of the `MoveNode()` method for two nodes. The last node of the cover (`i = 4`) is used for moving the whole rectangle, so there is only an automatic call for the `Move()` method. The first node (`i = 0`) covers the left border. The proposed move of the border (`dx`) gives the possible new width of the rectangle (`wNew`); by using the fixed ratio, the possible new height of this rectangle is calculated (`hNew`). These new sizes are checked against the allowed minimum sizes; if the changes can be made, then the `MoveBorder_Left()` method is called.

```
public override bool MoveNode (int i, int dx, int dy, Point ptM, MouseButtons btn)
{
    bool bRet = false;
    if (btn == MouseButtons .Left)
    {
        int wNew, hNew;
        if (i == 4)
        {
            Move (dx, dy);
        }
        else if (i == 0)      // on left side
        {
            wNew = rc .Width - dx;
            hNew = Convert .ToInt32 (wNew / fWHratio);
            if (wMin <= wNew && hMin <= hNew)
            {
                MoveBorder_Left (dx);
```

```
                bRet = true;
            }
        }
```

In the `MoveBorder_Left()` method the left border is really moved and the new width of rectangle is calculated. Then the new height of rectangle is calculated by using the predetermined ratio of the sides.

```
        private void MoveBorder_Left (int dx)
        {
            rc .X += dx;
            rc .Width -= dx;
            rc .Height = Convert .ToInt32 (rc .Width / fWHratio);
        }
```

There is one more interesting object in this form – the pair "control + comment". In one of the previous examples (**Form_Rectangles_StandardCase.cs**, **figure 3.1**) you can see two controls with comments, and though they look similar (what is the difference between one control with comment and two?), they belong to the different classes and behave differently.

In that previous example with two controls the whole group can be moved by any point. You can press the button either next to one of the controls or at any point of any comment and all four elements move synchronously. If you try the same with the pair in this **Form_Rectangles_FixedRatio.cs,** the result will depend on the mouse point:

- If you press the mouse next to the control, then the pair will move synchronously.
- If you press the mouse on the text (comment), only the text will move, but the control will stay at its position.
- If you press the text with the right button, you can rotate the text.

This pair "control + text" is an object of the `CommentedControl` class; this is one of the classes, widely used for design of forms. The behaviour of such objects is fully described by the three previous statements. Though this object looks very simple, it is the first example of the complex objects. The simplest of the complex objects! The main feature of all such objects is that their parts can be involved both in individual and synchronous (or related) movements. This involvement in both forms of movements requires a different type of registering with the mover. If you try to register the `CommentedControl` object in a standard way with `mover.Add()` or `mover.Insert()` methods, you will find out that the synchronous movement will work (if you press the button next to the control), but the comment will not move individually. To achieve all the movements that the `CommentedControl` class can provide, the special `IntoMover()` method of this class must be used for registering the objects. I will write more about the `CommentedControl` class and its use in the chapter *Control + text*.

When the **Form_Rectangles_FixedRatio.cs** is opened, all the objects are registered with the mover in such an order: button to switch the covers ON / OFF, the `CommentedControl` object, and then four rectangles (blue, yellow, green, and cyan).

```
        ccStrip .IntoMover (mover, 0);
        mover .Insert (0, btnCovers);
        mover .Add (rsBlue);
        mover .Add (rsYellow);
        mover .Add (rsGreen);
        mover .Add (rsCyan);
```

Later you can click any rectangle with the left button and bring this rectangle atop of others.

```
    private void OnMouseUp (object sender, MouseEventArgs e)
    {
        if (mover .Release ())
        {
            if (e .Button == MouseButtons .Left &&
                Auxi_Geometry .Distance (ptMouse_Down, e .Location) <= 3 &&
                mover .WasCaughtSource is Rectangle_FixedRatio)
            {
                PopupRectangle (mover .WasCaughtObject);
            }
        }
    }
```

Painting in the form is organized by going from the end of the mover's queue to its head and painting every element on the way. In one of the previous examples (**Form_Rectangles_StandardCase.cs, figure 3.1**) I have explained this process in details and showed that first the order of rectangles was changed in the `List` of rectangles, then the `RenewMover()` method was called to reorganize the mover's queue according to the new order of rectangles. However, in this **Form_Rectangles_FixedRatio.cs** neither a `List` of rectangles nor the `RenewMover()` method are present. Then how all this works? In this example I directly change the order of objects in the mover's queue.

```
private void PopupRectangle (int iInMover)
{
    while (iInMover > 0 && mover [iInMover - 1] .Source is Rectangle_FixedRatio)
    {
        mover .Reverse (iInMover - 1, 2);
        iInMover--;
    }
    Invalidate ();
}
```

There is neither deleting nor adding of objects in this form; the whole set of objects is small and simple, so I can do such a thing. But the technique described earlier, when you have a `List` of objects to deal with them and you use this `List` to populate the mover's queue, such technique is much more reliable, and I use it all the time.

# Rotation

> A lot of objects can be involved both in forward movement and rotation. I decided to interrupt the demonstration of standard elements for a discussion of rotation. After it the gallery of movable elements will continue, but already with their involvement in rotation.

The list of discussed standard elements in the previous chapters was opened by the solitary lines (**figure 2.2**). The lines in the **Form_Lines_Solitary.cs** can be rotated, but though the rotation of those lines looks OK, the code for that rotation is not absolutely correct. After discussion of the first example from this chapter you will easily understand, why the code in the **Form_Lines_Solitary.cs** has a flaw.

The common rule for all the movable objects: the design of their covers does not depend on whether these objects are going to be rotated or not. The covers are designed to provide the movements; the start of the forward movement and rotation are distinguished not by the touched place, but by the pressed mouse button. Certainly, it can be organized in a different way, when the places to start the forward movement and rotation would be different, but, from my point of view, it would be a wrong design, because then users would have to know and remember the difference between those areas. It is much better, when any object can be moved and rotated by any inner point. This will not demand from the users any extra knowledge about the screen objects, with which they deal in one program or another.

Though the cover of an object does not depend on its presumable involvement in rotation, there must be some pieces of code that make the rotation possible.

The first mandatory thing is an additional parameter which describes the angle of an object. Without such an angle it is impossible to define the position of object under rotation. For some objects such an angle exists among the parameters in any way and is used all the time. For others it is needed only for the period of rotation; especially for such objects this angle must be defined at the moment, when the rotation starts. Each class of rotated objects has its `StartRotation()` method, which includes the calculation of that angle. Because the rotation usually starts, when an object is caught by the right mouse button, then you will find in all further examples that this `StartRotation()` method is called from inside the `OnMouseDown()` method of the form.

Usually one angle is enough even for a complex object, because during the rotation the relative angles between the parts are not changed, so the positions of all the basic points are calculated according to this angle and the sizes. However, it depends on the object and in some cases it is easier to calculate the starting angles for several basic points and to change them synchronously throughout the rotation.

The rotation of an object can be started by catching it at any inner point and the center of rotation can be also changeable. What is really calculated at the starting moment is the difference between the angle from the mouse position to the center of rotation and that angle of an object. I call this difference between two angles a *compensation*. This compensation angle is never changed throughout the rotation; with the help of this compensation, the angle of an object can be calculated from the position of the cursor, and thus you get all the basic points and the view of an object at any moment.

`MoveNode()` method of an object is the place, where the real movement is checked and described, so an additional code in this method is the second mandatory thing to organize the rotation. The `MoveNode()` method has several parameters, which include the linear shift along both axes and the mouse position. It may look like one of these is redundant, but it is not so: my experience, based on design and use of many different movable objects, shows that the linear shifts along the axes are better used for forward movements, but for rotation the exact mouse position is the best and gives the most accurate and reliable results.

## *Circles*

File: **Form_Circles_Nonresizable.cs**
Menu position: *Graphical objects – Basic elements – Circles – Non-resizable*

This is going to be the first appearance of circles in this book. Trying to organize the examples of each chapter in an order from simple to complicated, I was surprised myself, when I realized that the simplest case of rotation would be a circle. This shape of objects was really a problem for a good resizing until I thought out the N-node covers (I will write about them further on), but if you take as an example a non-resizable circle, then it is the best object to talk about rotation.

When I began to design the class of circles for this example – `CircleRt` – I immediately understood one more amazing thing: it would be enough to have a single node cover for this class. I did not expect it; I was sure that after that first example with the primitive objects (**Form_Nodes.cs**, **figure 2.1**) I would not use such primitive covers anywhere else, but here we are, starting the exploration of the rotation and again using the most primitive objects. Though there is nothing strange at all, if you look at the situation according to the requests and rules of design. From the beginning I put one restriction on this class: the objects must be non-resizable, but only movable. If you have to implement both things, you cannot have less than two nodes in the cover; if you leave only movability, then a single node can be enough. In this case it does not matter that we need to organize two types of movement: forward movement and rotation. These two movements are started by two buttons, so one node is enough for both.

There is going to be a major difference between the `PrimitiveCircle` and the `CircleRt` classes. The first one was involved only in forward movement, which is obvious even for the monochromatic circles. But watching the monochromatic circle you cannot decide whether it is rotating or not, so the `CircleRt` must be multicolored.

```
public class CircleRt : GraphicalObject
{
    Point center;
    int radius;
    double angle;
    List<Color> clrs;
    double compensation;
```

Any object of the `CircleRt` class is described by its central point, radius, angle, and a set of colors. Visually the circle is divided into colored sectors; the starting angle of the first sector is considered as an angle of the circle.

The cover of such objects is primitive.

```
public override void DefineCover ()
{
    cover = new Cover (new CoverNode [] {
                      new CoverNode (0, center, radius, Cursors .SizeAll) });
}
```

The forward movement is simple, as only the central point must be moved.

```
public override void Move (int dx, int dy)
{
    center += new Size (dx, dy);
}
```

The `MoveNode()` method is also simple, as there is only one node in the cover.

```
public override bool MoveNode (int i, int dx, int dy, Point ptM, MouseButtons btn)
{
    bool bRet = false;
    if (btn == MouseButtons .Left)
    {
        Move (dx, dy);
        bRet = true;
    }
    else if (btn == MouseButtons .Right)
    {
        double angleMouse = Auxi_Geometry .Line_Angle (center, ptM);
        angle = angleMouse - compensation;
        bRet = true;
    }
    return (bRet);
}
```

The forward movement demonstrates nothing new: press a circle anywhere with the left button, the `MoveNode()` method automatically calls the `Move()` method, which calculates the new position of the central point.

Rotation starts, when the circle is pressed by the right button.

```
private void OnMouseDown (object sender, MouseEventArgs e)
{
    ptMouse_Down = e .Location;
    if (mover .Catch (e .Location, e .Button))
    {
```

```
            if (e .Button == MouseButtons .Right)
            {
                if (mover .CaughtSource is CircleRt)
                {
                    (mover .CaughtSource as CircleRt) .StartRotation (e .Location);
                }
            }
        }
    }
```

When the `CircleRt` object is pressed by the right button, the `StartRotation()` method of this class is called; the parameter of this method is the position of the cursor.

```
    public void StartRotation (Point ptMouse)
    {
        double angleMouse = Auxi_Geometry .Line_Angle (center, ptMouse);
        compensation = Auxi_Common .LimitedRadian (angleMouse - angle);
    }
```

The `StartRotation()` method calculates two things.

1. The angle of the line from the rotation center to the mouse cursor. The center of rotation is known – it is the center of a circle. The mouse position is passed as a parameter to this method. The angle from the rotation center to the mouse cursor is an angle of the line between two points; as I have to calculate such angle in many places of my application, this calculation uses one of the methods from the **MoveGraphLibirary.dll**; all the methods of this library are described in the **MoveGraphLibrary_Classes.doc** (see section *Programs and Documents* at the end of the book).
2. The compensation angle which is the difference between the angle to the mouse and the angle of a circle. When you press the mouse anywhere inside the circle to start the rotation, you visually estimate the position of the mouse in relation to the colored sectors (that was the reason for using multicolored circles; in monochromatic object you do not have this visual estimation). When you rotate a circle, you expect that those distances from the mouse to differently colored sectors would be unchanged. The positions of the colored sectors are defined by the angle of a circle, so the idea of rotation is to keep throughout the process the unchanged angle between the mouse and the angle of a circle. This is the compensation, which is not going to change throughout the whole rotation.

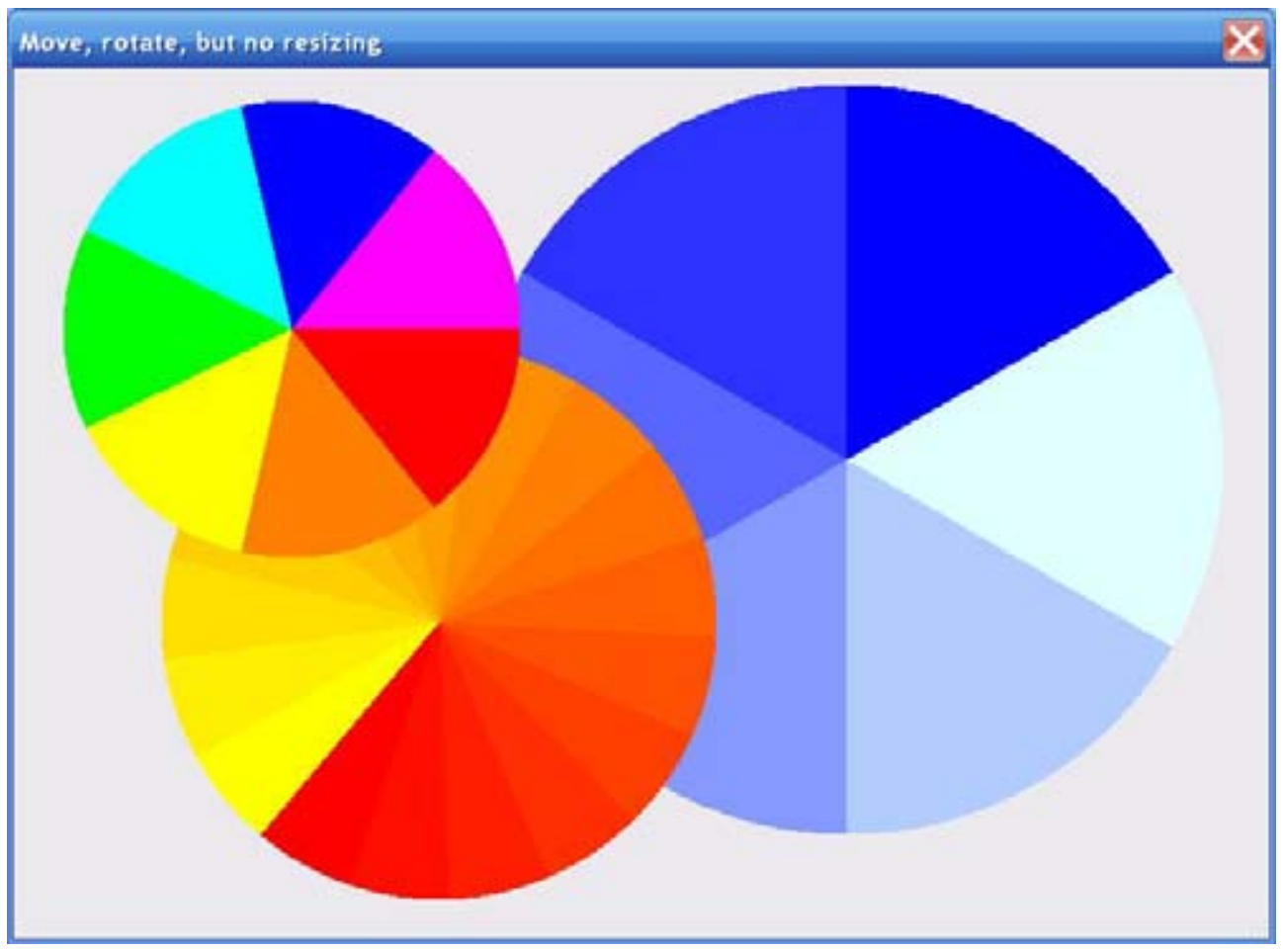

**Fig.4.1** Objects of the `CircleRt` class can not be resized, but can be moved forward and rotated.

The `compensation` is calculated at the starting moment of rotation and fixed for the whole time of it. The `MoveNode()` method uses this compensation to calculate the angle of a circle throughout the rotation. The mouse position is changing throughout the rotation, but the compensation is fixed, so for any mouse position it is easy to calculate the needed angle of a circle.

```
public override bool MoveNode (int i, int dx, int dy, Point ptM, MouseButtons btn)
{
    … …
    else if (btn == MouseButtons .Right)
    {
        double angleMouse = Auxi_Geometry .Line_Angle (center, ptM);
        angle = angleMouse - compensation;
```

This is the whole idea of rotation for the objects of any level of complexity. Some of the objects need and use only one angle, which allows to calculate everything; other objects may need several angles to calculate their different basic points. If there is one angle for an object, then you need to calculate a single compensation angle; if there are several angles to describe an object, then several compensation angles are needed. Everything else is absolutely the same.

- Calculate compensation angle(s) at the starting moment of rotation. The calculations are done in the `StartRotation()` method which has to receive the mouse position as a parameter.
- Inside the `MoveNode()` method use the current mouse position and the compensation angle(s) to calculate the basic point(s) of an object.

A couple of remarks on the **Form_Circles_Nonresizable.cs**. There are several objects of the `CircleRt` class in this form (**figure 4.1**); they can be moved, rotated, their order can be changed, but there is no `List` of such objects. I put the new circles directly into the mover's queue and perform all actions on the objects of this queue. I would not call this technique a widely used. On the contrary, I would never recommend such a thing, if there would be any complex objects in the form, but when you have to organize a form with a few simple objects, then it is possible to do it in such a way without any problems.

```
public Form_Circles_Nonresizable ()
{
    InitializeComponent ();
    mover = new Mover (this);
    mover .Add (new CircleRt (new Point (170, 160), 140, 0,
                Auxi_Colours .RainbowColorList));
    mover .Add (new CircleRt (new Point (260, 340), 170, -130,
                Auxi_Colours .SmoothColorsList (17, Color .Yellow, Color .Red)));
    mover .Add (new CircleRt (new Point (510, 240), 230, 30,
             Auxi_Colours .SmoothColorsList (6, Color .LightCyan, Color .Blue)));
}
```

The painting of the elements, while going along the mover's queue from the tail to the head, was already mentioned and demonstrated in the previous examples. In this example with the circles I have included the reordering of the objects directly into the `OnMouseUp()` method.

```
    private void OnMouseUp (object sender, MouseEventArgs e)
    {
        if (mover .Release ())
        {
            GraphicalObject grobj = mover .WasCaughtSource;
            if (e .Button == MouseButtons .Left &&
                grobj is CircleRt &&
                Auxi_Geometry .Distance (ptMouse_Down, e .Location) <= 3)
            {
                for (int i = mover .Count - 1; i > 0; i--)
                {
                    if (grobj .ID == mover [i] .Source .ID)
                    {
                        CircleRt circle = mover [i] .Source as CircleRt;
                        mover .RemoveAt (i);
                        mover .Insert (0, circle);
                        Invalidate ();
                        break;
                    }
                }
            }
        }
    }
```

If any circle is clicked by the left button and the distance between the mouse positions for the **`MouseDown`** and **`MouseUp`** events is greater than 3 pixels, then I consider it as a normal forward movement of a circle. If the distance is small, then it is a command to put the pressed circle on top of others. The last circle of the queue is always the lowest in a pile. Go through the objects of the mover's queue beginning from the end. If you find the circle with the same **id**, as the pressed one, and it is not already at the top, then it must be moved there. There is nothing else in the mover' queue except the circles. Remember the pressed circle, delete it from the queue, put the remembered circle into the head of the queue, and redraw everything.

## Segmented line

File: **Form_SegmentedLine.cs**
Menu position: *Graphical objects – Basic elements – Lines – Segmented line*

The previous example of rotation used an object (circle), which had one angle in description, so it was enough to calculate only one compensation angle. This new example demonstrates an object, which needs an array of compensation angles.

There are movable objects of four different classes in the **Form_SegmentedLine.cs** (**figure 4.2**).

- The familiar small button to switch the covers ON and OFF.
- The green text, which is an object of the `TextMR` class.
- The small red circle, which marks the rotation center, belongs to the `RotationCenter` class. It is a non-resizable circle, which can be moved to any place around the screen.
- The blue line, consisting of several segments. An object of the `SegmentedLine` class is based on an array of independently movable points; each pair of consecutive points is connected by a line. Because all these points (ends of segments) can be moved independently, then they must be defined at the moment of construction.

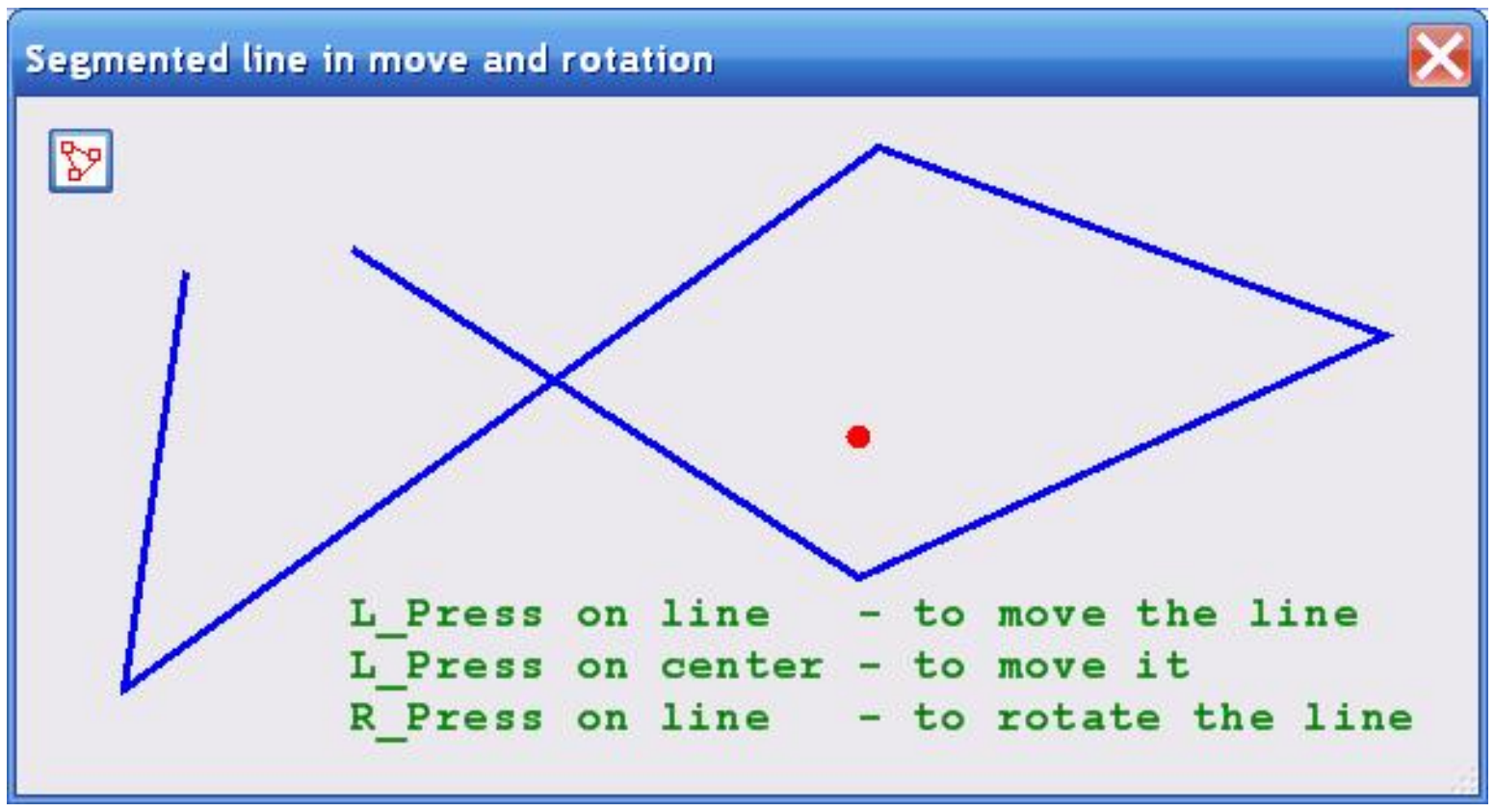

**Fig.4.2** Segmented line can be moved, reconfigured, and rotated

```
public SegmentedLine (Point [] pt, Pen penLine)
{
    pts = new PointF [pt .Length];
    for (int i = 0; i < pt .Length; i++)
    {
        pts [i] = pt [i];
    }
    pen = penLine;
    radius = new double [pts .Length];
    compensation = new double [pts .Length];
}
```

Two arrays are not needed for the forward movement of segmented line, but they are needed for a period of rotation between the **`MouseDown`** and **`MouseUp`** events. Because this is the main idea of this example, I decided to organize those arrays during the initialization of an object.

The cover of the segmented line consists of the small circular nodes on all the ends of segments and the strip nodes for each of the segments. The default width of the strip nodes is 6 pixels. To simplify the finding and moving of the end points, I set the radius of the circular nodes above them to 5 pixels.

```
public override void DefineCover ()
{
    CoverNode [] nodes = new CoverNode [pts .Length + (pts .Length - 1)];
    for (int i = 0; i < pts .Length; i++)
    {
        nodes [i] = new CoverNode (i, pts [i], 5);
    }
    for (int i = 0; i < pts .Length - 1; i++)
    {
        nodes [pts .Length + i] =
                        new CoverNode (pts .Length + i, pts [i], pts [i + 1]);
    }
    cover = new Cover (nodes);
}
```

Moving of the whole line means the synchronous movement of all the points.

```
    public override void Move (int dx, int dy)
    {
        Size size = new Size (dx, dy);
        for (int i = 0; i < pts .Length; i++)
        {
            pts [i] += size;
        }
    }
```

If you try to move the segmented line by the left button, then the reaction depends on the type of the caught node: each point (a circular node) can be moved individually, while any segment (a strip node) moves the whole line. The first `pts.Length` nodes of a cover are the circular nodes, so if the number of the caught node is less than `pts.Length`, then it is the movement of the point with the same number, as the node. If the number of the caught node is bigger, then it really does not matter, as it must be one of the segments (strip nodes), and any segment moves all of them synchronously, so the `Move()` method must be called.

```
public override bool MoveNode (int i, int dx, int dy, Point ptM, MouseButtons btn)
{
    bool bRet = false;
    if (btn == MouseButtons .Left)
    {
        if (i < pts .Length)
        {
            pts [i] += new Size (dx, dy);
        }
        else
        {
            Move (dx, dy);
        }
        bRet = true;
    }
```

Whenever I write the part of the `MoveNode()` method for the forward movement, I use the pair of parameters `(dx, dy)` for calculations. However, there is no law that you have to do the calculations with the help of only these parameters. There are situations, when the mouse position `(ptM)` looks better even for this part of the `MoveNode()`. The circular node over the point is small enough; you can ignore the possibility of the small difference between the point and the cursor at the starting moment of the movement and decide that the new position for the point is exactly at the mouse position. In such a case you can change one line of the code above to the

```
            pts [i] = ptM;
```

If you want to be absolutely correct, you can get and remember the shifts between the mouse cursor position and the exact point at the start of the movement `(xShift, yShift)`; these shifts can be used for exact calculations of the position for the point throughout the movement.

```
            pts [i] = new Point (ptM .X + xShift, ptM .Y + yShift);
```

The rotation starts, when the segmented line is pressed (caught) by the right mouse. At this moment the segmented line must be informed about the center of rotation through its `Anchor` property and the `StartRotation()` method must be called. As usual, this method gets one parameter – the mouse position.

```
    private void OnMouseDown (object sender, MouseEventArgs e)
    {
        if (mover .Catch (e .Location, e .Button))
        {
            GraphicalObject grobj = mover .CaughtSource;
            if (e .Button == MouseButtons .Right && grobj is SegmentedLine)
            {
                (grobj as SegmentedLine) .Anchor = center .Point;
                (grobj as SegmentedLine) .StartRotation (e .Location);
                bInRotation = true;
                Invalidate ();
            }
```

```
        }
    }
```

There is one significant change in the `SegmentedLine.StartRotation()` method in comparison with the previous example. Because of the independency of all the basic points of the segmented line, radius and compensation for each point must be calculated at the start of rotation and used throughout the rotation for their positioning. The segmented line is the classical example of an object that uses not a single compensation angle, but an array of such angles to organize a rotation.

When you press the segmented line to start the rotation, you have the center of rotation (`ptAnchor`), the mouse position (`ptMouse`), and the line, which is described by the array of points (`pts[]`). Draw a line (imaginary line) from the mouse cursor to the center of rotation; then draw another line (also an imaginary one) from one of the points to the center of rotation. You have this angle between the two lines and you have the distance of the point from the rotation center; these two values must be fixed for the whole duration of rotation. The two values are calculated for each point, so the `StartRotation()` method produces two arrays.

```
    public void StartRotation (Point ptMouse)
    {
        double angleMouse = -Math .Atan2 (ptMouse .Y – ptAnchor .Y,
                                          ptMouse .X – ptAnchor .X);
        for (int i = 0; i < pts .Length; i++)
        {
            radius [i] = Auxi_Geometry .Distance (ptAnchor, pts [i]);
            compensation [i] = Auxi_Common .LimitedRadian (angleMouse +
                 Math .Atan2 (pts [i] .Y - ptAnchor .Y, pts [i] .X - ptAnchor .X));
        }
    }
```

The position of each point throughout the rotation is calculated in that part of the `MoveNode()` method, which is associated with the movement by the right button. Calculations are identical for all the points, but use the two personal values for each of them. The current angle of the mouse is calculated prior to this. Knowing the mouse angle and the compensation angle for a particular point, you receive the real angle from the center of rotation to this point. Knowing the angle of the point and the distance from the rotation center, you get the position of the point. All the points change their positions, so the cover must be renewed by the `DefineCover()` method.

```
public override bool MoveNode (int i, int dx, int dy, Point ptM, MouseButtons btn)
{
    … …
    else if (btn == MouseButtons .Right)
    {
        double angleMouse = -Math.Atan2 (ptM.Y - ptAnchor.Y, ptM.X - ptAnchor.X);
        for (int j = 0; j < pts .Length; j++)
        {
            pts [j] = Auxi_Geometry .PointToPoint (ptAnchor,
                                   angleMouse - compensation [j], radius [j]);
        }
        DefineCover ();
        bRet = false;
    }
```

The rotation of the segmented line in my example can be started by pressing the line anywhere. However, there can be different opinions on this matter or different requests for its organization. For example, the rotation can be supposed to start only by the end points of segments or, on the contrary, only by the inner points of segments, but not by their ends. Such adjustments can be easily made by the small changes in the `OnMouseDown()` method and in the `MoveNode()` method.

Suppose that you want to allow rotation by the end points, but not by the segments. This means that rotation can be started only by the circular nodes; this adds one more check into the `OnMouseDown()` method.

```
    private void OnMouseDown (object sender, MouseEventArgs e)
    {
        if (mover .Catch (e .Location, e .Button))
        {
            GraphicalObject grobj = mover .CaughtSource;
```

```
        if (e .Button == MouseButtons .Right && grobj is SegmentedLine &&
            mover .CaughtNodeShape == NodeShape .Circle)
        {
```

The synchronous change of the `MoveNode()` method also includes an additional check, but this is the check of the nodes number, as only the first `pts.Length` nodes are circular.

```
public override bool MoveNode (int i, int dx, int dy, Point ptM, MouseButtons btn)
{
    … …
    else if (btn == MouseButtons .Right)
    {
        if (i < pts .Length)
        {
            double angleMouse = -Math.Atan2 (ptM.Y -ptAnchor.Y, ptM.X -ptAnchor.X);
```

These two possible changes are commented in the code of the **Form_SegmentedLine.cs**; if you turn these comments into the working code, then the segmented line will be rotated only by the end points of all the segments.

While the `SegmentedLine` object is in rotation, the circles for all the segment ends are painted. To determine, if these circles must be painted, there is an additional variable `bInRotation`, which changes its value to `true` inside the `OnMouseDown`() method, when the rotation starts, and returns back to `false` in the `OnMouseUp`() method. You can organize the same thing without any additional field, but with some help from mover. Comment all the places, where `bInRotation` is mentioned (four lines in the code) and change the checking in the `OnPaint`() method for such line (this line is left as a comment there).

```
    if (mover .Caught  &&  mover .CaughtSource is Polyline
                       &&  mover .Catcher == MouseButtons .Right)
```

There is one more object in the **Form_Polyline.cs** (**fig. 4.2**), which can be rotated, though you will not find any traces of it in the code. The text, which you can see on the screen, is an object of the `TextMR` class. Objects of this class can be used by themselves, as you see in this form; the class is also used as a base class for different types of comments. A `TextMR` object has a very simple cover, consisting of one polygonal node; rotation of such text goes around its middle point. Rotation can be started by pressing with the right button at any inner point of the text.

The `TextMR` objects and its derivatives are used so often that I automated the whole process of their forward movement and rotation; because of this automation no mentioning of these things are needed anywhere, but everything works smoothly, if an object is registered with a mover.

As I mentioned, objects of the `TextMR` class have a very primitive cover, consisting of one polygon (always rectangular) node; in the **Form_Polyline.cs** you can see it together with other covers. All classes, derived from the `TextMR`, have the same type of cover, but you will never see them. The cover for the `TextMR` class is left visible for the purpose of better explanation; the covers for this and all the derived classes do not need any visualization, because any text can be simply moved and rotated by any inner point. The explanation of a simple way to get rid of the visualization of cover is given further on in the *Appendix A. Visualization of covers*.

## *Rectangles*

File: **Form_Rectangles_WithRotation.cs**
Menu position: *Graphical objects – Basic elements – Rectangles – Move, resize, rotate*

Several classes of rectangles with the different types of resizing were demonstrated in the previous chapter, but none of them was involved in rotation. Here is one more class of rectangles – `RectangleRR`. The cover of this class is nearly identical to the `RectStandard` class with only some changes in the mouse cursor shape over part of the nodes.

When the objects of the `RectStandard` class, which you can see in the **Form_Rectangles_StandardCase.cs** (**figure 3.1**), have the nodes on their borders, then these nodes are used for resizing only along horizontal or vertical axis. For these directions the cursor in the form of `Cursors.SizeNS` or `Cursors.SizeWE` gives the best information about the possible movement. The `RectangleRR` class has exactly the same set of nodes, but the whole rectangle together with all its nodes can be turned on an arbitrary angle, so the use of those standard cursors is inappropriate. Instead, the `Cursors.Hand` cursor is used as a default parameter for circular and strip nodes. There is no need to mention any cursor in construction of the nodes for this cover, as the `Cursors.Hand` cursor is the default one both for circular and strip nodes.

Any object in rotation must have an `angle` parameter (field). Imagine a rectangle, which is positioned with its sides along the horizontal and vertical axes, like a blue rectangle at **figure 4.3**. Draw an imaginary horizontal line through the central point of rectangle; this line has an `angle = 0`. Now start rotating the rectangle; the line will rotate synchronously. There will be an angle between the horizontal line and the imaginary one; this is the current angle of this rectangle in rotation. If you start rotating the rectangle counterclockwise, then the angle is positive and growing. This is the way the angles are used in all the examples of this book and in all my applications. Angles are positive if measured anticlockwise from the given line; angles are negative if measured clockwise from the given line!*

In the code of the `RectangleRR` class you can see the variants for the resizable / non-resizable and rotatable / non-rotatable cases. These rectangles can be used in all of these variants; in other places objects of this class are used with an additional switch for each rectangle from one variant to another via the context menu, but in the **Form_Rectangles_WithRotation.cs** all the rectangles are resizable and rotatable all the time. There are two variants of covers for the `RectangleRR` class, as resizability requires more complex cover. For a non-resizable rectangle there is a single rectangular node to cover the whole area. Covers for resizable rectangles consist of nine nodes: four in the corners, four on the sides, and the same big node for the whole area (**figure 4.3**). The width of the strip nodes is set by default (6 pixels); the diameter of the circular nodes is enlarged to 12 pixels (`radiusCorner = 6`)

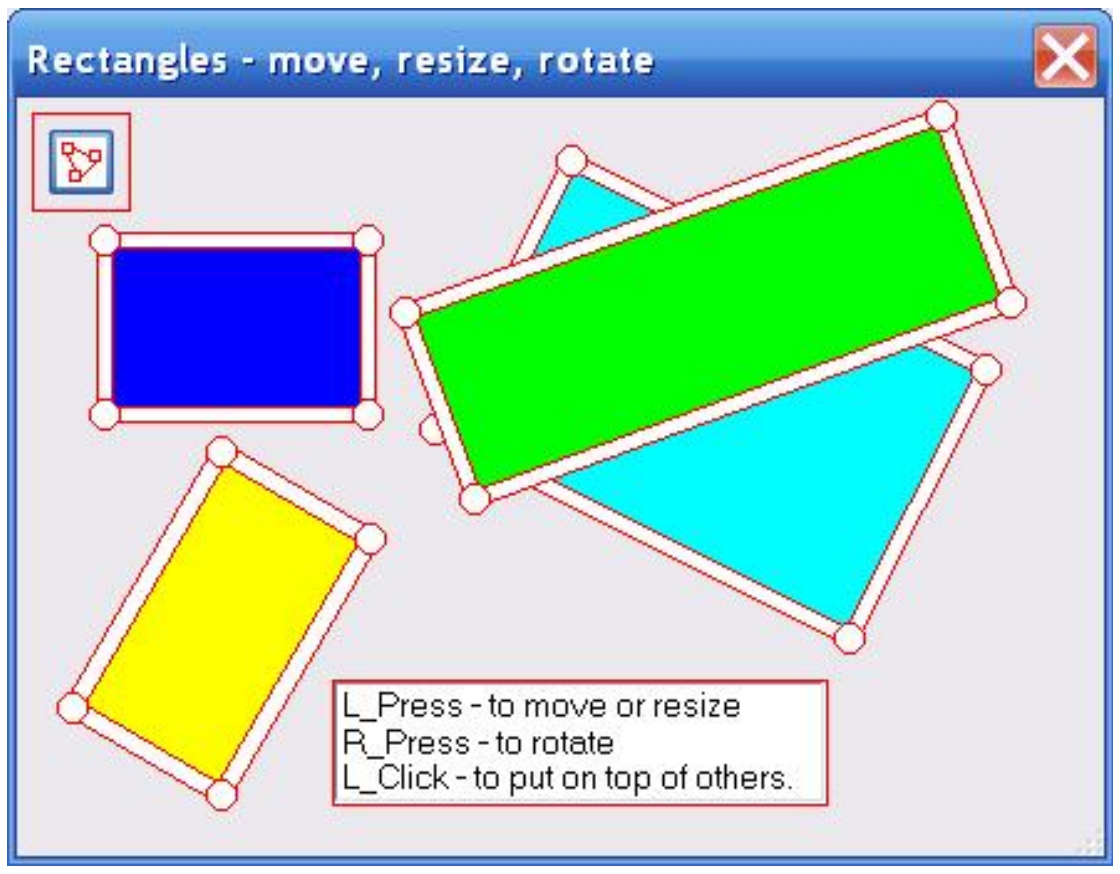

**Fig.4.3** Objects of the `RectangleRR` class

```
public override void DefineCover ()
{
    CoverNode [] nodes;
    if (bResize)
    {
        nodes = new CoverNode [9];
        for (int i = 0; i < 4; i++)
        {
            nodes [i] = new CoverNode (i, pts [i], radiusCorner);
        }
        for (int i = 0; i < 4; i++)
        {
            nodes [i + 4] = new CoverNode (i + 4, pts [i], pts [(i + 1) % 4]);
        }
        nodes [8] = new CoverNode (8, pts);
    }
    else
    {
        nodes = new CoverNode [] { new CoverNode (0, pts) };
    }
    cover = new Cover (nodes);
}
```

* I am sorry that I have to mention and even underline the statement, which is well known all round the world to everyone, who is older than 10 (in some countries even earlier), but when you have to deal with the Microsoft products like Visual Studio, you can expect to see a lot of interesting things. Many years ago, because it started from their very first version, the Microsoft decided to calculate the angles in the opposite way!!! I understand why it happened. The top left corner of the screen has the (0, 0) coordinate with the positive coordinates growing to the right and to the bottom of the screen. The Microsoft developers coded the standard formula for angle calculations, but forgot for a second (which turned into 25 years now) that for the whole world (outside the screen) the positive values on the Y axis are going not down, but up! Maybe the guys were sure that they would easily turn everyone into their faith, but certainly it did not work. The world continues to count the angles in the same way as it was doing for many many centuries; Microsoft is also very stubborn and never changed that old decision. The result? Throughout all these years millions of programmers around the world have to turn their brains inside out whenever they have to use some of the standard functions from the Microsoft's Math library.

Now you will be not surprised to see the strange sign in front of the calls to the `Math.Atan2()` method in my code. I have to remember all the time that this is one of their methods that return the result with the wrong sign and change it. The funniest side of this process is that only some methods return the wrong results, so you have to be even more alert.

Any RectangleRR object is initialized by a central point of the rectangle, two sizes, and an angle; this set of parameters allows to calculate the four corner points. The central point is used only at the moment of initialization; after the four corner points are calculated, only these points are used throughout the code. The central point will appear again, but especially for rotation between the **MouseDown** and **MouseUp** events.

```
public RectangleRR (PointF ptC, float w, float h, double angleDegree, Color color)
{
    w = Math .Max (minSide, Math .Abs (w));
    h = Math .Max (minSide, Math .Abs (h));
    angle = Auxi_Convert .DegreeToRadian (angleDegree);
    double radius = Math .Sqrt (w * w + h * h) / 2;
    double angle_plus = Math .Atan2 (h, w);
    pts = new PointF [4] {
           Auxi_Geometry .PointToPoint (ptC, angle + angle_plus, radius),
           Auxi_Geometry .PointToPoint (ptC, angle - angle_plus + Math.PI, radius),
           Auxi_Geometry .PointToPoint (ptC, angle + angle_plus + Math.PI, radius),
           Auxi_Geometry .PointToPoint (ptC, angle - angle_plus, radius) };
    brush = new SolidBrush (color);
}
```

The forward movement of any rectangle means the synchronous movement of its four corners.

```
        public override void Move (int dx, int dy)
        {
            SizeF size = new SizeF (dx, dy);
            for (int i = 0; i < 4; i++)
            {
                pts [i] += size;
            }
        }
```

The MoveNode() method in the part, which describes the resizing, reminds a bit the similar method for the RectStandard class, but is more complicated, because a rectangle may have an arbitrary angle. Here is the part of the MoveNode() method for one of the corner nodes.

```
public override bool MoveNode (int i, int dx, int dy, Point ptM, MouseButtons btn)
{
    bool bRet = false;
    double angleMouse;
    if (btn == MouseButtons .Left)
    {
        if (bResize)
        {
            double newD, newW, newH;
            PointF ptOpposite, ptA_oppositeside, ptB_oppositeside;
            switch (i)
            {
                case 0:
                    ptOpposite = pts [2];
                    angleMouse = Auxi_Geometry .Line_Angle (ptOpposite, ptM);
                    newD = Auxi_Geometry .Distance (ptOpposite, ptM);
                    newW = Math .Abs (newD * Math .Cos (angleMouse - angle));
                    newH = Math .Abs (newD * Math .Sin (angleMouse - angle));
                    if (newW >= minSide && newH >= minSide &&
                        RectangleNotInverted (i, ptM))
                    {
                        pts = new PointF [4] { ptM,
                          Auxi_Geometry.PointToPoint (ptM, angle + Math.PI, newW),
                          ptOpposite,
                          Auxi_Geometry.PointToPoint (ptM, angle-Math.PI/2, newH)};
                        DefineCover ();
                        bRet = true;
```

```
                    }
                    break;
```

It looks like this piece of code needs several words of explanation.

1. A small possible difference between the center of the circular node and the exact point, at which the node is pressed and caught, is ignored; the cursor point `ptM` is supposed to become the new point for the caught corner. This is the corner with `i = 0`. The opposite corner of rectangle has `i = 2`, so `ptOpposite` gets the actual point of that opposite corner.
2. `angleMouse` is the angle from the central point of rectangle to the current mouse position. The central point is on the line between two opposite corners, so I can use these two points to calculate the needed angle. The distance between two opposite corners gives the diagonal of rectangle. Both calculations use the methods from the **MoveGraphLibrary.dll**; all the methods are described in the **MoveGraphLibrary_Classes.doc** (see section *Programs and Documents* at the end of the book).
3. `(angleMouse – angle)` is the angle between the line from the central point to the caught corner (mouse position) and the angle of rectangle. Knowing this difference between two angles and the diagonal of rectangle, it is easy to calculate the proposed width (`newW`) and height (`newH`) of rectangle.
4. Check the proposed new sizes against the allowed minimum sides for rectangle. If none of the proposed sides is too small, then calculate the new points for the corners. One of these corners, which is opposite to the mouse, is not moving at all and retains its position (`ptOpposite`); another one is going to be at the mouse position (`ptM`), but two remaining corners must be calculated, which is easy to do. There is one more condition inside the checking; this one prevents the rectangle from being turned inside out.
5. If all the checks allow to move the corner, then the points of all four corners are saved and the cover is defined.

Calculations for moving the sides are slightly different, but only in details.

The previous explanation was about the resizing of an arbitrary turned rectangle; now it is time to look into the rotation process, which starts by the right button at any point of rectangle. When an object of the `RectangleRR` class is caught with the right mouse button, then the `StartRotation()` method of the caught object is called.

```
private void OnMouseDown (object sender, MouseEventArgs e)
{
    ptMouse_Down = e .Location;
    if (mover .Catch (e .Location, e .Button))
    {
        if (e .Button == MouseButtons .Right &&
            mover .CaughtSource is RectangleRR)
        {
            (mover .CaughtSource as RectangleRR) .StartRotation (e .Location);
        }
    }
}
```

The `StartRotation()` method gets the mouse position, as a parameter, and calculates three things.

1. The center of rotation, which is the central point of the pressed rectangle.
2. The angle from the rotation center to the mouse position.
3. The compensation angle, which is the difference between this angle to the mouse and the angle of an object.

```
public void StartRotation (Point ptMouse)
{
    center = Center;
    double angleMouse = Auxi_Geometry .Line_Angle (center, ptMouse);
    compensation = Auxi_Common .LimitedRadian (angleMouse - angle);
}
```

This compensation angle is not going to change throughout the rotation and is used to calculate the corner points in that part of the `MoveNode()` method, which deals with rotation. In the `RectangleRR.MoveNode()` method you can see not only the check of the pressed button, but also an additional check with the `bRotate` value. I have already mentioned that the same rectangles can be switched between being rotatable and not being rotatable via the commands of the context menu. This possibility is not used in the **Form_Rcetangles_WithRotation.cs;** here all the rectangles are always rotatable.

```
public override bool MoveNode (int i, int dx, int dy, Point ptM, MouseButtons btn)
{
    … …
    if (btn == MouseButtons .Left)
    {
        … …
    }
    else if (btn == MouseButtons .Right && bRotate)
    {
        angleMouse = Auxi_Geometry .Line_Angle (center, ptM);
        angle = angleMouse - compensation;
        double radius = Radius;
        double w = Auxi_Geometry .Distance (pts [0], pts [1]);
        double h = Auxi_Geometry .Distance (pts [0], pts [3]);
        double angle_plus = Math .Atan2 (h, w);
        pts = new PointF [4] {
    Auxi_Geometry .PointToPoint (center, angle + angle_plus, radius),
    Auxi_Geometry .PointToPoint (center, angle - angle_plus + Math .PI, radius),
    Auxi_Geometry .PointToPoint (center, angle + angle_plus + Math .PI, radius),
    Auxi_Geometry .PointToPoint (center, angle - angle_plus, radius) };
        DefineCover ();
        bRet = true;
```

Again, some comments on this code.

1. `angleMouse` is the angle from the central point of rectangle, which is also the rotation center, to the current mouse position.
2. Knowing the `compensation` which was calculated at the starting of rotation, it is easy to achieve the angle of rectangle.
3. `radius` is the half of the diagonal.
4. The `pts[]` array contains the corner points. Distance between `pts[0]` and `pts[1]` is the width of rectangle; distance between `pts[0]` and `pts[3]` is its height.
5. `anglePlus` is the angle between the bottom of rectangle and its diagonal.
6. Knowing the central point, half of the diagonal (`radius`), and two sizes, it is easy to calculate the points for all four corners.
7. All the basic points (four corners) change their location, so the cover must be redefined.

Certainly, the dimensions of rectangle are not going to change throughout the rotation, so they can be also calculated inside the `StartRotation()` method and saved in some additional fields.

<u>Short conclusion for the *Rotation* chapter.</u>

I have shown three examples with the rotation of different objects. Many more different objects will be involved in rotation in the further examples, but the rules are always the same.

- To be involved in rotation, an object has to have an `angle` field. If an object has several independently moving points, then each of them has to have its personal angle.
- Rotation is started, when an object is caught by the right button. At this moment the `StartRotation()` method of an object must be called; this method gets the mouse position as a parameter.
- The `StartRotation()` method calculates the difference between the angle from the rotation center to the mouse and the angle of an object. This difference is called `compensation` and is not going to change throughout the rotation. If there are several independent points in the object, then for each of them the compensation is calculated in the similar way.

- The position of an object throughout the rotation is calculated in that part of the `MoveNode()` method, which is associated with the right button. For any current mouse position, the compensation allows to calculate the angle of the basic point. If there is a set of such points with the personal compensation angles, then all of these points are calculated in exactly the same way from the current mouse angle.

Reminder. I hope that now you can easily understand, what was wrong with the rotation of the solitary lines in the **Form_Lines_Solitary.cs** (**figure 2.2**).

# Texts

Texts play a very important role in the visualization of information. Depending on the request, texts can be involved in individual movements, they can influence each others movements, or they can be the part of some complex objects. This chapter is about the individual movements of texts and some aspects of their related movements.

## *TextM – the simplest class of movable texts*

At least one third of this book is about moving and resizing of the abstract figures. It started with the lines, the next were rectangles. After this chapter there will be polygons and further on to more and more interesting objects. The idea of such an order of explanation is obvious: you might have in your programs the objects of absolutely different shapes; you need to know, how to make them movable and resizable. On the other hand, a lot of people prefer to see the abstract ideas applied to the real objects, with which they need to deal in their programs every day. I cannot think out more often used objects in our programs than texts.

Let us not discuss those texts (better to say *documents*), which come out as a result of work of some text editor. I am writing here about those short (very often it is a single word) or middle size texts that are used as comments or explanations to other objects in very many applications. It can be an explanation to a parameter, which user has to type in; it can be an explanation or a title to some `List`; it can be a comment on a plot; it can be some short information about the features of an object or a form.

In the last role the texts started to appear in the examples of this book even before any explanation: you can see it in the *Introduction* at **figure I.1**. This figure shows the view of the first (main) form, when the Demo application is just started, so the movable text is among the very first objects that I demonstrate. That text, which you see in the **Form_Main.cs**, is an object of the `TextM` class. The information looks like a colored panel with a text on it. The panel can be moved around the screen by any point, but it is not resizable and you cannot rotate it. I do not think that all the texts must be rotated, though a lot of them throughout the Demo application are rotatable. But if I need to show some information in a simple standard way, when the text is painted horizontally and easy to read, then I usually design it as a `TextM` object. As the same type of information can be seen in many other forms of the Demo application, let us look into the design of this class.

The full set of parameters for the `TextM` constructor includes location (top left corner), text, font, color of text, two Boolean parameters, which regulate the drawing of the frame and background area, and the color for this area.

```
public TextM (Form form, Point ptLeftTop, string txt,
          Font fnt, Color clr, bool show_frame, bool show_background, Color back)
```

Of all these parameters only the first three are mandatory; others can be declared in different variations or omitted. If any of the last five parameters are not specified, then they receive the default values: font and color of the text are received from the form (`form.Font` and `form.ForeColor`), the frame is shown; the background is also painted, using the `SystemColors.Control` color for it. This background color is usually the same as the form has, but the area is not transparent and there is a frame around it, so the area of thus shown text can be easily distinguished from the surrounding form.

The area of text is a rectangle, calculated with one of the methods from the **MoveGraphLibrary.dll**. Text can consist of any number of lines; the width of the resulting rectangle is equal to the longest line.

```
protected void CalcSizes (Form form, string txt, Font fnt)
{
    SizeF sizef = Auxi_Geometry .MeasureString (form, txt, fnt);
    nW = Convert .ToInt32 (sizef .Width) + wAdd;
    nH = Convert .ToInt32 (sizef .Height) + hAdd;
}
```

The text is going to be moved by any inner point. There is no resizing, so the cover of such object can be constructed of a single rectangular node, covering exactly the needed area. One of the standard `Cover` constructors works perfectly in this situation.

```
public override void DefineCover ()
{
    cover = new Cover (Area, Resizing .None);
}
```

It is impossible to think out an object, which would have simpler methods for moving than this `TextM` class. There is only one basic point for an object – the top left corner of rectangle; this point must be changed in the `Move()` method.

```
public override void Move (int dx, int dy)
{
    ptLT += new Size (dx, dy);
}
```

There is only one node in the cover. The object can be moved by the left button, so in this case the `MoveNode()` method has to call the `Move()` method and that is all.

```
public override bool MoveNode (int i, int dx, int dy, Point ptM, MouseButtons btn)
{
    bool bRet = false;
    if (btn == MouseButtons .Left)
    {
        Move (dx, dy);
        bRet = true;
    }
    return (bRet);
}
```

The cover is designed, when a movable text is initialized. Are their situations when the cover of the existing text has to be changed? Yes, the single node must always cover the whole area of the text and these two areas must be equal; whenever the area of text changes, the `DefineCover()` method must be called. There are three causes for such an action:

- Change of the text.
- Change of the font.
- Change of location. This is not the move of the text by a mouse, but applying the new value through the `Location` property.

All these changes can happen after using an appropriate property. For example, here is the way to change the font.

```
public Font Font
{
    get { return (font); }
    set
    {
        m_font = value;
        CalcSizes (form, text, font);
        DefineCover ();
    }
}
```

The `TextM` class has all the needed properties and methods to use this class in many different situations; the class is described in the **MoveGraphLibrary_Classes.doc** (see information in the section *Programs and Documents*). In the accompanying Demo application the class is used to show a short information on the available commands in several of the forms. Further on I will demonstrate in a couple of forms and discuss the type of similar information that can be opened and closed at any moment by a user. That would be the `ClosableInfo` class, derived from the `TextM` class.

## Individual movements

File: **Form_TextsIndividual.cs**
Menu position: *Graphical objects – Texts – Individual movements*

In the previous chapter in the example of the segmented line (**Form_SegmentedLine.cs**, **figure 4.2**), I have demonstrated the use of the movable / rotatable text of the `TextMR` class. This class is included into the **MoveGraphLibrary.dll** and nearly all the work with this class is automated. This was done, because the movable and rotatable texts are used very often; the automation allows to get rid of the same lines of code in many places. On the other hand, it would be interesting to look into some details of designing such texts, so here is the **Form_TextsIndividual.cs** (**figure 5.1**), which uses not any predetermined class from the **MoveGraphLibrary.dll**, but the `MoveableText` class, which is developed in the same file.

The design of cover for any object is determined by the set of requirements to this class.

- The text may have multiple lines.
- The text can be moved and rotated by any point. The sensitive area is not combined of the rectangles for each line, but is a single rectangle with the width determined by the longest line.
- The size of this solitary rectangle is determined only by the used font.
- An object is not resizable by the mouse. The size of an object can change only as the result of using another font or changing the text itself.
- An object is always rotated around the central point of the sensitive rectangle.

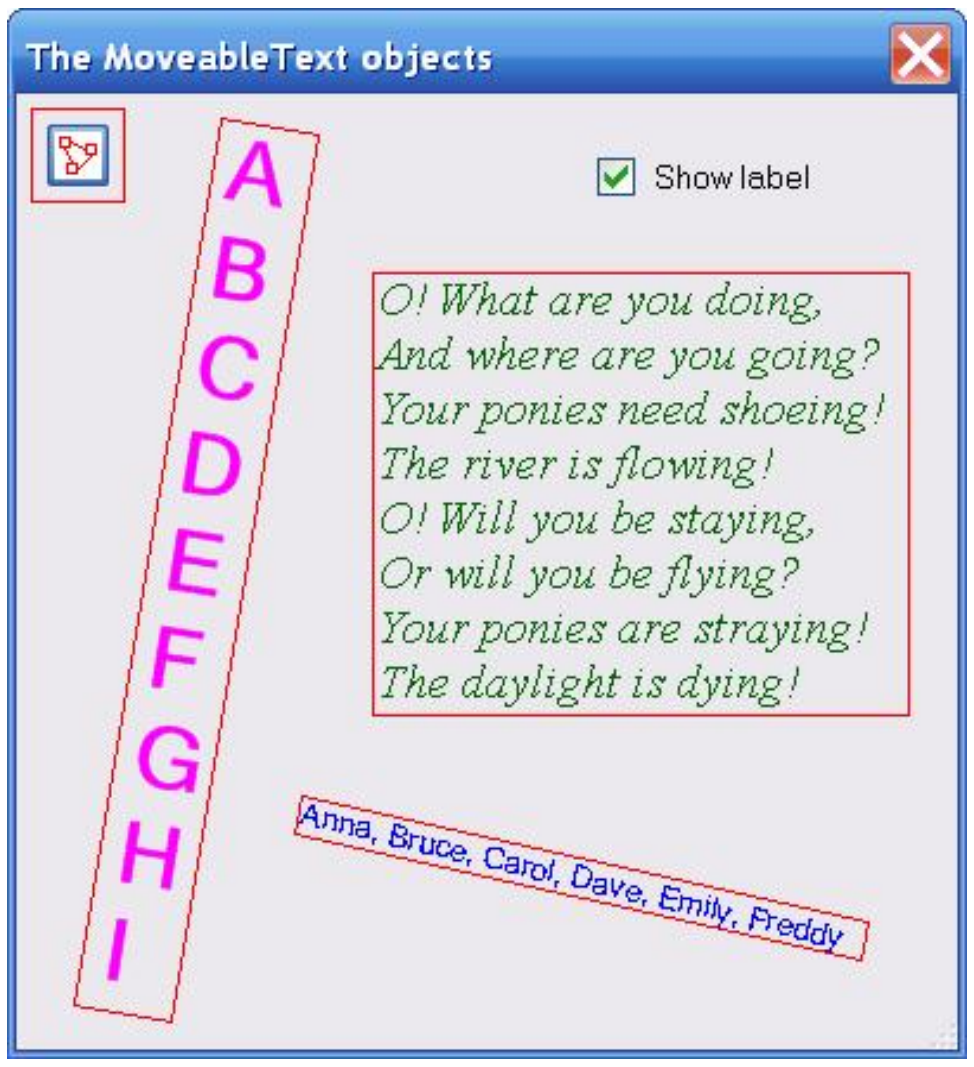

**Fig.5.1** Texts of the `MoveableText`

This is not the only possible set of requirements for design of movable texts. If you change any of these requirements, you will probably find that you would have to change the cover design. But in this example let us work with the proposed set of requirements.

The movable text can be initiated with a set of parameters, of which only the first three are mandatory; others can get the default values.

```
public MoveableText (Form form,                // the form, in which this text is used
                     Point ptAnchor,           // the central point
                     string txt, …)            // the text itself
{
    formParent = form;
    ptMiddle = ptAnch;
    text = CheckedText (txt);
    SizeF sizef = Auxi_Geometry .MeasureString (form, text, fnt);
    nW = Convert .ToInt32 (sizef .Width);
    nH = Convert .ToInt32 (sizef .Height);
    font = fnt;
    angle = Auxi_Common .LimitedRadian (Auxi_Convert .DegreeToRadian (ang_Deg));
```

The sizes of the text are calculated on the basis of the used font. By using these sizes and the angle of text, the corners of the rectangular area are calculated with the help of the `Auxi_Geometry.TextCorners()` method. The cover always consists of a single node, but the behaviour of this node depends on either the text is movable or not. It looks a bit strange first to develop a whole theory of turning any screen object into movable / resizable and then to think about turning such movable objects into unmovable and to design a special addition for this purpose, but… I will write much more about this situation later, while describing the design of user-driven applications. The reason for the discussion of turning movable objects into unmovable is the necessity of such actions from time to time. The main principle of the user-driven applications is that user has the total control of any application and decides at any moment about the main features of any object. It is possible that at some moment user would prefer to turn otherwise movable objects into unmovable; for this situation an object must be ready to switch its behaviour.

The turn of any object into unmovable does not mean that it is simply excluded from the mover's queue and becomes invisible for the mover. No, it is still sensitive, and with the mover's help it can be found, for example, to change its color, font, or do something else, but it becomes unmovable. Whether it is movable or unmovable, the cover of the `MoveableText` object consists of a single node; the difference is only in the behaviour of this node

- For an unmovable text, I use the `Behaviour.Frozen` value for behaviour and `Cursors.Default` as the shape of cursor.
- For a movable text, I do not specify these parameters, so they get the default values for any polygonal node: `Behaviour.Moveable` and `Cursors.SizeAll`.

```
public override void DefineCover ()
{
   Point [] pts = Auxi_Geometry.TextCorners (nW, nH, angle, ptMiddle, TextBasis.M);
   CoverNode [] nodes = new CoverNode [1];
```

```
    if (Movable)
    {
        nodes [0] = new CoverNode (0, pts);
    }
    else
    {
        nodes [0] = new CoverNode (0, pts, Behaviour .Frozen, Cursors .Default);
    }
    cover = new Cover (nodes);
}
```

The position of the text is determined by a single point, so for the forward moving it is enough to move this point for a specified number of pixels.

```
        public override void Move (int dx, int dy)
        {
            ptMiddle += new Size (dx, dy);
        }
```

The `MoveNode()` method is also very simple, as there is only one node in the cover and no restrictions on any movement at all. For the forward movement, the `Move()` method is called from inside the `MoveNode()` method.

```
public override bool MoveNode (int i, int dx, int dy, Point ptM, MouseButtons btn)
{
    bool bRet = false;
    if (btn == MouseButtons .Left)
    {
        Move (dx, dy);
        bRet = true;
    }
```

In the **Form_TextsIndividual.cs** (**figure 5.1**), you can see three different objects of the `MoveableText` class. The interesting things with all of them begin in the `OnMouseDown()` method of the form.

```
    private void OnMouseDown (object sender, MouseEventArgs e)
    {
        if (mover .Catch (e .Location, e .Button))
        {
            GraphicalObject grobj = mover .CaughtSource;
            if (e .Button == MouseButtons .Right && grobj is MoveableText)
            {
                MoveableText mt = grobj as MoveableText;
                mt .StartRotation (e .Location);
                mt .AngleLabel (checkboxLabel .Checked);
            }
        }
    }
```

When any `MoveableText` object is caught by the right button, then it means that the rotation is started. At this moment two methods of this class are called. The first one is called `StartRotation()`; it calculates the compensation angle between the angle to the point where the mouse have caught the object and the angle of object. The use of this compensation was already explained earlier in the chapter *Rotation*. If the parameter of the second method – `AngleLabel()` – is set to true, then this method organizes a special label to show the changing angle of the text throughout the rotation.

Any textual information can be shown either as graphical object or as a text in a control. When I know that there are no controls in the form, then I prefer to show the needed information as a graphical object. But any controls are always shown on top of all the graphical objects, so, if there are any controls in the form, they can block this graphical information from view. The `MoveableText` class is designed to be used in any situation with any combination of objects in view, so it is better to show an auxiliary information related to the rotation as a `Label`.

Texts themselves can be positioned at any place on the screen; while the text is in rotation, the information about its angle must be shown somewhere very close to this text; in such way it would be easier to look simultaneously at the text and the auxiliary information. The `AngleLabel()` method calculates the appropriate position for a label

`(cxLabel, cyLabel)`, creates the `Label` with the needed information and puts it at the top of the set of controls of the form; thus the new label appears above anything else. Position of the text in rotation is described by its central point; the label with the information about the angle appears at the level of this central point; whether it is to the left or to the right of the text depends on which side of the form (left or right) the text itself is shown.

```
public void AngleLabel (bool bShowLabel)
{
    if (bShowLabel)
    {
        … …
        labelAngle = Auxi_Common .CreateInfoLabel (new Point (cxLabel, cyLabel),…
        formParent .Controls .Add (labelAngle);
        labelAngle .BringToFront ();
    }
}
```

The rotation of the text is described in the second half of its `MoveNode()` method; this technique was discussed in the chapter *Rotation*. .The compensation angle was calculated in the `StartRotation()` method at the beginning of rotation; now the current mouse position gives the angle to the mouse and together with the compensation angle it gives the current angle of the text.

```
public override bool MoveNode (int i, int dx, int dy, Point ptM, MouseButtons btn)
{
    … …
    else if (btn == MouseButtons .Right)
    {
        double angleMouse = -Math.Atan2 (ptM.Y - ptMiddle.Y, ptM.X - ptMiddle.X);
        Angle = angleMouse - compensation;
        bRet = true;
    }
    return (bRet);
}
```

The `MoveableText.Angle` property sets the new angle of the text. The new angle means that the rectangular area of the text has turned, so the cover has to be updated by calling the `DefineCover()` method.

```
        public double Angle
        {
            get { return (angle); }
            set
            {
                angle = value;
                DefineCover ();
                UpdateLabelAngle ();
            }
        }
```

The angle is set to the new value; the cover is defined according to this new angle. Then the method to update the label is called; this method will work only if the label was created before.

```
private void UpdateLabelAngle ()
{
    if (labelAngle != null)
    {
        labelAngle .Text =
           Convert .ToInt32 (Auxi_Common.LimitedDegree (Angle_Degree)).ToString ();
        labelAngle .Update ();
    }
}
```

When any `MoveableText` object is released by mover, then the `Release()` method of this object must be called.

```
private void OnMouseUp (object sender, MouseEventArgs e)
{
    if (mover .Release ())
    {
        if (mover .WasCaughtSource is MoveableText)
        {
            (mover .WasCaughtSource as MoveableText) .Release ();
        }
    }
}
```

This `MoveableText.Release()` method is doing anything at all only if the label with the information was created before; in such case the label is deleted from the set of controls. Thus the label exists, if required, only between the **`MouseDown`** and **`MouseUp`** events, and disappears after it without any trace.

```
public void Release ()
{
    if (labelAngle != null)
    {
        formParent .Controls .Remove (labelAngle);
        labelAngle = null;
    }
}
```

The `MoveableText` class demonstrates the design of the movable and rotatable texts; it also includes the possibility of showing the currently changing angle throughout the rotation of a text. The class is nearly an exact copy of the `TextMR` class, which can be found in the **MoveGraphLibrary.dll**. This `TextMR` class is widely used in all my programs, but you will never see any calls to its `StartRotation()` method or any mentioning of the labels. At the same time all those numerous texts can be moved and rotated, and the labels are shown, if required. The explanation of work without mentioning is simple: I do not want to repeat again and again all the calls to these methods in every form, so the work with the `TextMR` class is automated inside the `Mover` class. For this particular class of texts, mover knows how to create a label, how to change the information in this label, and when to delete the label. Everything is organized exactly in the same way, as it is shown for the `MoveableText` class, but everything is hidden.

Several lines ahead I mentioned that those labels with an angle "*are shown, if required*". Where and how? The labels have to be shown only starting from the **`MouseDown`** event and only in case of the text caught by the mover. Mover is responsible for organizing a label and dealing with it, so it would be natural to inform the mover about the need for a label at the moment, when a text can be caught. This is done by using the special form of the `Mover.Catch()` method, which has an additional parameter.

```
mover .Catch (Point ptMouse, MouseButtons catcher, bool bShowAngle)
```

This version of the method is often used for the forms with the textual comments. The majority of these comments are derived from the `TextMR` class; the moving and rotation of all these objects are used so often that they are automated and do not need to be mentioned anywhere in the code. However, throughout the rotation of any comment, its angle can be shown nearby, and this is regulated by the third parameter of the method. Usually, this parameter is controlled by the user and can be changed at any moment, for example, via some context menu.

Mover can easily check the type of the caught object. If it is a text and the parameter of the `Catch()` method is set to `true`, then mover organizes a label, shows there the changing angle throughout the rotation, and deletes the label, when the text is released. It is all mover's responsibility; everything works correctly; you do not need to think about it. Certainly, if for any reason you are not satisfied with the mover's demonstration of the angle, then do not use this version of the `Catch()` method and organize the showing of angle yourself. I have demonstrated the technique on the example of the `MoveableText` class; you can do it in a similar way or any other.

The `TextMR` class is also used as a base class for many different types of comments for complex objects, included into the **MoveGraphLibrary.dll** and used for different types of plotting and construction of different applications. You will see the classes, derived from the `TextMR` class, in nearly all the further chapters of the book. All those classes will be involved in forward movement and rotation without any mentioning in the code, because they all inherit their nice features from the `TextMR` class.

## *Related movements*

File: **Form_SpottedTexts.cs**
Menu position: *Graphical objects – Texts – Related movements*

Texts in applications are often involved not only in individual movements, but also in different related movements with other objects. Later on I will write about the plotting for scientific / engineering programs and also about bar charts. These types of plotting have scales with the sets of textual or numerical information positioned along those scales. For changing the parameters of scales, some tuning forms are used; these forms include a sample of text. While this sample is moved or rotated, all the textual elements along the associated scale move or rotate in the similar way, so a move of a single element (sample) allows to position the textual information of the scale quickly and identically. Let us design a class of textual information and see, how this tuning by a sample can work.

There are 13 movable texts in the **Form_SpottedTexts.cs** (**figure 5.2**). Though all these objects belong to the same `SpottedText` class, only one of them is always shown as a really spotted text; the spots on other objects (*months*) can be switched ON / OFF with the help of a special check box. This form demonstrates a very interesting situation, when the objects of the same class behave in different ways and even can be moved in different ways.

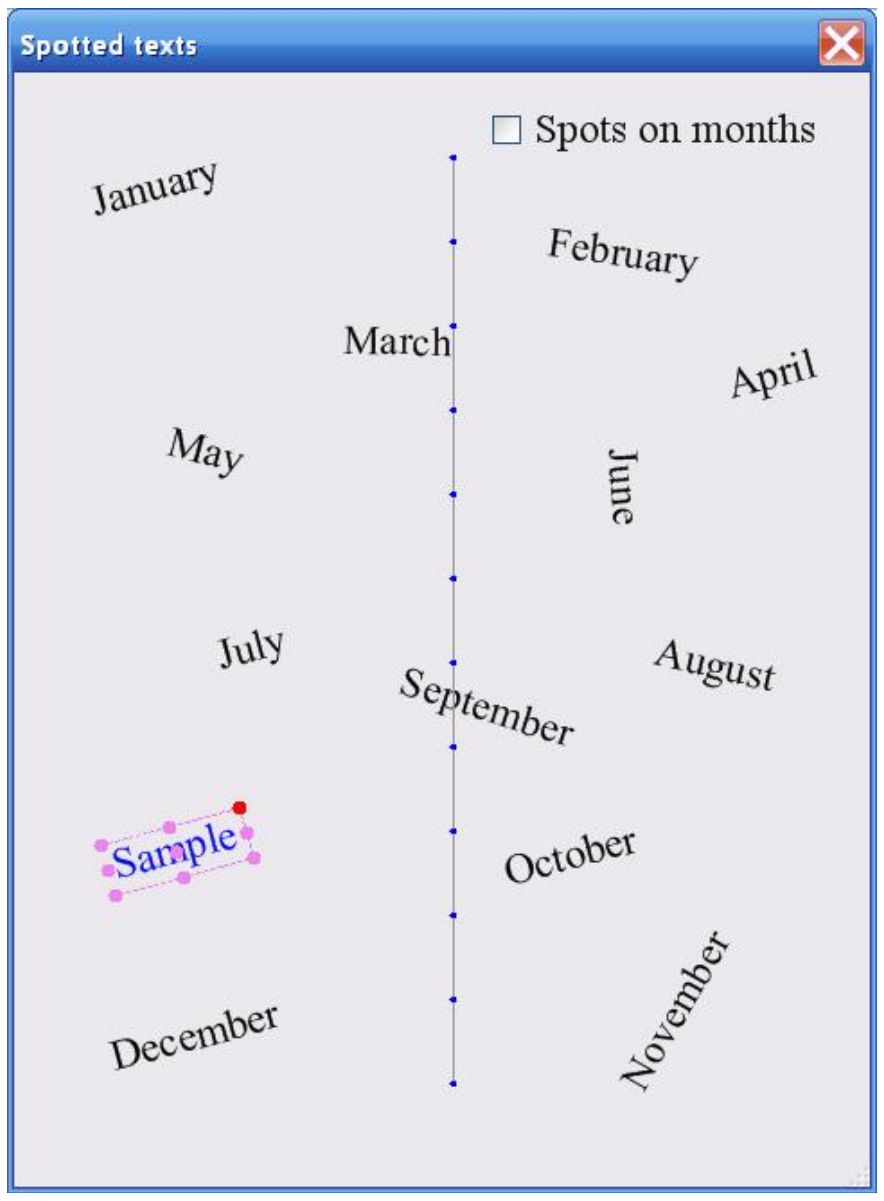

**Fig.5.2** The `SpottedText` objects

Before starting to play with these texts, several words about their geometry and special spots of their area. Any text occupies some rectangular area on the screen; the sizes of this rectangle are determined by the text itself and the used font. Even for a text, consisting of several lines, I use a single rectangular area with the width, determined by the longest line. Consider some horizontally written text, which occupies a rectangular area. This rectangle has four corners, four points in the middle of the sides, and the central point; the total number of specially mentioned points is nine. The `TextBasis` enumeration includes exactly nine members to work with these special points. I hope that eight of their names are obvious; the `M` member is associated with the **m**iddle.

```
public enum TextBasis { NW, N, NE, W, M, E, SW, S, SE };
```

The **MoveGraphLibrary.dll** includes several methods to draw the texts and get the coordinates of the texts; these methods use this `TextBasis` enumeration. For example, one of the methods allows to draw the text, anchoring it on the specified point by any member of this enumeration. To draw the text "January", which you can see in the top left corner of **figure 5.2**, you can use a line of code, shown below; the specified point will be in the bottom left corner of the letter **J**. Some of the parameters specify the font, angle, and color. The last two parameters specify the point in the form and the special point of the text, which must be associated with this real point.

```
Auxi_Drawing .DrawText (grfx, "January", Font, 0.3, ForeColor,
                        new Point (40, 80), TextBasis .SW);
```

Another very useful method can be seen in the constructor of the `SpottedText` class; it returns the set of those special points for a text with the known sizes, angle, position, and the special point of the text, associated with this position.

```
public SpottedText (Form form, Point ptAnch, TextBasis basicpoint, string txt,
                    Font fnt, double ang_Deg, Color clr)
{
    formParent = form;
    text = CheckedText (txt);
    sizef = Auxi_Geometry .MeasureString (form, text, fnt);
    basis = basicpoint;
    pts = Auxi_Geometry .TextGeometry_Degree (sizef, ang_Deg, ptAnch, basis);
    font = fnt;
```

The array of points, returned by the `Auxi_Geometry.TextGeometry_Degree()` method is filled by the points according to the `TextBasis` enumeration, so the first three points are from the top line, then the three points from the middle line, and the last three are from the bottom of the text (see the word *Sample* with the dots at **figure 5.2**). These nine points are used for the cover design in the `SpottedText` class.

Cover of any SpottedText object consists of 10 nodes: nine of them are the small circular nodes in the special places of the text's area; the last node is a polygon, covering the whole area of text. Those nine points are calculated by one of the methods from the **MoveGraphLibrary.dll** and include four corners, four points in the middle of each side, and the central point of the rectangular area. The previous explanation about the TextBasis enumeration and the array of points together with the code of the DefineCover() method makes it obvious that the first nine nodes of the cover go in parallel with the members of the TextBasis enumeration, which makes the work with the SpottedText objects easier, as the number of a node can be cast into the value of enumeration and wise verse.

```
public override void DefineCover ()
{
    CoverNode [] nodes = new CoverNode [10];
    for (int i = 0; i < 9; i++)
    {
        nodes [i] = new CoverNode (i, pts [i], nRadius);
    }
    nodes [9] = new CoverNode (9, Frame);
    cover = new Cover (nodes);
}
```

I have to admit that 10 nodes are used in this cover not because they provide different types of movement, but because they simplify my explanation and visualization of some features. The same movement can be organized on a cover of a single rectangular node, but then I will have to organize the similar set of nine points and keep track of them; mover takes this burden from me. As all 10 nodes are used for the same kind of movement, then for forward movement there is no need in checking the number of the caught node; the MoveNode() method is very simple.

```
public override bool MoveNode (int i, int dx, int dy, Point ptM, MouseButtons btn)
{
    bool bRet = false;
    if (btn == MouseButtons .Left)
    {
        Move (dx, dy);
    }
    else if (btn == MouseButtons .Right && i != (int) basis)
    {
        double angleMouse =
                 -Math .Atan2 (ptM .Y - AnchorPoint .Y, ptM .X - AnchorPoint .X);
        Angle = angleMouse - compensation;
        bRet = true;
    }
```

The rotation of these texts is organized in nearly a standard way, but with one small addition, or better to call it exception. Each text has nine special points; each one of them can be used as the center of rotation. The currently selected center is painted in different color (red); others are violet. The rotation can be started, when the right button is pressed at any point, except the area of the node that covers the rotation center; for this purpose there is an additional comparison of the caught node with the number of the current center point in the second half of the MoveNode() method.

```
else if (btn == MouseButtons .Right  &&  i != (int) basis)
```

All 13 texts in the **Form_SpottedTexts.cs** – 12 *months* plus a *sample* - belong to the SpottedText class and are supposed to work in the same way, but if you try to move them forward, you will find something interesting: the names of the months can be moved by any inner point, but the *Sample* can be moved only by pressing one of the colored spots. To organize such a strange difference in their behaviour, I added several lines of code into the OnMouseDown() method. First, on catching any SpottedText object, I check (via the unique identification number), which of the texts was caught.

```
private void OnMouseDown (object sender, MouseEventArgs e)
{
    if (mover .Catch (e .Location, e .Button))
    {
        GraphicalObject grobj = mover .CaughtSource;
        if (e .Button == MouseButtons .Left)
        {
            if (grobj is SpottedText)
            {
```

```
int iNode = mover .CaughtNode;
if (grobj .ID == sample .ID)
{
    … …
else
{
    if (iNode < Enum .GetNames (typeof (TextBasis)) .Length)
    {
        (grobj as SpottedText) .TextBasis = (TextBasis) iNode;
        Invalidate ();
    }
}
```

If it is not the *sample* (so, it is one of the *months*), then an object will be simply moved according to the mentioned `MoveNode()` method. The only thing, which is done here, is the setting of the new rotation center, if any small node was pressed. To be any of the small nodes, the number of the caught node must be less than the number of members in the `TextBasis` enumeration.

```
if (iNode < Enum .GetNames (typeof (TextBasis)) .Length)
{
    (grobj as SpottedText) .TextBasis = (TextBasis) iNode;
    Invalidate ();
}
```

If the *sample* was caught, then the reaction depends on the node, by which it was caught:

- If it is one of the circular nodes (colored spots), then all the *months* are lined at their initial places and their lining is defined by the pressed spot of the *Sample* (**figure 5.3**).
- If the pressed point is not in any of the colored spots, then not only the months ignore any command and continue to stay freely wherever they want, but the *sample* itself is released from mover and is not going to move anywhere. Throughout the whole Demo application, it is the only one of two examples of using `mover.Release()` method anywhere outside the `OnMouseUp()`, but I have never said that it can be used only inside the **MouseUp** event. You can decide yourself, which events to use for catching and releasing objects. I prefer to use the `MouseDown` and `MouseUp` events, but there is no strict law about it.

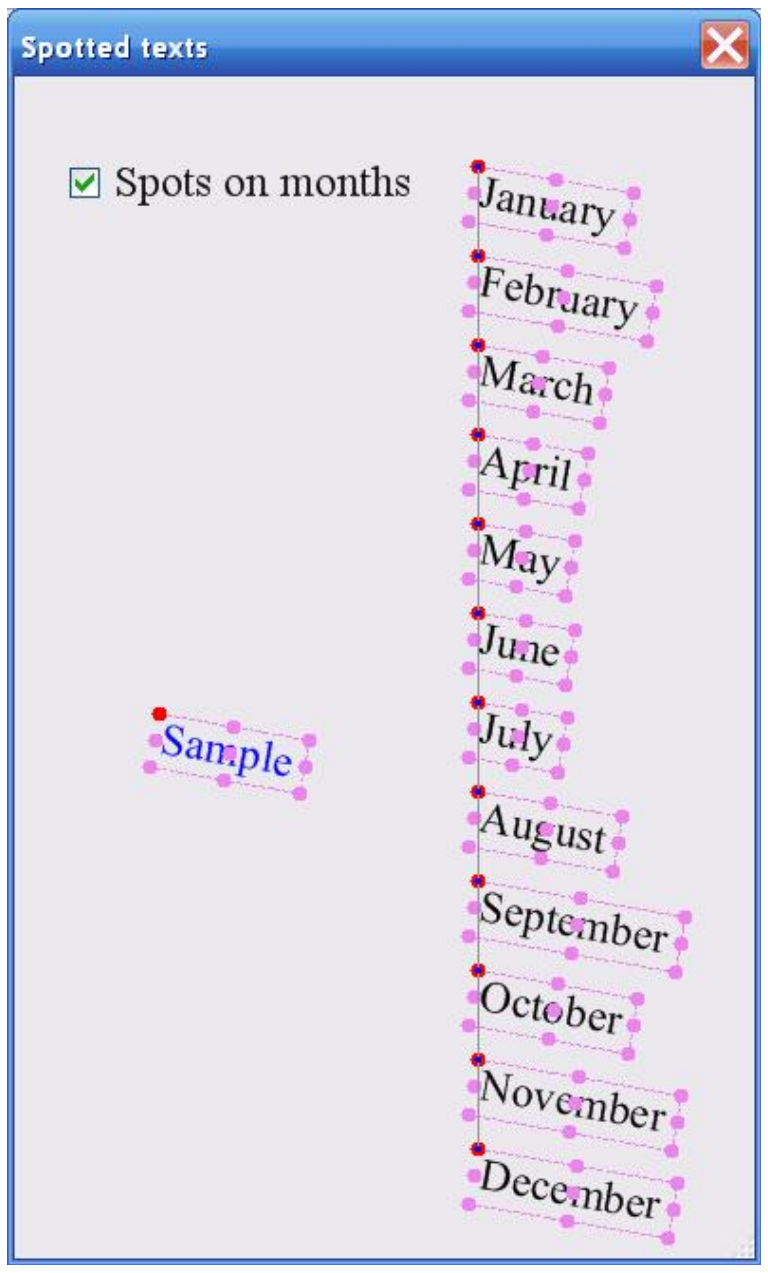

**Fig.5.3** All *months* are lined, if any special spot of the *sample* is pressed

```
void OnMouseDown (object sender, MouseEventArgs e)
{
    … …
    if (grobj .ID == sample .ID)
    {
        if (iNode < Enum .GetNames (typeof (TextBasis)) .Length)
        {
            sample .TextBasis = (TextBasis) iNode;
            for (int i = 0; i < month .Length; i++)
            {
                month [i] .Relocate (ptAnchor [i], sample .TextBasis,
                                     sample .Angle_Degree);
            }
            Invalidate ();
        }
        else
        {
            mover .Release ();
        }
    }
```

**Figure 5.2** makes it obvious that any *month* can be rotated individually and such rotation has no effect on other objects. The rotation of texts was already discussed in the previous chapter; rotation of the SpottedText objects starts in exactly the same way.

```
private void OnMouseDown (object sender, MouseEventArgs e)
{
    if (mover .Catch (e .Location, e .Button))
    {
        GraphicalObject grobj = mover .CaughtSource;
        if (e .Button == MouseButtons .Left)
        … …
        else if (e .Button == MouseButtons .Right)
        {
            if (grobj is SpottedText)
            {
                (grobj as SpottedText) .StartRotation (e .Location);
            }
        }
    }
}
```

It does not matter whether it is one of the *months* or a *Sample* is pressed by the right button; the StartRotation() method for the pressed object is called. Each object can be rotated around one of nine special points; the compensation angle depends on the currently selected rotation center for the pressed object.

```
public void StartRotation (Point ptMouse)
{
    Point ptAnchor = pts [(int) basis];
    double angleMouse = -Math .Atan2 (ptMouse .Y - ptAnchor .Y,
                                      ptMouse .X - ptAnchor .X);
    compensation = Auxi_Common .LimitedRadian (angleMouse - angle);
}
```

When any object is pressed by the right button to start its rotation, there is no difference in behaviour. But when you start the rotation itself, then you can see the difference between the rotation of any *month* or a *Sample*. Any month is turning around in an ordinary way. But when the *Sample* is rotated, then all the *months* immediately get the same angle and rotate synchronously.

```
private void OnMouseMove (object sender, MouseEventArgs e)
{
    if (mover .Move (e .Location))
    {
        GraphicalObject grobj = mover .CaughtSource;
        if (grobj is SpottedText &&
            grobj .ID == sample .ID &&
            e .Button == MouseButtons .Right)
        {
            for (int i = 0; i < month .Length; i++)
            {
                month [i] .Angle = sample .Angle;
            }
        }
        Invalidate ();
    }
}
```

In the **Form_SpottedTexts.cs** you can see only the mechanism of synchronous rotation, in which a set of related objects copy the angle from the single sample. In the plotting and tuning forms that I mentioned at the beginning of this section, the same idea is used not only for rotation, but for forward movement also: sample has its own anchor point from which the shifts are calculated; when the sample is moved around, all the related texts get the identical shifts from their anchor points.

# Polygons

The discussion of the simple elements was interrupted only to describe the rotation procedure, because a lot of elements, which I am going to demonstrate further on, can be involved both in forward movement and rotation. Let us continue with the regular polygons and some interesting objects, into which the regular polygons can be transformed.

## *Regular polygons*

File: **Form_RegPoly_Variants.cs**
Menu position: *Graphical objects – Basic elements – Polygons – Regular polygons*

In case you need to draw a regular polygon, you have to declare several parameters: center of a polygon; radius for the vertices, number of vertices, angle for the first vertex, direction for the consecutive vertices, and the color to fill the figure. These are the parameters that are used for construction of the `RegPoly_Variants` objects in the **Form_RegPoly_Variants.cs**.

```
public RegPoly_Variants (PolygonType polytype, PointF center, float rad,
                        int vertices, double angleDegree, Color clr)
```

Only for initialization an angle is declared in degrees; I think it is more natural for users to describe the position around the circle in degrees, but for all the inner calculations further on it is easier to deal with radians. Calculation of all the vertices is easy, but I have to do it so often in many of my examples and in different applications that there is the method `Auxi_Geometry.RegularPolygon()` to get the vertices of the regular polygon.

```
public static PointF [] RegularPolygon (PointF ptCenter, double radius,
                                        int nVertices, double fAngle)
```

In many cases I use this method from the **MoveGraphLibrary.dll**, but the `RegPoly_Variants` class uses its own method, which is identical. The consecutive vertices of my polygons are going counterclockwise.

```
protected PointF [] Vertices
{
    get
    {
        PointF [] pts = new PointF [nVertices];
        double ang;
        for (int i = 0; i < nVertices; i++)
        {
            ang = angle + 2 * Math .PI * i / nVertices;
            pts [i] =
            new PointF (ptC .X + Convert .ToSingle (radius * Math .Cos (ang)),
                        ptC .Y - Convert .ToSingle (radius * Math .Sin (ang)));
        }
        return (pts);
    }
}
```

If the regular polygons in the **Form_RegPoly_Variants.cs** are shown without covers, then all these polygons look similar and differ only by color. When the visualization of covers is switched ON, it makes obvious that there are three different types of covers (**figure 6.1**). At the same time all these regular polygons belong to the same class `RegPoly_Variants`, so the representatives of the same class may have different covers and, as a result, can be moved and resized differently from each other. Constructor of the `RegPoly_Variants` class has one parameter, which belongs to the `PolygonType` enumeration and determines the initial type of resizing. Later any polygon can change its type of resizing; to do this, click a polygon with the right button and choose in the opened menu one of three possibilities of resizing.

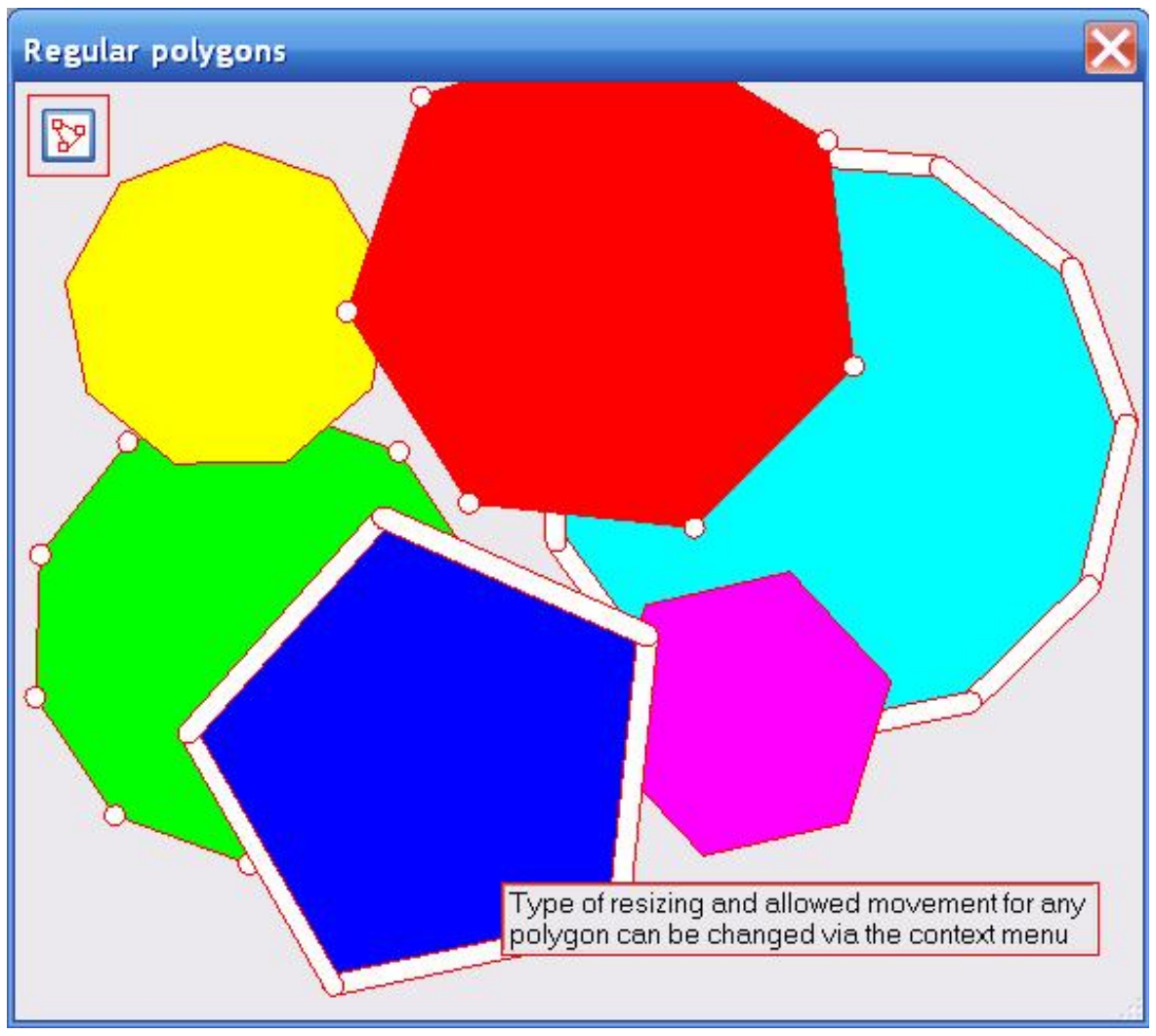

**Fig.6.1** Regular polygons with different covers

The names of the members of the PolygonType enumeration clearly inform about the resizing, associated with each of them.

```
public enum PolygonType { NonResizable, ZoomByVertices, ZoomByBorder };
```

The first type of regular polygons is only movable by any inner point, bit not resizable at all; yellow and violet polygons from the picture are of this type. The cover of such polygon consists of a single polygonal node equal to its area.

```
public override void DefineCover ()
{
    switch (type)
    {
        case PolygonType .NonResizable:
            cover = new Cover (new CoverNode [] { new CoverNode (0, Vertices) });
            break;
```

The next type of polygons is resizable by vertices; red and green polygons from the picture are of such a type. For this purpose each vertex is covered by a small circular node. The movability of the whole polygon is provided exactly in the same way, as in the previous case, by the big polygonal node.

```
public override void DefineCover ()
{
    CoverNode [] nodes;
    PointF [] pts;
    switch (type)
    {
        … …
        case PolygonType .ZoomByVertices:
            nodes = new CoverNode [nVertices + 1];
            pts = Vertices;
            for (int i = 0; i < nVertices; i++)
            {
                nodes [i] = new CoverNode (i, pts [i], 5);
            }
            nodes [nVertices] = new CoverNode (nVertices, Vertices);
            cover = new Cover (nodes);
            break;
```

Though the resizing of polygons by vertices works correctly and it is not a problem for users to start the resizing by any vertex, but there is one more variant, which fits with the standard design of movable / resizable objects even better. The most common (and expected by users!) solution is the resizing of an object by any border point. This is the third variant of regular polygons in the **Form_RegPoly_Variants.cs;** blue and cyan polygons on the figure have such covers. In these polygons, there are no circular nodes above the vertices, but each segment of perimeter between the two consecutive vertices is covered by a strip node. Certainly, all the vertices are covered by these strips, so any point of perimeter can be used for resizing in identical way. Here is the design of cover for such polygons.

```
public override void DefineCover ()
{
    switch (type)
    {
        case PolygonType .ZoomByBorder:
            nodes = new CoverNode [nVertices + 1];
            pts = Vertices;
            for (int i = 0; i < nVertices; i++)
            {
                nodes [i] = new CoverNode (i, pts [i], pts [(i+1) % nVertices], 5);
            }
            nodes [nVertices] = new CoverNode (nVertices, Vertices);
            cover = new Cover (nodes);
            break;
```

We have polygons with different types of resizing, but the code of the form does not have a single place, where the exact type of resizing is checked or analysed. So, what is the place in the code, where the difference in covers plays its role? Any movement of any object starts in its `MoveNode()` method, so the first place to look at is this method for the `RegPoly_Variants` class. There are three cases in this method corresponding to different types of covers.

For non-resizable polygons, there is an automatic call for `Move()` method; the whole object is moved.

```
public override bool MoveNode (int i, int dx, int dy, Point ptM, MouseButtons btn)
{
    bool bRet = false;
    if (btn == MouseButtons .Left)
    {
        switch (type)
        {
            case PolygonType .NonResizable:
                Move (dx, dy);
                bRet = true;
                break;
```

For polygons resized by vertices the decision is based on the number of the node. If it is the last node (the one, which covers the whole object), then the `Move()` method is called; all other nodes are responsible for resizing, and in this case it does not matter, which one of them is caught. The nodes around each vertex are small, so if any "vertex" is caught, then the mouse at that moment is exactly at the vertex or very close to it. Because of this closeness, the new radius for the vertices is simply calculated as the distance between the mouse and the center of an object.* If this distance exceeds the minimum allowed radius, then the radius of vertices is changed thus resizing the polygon.

```
            case PolygonType .ZoomByVertices:
                if (i < nVertices)
                {
                    distance = Auxi_Geometry .Distance (ptC, ptM);
                    if (distance >= minR)
                    {
                        radius = Convert .ToInt32 (distance);
                    }
                }
                else
                {
                    Move (dx, dy);
                }
                break;
```

For the polygons, resized by any border point, the procedure is a bit more complicated. The strip node, connecting two consecutive vertices, can be grabbed for moving at any point; the distance between this point and the center can significantly vary and can differ a lot from the radius of vertices, so this distance cannot be substituted as the new radius. But it can be used for calculations of the new radius, if multiplied by the special *scaling* coefficient. This coefficient is not going to change throughout the resizing, so it has to be calculated once at the moment, when an object is caught for resizing. The calculation of this coefficient is done by the special method of the `RegPoly_Variants` class.

```
        public void StartScaling (Point ptMouse)
        {
            if (type == PolygonType .ZoomByBorder)
            {
                scaling = radius / Auxi_Geometry .Distance (ptC, ptMouse);
            }
        }
```

As I mentioned, this coefficient is calculated once, when an object is caught for resizing, so this method must be called from the `OnMouseDown()` method of the form.

* It is very easy to take into consideration the difference between the current radius and the distance from mouse to the center, but this difference never exceeds a few pixels, so I ignored this difference. If you want the higher accuracy, you can achieve it by adding a couple of lines into the code.

```
private void OnMouseDown (object sender, MouseEventArgs e)
{
    ptMouse_Down = e .Location;
    if (mover .Catch (e .Location, e .Button))
    {
        if (mover .CaughtSource is RegPoly_Variants)
        {
            (mover .CaughtSource as RegPoly_Variants) .StartScaling (e .Location);
        }
    }
    ContextMenuStrip = null;
}
```

Now, when you know everything about the origin and calculation of the scaling coefficient, it is easier to understand its use in the `MoveNode()` method for the polygons of the appropriate type.

```
public override bool MoveNode (int i, int dx, int dy, Point ptM, MouseButtons btn)
{
    bool bRet = false;
    if (btn == MouseButtons .Left)
    {
        switch (type)
        {
            case PolygonType .ZoomByBorder:
                if (i < nVertices)
                {
                    distance = Auxi_Geometry .Distance (ptC, ptM);
                    if (distance * scaling >= minR)
                    {
                        radius = Convert .ToInt32 (distance * scaling);
                    }
                }
                else
                {
                    Move (dx, dy);
                }
```

As in the previous case of the polygons resized by vertices, the decision is based on the number of the caught node. The last node in the cover is responsible for moving the whole object, so the attempt to move this node calls the `Move()` method. All other nodes are responsible for resizing, and again it does not matter, which one of them is caught, but the scaling coefficient must be included into calculations.

There can be different variants of doing the same things. For example, in the shown code the decision about the movement is based on the number of the node, but it can be also based on its shape. For the polygons, resized by vertices, it will look like this.

```
        if (cover .GetNodeShape (i) == NodeShape.Circle) {
            … …
        } else {
            Move (dx, dy);
```

The `StartScaling()` method in the shown code is called for any polygon; then inside the method there is an additional check by the type of the cover. But the method is needed only, when the strip node is caught, so the checking of the type of the pressed node can be added into the `OnMouseDown()` method of the form prior to calling the `MoveNode()` method of the polygon. Usually, there are several different ways of doing the same things, and it is up to the developer to decide on one or another.

The regular polygons of the `RegPoly_Variants` class can change not only the type of cover, but also the allowed movement. By default any of them can be moved freely around the screen, but the same context menu on each polygon allows to limit its movement either to horizontal or to vertical direction. The cover of an object on such an action is not affected at all, but the `Move()` method, which is responsible for moving the whole polygon, has three different options. Any polygon is grabbed and moved by the mouse in the same way, or better to say that the mouse is moved in the same way, but, depending on the type of the allowed movement, either both of the parameters are used by the RegPoly_Variants.Move() method or only one of them.

```
public override void Move (int dx, int dy)
{
    switch (movetype)
    {
        case MovementAllowed .Any:          ptC += new Size (dx, dy);    break;
        case MovementAllowed .Horizontal:   ptC .X += dx;                break;
        case MovementAllowed .Vertical:     ptC .Y += dy;                break;
    }
}
```

There is no visual indication on any polygon of the type of movement, which is currently allowed for this object. If you want, you can easily add such an indication, for example, somewhere in the center of a polygon. For example, there can be an arrow Left – Right, or Up – Down, or both of them. But I do not think that such an indication is needed. I will return to the discussion of changing the movability of objects further on in the book.

I am not going to delete or rewrite the previous several lines, but at the last moment I added a special type of visualization, which informs about the possible movement for each polygon. The node which is responsible for moving the whole object is always the same: it is the polygonal node, which covers the whole object. It is always the last node regardless of the cover type and I can do the same thing by checking the number of node. It is the only polygonal node in all three variants of covers, so this part of code is based on the shape of nodes. By default the polygonal nodes get the standard cursor Cursors.SizeAll. I can change the cursor for this particular node depending on the allowed movement; I will do it in two of three cases. This piece of code is at the end of the RegPoly_Variants.DefineCover() method and commented, but if you turn this commented part into the working code, you will see the difference in cursors for different types of allowed movement.

```
        public override void DefineCover ()
        {
            … …
            if (movetype == MovementAllowed .Horizontal)
            {
                cover .SetCursor (NodeShape .Polygon, Cursors .SizeWE);
            }
            else if (movetype == MovementAllowed .Vertical)
            {
                cover .SetCursor (NodeShape .Polygon, Cursors .SizeNS);
            }
```

Regular polygons in the **Form_RegPoly_Variants.cs** are constructed with different types of covers. Later the cover of any polygon can be changed to a different type, which means the changing of resizing for this polygon. Also the order of polygons can be changed, as any of them can be put on top of all others by a simple click. Let us look at the details of all these things.

The set of polygons is organized, when the form is constructed. There are more polygons in the form, than are shown at **figure 6.1**; when the form is opened, nine polygons are constructed with the number of vertices changing from three to 12.

```
    List<RegPoly_Variants> polygons = new List<RegPoly_Variants> ();

    private void NewSet ()
    {
        polygons .Clear ();
        for (int i = 3; i < 12; i++)
        {
            … …
            polygons .Add (new RegPoly_Variants ((PolygonType) (i % 3), pt, rad, i,
                                                  angleDeg, clr));
        }
    }
```

All objects are included into the mover's queue; first the button to switch covers ON and OFF, then the short information, and then all the polygons in the same order, as they are included into the List.

```
        private void RenewMover ()
        {
            mover .Clear ();
            foreach (RegPoly_Variants poly in polygons)
            {
                mover .Add (poly);
            }
            mover .Insert (0, text);
            mover .Insert (0, btnCovers);
        }
```

Button and information never change their places in the mover's queue, but polygons can be easily reordered. It is enough to click any polygon with the left button to put it on top of all others. If the distance between the points of pressing and releasing the left button is greater than three pixels, then it is considered as an ordinary forward moving of the pressed polygon; the lesser distance is considered as a request to move the pressed polygon into the head of the `List`.

```
    private void OnMouseUp (object sender, MouseEventArgs e)
    {
        ptMouse_Up = e .Location;
        if (mover .Release ())
        {
            GraphicalObject grobj = mover .WasCaughtSource;
            if (grobj is RegPoly_Variants &&
                Auxi_Geometry .Distance (ptMouse_Down, ptMouse_Up) <= 3)
            {
                if (e .Button == MouseButtons .Left)
                {
                    PopupPolygon (grobj .ID);
                }
```

Identification of the polygon to be moved ahead is easily done with the **id** of the pressed polygon; it is enough to look through the whole `List` and check the **id** of the polygons one after another. Change of order among the movable objects requires the renewal of the mover's queue, which is done by the `RenewMover()` method.

```
        private void PopupPolygon (long id)
        {
            for (int i = polygons .Count - 1; i > 0; i--)
            {
                if (id == polygons [i] .ID)
                {
                    RegPoly_Variants poly = polygons [i];
                    polygons .RemoveAt (i);
                    polygons .Insert (0, poly);
                    RenewMover ();
                    Invalidate ();
                    break;
                }
            }
        }
```

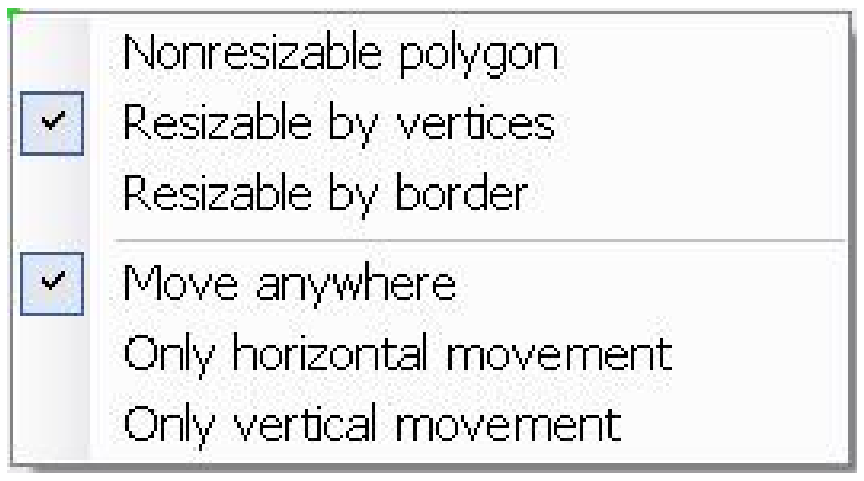

**Fig.6.2** Menu on polygons to change the resizing and allowed movement

The change of cover, which means the change of the resizing type, and the change of the allowed movement for any polygon are organized via the context menu (**figure 6.2**).

```
        private void OnMouseUp (object sender, MouseEventArgs e)
        {
            … …
                else if (e .Button == MouseButtons .Right)
                {
                    Identification (grobj .ID);
                    if (iPolyTouched >= 0)
                    {
                        ContextMenuStrip = contextMenuOnPoly;
```

```
                    }
                }
```

The `Identification()` method goes through the `List` of polygons and finds the touched one by comparison of **id** numbers. Three commands from the menu allow to change the resizing type of the polygon, which means the change of its cover. There is a possibility that at this moment the visualization of covers in the form is switched ON; in this case the repainting of the form is needed; I put the repainting in any case without checking the current state of visualization for covers.

```
        private void Click_miResizableByBorder (object sender, EventArgs e)
        {
            polygons [iPolyTouched] .PolygonType = PolygonType .ZoomByBorder;
            Invalidate ();
        }
```

Three other commands of menu allow to change the possibility of movement. This feature is not shown in any way, so the repainting is not needed.

```
        private void Click_miVerMove (object sender, EventArgs e)
        {
            polygons [iPolyTouched] .MovementType = MovementAllowed .Vertical;
        }
```

None of these commands change the order of movable objects, so there is no need to call the `RenewMover()` method.

## *Regular polygons that can disappear*

File: **Form_RegPoly_Disappear.cs**
Menu position: *Graphical objects – Basic elements – Polygons – Regular polygons that can disappear*

In all the real applications you often have a set of different objects on the screen. The movability of all these objects and the possibility of their resizing allow to design an absolutely new type of programs, in which users make ALL the decisions. Among these decisions are the commands on erasing the unneeded objects, which can be done in two possible ways. Objects can be deleted via the direct command, for example, by selection of a context menu line, or they can be deleted by squeezing them to a tiny size, if such an interpretation of their shrinking is put into code. Such method of deleting objects was already demonstrated with rectangles; let us do the same with the regular polygons.

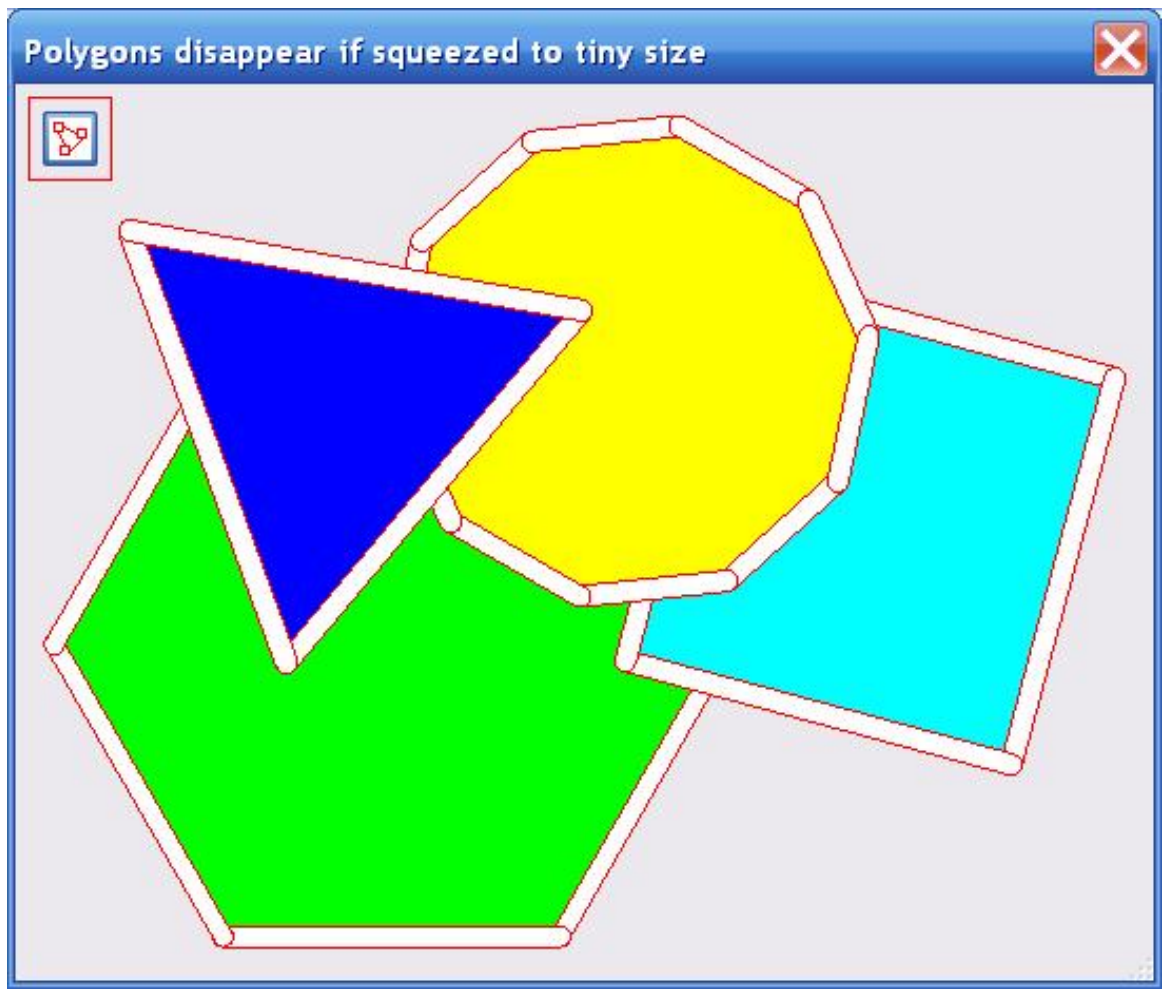

**Fig.6.3** These polygons can disappear, if squeezed to tiny size

The **Form_RegPoly_Disappear.cs** is initialized with several regular polygons in view (**figure 6.3**). Each polygon is of the `RegPoly_Disappear` class, which is derived from the `RegPoly_Variants` class. Though the base class has three different types of resizing, the derived class uses only one of them, defined in its constructor. A polygon with a possibility of disappearance can be resized by any border point.

```
public class RegPoly_Disappear : RegPoly_Variants
{
    static int nRadiusDisappearance = 5;

    public RegPoly_Disappear (PointF ptC, float rad, int vertices,
                              double angleDegree, Color color)
        : base (PolygonType .ZoomByBorder, ptC, rad, vertices, angleDegree, color)
    {
    }
```

The derived class has nearly nothing of its own; the only additional field is the radius of disappearance. If an object is resized and released, when the radius of vertices is less than the radius of disappearance, which is set to 5 pixels, then this object must go out.

Describing the `MoveNode()` method in the base class (`RegPoly_Variants` class), I mentioned that the decision on the particular movement can be based not only on the number of node, but on its shape; this technique is used in the `RegPoly_Disappear` class.

```
public override bool MoveNode (int i, int dx, int dy, Point ptM, MouseButtons btn)
{
    bool bRet = false;
    if (btn == MouseButtons .Left)
    {
        if (cover [i] .Shape == NodeShape .Strip)
        {
            double distanceMouse = Auxi_Geometry .Distance (Center, ptM);
            if (distanceMouse * scaling > 1)
            {
                radius = Convert .ToInt32 (distanceMouse * scaling);
            }
        }
        else
        {
            Move (dx, dy);
        }
        bRet = true;
    }
    return (bRet);
}
```

The decision of whether to erase a polygon or not is made inside the `OnMouseUp()` method; the process consists of two major parts. First, several conditions must be fulfilled for this to occur. Second, if the conditions are fulfilled, then an object to delete must be identified.

```
    private void OnMouseUp (object sender, MouseEventArgs e)
    {
        int iWasCaught, iNode;
        NodeShape shapeNode;
        if (mover .Release (out iWasCaught, out iNode, out shapeNode))
        {
            if (e .Button == MouseButtons .Left)
            {
                GraphicalObject grobj = mover [iWasCaught] .Source;
                if (grobj is RegPoly_Disappear)
                {
                    RegPoly_Disappear poly = grobj as RegPoly_Disappear;
                    if (shapeNode == NodeShape .Strip &&
                        poly .Radius < RegPoly_Disappear .RadiusDisappearance)
                    {
                        long id = grobj .ID;
                        for (int i = polygons .Count - 1; i >= 0; i--)
                        {
                            if (id == polygons [i] .ID)
                            {
                                polygons .RemoveAt (i);
```

As you can see from this code, there is a long list of nested **`if`** statements, which means a set of conditions to fulfil.

1. The erasing of any polygon can happen, only when an object is released; mover gives this information as a return value of its `Release()` method. The method is used in such a version which allows to get the number of object and the shape of the released node; both parameters are used for further checking.

```
if (mover .Release (out iWasCaught, out iNode, out shapeNode))
```

2. The erasing can happen only after resizing, which is done with the left button.

```
if (e .Button == MouseButtons .Left)
```

3. The released object must be a polygon.

```
GraphicalObject grobj = mover [iWasCaught] .Source;
if (grobj is RegPoly_Disappear)
```

4. A polygon has to be deleted only as a result of resizing; the resizing is done only by the strip nodes covering the border, but not by the polygonal node which is responsible for moving. Thus there is the checking of shape of a node.

```
if (shapeNode == NodeShape .Strip &&
```

5. The radius of vertices at the moment of release must be less than the predefined radius for disappearance.

```
poly .Radius < RegPoly_Disappear .RadiusDisappearance)
```

When all the conditions are fulfilled, then the polygon to erase must be found in the list of polygons. Identification is made with the **id** number, which is obtained from the released object.

```
long id = grobj .ID;
for (int i = polygons .Count - 1; i >= 0; i--)
{
    if (id == polygons [i] .ID)
```

From a lot of further examples you will see that this is a pretty standard set of checks and the standard procedure for erasing an object.

## Polygons which are always convex

File: **Form_Polygons_Convex.cs**
Menu position: *Graphical objects – Basic elements – Polygons – Always convex polygons*

Nodes, which are used for moving / resizing of objects, can be of three different types. One of these types is polygon, and the only condition is that it must be convex. In all the previous examples, the polygonal nodes were either rectangles or regular polygons. But there is no requirement that the polygonal node must be of a "simple" or "standard" form; it can be arbitrary, until it is convex. Let us look at the example of such arbitrary polygonal nodes. The objects are initialized in the form of regular polygons, but this is only for the easiness of initialization. After it polygons can be reconfigured in an arbitrary way (**figure 6.4**), but the program is supposed to keep them convex and not allow to break the rule.

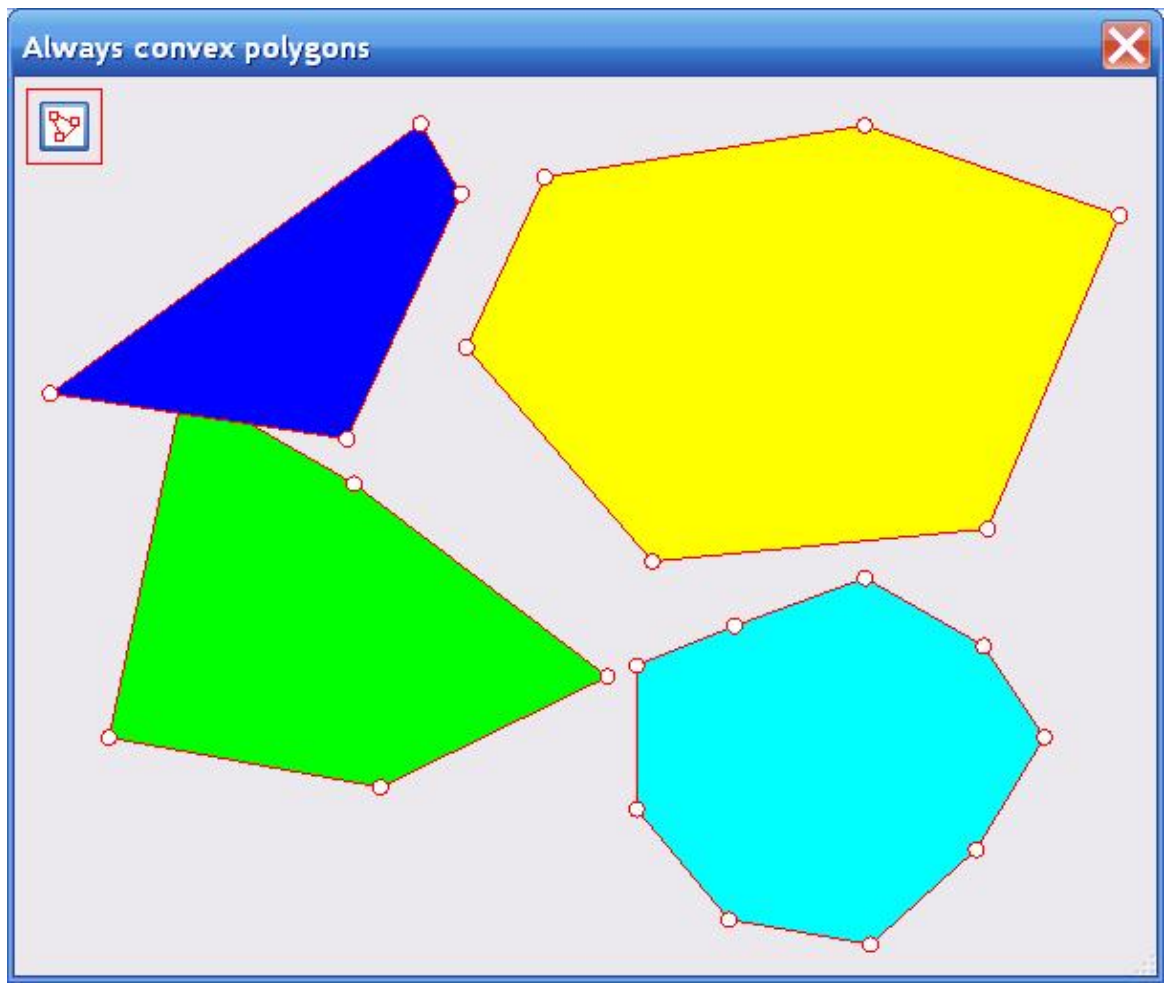

**Fig.6.4** Always convex polygons from the **Form_Polygons_Convex.cs**

The only way to define a convex polygon is to have the points of all its vertices. This is going to be the only needed field for such an object plus a brush to paint it.

```
public class Polygon_Convex : GraphicalObject
{
    PointF [] pts;
    SolidBrush brush;
```

The cover for the objects of the `Polygon_Convex` class is the same, as for one of the types of the `RegPoly_Variants` class: each vertex is covered with a small circular node, which allows to move this particular vertex; the whole area of object is covered by a single polygonal node. The radius of the circular nodes over the vertices is slightly increased from the default value; this makes the moving of vertices a bit easier.

```
public override void DefineCover ()
{
    int nVertices = pts .Length;
    CoverNode [] nodes = new CoverNode [nVertices + 1];
    for (int i = 0; i < nVertices; i++)
    {
        nodes [i] = new CoverNode (i, pts [i], 4);
    }
    nodes [nVertices] = new CoverNode (nVertices, pts);
```

```
        cover = new Cover (nodes);
    }
```

The difference of these objects from the regular polygons is in their `Move()` and `MoveNode()` methods. For the regular polygons, it is enough to move the central point, and the whole object is moved; for the `Polygon_Convex` class, all the vertices must be moved, because their positions are arbitrary and not defined by one or another parameter.

```
    public override void Move (int dx, int dy)
    {
        Size size = new Size (dx, dy);
        for (int i = 0; i < pts .Length; i++)
        {
            pts [i] += size;
        }
    }
```

The positions of vertices are not defined by any parameters, but these positions are partly restricted by the neighbouring vertices. The `MoveNode()` method is the place, where the possibility of each proposed reconfiguration is checked and either allowed or not. The methods, which are used in this method to determine the possibility of movement, are from the **MoveGraphLibrary.dll**; they are described in the **MoveGraphLibrary_Classes.doc** (see *Programs and Documents*). The allowed movements of vertices are those that still keep the area of a polygon as a convex polygon; to check this situation, the new position of a vertex must be compared not only with the two neighbouring vertices, but some comparison with the vertices, which are farther away along the perimeter, is needed.

To avoid special cases of slightly different checking, I set the minimum number of vertices in the `Polygon_Convex` class to four. Each vertex has its own number; vertices (and their numbers) produce an infinitive loop, but if you write them in a line, the numbers will look like this

```
            jPrevPrev     jPrev     i     jNext     jNextNext
```

It does not matter that for a polygon with a small number of vertices the numbers on two ends of this row are the same. The above mentioned numbers are used in the code of the `MoveNode()` method; `ptNew` is the supposed position of the caught node (vertex), if all the checks allow this move. I think that now the code is not difficult to understand.

```
public override bool MoveNode (int i, int dx, int dy, Point ptM, MouseButtons btn)
{
    bool bRet = false;
    if (btn == MouseButtons .Left)
    {
        if (i < pts .Length)    // variant   cover [i] .Shape == NodeShape .Circle
        {
            PointF ptNew = new PointF (pts [i] .X + dx, pts [i] .Y + dy);
            int nVertices = pts .Length;
            int jNext = (i + 1) % nVertices;
            int jNextNext = (i + 2) % nVertices;
            int jPrev = (i + nVertices - 1) % nVertices;
            int jPrevPrev = (i + nVertices - 2) % nVertices;
            if (Auxi_Geometry .Distance (pts [jPrev], ptNew) > minSide &&
                Auxi_Geometry .Distance (pts [jNext], ptNew) > minSide &&
                Auxi_Geometry .OppositeSideOfLine (pts [jPrev], pts [jNext],
                                                   ptNew, pts[jNextNext])     &&
                Auxi_Geometry .SameSideOfLine (pts [jPrevPrev], pts [jPrev],
                                               ptNew, pts[jNext])             &&
                Auxi_Geometry .SameSideOfLine (pts [jNextNext], pts [jNext],
                                               ptNew, pts [jPrev]))
            {
                pts [i] = ptNew;
                bRet = true;
            }
```

Two methods from the **MoveGraphLibrary.dll** - `Auxi_Geometry.OppositeSideOfLine()` and `Auxi_Geometry.SameSideOfLine()` – are very helpful in the needed checking. Both methods work with a segment of line, determined by two points, and check, if other two points are on the opposite sides of this line or on the same side.

This are exactly the checks, which are needed to decide, if the proposed move of the caught vertex is still going to keep the polygon convex.

## *Triangles*

File: **Form_Triangles.cs**
Menu position: *Graphical objects – Basic elements – Triangles*

With all the previous examples I tried to move from the simplest to more and more complicated figures. Let us change the trend, make a step back, and look at the simple triangles. These objects of the `Triangle` class can be reconfigured by any vertex. What is more, there are no restrictions on these movements, so any object can be reconfigured into any other triangle, into a line, or even into a single dot. With such transformations, there can be some strange or even amusing results: a triangle can disappear.

When three vertices of such object are placed on a straight line then, depending on the position of the middle vertex, you can see a thin line, or a collection of separate dots, or even nothing. A triangle also disappears from view, when all three vertices are moved into the same point; only the visualization of cover can help in finding such an object.

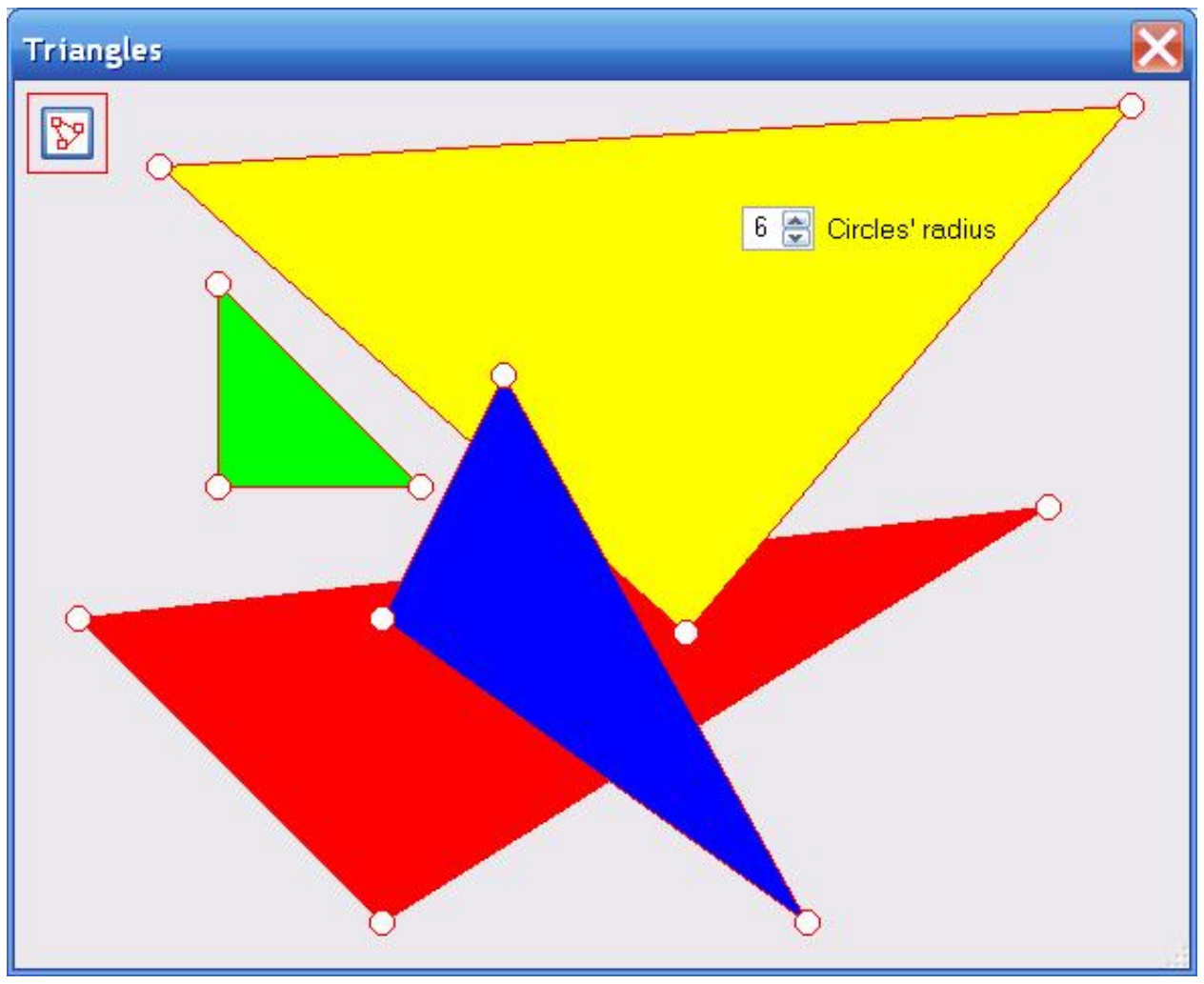

**Fig.6.5** Triangles in the **Form_Triangles.cs**

The cover design for these triangles is obvious from **figure 6.5**: three small circular nodes on vertices and the polygonal node, covering the whole area of an object.

```
public override void DefineCover ()
{
    CoverNode [] nodes = new CoverNode [4];
    for (int i = 0; i < 3; i++)
    {
        nodes [i] = new CoverNode (i, pts [i], radiusPoint);
    }
    nodes [3] = new CoverNode (3, pts);
    cover = new Cover (nodes);
}
```

All three vertices are the basic points of a rectangle, so all of them must be moved by the `Move()` method.

```
public override void Move (int dx, int dy)
{
    SizeF size = new SizeF (dx, dy);
    for (int i = 0; i < pts .Length; i++)
    {
        pts [i] += size;
    }
}
```

There are no restrictions on movement of any node. The nodes on vertices can be relocated in an arbitrary way; so a triangle can be reconfigured in any possible way. Squeezing into a line or a dot is possible, but there is one consequence of such transformation: a triangle becomes temporarily unmovable. The understanding of this fact is easier, when the visualization of cover is switched ON.

If you have three vertices on a line, you see the three round nodes at these points and a line, connecting them. As a result, there is no real area, by which an object can be moved to another place. It is not the problem of a polygonal node, defined by three points, becoming non-convex. Any three points, not placed on the same line, define a convex polygon. But when those three points are on the same line, then there is no polygon at all; the area of such a polygon is null; mover cannot detect it, and mover cannot call the `Move()` method in such situation, because it does not detect any node.. When the three vertices are on the same line, they can be moved individually, but not together. Only when one of the vertices is moved from the same line with others, the object will gain some area, by which it can be moved again.

Three circular nodes can be even moved exactly to the same point; in such a case they look like a single circle. Or you can slightly move the nodes apart, so that they overlap without gap in the middle. Though on switching OFF the visualization of cover you see a small rectangle, but it cannot be moved to another position, because those nodes on vertices precede the polygonal node and entirely block it. Only when the vertices are moved apart enough to have some area in between, which is not covered by them, only then an object becomes movable again.

This curious situation is seen much better, if the circular nodes are enlarged; there is a control in the form, which allows to change the radius of nodes in the [1, 20] range. Each change of the radius of nodes causes the redefinition of the covers, but there is no need to renew the mover's queue, as such operation neither changes the number of movable objects nor their order.

```
private void ValueChanged_numericRadius (object sender, EventArgs e)
{
    radius = Convert .ToInt32 (numericUD_Radius .Value);
    foreach (Triangle elem in triangles)
    {
        elem .Radius = radius;
    }
    Invalidate ();
}
public int Radius
{
    get { return (radiusPoint); }
    set
    {
        radiusPoint = Math .Abs (value);
        DefineCover ();
    }
}
```

## *Chatoyant polygons*

File: **Form_ChatoyantPolygons.cs**
Menu position: *Graphical objects – Basic elements – Polygons – Chatoyant polygons*

With all the previous examples, the name of a section tells about the form of inner objects and the special features of their resizing. In the name of this section you can see the word *polygons*, but no more details about the resizing. This is because I cannot explain in a word or two, what can be done with these polygons; they can be transformed into figures, which I would not even imagine before I thought out such reconfiguring.

The objects of the `ChatoyantPolygon` class combine some features from the regular polygons and from the triangles, which were shown in the previous examples.

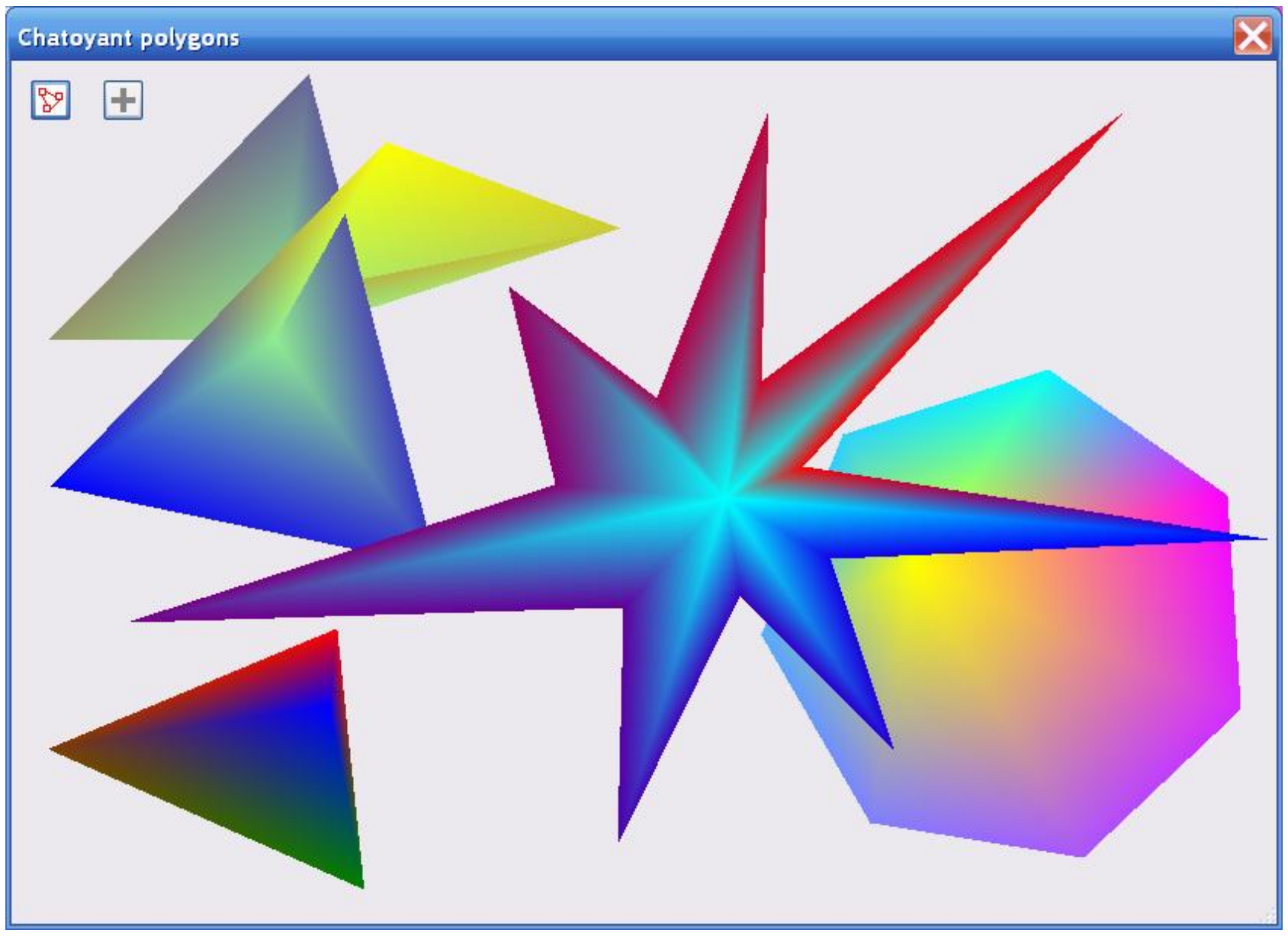

**Fig.6.6** Chatoyant polygons

**Figure 6.6** shows several of the `ChatoyantPolygon` objects in the **Form_ChatoyantPolygon.cs**. The figure does not show the covers of these objects, but gives an idea of the possible variety of such objects. **Figure 6.7** shows one `ChatoyantPolygon` object with its cover, which makes further explanation easier for understanding.

Each `ChatoyantPolygon` object has a central point and a set of at least three vertices.

- The central point is the place for a circular node.

- Each vertex is also covered by a circular node.
- Every two consecutive vertices are connected by a strip node; the last vertex is connected with the first one, so each vertex is connected with two neighbours.
- Each pair of two consecutive vertices and the central point form a polygonal (triangular) node.

The cover of the `ChatoyantPolygon` class uses the nodes of all three possible types to organize moving, resizing, and reconfiguring of the objects. The easiest way to initialize such an object is to do it in the form of a regular polygon, but it can be done also in other ways.* When the `ChatoyantPolygon` object is initialized in the form of a regular polygon, then the central point is really the central, but generally this is only the point of rotation for the whole object and nothing more; this "central" point can be moved anywhere and can be either inside the polygon or outside. When such object is initially organized in the form of a regular polygon, then the strip nodes cover the perimeter of an object; the union of the polygons (triangles) covers the whole area of an object. Later, when any point associated with a circular node can be moved around the screen to an arbitrary place, a polygon quickly changes its shape and the strips are not on the perimeter any more, but an object can be still moved, reconfigured, and zoomed in the same way. In the further explanation, whenever I write *central point*, it means the point, which was originally central at the moment of initialization, but can be now anywhere; the same thing about vertices.

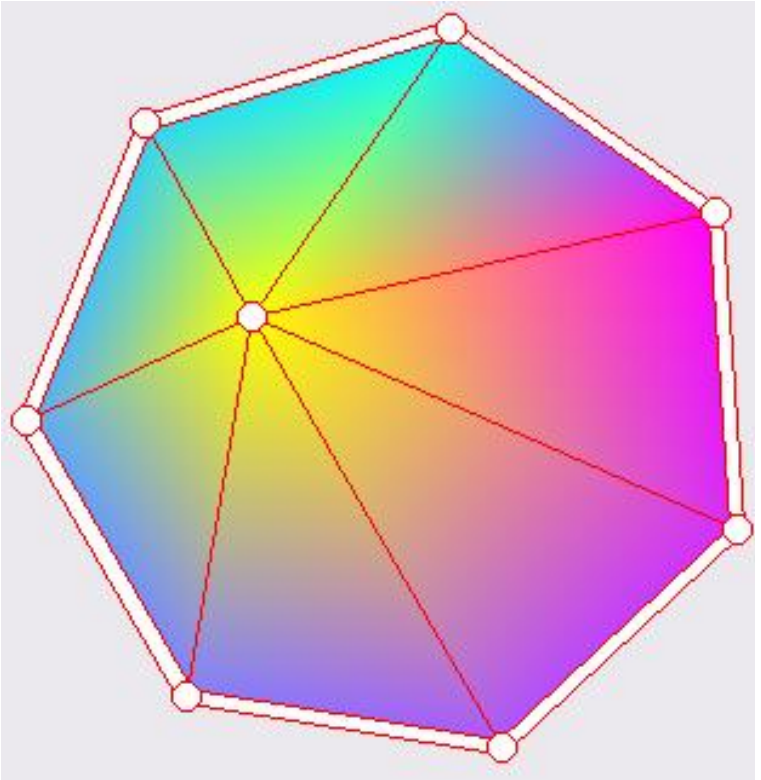
**Fig.6.7** Polygon's cover

Circular nodes are used for individual movements of the points, with which they are associated, regardless of whether it is a vertex or a center. This is the way for reconfiguring a polygon, and though often born as regular polygons, these objects can be quickly turned into very strange figures.

Strip nodes are used for zooming.

Polygonal nodes (triangles) are used for moving the whole objects.

For originally regular N-gon, the number of nodes is 3 * N + 1, so a polygon with 12 vertices has 37 nodes. It looks like there must be a lot of code writing for such object, especially in the `MoveNode`() method, but in reality it is not so, because the nodes of the same group have similar behaviour. As often done with the covers containing different types of nodes, the smaller nodes are included into the cover before the bigger ones, so here is the order of nodes for the `ChatoyantPolygon` class:

1. N circular nodes on the vertices.
2. One circular node on the center.
3. N strip nodes along the border; the strips connect the neighboring vertices.
4. N triangular nodes for the inner area.

```
public override void DefineCover ()
{
    CoverNode [] nodes = new CoverNode [3 * VerticesNum + 1];
    for (int i = 0; i < VerticesNum; i++)
    {
        nodes [i] = new CoverNode (i, ptVertices [i], 6);
    }
    nodes [VerticesNum] = new CoverNode (VerticesNum, center, 6);
    int k0 = VerticesNum + 1;
    for (int i = 0; i < VerticesNum; i++)
    {
        nodes [k0 + i] = new CoverNode (k0 + i, ptVertices [i],
                                        ptVertices [(i + 1) % VerticesNum]);
    }
    k0 = 2 * VerticesNum + 1;
    for (int i = 0; i < VerticesNum; i++)
    {
```

* There is no requirement for polygon to be convex and the central point can be placed anywhere. You can take any set of points (N >= 3) and declare them to be the vertices. In such way the convex polygon in the **Form_Main.cs** is constructed.

```
                PointF [] pts = new PointF [3] { ptVertices [i],
                                     ptVertices [(i + 1) % VerticesNum], center };
                nodes [k0 + i] = new CoverNode (k0 + i, pts);
            }
            cover = new Cover (nodes);
        }
```

Such a polygon with N vertices has (N + 1) independently movable basic points, so the `Move()` method has to change synchronously these (N + 1) points.

```
        public override void Move (int dx, int dy)
        {
            Size size = new Size (dx, dy);
            for (int i = 0; i < VerticesNum; i++)
            {
                ptVertices [i] += size;
            }
            center += size;
        }
```

Any movement of a polygon starts, when a polygon is pressed with a mouse. Depending on the pressed button and the caught node, it can be a reconfiguring, a resizing, a forward movement of the whole figure, or rotation. All four possibilities are demonstrated by these polygons.

```
    private void OnMouseDown (object sender, MouseEventArgs e)
    {
        ptMouse_Down = e .Location;
        if (mover .Catch (e .Location, e .Button))
        {
            GraphicalObject grobj = mover .CaughtSource;
            if (grobj is Polygon_Chatoyant)
            {
                if (e .Button == MouseButtons .Left)
                {
                    (grobj as Polygon_Chatoyant) .StartScaling (e .Location,
                                                          mover .CaughtNodeShape);
                }
                else if (e .Button == MouseButtons .Right)
                {
                    (grobj as Polygon_Chatoyant) .StartRotation (e .Location);
                }
            }
        }
    }
```

The real movements are described by the `MoveNode()` method of the polygons, but two types of movement require some preliminary actions.

All the vertices can move individually without affecting each other, so there is no way to describe the positions of all of them by one or two general parameters. Instead, the position of each vertex can be defined by the distance and the angle from the central point; all these parameters must be calculated before rotation or zooming is started.

```
        private void VerticesArrays ()
        {
            for (int i = 0; i < VerticesNum; i++)
            {
                radii [i] = Auxi_Geometry .Distance (center, ptVertices [i]);
                angles [i] = Auxi_Geometry .Line_Angle (center, ptVertices [i]);
            }
        }
```

We have already explored several examples of rotation and know the main rule of such movement. Before starting any rotation, the compensation angle must be calculated. Because in these polygons there are independently movable vertices, then the compensation for each of them must be calculated.

```
public void StartRotation (Point ptMouse)
{
    VerticesArrays ();
    double angleMouse = Auxi_Geometry .Line_Angle (center, ptMouse);
    for (int i = 0; i < VerticesNum; i++)
    {
        compensation [i] = Auxi_Common .LimitedRadian (angleMouse - angles [i]);
    }
}
```

For the same reason (independently movable vertices) not one scaling coefficient, but a whole array of coefficients for all vertices must be calculated at the moment, when zooming starts.

```
    public void StartScaling (Point ptMouse, NodeShape nodeshape)
    {
        if (nodeshape == NodeShape .Strip)
        {
            VerticesArrays ();
            double distanceMouse = Auxi_Geometry .Distance (center, ptMouse);
            for (int i = 0; i < VerticesNum; i++)
            {
                scaling [i] = radii [i] / distanceMouse;
            }
        }
    }
```

With all the preparations finished, we can look now into the `MoveNode()` method, which is responsible for all the movements. In the `Polygon_Chatoyant` class the type of movement initiated by pressing any node is determined by the number of node; the order of nodes was mentioned above and can be seen from the `DefineCover()` method.. If one of the vertices is moved, then it is only an individual relocation of this vertex and nothing else.

```
public override bool MoveNode (int i, int dx, int dy, Point ptM, MouseButtons btn)
{
    if (btn == MouseButtons .Left)
    {
        if (i < VerticesNum)
        {
            ptVertices [i] += new Size (dx, dy);
        }
```

If it is a central point, then the center is moved.

```
        else if (i == VerticesNum)
        {
            center += new Size (dx, dy);
        }
```

If it is any of the triangular nodes, then the whole object is moved by calling the `Move()` method.

```
        else if (i >= 2 * VerticesNum + 1)
        {
            Move (dx, dy);
        }
```

Otherwise it can be only one of the strip nodes; in this case the set of scaling coefficients is used for zooming.

```
        else
        {
            double distanceMouse = Auxi_Geometry .Distance (center, ptM);
            if (distanceMouse > 25)
            {
                for (int j = 0; j < VerticesNum; j++)
                {
                    ptVertices [j] = Auxi_Geometry.PointToPoint (center, angles[j],
                                                     distanceMouse * scaling [j]);
```

```
                }
            }
        }
```

If the move was started by the right button, then regardless of the caught node it is a rotation. Previously calculated radius and compensation for each vertex are used to calculate the new position.

```
    else if (btn == MouseButtons .Right && bRotate)
    {
        double angleMouse = -Math .Atan2 (ptM .Y - center .Y, ptM .X - center .X);
        for (int j = 0; j < VerticesNum; j++)
        {
            ptVertices [j] = Auxi_Geometry .PointToPoint (center,
                       Auxi_Common .LimitedRadian (angleMouse - compensation [j]),
                                                                   radii [j]);
        }
        DefineCover ();
        return (true);
    }
```

In the first line of this piece of code you can see that not only the pressed button is checked, but there is also another check for the value of the `bRotate` field. This field is assigned a `true` value in the constructor of the `Polygon_Chatoyant` class, the value is not changed anywhere, so all these polygons are rotatable all the time. However, there is a `Rotatable` property, which allows to change this value. You can organize an interesting addition to this **Form_ChatoyantPolygon.cs**. Add a small menu of one line to change the possibility of rotation for each polygon. If you need any example for this work, you can look into the **Form_RegPoly_Variants.cs**, which uses similar type of context menu. Via the menu you can declare each chatoyant polygon either rotatable or not.

One more remark about zooming. In the current version the `StartScaling()` method gets two parameters, of which the second one is the shape of a node. Zooming is supposed to start only with the strip nodes, so this method starts with the check of the shape. You can easily move this check one level up into the `OnMouseDown()` method and call the `StartScaling()` method only for strips. In this way the `StartScaling()` method will have only one parameter and will look similar to the `StartRotation()` method. I even put the needed lines of code into the `OnMouseDown()` method, but commented them.

The example with the chatoyant polygons is the first to demonstrate that with the help of those primitive nodes not only the simple objects can be moved, but much more complicated.

There is one small button with the plus sign on it in the **Form_ChatoyantPolygons.cs** (**figure 6.6**); this button allows to add more polygons into the form. On clicking this button, you open the **Form_AddChatoyantPolygon.cs** (**figure 6.8**). Everything is movable in this form and all the changes, made by user, are saved on closing the form and used on next opening.

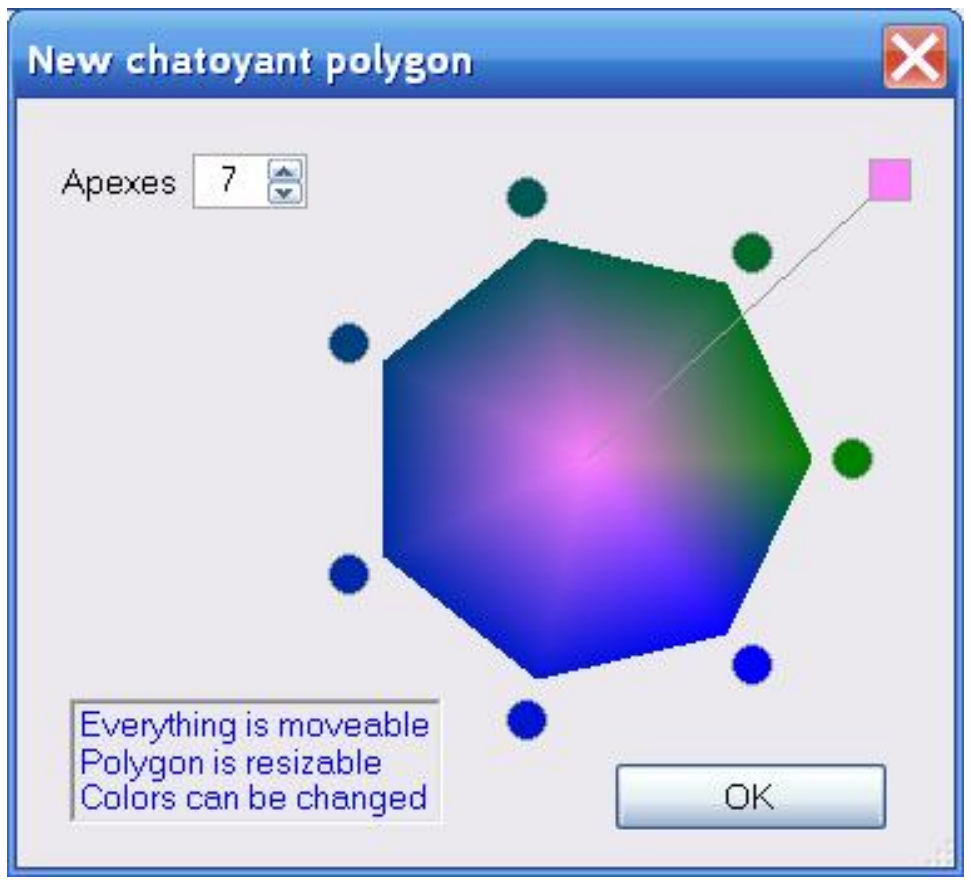

**Fig.6.8 Form_AddChatoyantPolygon.cs**

The dominant object in this form is the regular chatoyant polygon; this polygon is resizable by any border point. Small circles next to the vertices are for setting the colors in the vertices. Click the circle and the standard `ColorDialog` will be opened. The small square in the corner is for the same purpose, but allows to change the color in the central point. Small circles always stay next to the vertices; that square can be moved (by its border) to any place. The control in the top left corner allows to change the number of vertices in the polygon. Control with its comment is an object of the `CommentedControl` class, so the comment can be placed anywhere in relation to that control. When I write that some element is in the top left corner or something like this, it means only the position of the element on this figure. All the objects can be moved to any place, so the view of your variant of this form can be absolutely different, but the form will still produce the new chatoyant polygon with the number of vertices and the colors determined by you. Certainly, this will happen only after you click the OK button. Regardless of the place, where you decide to put this button.

This form is the first and very simple example of the idea that I unite under the term *user-driven applications*. The only purpose of the **Form_AddChatoyantPolygon.cs** is to prepare the new polygon by declaring the number of vertices and the colors in all the basic points. The form contains several objects to fulfil the job. It does not matter where I, as a designer, prefer to put all these elements and where you move them. The form works regardless of how you rearrange it. We will return to the discussion of the user-driven applications much later.

## Dorothy's house

File: **Form_DorothyHouse.cs**
Menu position: *Graphical objects – Basic elements – Polygons – Dorothy's house*

All the examples from the beginning of the book and up till this one were of an abstract type. This example is especially for those, who want to see something real.

I used the examples with different houses in nearly all of my previous demo applications. From the programming point of view, these examples are similar to demonstration of rectangles and polygons, but maybe they are making the set of examples more interesting. It is possible that the **Form_DorothyHouse.cs** is simpler than the previous example with the chatoyant polygons, but it makes the set of examples more diverse.

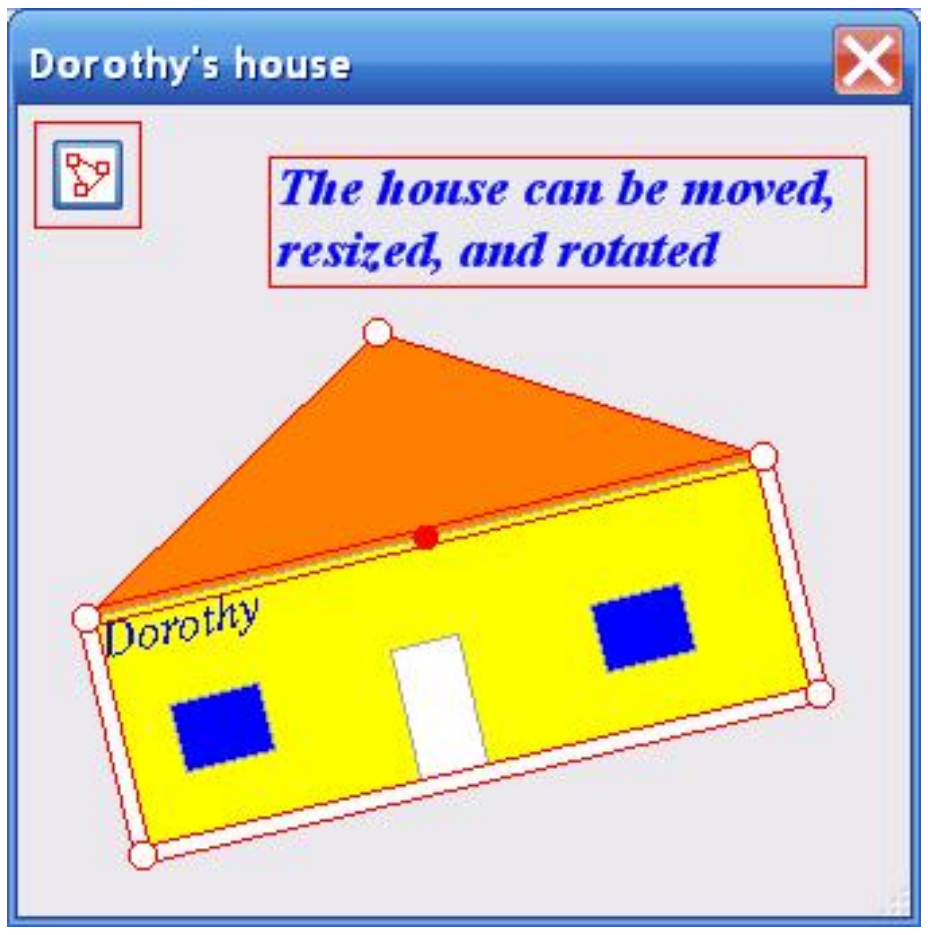

**Fig.6.9 Form_DorothyHouse.cs**

The house, turning around some point, looks like a flying house, and I think that the most famous of all the flying buildings was the Dorothy's house, so I called this class `DorothyHouse` and put the name of the owner on the building.

**Figure 6.9** shows the house together with its cover. There is a small circular node at the edge of the roof, which allows to change the height of the roof. I cannot remember seeing in Kansas a single house with the ridge moved anywhere from the center, so in this example the roof top can be moved only up or down, but not to the sides. The `DorothyHouse` class has one specific rule of resizing: when either left or right side of the house is moved, the house changes symmetrically on both sides. Because of this rule, it is easier to keep in the fields the positions of all five basic points, which belong to the perimeter of the house. There is also one more point, painted as a small red circle, around which the house can be rotated. The decision on the center of rotation affects the equations, used in calculations, but they are always at the same level of simplicity. Anyway, in this example the house can be rotated around the middle point of the ceiling.

```
    public class DorothyHouse : GraphicalObject
    {
        Point ptAnchor;                    // rotation center; middle of the ceiling
        Point ptLT, ptRT, ptRB, ptLB;      // four corners of the building
        Point ptRidge;                     // top of the roof
        double angle;                      // radians
        double width;                      // width of the house
        double height;                     // height of the house
        double roof;                       // height of the roof
```

The angle of the house is calculated as an angle of the ceiling.

The cover for a `DorothyHouse` object consists of 10 nodes: five circular nodes on the vertices, four strips along the sides of the rectangular part, and the polygonal node (pentagon) to cover the whole area. Because the house can be rotated and the direction, in which each side can be moved, depends on the current angle of the house, I changed the cursor over the strip nodes to `Cursors.Hand`.

```
public override void DefineCover ()
{
    CoverNode [] nodes = new CoverNode [4 + 1 + 4 + 1];
    int nr = 5;
    nodes [0] = new CoverNode (0, ptLT, nr);
    nodes [1] = new CoverNode (1, ptRT, nr);
    nodes [2] = new CoverNode (2, ptRB, nr);
    nodes [3] = new CoverNode (3, ptLB, nr);
    nodes [4] = new CoverNode (4, ptRidge, nr);
    nodes [5] = new CoverNode (5, ptLT, ptLB);
    nodes [6] = new CoverNode (6, new PointF [] { ptLT, ptRT }, Cursors .Hand);
    nodes [7] = new CoverNode (7, ptRT, ptRB);
```

```
    nodes [8] = new CoverNode (8, ptLB, ptRB);
    nodes [9] = new CoverNode (9, new PointF [] {ptLT, ptRidge, ptRT, ptRB, ptLB});
    cover = new Cover (nodes);
}
```

You can see from the code of the `DefineCover()` method that one node on the borders of rectangular part of a house is defined differently from three others. And from **figure 6.9** you can see that this strip node is visualized differently, when cover is painted. The small change in the definition of one strip node was done purposely, so that it would be possible to see the cover and the red spot, marking the center of rotation.

When a node is defined by a single point, it is organized as a circular node; when visualized, such node is filled with white. When a node is defined by two points, it is organized as a strip and is also filled with white on visualization. Both types of nodes have the default cursor `Cursors.Hand`. The definition of a node by an array of points is the standard way to define a polygonal node; by default such node gets the `Cursors.SizeAll` cursor. However, if an array of points has less than three points, the polygonal node cannot be constructed; in such case the result depends on the number of points in the provided array. If there is only one point in array, then a circular node is constructed; if there are two points, then a strip node. But in both of these cases the node still gets the default cursor of a polygonal node and is not filled on visualization. The second thing is just what I need, because I do not want the strip node on the ceiling to close the view of the red circle. But I have to declare another cursor; the `Cursors.Hand`, which is used for the nodes on all other sides of the house. That was the reason, why one node of the house got the special design.

For the forward movement of the house, all the basic points must be moved synchronously.

```
    public override void Move (int dx, int dy)
    {
        Size size = new Size (dx, dy);
        ptLT += size;
        ptRT += size;
        ptRB += size;
        ptLB += size;
        ptRidge += size;
        ptAnchor += size;
    }
```

Movement of each node is described in the `MoveNode()` method. The movement of the polygonal node ( `i = 9` ) results in moving of the whole house, so in this case the `Move()` method must be called. An attempt to move any other node is described in similar way. First, the current mouse position must be checked. There are no restrictions on the mouse movements, but there are restrictions on the house sizes. For example, the mouse can grab the side of the house and then there can be an attempt to move the mouse across the opposite side of the house. Without any restrictions, this will turn the house inside out. This is not allowed, so the change of the house is stopped somewhere on the way, though the cursor can continue its movement. The checking of the possibility for the resizing is based on whether the mouse and another basic point are on the same side of the line or on different sides. The selection of these point and line depends on the number of the caught node. Here is the part of the `MoveNode()` method for the case of moving the floor ( `i = 8` ).

```
public override bool MoveNode (int i, int dx, int dy, Point ptM, MouseButtons btn)
{
   if (btn == MouseButtons .Left)
   {
      switch (i)
      {
         … …
         case 8:          // floor
             if (!Auxi_Geometry .SameSideOfLine (ptLT, ptRT, ptM, ptRidge))
             {
                distToCeiling =
                           Auxi_Geometry .DistanceToLine (ptM, ptLT, ptRT, out pt);
                height = Math.Min (Math.Max (minHeight, distToCeiling), maxHeight);
                CalculateBasicPoints ();
             }
             break;
```

When the floor is moved, the mouse point and the upper point of the roof (`ptRidge`) have to be on the opposite sides of the ceiling (`ptLT, ptRT`). If this condition is satisfied, then the distance between the cursor and the ceiling is calculated.

```
        distToCeiling = Auxi_Geometry .DistanceToLine (ptM, ptLT, ptRT, out pt);
```

House has a range for changing its height, so this distance between the mouse and the ceiling can be used as the new height of the house only if it is inside this range.

```
        height = Math.Min (Math .Max (minHeight, distToCeiling), maxHeight);
```

On any resizing of the house, nearly all the basic points take the new positions; the new points are calculated by the `CalculateBasicPoints()` method.

```
    private void CalculateBasicPoints ()
    {
        ptLT = Auxi_Geometry .PointToPoint (ptAnchor, angle + Math .PI, width / 2);
        ptRT = Auxi_Geometry .PointToPoint (ptLT, angle, width);
        ptLB = Auxi_Geometry .PointToPoint (ptLT, angle - Math .PI / 2, height);
        ptRB = Auxi_Geometry .PointToPoint (ptRT, angle - Math .PI / 2, height);
        ptRidge = Auxi_Geometry.PointToPoint (ptAnchor, angle + Math.PI / 2, roof);
    }
```

This method uses the `Auxi_Geometry.PointToPoint()`, which calculates the second point, if the first point is known and the angle from the first point to the second.

Rotation of the `DorothyHouse` object is similar to rotation of rectangle, described in the chapter *Rotation*. Rotation can be started by pressing the house with the right button at any inner point.

```
    private void OnMouseDown (object sender, MouseEventArgs e)
    {
        if (mover .Catch (e .Location, e .Button))
        {
            if (e .Button == MouseButtons .Right)
            {
                GraphicalObject grobj = mover .CaughtSource;
                if (grobj is DorothyHouse)
                {
                    (grobj as DorothyHouse) .StartRotation (e .Location);
                }
```

The `DorothyHouse.StartRotation()` method calculates the compensation angle between the angle to the mouse (from the center of rotation) and the angle of the house. The angle to the mouse is calculated from the rotation center, for which the middle of the ceiling was chosen.

```
    public void StartRotation (Point ptMouse)
    {
        double angleMouse = -Math .Atan2 (ptMouse .Y – ptAnchor .Y,
                                          ptMouse .X – ptAnchor .X);
        compensation = Auxi_Common .LimitedRadian (angleMouse - angle);
    }
```

This compensation is fixed for the whole time of rotation (between the **`MouseDown`** and **`MouseUp`** events) and allows to calculate the angle of the house depending on the current angle of the mouse throughout the rotation. The angle of the house, in its turn, is used to calculate all the basic points.

```
public override bool MoveNode (int i, int dx, int dy, Point ptM, MouseButtons btn)
{
    … …
    else if (catcher == MouseButtons .Right)
    {
        double angleMouse = -Math.Atan2 (ptM.Y - ptAnchor.Y, ptM.X - ptAnchor.X);
        angle = angleMouse - compensation;
        CalculateBasicPoints ();
    }
    DefineCover ();
```

An interesting feature of the `DorothyHouse` class is that its `DefineCover()` method is called from inside the `MoveNode()` method not only for rotation, but for the forward movement also. This is the easiest way to work with this class, because of that small addition in design (the mirror movement of both sides on resizing) the movement of any node causes the relocation of nearly all other nodes; in such case it is much easier to redefine the cover.

The DorothyHouse class demonstrates again one important thing: there is nothing special in the design of covers for the objects, involved in rotation. The same objects can be involved in rotation or not; this is provided by adding several lines of code into the OnMouseDown() method of the class, but not by anything special in its cover design. Certainly, if the objects of the class are going to be rotated, then some sort of the StartRotation() method must be written.

# Curved borders. N-node covers

Objects can be resized by their borders; the standard way of doing it is to press the border at any point and move it. In the case of straight borders the solution is obvious: cover a straight segment either by a strip node or by a narrow rectangular node. But not all the objects have the borders, consisting of the straight lines. For resizing of the objects with the curved borders the special covers are used, which are called *N-node covers*.

File: **Form_NnodeCovers.cs**
Menu position: *Graphical objects – N-node covers*

At **figure 7.1** you can see three graphical objects with the curved borders: circle, rounded strip, and ring. Two of these shapes have already appeared at **figure 2.1** in the chapter *The first look at nodes and covers*. In the first example of this book (**Form_Nodes.cs**), a circle and a strip had the covers consisting of a single node; those objects were used only for moving. Though the shape is identical, in the **Form_NnodeCovers.cs** the objects are used both for moving and resizing, so their covers must be different. The form of an object does not absolutely determine the design of cover; the cover depends on what you want to do with this object. The covers of the objects from **figure 7.1** reflect my view on how these objects must be moved and resized.

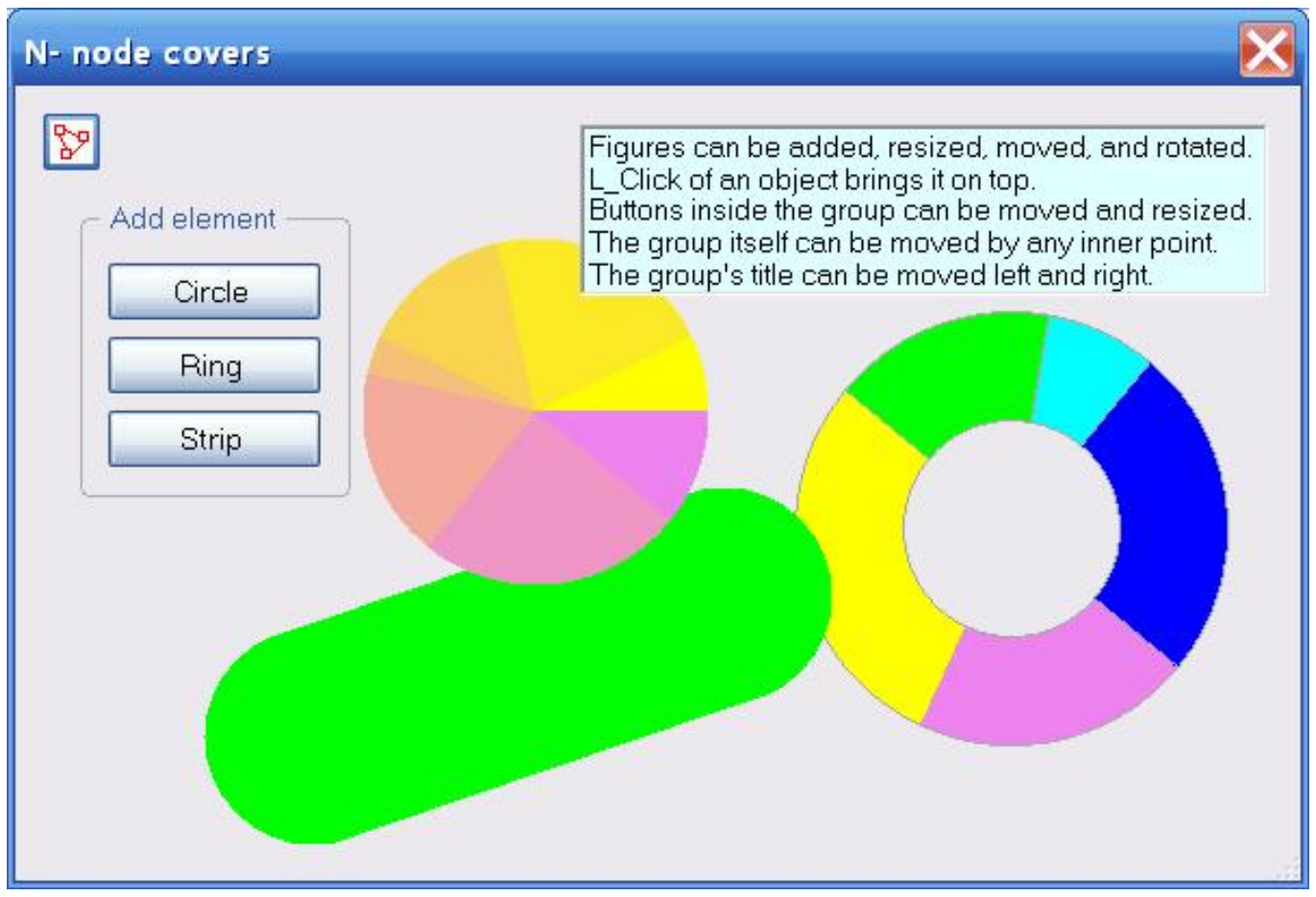

**Fig.7.1** Objects with the curved borders

<u>Circle</u> must be moved by any inner point and resized by any border point. There must be a minimum radius restriction, so that a circle would not disappear from view as a result of an unrestricted squeezing.

<u>Ring</u> must be moved by any inner point and resized by any point of its inner and outer borders. There must be a restriction on the minimum inner radius, so that a ring would not turn into circle. Also, there must be a restriction on the minimum width of a ring, so that a ring would not disappear.

<u>Strip</u> must be moved by any inner point and there must be two different types of resizing. By moving any point of the curved part of border, the length of a strip can be changed, but its width is not affected. On catching the border along its straight parts, the width of a strip can be changed. Throughout such change of the width, the length of the straight part of the strip stays unchanged, but the radius of semicircles is equal to half of the strip's width, so the overall length is changing.

What was the common feature for the covers, which allowed the resizing of all the previous examples with the straight borders? The borders were covered with the narrow nodes. Those nodes were wide enough to find and grab them easily, but at the same time they were narrow, so the real point of the catch was never far away from the nearest border point. The tiny difference between those two points allowed me to ignore it and to identify the mouse point with the border point.

To resize the objects with the curved borders, I use exactly the same technique: I have to cover the arbitrary curved border with the narrow sensitive area, which allow me to grab the border at any point and at the same time allow to identify the point of the mouse press with the nearest border point. I cannot cover such curved border with a single node, so I use a big amount of small nodes; the united area of all these nodes works like a sensitive strip. The covers, which use the proposed technique of covering the borders with the big number of small nodes, are called ***N-node covers***. How to place those nodes and which form they must be – these things depend on each developer and his preferences. Users do not see these details, they only see if the objects can be resized easily and without mistakes, or are there some problems on resizing.

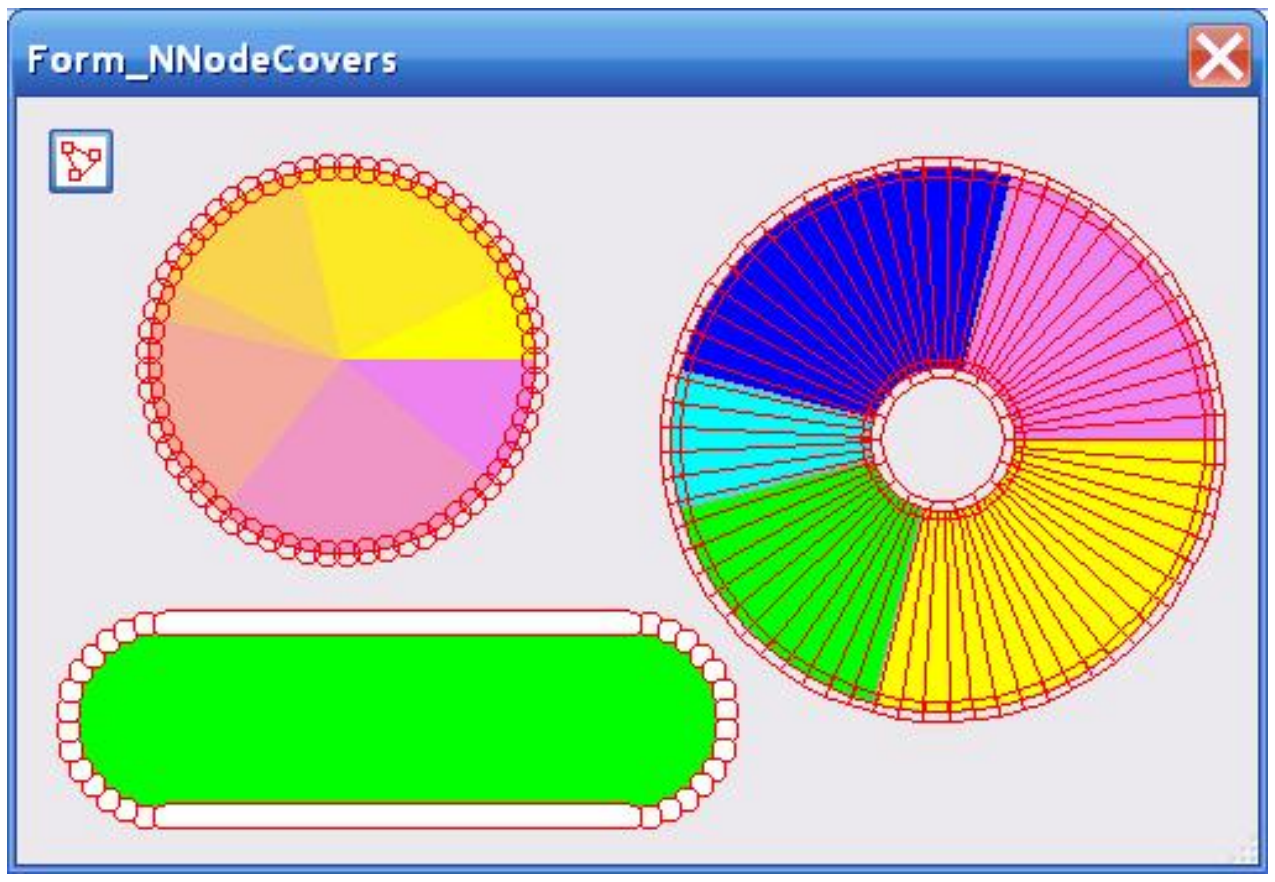

**Fig.7.2** Objects with their covers

The most common technique, which I prefer to use, is to cover the curved border with a set of overlapping circular nodes. The neighbouring nodes overlap, so that there would be no gaps along the border and no such places, where the width of a sensitive strip diminishes to a single point. The advantage of using a set of overlapping circles is in the simplicity of their calculation (positioning), but this technique is not the only possible solution; the borders of the rings in the **Form_NnodeCovers.cs** are covered by a set of polygonal nodes. Each small polygonal node has a form of a trapezoid with two sides placed along the radiuses. The trapezoids do not overlap, but stay side by side; all the trapezoids have the same height, so the width of such strip is the same along the entire border. **Figure 7.2** demonstrates the objects together with their covers.

Circle (of the `CircleNR` class) has the simplest cover of these objects. The first thing to do is to calculate the number of circular nodes, which are needed to cover the perimeter. I decided about the 5 pixels radius for all the small nodes (`nrSmall`) and the 8 pixels distance between the neighbours (`distanceNeighbours`). This produces a bended strip of at least 6 pixels width along the perimeter of a circle; my experience shows that a strip of 3 pixels on each side of a line makes the moving of such a line easy enough.

```
private void NodesOnCircle ()
{
    nNodesOnCircle = Convert.ToInt32 ((2 * Math.PI * radius) / distanceNeighbours);
}
```

The cover consists of the predefined number of small circular nodes (`nNodesOnCircle`), which are used for resizing, and a big circular node, which is used for moving the whole object. As usual, the biggest node must the last one in the cover, but there is one more reason to put this big circular node at the end. If you change the order of nodes and put the big circular node at the head, then it will close that half of each small node, which is inside the border; the border would be sensitive only from outside, which would significantly worsen the resizing.

```
public override void DefineCover ()
{
    CoverNode [] nodes = new CoverNode [nNodesOnCircle + 1];
    for (int i = 0; i < nNodesOnCircle; i++)
    {
        nodes [i] = new CoverNode (i, Auxi_Geometry .PointToPoint (center,
                                 2 * Math .PI * i / nNodesOnCircle, radius), nrSmall);
    }
    nodes [nNodesOnCircle] = new CoverNode (nNodesOnCircle, center,
                                          Convert .ToInt32 (radius), Cursors .SizeAll);
    cover = new Cover (nodes);
    cover .SetClearance (false);
}
```

All the nodes in this cover are circular, so the decision on particular movement started by catching one or another node cannot be based on its shape, but only on the number. This is easy enough to do, as only the last node in the cover is responsible for moving, but all others – for resizing, and it does not matter, which of these small circles is pressed to start the movement. The only thing, which is done on moving any of the small nodes, is to calculate the new radius.

```
public override bool MoveNode (int i, int dx, int dy, Point ptM, MouseButtons btn)
{
    bool bRet = false;
    if (btn == MouseButtons .Left)
    {
        if (i == nNodesOnCircle)
        {
            Move (dx, dy);
        }
        else
        {
            int nRadNew = Convert .ToInt32 (Auxi_Geometry .Distance (center, ptM));
            if (nRadNew != radius && nRadNew >= nMinRadius)
            {
                radius = nRadNew;
                bRet = true;
            }
        }
    }
    else if (btn == MouseButtons .Right)
    {
```

```
        double angleMouse = Auxi_Geometry .Line_Angle (center, ptM);
        angle = angleMouse - compensation;
        bRet = true;
    }
```

The forward movement of a circle and its rotation were already discussed, so there is nothing new in these parts.

Ring (of the `RingNR` class) has more complicated cover than circle, because there are two resizable borders instead of one and also because the area of a ring has to be covered by a set of nodes to provide the forward movement of the ring. It is easy to cover both the inner and outer borders by two sets of small circular nodes, as was done with the circle, but in order to show that the overlapping small circles is not the only possible solution, I replaced those small nodes with two sets of polygons - trapezoids. Each trapezoid occupies the 10 pixels part of the border, on which it is placed, and spreads for 4 pixels outside and inside of the curve (so the height of each trapezoid is 8 pixels). Based on the standard width of each trapezoid `(width)`, the needed numbers of them to cover both the inner `(nNodesOnInner)` and outer `(nNodesOnOuter)` borders of the ring are calculated.

The area of the ring itself is also covered with a set of trapezoids. They have to cover the area of a ring without any gaps, so they can be calculated in different ways. To decrease the number of calculations, their sides are placed at the same radii, as the sides of the small trapezoids on the outer border, so the number of nodes to cover the area of a ring `(nNodesPoly)` is equal to the number of nodes on the outer border. Thus the number of nodes in all three mentioned sets depends on two radiuses (outer and inner) and is calculated by the `NodeNumbers()` method.

```
        private void NodeNumbers ()
        {
            nNodesOnOuter = Convert .ToInt32 ((2 * Math .PI * rOuter) / width);
            nNodesOnInner = Convert .ToInt32 ((2 * Math .PI * rInner) / width);
            nNodesPoly = nNodesOnOuter;
        }
```

The `RingNR.DefineCover()` method is a bit long to include it here, but the code is very simple; each polygonal node (trapezoid) is described by the four corner points, which are easily calculated from the angle and the known radius. The most important thing is the order of nodes in the cover:

1. Nodes on the outer border.
2. Nodes on the inner border.
3. Nodes to cover the area of a ring.

All the nodes in this cover are polygonal, so the decision on particular movement, started by one or another node, cannot be based on its shape but only on the number. The `RingNR.MoveNode()` method has to include more checking, as the moving of the outer border has to be checked against one restriction (minimum allowed width of the ring) and the proposed movement of the inner border has to be checked against the same minimum allowed width, but also against the minimum allowed radius of this border. Though there can be a lot of nodes in the cover, the `MoveNode()` method is simple, as the reaction on moving a node is not specific for each of them, but only depends on which group this node belongs to. Here is the part of the method for forward movement and resizing.

```
public override bool MoveNode (int i, int dx, int dy, Point ptM, MouseButtons btn)
{
    bool bRet = false;
    if (btn == MouseButtons .Left)
    {
        if (i >= nNodesOnOuter + nNodesOnInner)
        {
            Move (dx, dy);
        }
        else if (i >= nNodesOnOuter)
        {
            int newInner = Convert.ToInt32 (Auxi_Geometry .Distance (center, ptM));
            if (minInner <= newInner && newInner <= rOuter - minWidth)
            {
                rInner = newInner;
                bRet = true;
            }
```

```
        }
        else
        {
            int newOuter = Convert.ToInt32 (Auxi_Geometry.Distance (center, ptM));
            if (rInner + minWidth <= newOuter)
            {
                rOuter = newOuter;
                bRet = true;
            }
        }
    }
```

The forward movement of a ring and its rotation are absolutely identical with the case of a circle. A ring has to have an angle, an identical `StartRotation()` method to calculate the compensation at the start of rotation, and the same technique of using this compensation angle to calculate the real angle of a ring throughout the rotation..

Strip (of the `StripNR` class) has more variations in resizing and includes different types of nodes into its cover. Usually the nodes are included into a cover beginning from the small ones, but here I changed the order slightly and put the strip nodes on the straight parts of border ahead of the circular nodes, which cover the curves. This cover includes not a solitary circular node, which can be a problem to find, but a set of such nodes, which cover a good piece of a border, so it really does not matter whether those circular nodes are at the head of a cover or a bit later. The small nodes on the curves of the border are placed in the same way, as they were placed on the border of a circle: each small node has a radius of 5 pixels and is placed not more than 8 pixels from the neighbour. Thus we receive the order of nodes in the cover of a strip.

1. Two strip nodes along to straight parts of the border.
2. Two sets of circular nodes along the curved parts of the border.
3. One big strip node to cover the whole area.

```
    public override void DefineCover ()
    {
        PointF [] pts = CornerPoints ();
        CoverNode [] nodes;
        int nNodes, k0;
        double small_angle = Math .PI / (nNodesOnHalfCircle - 1);
        nNodes = 2 + 2 * nNodesOnHalfCircle + 1;
        nodes = new CoverNode [nNodes];
        nodes [0] = new CoverNode (0, pts [0], pts [1], nrSmall);
        nodes [1] = new CoverNode (1, pts [2], pts [3], nrSmall);
        k0 = 2;
        for (int i = 0; i < nNodesOnHalfCircle; i++)
        {
            nodes [k0 + i] = new CoverNode (k0 + i,
                Auxi_Geometry .PointToPoint (ptC0,
                       angle + Math .PI / 2 + i * small_angle, radius), nrSmall);
        }
        k0 += nNodesOnHalfCircle;       // 2 + nNodesOnHalfCircle
        for (int i = 0; i < nNodesOnHalfCircle; i++)
        {
            nodes [k0 + i] = new CoverNode (k0 + i,
                Auxi_Geometry .PointToPoint (ptC1,
                       angle - Math .PI / 2 + i * small_angle, radius), nrSmall);
        }
        k0 = nodes .Length - 1;
        nodes [k0] = new CoverNode (k0, ptC0, ptC1, radius, Cursors .SizeAll);
        cover = new Cover (nodes);
        cover .SetNodeClearance (nodes .Length - 1, false);
    }
```

There are two restrictions on the minimum allowed sizes to avoid an accidental disappearance of a strip: the minimum radius of semicircles at the ends is 12 pixels (this radius determines the minimum width of a strip as 24 pixels) and the minimum length of the straight part is 20 pixels. As usual, these restrictions are taken into consideration inside the `MoveNode()` method, when one or another node is caught for moving.

The **Form_NnodeCovers.cs** is populated by different objects with the N-node covers. Part of what is going in the `OnMouseDown()` and `OnMouseUp()` methods of the form is similar to all the previous examples, but there are also absolutely new features, which originate from this special type of covers. First, let us look at more familiar parts.

```
private void OnMouseDown (object sender, MouseEventArgs e)
{
    ptMouse_Down = e .Location;
    if (mover .Catch (e .Location, e .Button))
    {
        GraphicalObject grobj = mover .CaughtSource;
        if (e .Button == MouseButtons .Left)
        {
            if (grobj is StripNR)
            {
               (grobj as StripNR).StartLengthChange (e.Location, mover.CaughtNode);
            }
        }
        else if (e .Button == MouseButtons .Right)
        {
            if (grobj is ElementNR)
            {
                (grobj as ElementNR) .StartRotation (e .Location);
            }
        }
    }
}
```

The resizable strip can be looked at as a rectangle with the semicircular additions on two opposite sides of it. When any small node on the curve is moved, then only the length of the straight part of this strip is changed. This change goes in such a way that the opposite end of the strip is not moved at all, but the curve, which was pressed, can be moved one way or another. The different points of the curve are at different distances from the rectangular part of the same strip; the distance between the mouse and the rectangular part is not supposed to change throughout this resizing, so, when any of those small nodes is caught for forward movement, then at this starting moment of resizing the distance between the mouse and the straight part of the strip must be calculated. This is done by calling the `StripNR.StartLengthChange()` method. The call to this method is similar to the use of the `Polygon_Chatoyant.StartScaling()` method, which was mentioned in the section about the chatoyant polygons. The `StartLengthChange()` method has to receive the number of the node as the second parameter, because there are two ends of the strip; this number allows to identify which of the ends is moved.

```
    public void StartLengthChange (Point ptMouse, int iNode)
    {
        PointF [] pts = CornerPoints ();
        if (iNode < 2 + nNodesOnHalfCircle)
        {
            fStartingDistanceToRect =
                           Auxi_Geometry .DistanceToLine (ptMouse, pts [1], pts [2]);
        }
        else
        {
            fStartingDistanceToRect =
                           Auxi_Geometry .DistanceToLine (ptMouse, pts [0], pts [3]);
        }
    }
```

First, the corners of the rectangular part of the strip are calculated. After it, the distance is calculated between the mouse and the nearest side of that rectangle. The number of the node is essential for this part of calculations.

Now let us look at what is really unique for the N-node covers. The biggest difference between the N-node covers and the covers from all the previous examples is not in the number of nodes, but in the fact that the number of nodes in the cover can vary. For all the previous examples only the shape of an object or its required resizing determined the number of nodes in the cover. With the examples in the **Form_NnodeCovers.cs** it is different: the number of nodes is determined by the size of object; this can become a big problem, even a disaster, if you try to use such covers in the same way as others.

For all the previously demonstrated objects with the possibility of resizing there were no restrictions on calling the `DefineCover()` method on any change of the sizes. It is done, for example, in all the classes for different polygons. When the number of nodes in the cover is fixed for the whole class (or even for an object with the fixed number of vertices) and does not depend on the size of the particular object, then the `DefineCover()` method can be used at any moment.

When an object is caught by the mover, it is caught by one or another node of the cover. Movement of the node is translated into the movement of an object by the `MoveNode()` method, in which the exact type of movement (forward movement of the whole object, resizing or reconfiguring) is usually determined according to the identification number of the caught node. If the number of nodes in the cover is changed during the period of time, when an object is grabbed by the mover, then you can expect a lot of strange situations. The number of the caught node is determined at the beginning of any movement and is not going to change in any way until the end of the movement. But if the total number of nodes in the cover is changing throughout the resizing and the whole set of nodes consists of nodes with different types of movements (for example, the `RingNR` class has three such groups), then the specified number can easily move from one group to another or even go outside the whole range of numbers for nodes. In the first case the type of movement will simply change somewhere on the way; this is definitely an unexpected thing for anyone; in the second case the program can crash.

To understand the first situation, let us look once more at the `RingNR` class and remember the order of nodes in its cover: first the nodes on the outer border, then the nodes on the inner border, and then the nodes for the whole area. Suppose that you pressed the inner border and began to squeeze the hole. With the diminishing hole, less and less nodes are needed to cover the inner border. If you would be constantly changing the cover according to the changing size, then at some moment the specified number of the caught node will move into the set of nodes that cover the whole area of the ring and instead of squeezing the hole you will see the movement of the ring. It is definitely not the expected thing. The rule of N-node covers prevents you from running into such a situation.

**The rule of N-node covers.** The `DefineCover()` method cannot be called from inside the `MoveNode()` method, and though the cover has to be changed because the sizes of an object have changed, but the call to the `DefineCover()` method must be postponed until the release of an object. And before this call is made the new number of nodes must be calculated.

The redesign of cover for a previously caught object is done, when this object is released. At this moment the new number of needed nodes is calculated according to the new sizes and the `DefineCover()` method is called.

```
private void OnMouseUp (object sender, MouseEventArgs e)
{
    int iWasCaught;
    if (mover .Release (out iWasCaught))
    {
        GraphicalObject grobj = mover [iWasCaught] .Source;
        if (e .Button == MouseButtons .Left)
        {
            if (grobj is ElementNR)
            {
                RedefineCover (iWasCaught);
                if (dist <= 3)
                {
                    PopupFigure (grobj .ID);
                }
            }
        }
    }
}
```

Based on the number of the caught object `(iWasCaught)`, the class of the caught object is determined and the `RedefineCover()` method of this particular class is called.

```
private void RedefineCover (int iObj)
{
    GraphicalObject grobj = mover [iObj] .Source;
    if (grobj is RingNR)
    {
```

```
                (grobj as RingNR) .RedefineCover ();
                Invalidate ();
            }
            else if (grobj is CircleNR)
            {
                (grobj as CircleNR) .RedefineCover ();
                Invalidate ();
            }
            else if (grobj is StripNR)
            {
                (grobj as StripNR) .RedefineCover ();
                Invalidate ();
            }
        }
```

This is a normal way of calling the `RedefineCover()` method for the caught object, but here the order of an object is sent as a parameter; then the type of an object is determined. Instead, the direct passing of the element can be used.

```
        private void OnMouseUp (object sender, MouseEventArgs e)
        {
            … …
                    if (grobj is ElementNR)
                    {
                        RedefineCover (grobj);
```

Certainly, the small change of the `RedefineCover()` method would be also needed.

```
        private void RedefineCover (GraphicalObject grobj)
        {
            if (grobj is RingNR)
            {
                (grobj as RingNR) .RedefineCover ();
                Invalidate ();
            }
```

Any class, which uses the N-node cover, must have the `RedefineCover()` method. The method is always very simple: calculate the new number of nodes according to the new sizes and call the `DefineCover()` method. Here is the `RedefineCover()` method for the `RingNR` class.

```
        public void RedefineCover ()
        {
            NodeNumbers ();
            DefineCover ();
        }
```

Is the mentioned rule of the N-node covers a mandatory one or are there any exceptions? Yes, there are exceptions. You have to understand those two cases, when the rule has to be used:

1. When the number of the caught node can move from the group of nodes with one behaviour to the group of nodes with different behaviour.
2. When the number of the caught node can move outside the range of nodes.

In all other cases the `DefineCover()` method can be called from inside the `MoveNode()` method without any problems. For example, you can see such thing in the `StripNR` class. The resizing of a strip with the requirement for another number of nodes can occur only on moving the two strip nodes on the border, but these nodes never change their numbers (0 or 1); the caught strip node will have the same number in any new cover. (That is why I put these strip nodes ahead of the circular nodes in the cover.) The resizing caused by moving the circular nodes, which cover the curves, does not change the number of nodes, so again the number of the caught node will be the same in the new cover as at the starting moment of such movement. Thus there are no dangerous situations with calling the `DefineCover()` method from any part of the `StripNR.MoveNode()` method.

For the `CircleNR` and `RingNR` classes the situation is different: you cannot call the `DefineCover()` method from inside the `MoveNode()` method, but only after the release of the caught object.

# Transparent nodes

It can be a very big problem to design the covers for some objects with the holes or for objects with the unusual form. For some of them the standard way of covering the area of object by a set of ordinary nodes can demand a lot of painstaking work with equations. Surprisingly, the covers for the same objects can be easily organized with the help of the transparent nodes.

Ring was one of the examples, which I used to explain the idea of the *N-node covers*. As you can see at **figure 7.2**, covers for the objects of the `RingNR` class consist of three different sets: nodes on the outer border, nodes on the inner border, and the polygonal nodes to cover the area of a ring. The first two sets of nodes in such cover provide the resizing of a ring. Suppose that you want to develop the movable but non-resizable rings. At first it looks like you can take the cover for the `RingNR` class, get rid of all the border nodes, leave only an array of polygons to cover the area of a ring, and everything is going to be OK.

```
public override void DefineCover ()
{
    PointF [] pts = new PointF [4];
    CoverNode [] nodes = new CoverNode [nPoly];
    pts [0] = Auxi_Geometry .PointToPoint (center, 0, rInner);
    pts [1] = Auxi_Geometry .PointToPoint (center, 0, rOuter);
    for (int i = 0; i < nPoly; i++)
    {
        double angle = 2 * Math .PI * (i + 1) / nPoly;
        pts [2] = Auxi_Geometry .PointToPoint (center, angle, rOuter);
        pts [3] = Auxi_Geometry .PointToPoint (center, angle, rInner);
        nodes [i] = new CoverNode (i, pts);
        pts [0] = pts [3];
        pts [1] = pts [2];
    }
    cover = new Cover (nodes);
}
```

In reality it is not so simple. The problem is that the united area of all those trapezoids is not equal to the area of a ring. It was the same in the previous example of a ring, but there the small discrepancies were covered by the nodes along the borders. If we cover the area of a ring with the same trapezoids with their corners placed on two circles, we receive some problems on both borders:

- There will be small non-sensitive areas along the outer border.
- Small parts of the hole will become sensitive, though they do not belong to the ring.

The difference between the exact area of a ring and the area of a cover is not big; this difference also decreases, if the number of polygons is increased. You can ignore this problem because with the big number of nodes the discrepancy between the circle and its approximation by a set of polygons (united area of all those trapezoids) can be really small, but this is not a good-looking solution. Much better and really good solution can be achieved with the use of transparent node. Such cover consists of only TWO nodes!

```
public override void DefineCover ()
{
    CoverNode [] nodes = new CoverNode [2];
    nodes [0] = new CoverNode (0, center, rInner, Behaviour .Transparent);
    nodes [1] = new CoverNode (1, center, rOuter, Cursors .SizeAll);
    cover = new Cover (nodes);
}
```

This is really an elegant cover, consisting of two circular nodes. One of them – the bigger one – covers the whole area of a ring together with the circular hole inside. If this node would be left without any additions, then the ring would become movable not only by any point of its area, which is correct and expected, but also by any point of its hole, which is definitely a mistake. To solve this problem, the second circular node is used, which covers the hole and "burns out" this part of the bigger sensitive area. To do such a thing, this node must have the parameter `Behaviour.Transparent` and must be included into the cover ahead of the bigger node. The order of nodes is extremely important in this case.

I have mentioned not once that mover analyses each cover node after node according to their order in the cover. When the mover sees the node with the `Transparent` behaviour, it not only ignores this node, but also skips all further nodes of the same cover. Beginning from this node, an object becomes transparent for the mover and the mover looks for something farther on in its queue of objects to grab and move. Two facts contribute to an extraordinary importance and, at the first glance, a strange use of transparent nodes: the invisibility of nodes and nonequivalence of the area of object and area of its cover.

For the majority of objects these two areas can be equivalent or nearly the same. To make an object movable by any inner point, its whole area must be covered by some set of nodes. If an object is movable but not resizable, then the sensitive area is limited by the area of object, so the area of cover can be equivalent to the area of object. When an object is resizable, then the resizing is usually done by the border points, in which case the border is covered by the sensitive strip, so the area of cover becomes slightly wider than the object itself. All this is used in the simple and most obvious cases; for the complicated cases of nontrivial areas only the transparent nodes can help.

In some situations, the use of transparent nodes allows to change the covers with a lot of nodes (and a lot of their calculations) into really simple; the above shown change of cover for a ring is of such a case. Further on I will show an example with a crescent; which is remarkable from two points. First, it shows a power of using the transparent nodes. Second, it is an example of an object, in which the area of object (visible or sensitive area) differs so much from the area of cover that it would be difficult even to imagine.

Pay attention to one aspect of using the transparent nodes: the shape of cursor above this node is never declared and will be set to `Cursors.Default`. Certainly, if the mover looks through the transparent node and finds some non-transparent node from another object underneath, then the cursor takes the shape that is defined by that node at the lower level.

## Rings

File: **Form_Rings.cs**
Menu position: *Graphical objects – Basic elements – Rings – Multicolored*

The first example of an object using a transparent node in its cover is going to be a ring of the `RingT` class. The resizing of such rings is provided by two sets of small nodes, covering the inner and the outer borders (**figure 8.1**). It is very similar to the covers of the `RingNT` class, which can be seen at **figure 7.2**, only in the new example the border nodes have the circular shape. The whole cover consists of such nodes and in such an order.

1. Nodes on the outer border.
2. Nodes on the inner border.
3. Circular node, covering the hole.
4. Big circular node up to the outer border.

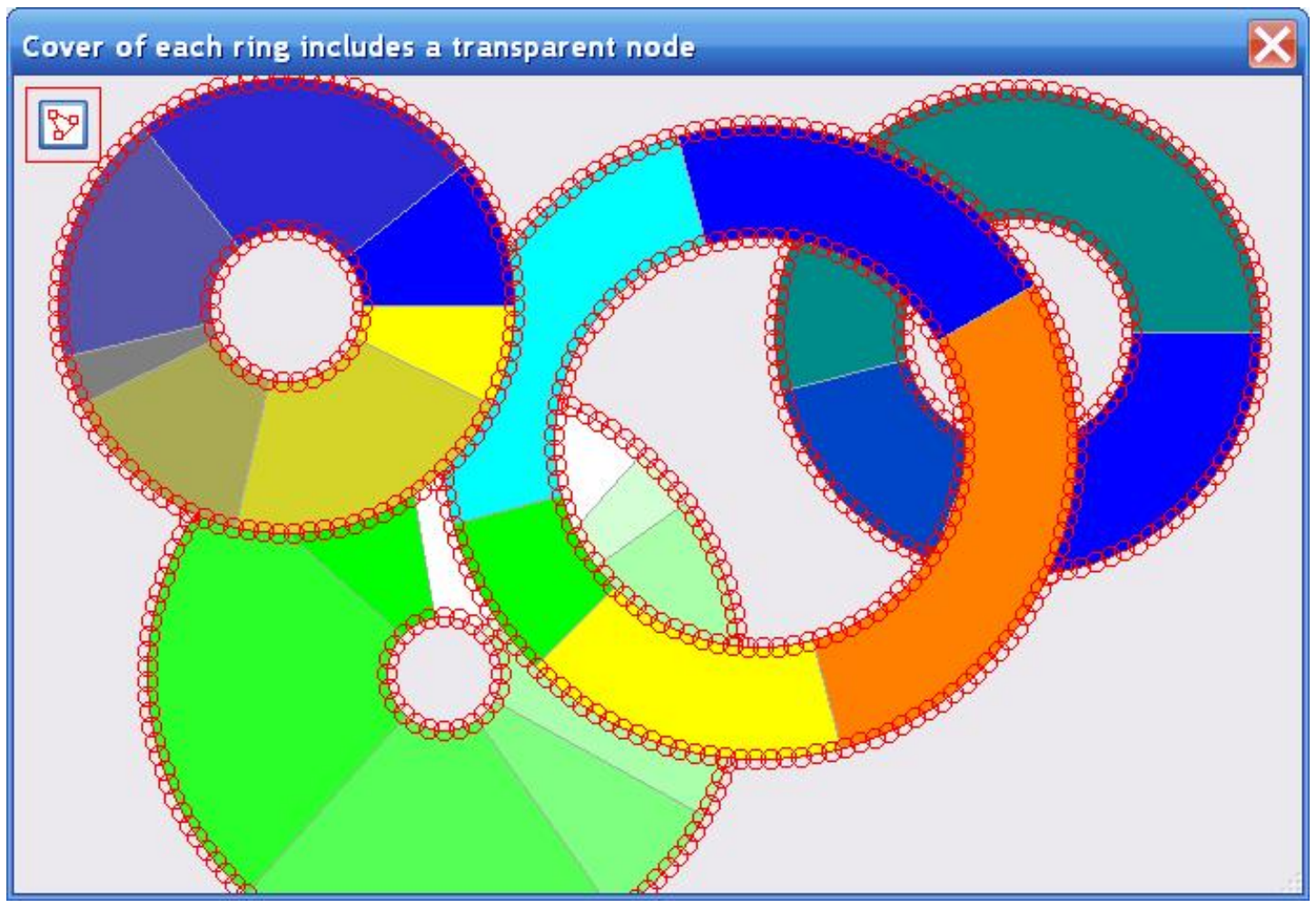

**Fig.8.1** Rings with the covers, using a transparent node

Each small node on a border has a radius of 5 pixels (`nrSmall`); the neighbouring nodes are placed with a step of 8 pixels (`distanceNeighbours`). Knowing both radii of a ring, it is not a problem to get the needed number of those small nodes.

```
private void NodesOnBorders ()
{
    nNodesOnOuter = Convert.ToInt32 ((2 * Math.PI * rOuter) / distanceNeighbours);
    nNodesOnInner = Convert.ToInt32 ((2 * Math.PI * rInner) / distanceNeighbours);
}
```

When the number of nodes is calculated and the order of nodes is known, there are no problems with the cover design.

```
    public override void DefineCover ()
    {
        CoverNode [] nodes = new CoverNode [nNodesOnOuter + nNodesOnInner + 2];
        for (int i = 0; i < nNodesOnOuter; i++)
        {
            nodes [i] = new CoverNode (i, Auxi_Geometry .PointToPoint (center,
```

```
                        2 * Math .PI * i / nNodesOnOuter, rOuter), nrSmall);
        }
        int k = nNodesOnOuter;
        for (int i = 0; i < nNodesOnInner; i++)
        {
            nodes [k+i] = new CoverNode (k + i, Auxi_Geometry.PointToPoint (center,
                                   2 * Math .PI * i / nNodesOnInner, rInner), nrSmall);
        }
        k += nNodesOnInner;
        nodes [k] = new CoverNode (k, center, Convert .ToInt32 (rInner),
                                              Behaviour .Transparent);
        nodes [k + 1] = new CoverNode (k + 1, center, Convert .ToInt32 (rOuter),
                                                      Cursors .SizeAll);
        cover = new Cover (nodes);
        cover .SetClearance (false);
    }
```

In all other aspects it is a classical example of the N-node cover with the call for updating the cover of the ring at the moment of its release. This redefinition of cover is required only after resizing but not after an ordinary movement of a ring around the screen, so the call to the `RedefineCover()` method is preceded by an additional check of the number of released node. The check is easy as only the two last nodes must be excluded from starting the redefinition of cover.

```
        private void OnMouseUp (object sender, MouseEventArgs e)
        {
            int iWasCaught, iNode;
            if (mover .Release (out iWasCaught, out iNode))
            {
                GraphicalObject grobj = mover [iWasCaught] .Source;
                if (e .Button == MouseButtons .Left)
                {
                    if (grobj is RingT && iNode < grobj .NodesCount - 2)
                    {
                        (grobj as RingT) .RedefineCover ();
                        Invalidate ();
                    }
```

## *Regular polygons with circular holes*

File: **Form_RegPoly_CircleInside.cs**
Menu position: *Graphical objects – Basic elements – Polygons – Regular polygons with circular holes*

The next example demonstrates the regular polygons with the circular holes (class `RegPoly_CircleInside`). The resizing of an object is provided by the strip nodes, covering the perimeter (**figure 8.2**). The border of a hole is covered by a set of circular nodes, which provide the resizing of this hole. In a similar way to the previous example of a ring, the circular transparent node precedes the big polygonal node, which covers the whole object together with the hole. Here is the order of nodes in the cover of this class.

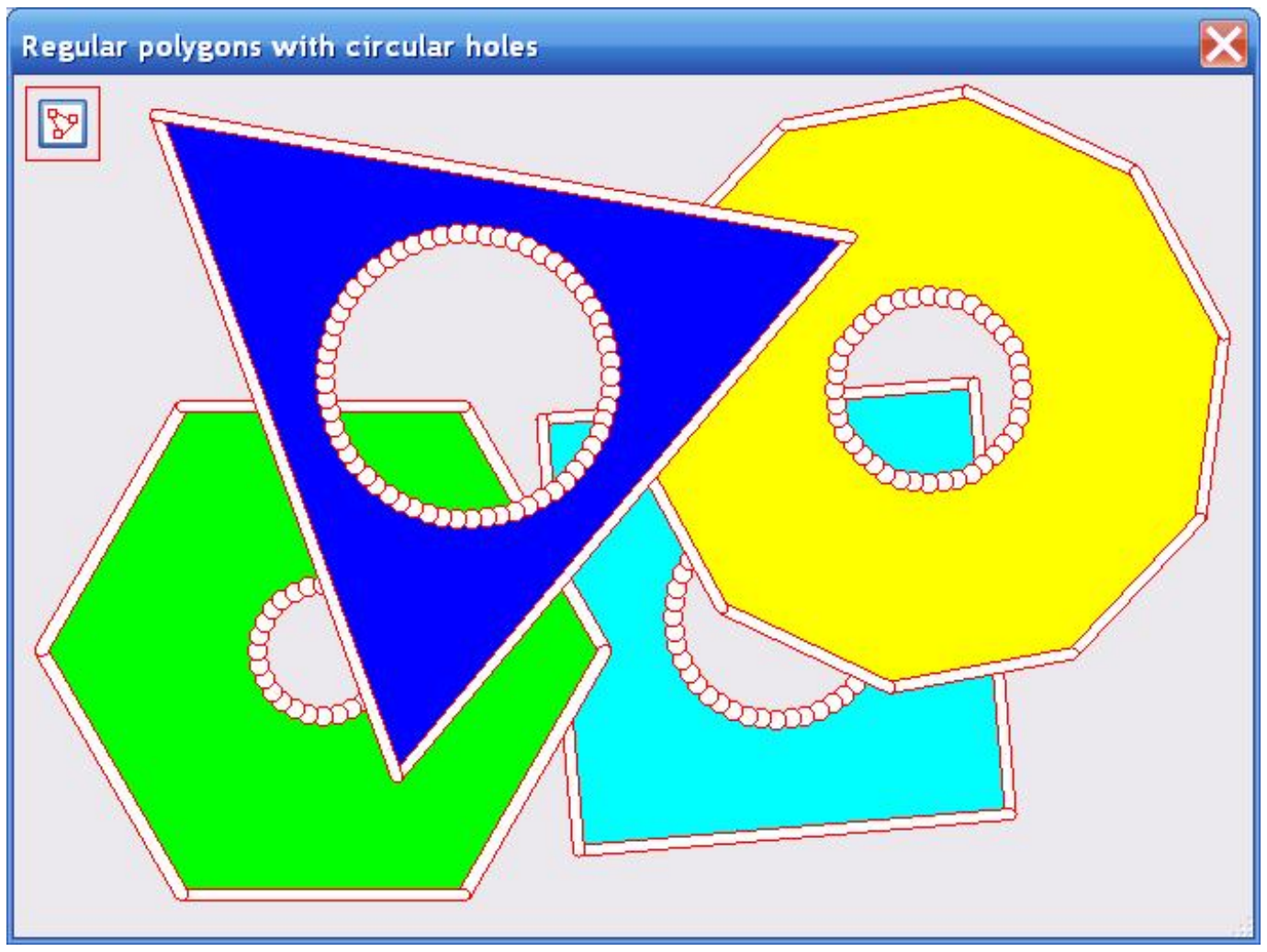

**Fig.8.2** Regular polygons with the circular holes

1. Small circular nodes, covering the border of a hole. The number of nodes depends on the circumference.
2. Strip nodes along the outer border; the number of strips is equal to the number of vertices.
3. Circular node, covering the hole.
4. Big polygonal node in the form of a regular polygon up to the outer border.

```
public override void DefineCover ()
{
    PointF [] pts = Vertices;
    CoverNode [] nodes = new CoverNode [nNodesOnCircle + nVertices + 2];
    for (int i = 0; i < nNodesOnCircle; i++)
    {
        nodes [i] = new CoverNode (i, Auxi_Geometry .PointToPoint (center,
                       2 * Math .PI * i / nNodesOnCircle, radiusCircle), nrSmall);
    }
    for (int i = 0, j = nNodesOnCircle; i < nVertices; i++, j++)
    {
        nodes [j] = new CoverNode (j, pts [i], pts [(i + 1) % nVertices]);
    }
    int k = nNodesOnCircle + nVertices;
    nodes [k] = new CoverNode (k, center, Convert .ToInt32 (radiusCircle),
                                      Behaviour .Transparent);
    nodes [k + 1] = new CoverNode (k + 1, pts);
    nodes [k] .Clearance = false;       // nodes .Length - 2
    cover = new Cover (nodes);
}
```

There are two restrictions on resizing such objects: minimum allowed radius of a circle and the minimum width between the circle and the outer border. These restrictions are taken into consideration, while trying to move one or another node, but this depends on the particular node.

When the circular hole is resized, then both of the restrictions are used, as they define the range in which the radius of circle can change. The hole can be resized only by the small nodes on its circumference; the nodes are so small that the difference between the mouse position and the circular border at the initial moment is ignored; the distance from the mouse cursor to the center of a circle is used as the new radius of a hole.

```
public override bool MoveNode (int i, int dx, int dy, Point ptM, MouseButtons btn)
{
    if (catcher == MouseButtons .Left)
    {
        double distanceMouse = Auxi_Geometry .Distance (center, ptM);
        if (i < nNodesOnCircle)
        {
            // inner resizing
            if (minCircleRadius <= distanceMouse &&
                distanceMouse <= radiusVertices * Math .Cos (Math .PI / nVertices)
                                                              - minCircleToSide)
            {
                radiusCircle = Convert .ToSingle (distanceMouse);
                bRet = true;
            }
        }
```

When the outer border is moved, then only the proposed distance between the circle and the outer border must be checked. The strips, which cover the outer border, can be caught for moving by any point. The distance from the mouse cursor to the center of an object is transformed into the new radius for all the vertices with the help of the scaling coefficient that is calculated at the starting moment of such resizing; this technique was explained in the first example with the regular polygons (**Form_RegPoly_Variants.cs**, **figure 6.1**). Because the resizing of a hole is started by the circular nodes and the moving of the outer border is started by the strip nodes, the decision about the particular type of resizing can be based either on the number of the caught node or its shape.

```
public override bool MoveNode (int i, int dx, int dy, Point ptM, MouseButtons btn)
{
    if (catcher == MouseButtons .Left)
    {
        … …
        else if (cover .GetNodeShape (i) == NodeShape .Strip)
        {
            double radiusNew = distanceMouse * scaling;
```

```
            if (radiusNew >= (radiusCircle +
                                  minCircleToSide) / Math .Cos (Math .PI / nVertices))
            {
                radiusVertices = Convert .ToSingle (radiusNew);
                bRet = true;
            }
        }
        else
        {
            Move (dx, dy);
        }
```

When a mover senses the `Transparent` node, it skips the whole cover of this object, so in this case the `MoveNode()` method is not called. If the `MoveNode()` method is called and the caught node is neither a circle nor a strip, then the only left option is the polygonal node which is used to move the whole object. Thus the `Move()` method is called.

## *Convex polygons with polygonal holes*

File: **Form_ConvexPoly_RegPolyInside.cs**
Menu position: *Graphical objects – Basic elements – Polygons – Convex polygons with polygonal holes*

Objects from **figure 8.3** represent the convex polygons with the holes in the form of regular polygons (class `ConvexPoly_RegPolyInside`). To simplify the construction of such objects, they are initialized as regular polygons inside and outside; then the outer border can be reconfigured, but must always keep the form of a convex polygon. Another restriction is the minimum allowed distance between the hole and the outer border.

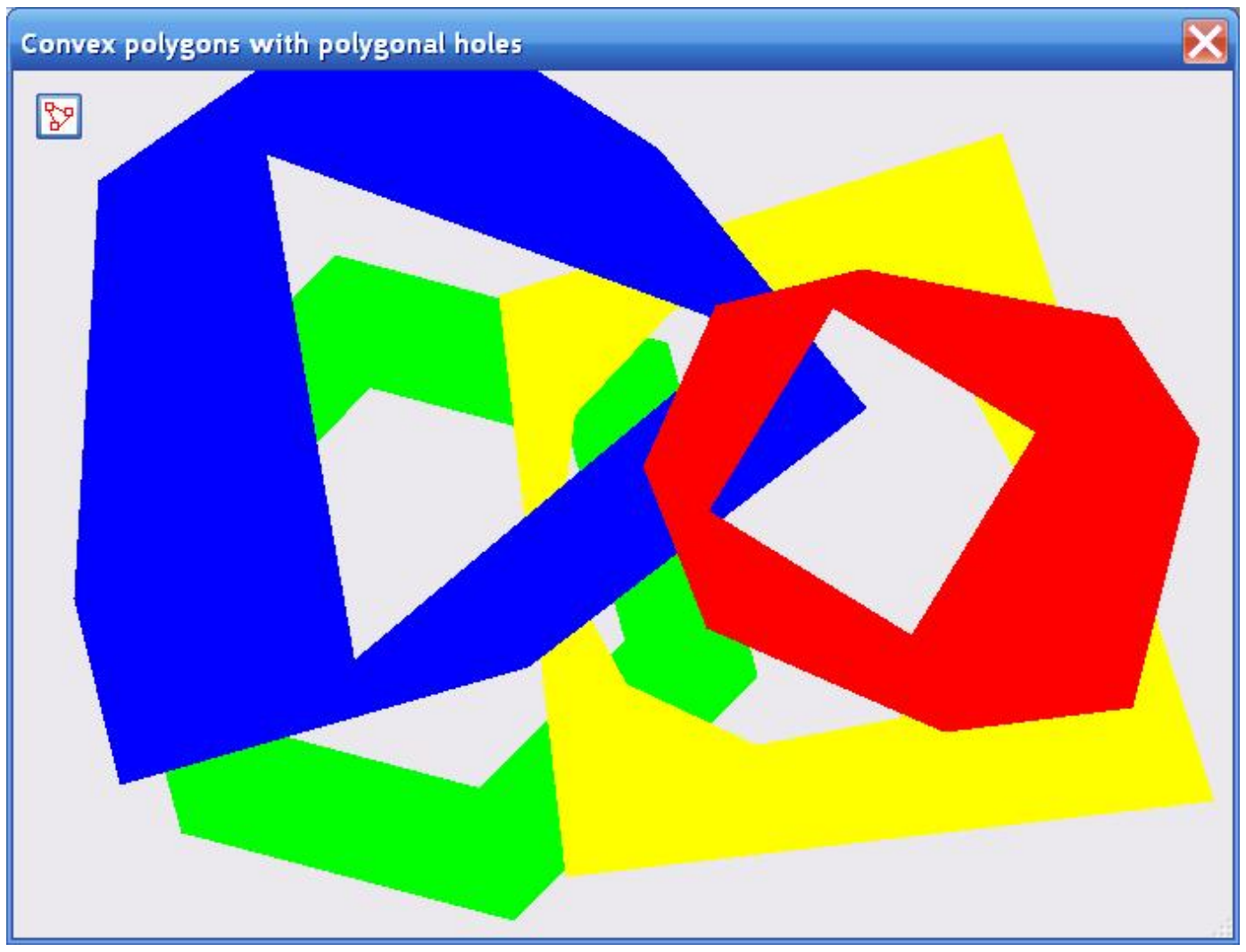

**Fig.8.3** Convex polygons with the polygonal holes

Both borders can be used for resizing, so overall the cover of each object consists of five different groups of nodes, which are included into the cover in such an order.

1. A set of strip nodes cover the segments of the inner polygon. The number of strips is equal to the number of vertices in this polygon.
2. A set of circular nodes are positioned on the vertices of the outer border.
3. A set of strip nodes cover the segments of the outer border. The number of strips is equal to the number of vertices in the outer polygon.
4. Polygonal transparent node, covering the hole.
5. Big polygonal node in the form of a convex polygon up to the outer border.

As both borders (inner and outer) are covered by the strip nodes (**figure 8.4**), then the decision about the particular movement inside the `MoveNode()` method of this class cannot be based on the shape of a node but only on its number. The inner border is always a regular polygon and the outer border is a convex polygon; the resizing started on the borders use the same idea of scaling, but with different variations, which are defined by the number of the caught node.

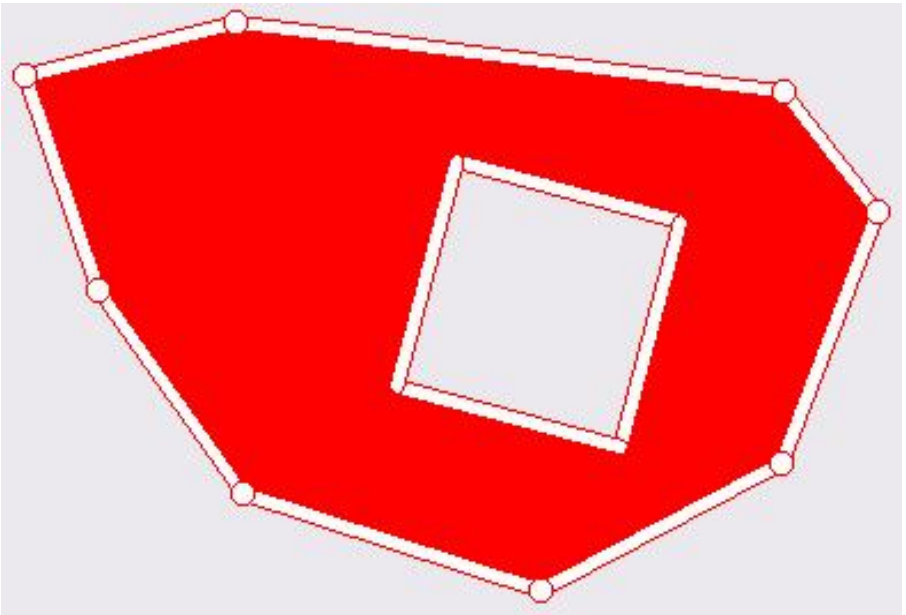

**Fig.8.4** A `polygon` with its cover

```
void OnMouseDown (object sender, MouseEventArgs e)
{
    ptMouse_Down = e .Location;
    if (mover .Catch (e .Location, e .Button))
    {
        GraphicalObject grobj = mover .CaughtSource;
```

```
        if (grobj is ConvexPoly_RegPolyInside)
        {
            ConvexPoly_RegPolyInside poly = grobj as ConvexPoly_RegPolyInside;
            if (e .Button == MouseButtons .Left)
            {
                if (mover .CaughtNodeShape == NodeShape .Strip)
                {
                    poly .StartScaling (e .Location, mover .CaughtNode);
                }
            }
            else if (e .Button == MouseButtons .Right)
            {
                poly .StartRotation (e .Location);
            }
```

All the vertices of the hole in the form of a regular polygon stay at the same distance from the center of an object, so it is enough to have a single scaling coefficient for all of them.

```
    public void StartScaling (Point ptMouse, int iNode)
    {
        double distanceMouse = Auxi_Geometry .Distance (center, ptMouse);
        if (iNode < nVertices_Inner)        // on inner border
        {
            scaling_Inner = radiusVertices_Inner / distanceMouse;
        }
        else                // on outer border
        {
            for (int i = 0; i < ptsOut .Length; i++)
            {
                angle_Outer [i] = Auxi_Geometry .Line_Angle (center, ptsOut [i]);
                scaling_Outer [i] = Auxi_Geometry .Distance (center,
                                                     ptsOut [i]) / distanceMouse;
            }
        }
    }
```

The outer border is a convex polygon with the semi-independent positioning of vertices. They are semi-independent because they have to produce a convex polygon at any moment, but from the point of zooming they are absolutely independent, so for them the whole set of scaling coefficients must be calculated (one coefficient per vertex). In addition to the scaling coefficients, the set of angles must be also calculated (one per vertex); throughout the scaling, the angles of the outer vertices do not change and all those vertices are moving along their personal radii.

Similar thing happens, when the rotation of any `ConvexPoly_RegPolyInside` object is started. For the inner regular polygon it is enough to calculate a single compensation angle. For the outer convex polygon two arrays must be calculated: one includes radii for all the vertices, another – personal compensation angles.

```
    public void StartRotation (Point ptMouse)
    {
        double angleMouse = Auxi_Geometry .Line_Angle (center, ptMouse);
        compensation_Inner = Auxi_Common .LimitedRadian (angleMouse - angle_Inner);
        for (int i = 0; i < ptsOut .Length; i++)
        {
            radius_Outer [i] = Auxi_Geometry .Distance (center, ptsOut [i]);
            compensation_Outer [i] = Auxi_Common .LimitedRadian (angleMouse –
                                  Auxi_Geometry .Line_Angle (center, ptsOut [i]));
        }
    }
```

Any move of any part of the inner and outer borders can violate the restriction on minimum allowed distance between two borders, so before any of these movements are allowed, the `EstimatedWidth()` method is called.

## *Crescent*

File: **Form_Crescent.cs**
Menu position: *Graphical objects – Basic elements – Crescent*

Previous examples of this chapter show the objects of different shapes, but they are similar in the way they use the transparent nodes in their covers: those nodes are used to "burn out" a hole in the area of an object, so each transparent node is smaller than the object and entirely inside its area. Here is an absolutely different type of object and a different way of using the transparent node. It is also an example in which it becomes obvious that the area of cover can differ so much from the area of an object that it is difficult to imagine the bigger discrepancy. **Figure 8.5** shows an object of the `Crescent` class together with its cover and some additional information, helpful for explanations.

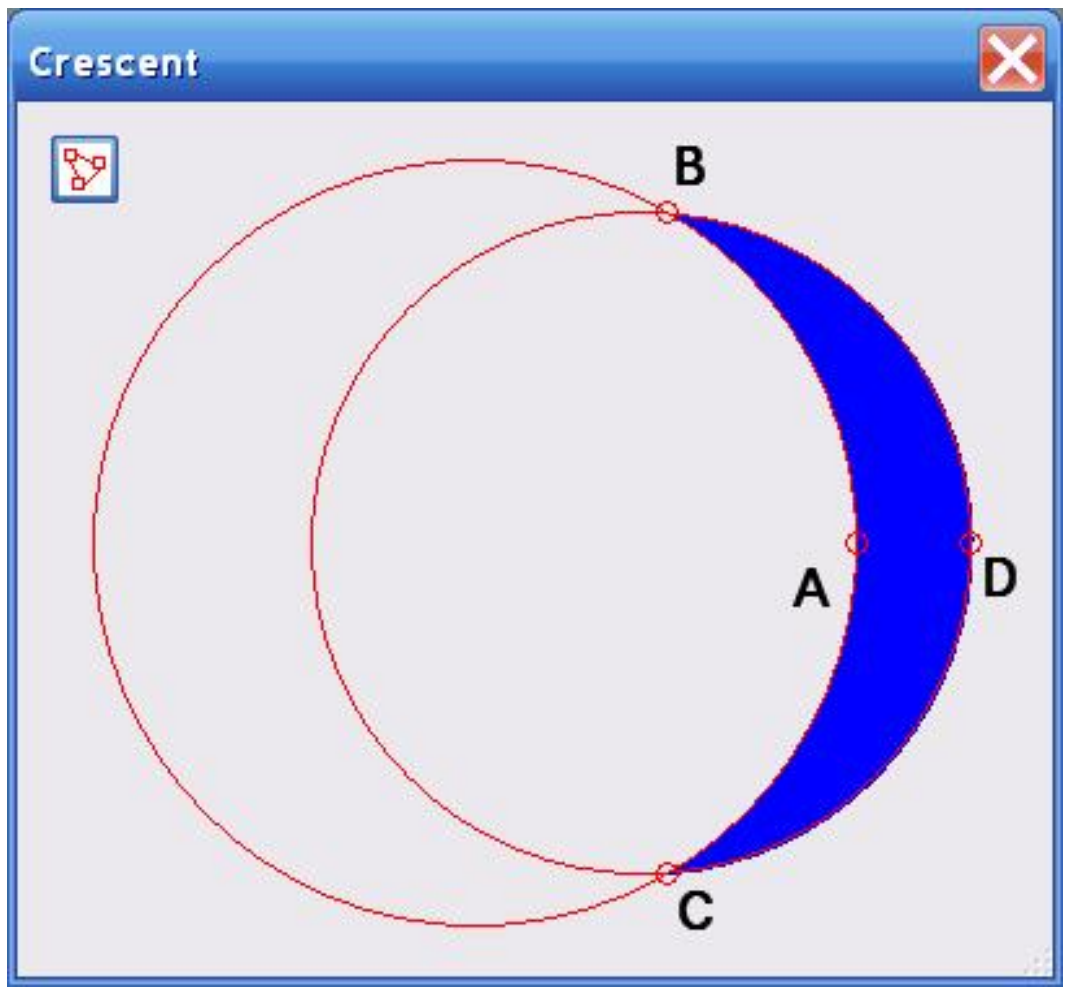

**Fig.8.5** `Crescent` object with its cover

The task looks simple enough: to have a crescent that can be moved, rotated, and resized. Moving and rotation can be organized by using only two circular nodes; at the figure they are shown by the red circles. To make a crescent resizable, I had first to decide, in which way it would be "natural" to do it, so I came to such a conclusion:

- Two small areas on both sides of the widest part of the crescent (points A and D) would be enough to change the width of an object.
- The tips of the horns (points B and C) are the good places to change the height of crescent.

At **figure 8.5** you can see four small circular nodes in the mentioned points, so the whole cover consists of six nodes.

```
public override void DefineCover ()
{
    CoverNode [] nodes = new CoverNode [] { new CoverNode (0, ptA, 4),
                                            new CoverNode (1, ptD, 4),
                                            new CoverNode (2, ptB, 4),
                                            new CoverNode (3, ptC, 4),
                new CoverNode (4, ptCenterInner, Convert .ToSingle (rInner),
                                 Behaviour .Transparent),
                new CoverNode (5, ptCenterOuter, Convert .ToSingle (rOuter),
                                 Cursors .SizeAll) };
    cover = new Cover (nodes);
    cover .SetClearance (false);
}
```

The cover of this object is not in compliance with the standard rule of resizing an object by any border point. This is done not because of any possible problem of covering such a border, but simply because I cannot understand myself, what would I expect on pulling a border at any other border points. So I decided that four sensitive points with more expected reaction than at other border points would be enough for resizing a crescent.

Crescent has two parts of its border: inner and outer. Both sides are the parts of some circle, so they are defined by two radii. The transparent circle, which defines the inner border, has bigger radius than the circle, which forms the outer border. The bigger circle covers and "burns out" the significant part (always more than half) of the second circle, so I always have a crescent.

What are the limitations on resizing such crescent?

<u>Point A</u> can be moved along the AD line one way or another. If it is moving away from point D, then the crescent is becoming wider. It will stop being a crescent, when point A crosses the BC line; I put the limit of moving point A in this direction at 3 pixels before such crossing, so that crescent will not lose its shape. When point A is moving towards point D, then the crescent becomes thinner. I want always to have some space between the nodes on these two points so that by its middle part the crescent can be moved around the screen. Thus I put the limit on the minimum width of a crescent, which is slightly bigger than the diameter of those nodes.

It is not so easy to explain some of the limits, which you can see in the `MoveNode()` method for moving the points B and C. I made some estimates and put it so that a crescent would always have the sharp horns. Anyway, this crescent is only an example to show the use of transparent nodes in design, moving, and resizing of not very often used, but interesting objects.

An interesting remark not on the design of the `Crescent` class, but more on the overall design and use of movable objects. When this **Form_Crescent.cs** was already working and I needed to prepare an illustration for this example, I understood that it would be nice to have those letters on the figure, which make my explanation easier to write and understand. One way would be to grab the picture, put it into some other program (*Paint*, for example), and add those letters there. I am not too much familiar with the *Paint* program, so it would be not too easy for me to put the letters exactly at the places, where I would like them to be. And what to do, if I do not like the first result and would like to move those letters? I am sure there are some ways to do it, but I am not familiar with them.

But I have everything I need at my own disposal in the same Demo program. So I added into the **Form_Crescent.cs** four simple objects of the `TextM` class, included them into the mover's queue, and received four movable letters, which I can put at any needed position and move at any moment in any way I want. They are still there in the code, but I have commented those lines. In case of need (another figure?), I can return them back into mover's queue.

# *Sector of a circle*

Sector of a circle is another type of object that can demonstrate an interesting example of using transparent nodes. There is one limitation on all the examples of this section: the angle of a sector must not exceed 180 degrees. If by the requirements of the particular task you need to work with bigger sectors, then similar, but slightly different covers must be used. But that would be another task. Here we are dealing with the sectors, which do not exceed 180 degrees.

## Non-resizable sectors

File: **Form_CircleSector_Nonresizable.cs**
Menu position: *Graphical objects – Basic elements – Sector of a circle – Non-resizable*

The first examples of the sectors are going to be non-resizable; objects of the `CircleSector_Nonresizable` class are only movable. This is a very rare type of objects in my programs, but while going from one example of sectors in this chapter to another, you will see that they differ in that part of cover, which provides the resizing, but use exactly the same algorithm of organizing their forward movement and rotation. These two things are designed on the basis of using two transparent nodes, so this design can be explained even with the movable, but non-resizable sectors.

Any sector is defined by its central point, radius, and two angles. The first of them – `angleStart` – defines one of the sides; the second – `angleSweep` – defines the angle from that first side to another. Positive angles are measured counterclockwise from the given line.

Sector is some part of a circle, so it is natural to try using the circular node for moving of such an object. But the part of the circular node which is outside the sector must be not used for moving an object; it must be "burned out"; this is done by two polygonal transparent nodes (rectangles), which are positioned next to the sides of a sector. The sizes of these rectangular nodes are calculated in such a way that each one closes and eliminates half of the full circle. Part of the circle is eliminated by both of the polygonal nodes, but it does not matter. Whenever a mover sees a transparent node at any point, there is no analysis on the further part of this cover; it is ignored regardless of what is there behind that transparent node. The transparent nodes are used to exclude some parts of another node, so the transparent nodes must be placed in the cover ahead of the node, from which they have to cut out some part.

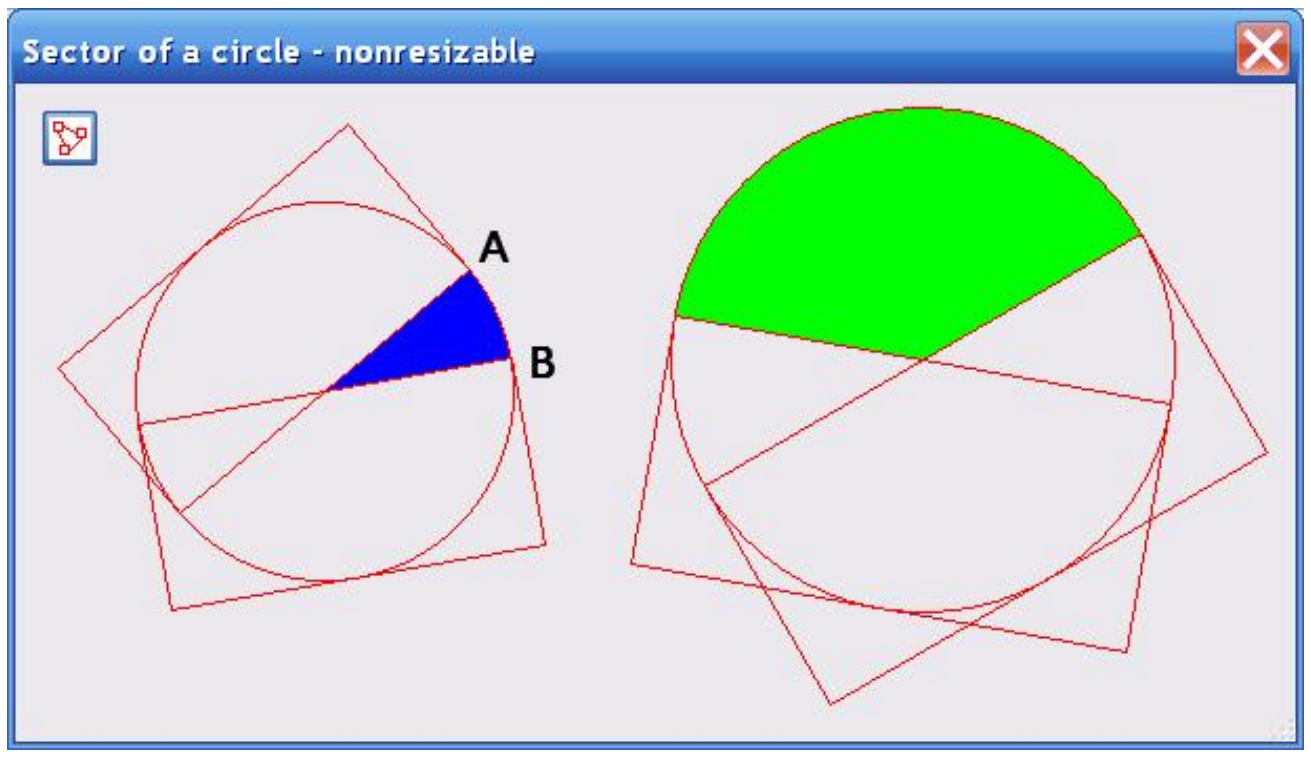

**Fig.8.6** Covers for non-resizable sectors

Point A and B mark to ends of the arc; these are the points, where two sides of sector meet with the circle. I need point B to be clockwise from point A, as shown at **figure 8.6**. First the angles to points A and B are defined; this depends on the sign of the `angleSweep`. Then the corners of two rectangles are calculated. The rectangles are based on the sides A and B; the corners of rectangles are going to be the points of the polygonal nodes.

```
public override void DefineCover ()
{
    PointF ptA, ptB;
    double angleA, angleB;
```

```
    if (angle >= 0)
    {
        angleB = angleStart;
        angleA = angleStart + angleSweep;
    }
    else
    {
        angleA = angleStart;
        angleB = angleStart + angleSweep;
    }
    ptA = Auxi_Geometry .PointToPoint (center, angleA, radius);
    ptB = Auxi_Geometry .PointToPoint (center, angleB, radius);
    PointF [] ptsA = new PointF [] {ptA,
                Auxi_Geometry .PointToPoint (ptA, angleA + Math .PI / 2, radius),
                Auxi_Geometry .PointToPoint (center, angleA + 3 * Math .PI / 4,
                                             radius * Math .Sqrt (2)),
                Auxi_Geometry .PointToPoint (center, angleA - Math .PI, radius) };
    PointF [] ptsB = new PointF [] {ptB,
                Auxi_Geometry .PointToPoint (center, angleB - Math .PI, radius),
                Auxi_Geometry .PointToPoint (center, angleB - 3 * Math .PI / 4,
                                             radius * Math .Sqrt (2)),
                Auxi_Geometry .PointToPoint (ptB, angleB - Math .PI / 2, radius) };
    CoverNode [] nodes = new CoverNode [3] {
                       new CoverNode (0, ptsA, Behaviour .Transparent),
                       new CoverNode (1, ptsB, Behaviour .Transparent),
                       new CoverNode (2, center, radius, Cursors .SizeAll) };
    cover = new Cover (nodes);
    cover .SetClearance (false);
}
```

This is nearly all that is needed to move such a sector. The cover consists of three nodes, but two of them are transparent, so if such an object is caught by the mover, then it can be done only by the sector, which is the part of the circular node. Thus there is even no need for checking the number or the shape of the node inside the `MoveNode()` method; if any node is caught, it can be the only node by which this object can be moved. And then the `Move()` method must be called.

```
public override bool MoveNode (int i, int dx, int dy, Point ptM, MouseButtons btn)
{
    bool bRet = false;
    if (btn == MouseButtons .Left)
    {
        Move (dx, dy);
    }
    else if (btn == MouseButtons .Right)
    {
        double angleMouse = Auxi_Geometry .Line_Angle (center, ptM);
        angleStart = angleMouse - compensation;
        bRet = true;
    }
    return (bRet);
}
```

Rotation is organized in a classical way with a `StartRotation()` method to calculate a compensation angle at the starting moment and using this fixed compensation throughout the full time of rotation. Compensation is the difference between the angle to the mouse and the angle of an object. In case of a sector, the `angleStart` plays the role of the object's angle.

## Sectors resized only by arc

File: **Form_CircleSector_ResizableArc.cs**
Menu position: *Graphical objects – Basic elements – Sector of a circle – Resizable arc*

The next example of a sector (`CircleSector_ResizableArc` class) has one improvement: it can be resized by any point of its arc. The angle between two sides is still fixed from the moment of initialization and cannot be changed (yet!), but the radius of the sector can be changed. To do this, the arc is covered by a set of small circular nodes – the classical type of the N-node cover.

As seen from **figure 8.7**, the same pair of transparent nodes is used to let sensitive the needed part of the big circular node, but there is also a set of small nodes along the arc. These nodes are included into the cover as the first elements. The number of small circular nodes on the arc (`nNodesOnArc`) depends on the length of the arc and the distance between the neighbours (`distanceNeighbours = 8`), but the number of these circles cannot be less than two.

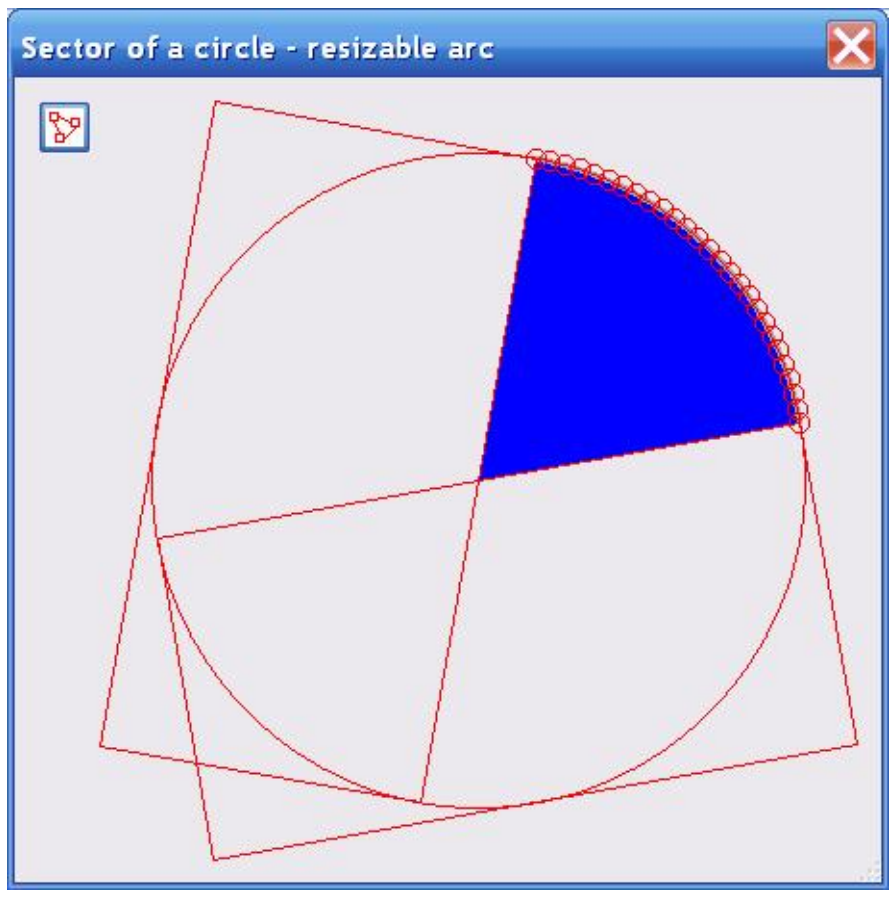

**Fig.8.7** Sector can be resized by the arc

```
    private void NodesOnArc ()
    {
        nNodesOnArc = Math .Max (Convert .ToInt32 (Math .Abs (angleSweep) * radius
                                                   / distanceNeighbours) + 1, 2);
    }
```

The first part of the `DefineCover()` method is identical to the previous example. The second part is slightly different, as all the small nodes must be included into the cover ahead of three big nodes.

```
public override void DefineCover ()
{
    … …
    PointF [] ptsB = new PointF [] {… …
    CoverNode [] nodes = new CoverNode [nNodesOnArc + 2 + 1];
    double delta = angleSweep / (nNodesOnArc - 1);
    for (int i = 0; i < nNodesOnArc; i++)
    {
        nodes [i] = new CoverNode (i, Auxi_Geometry .PointToPoint (center,
                                         angleStart + i * delta, radius), nrSmall);
    }
    nodes [nNodesOnArc] = new CoverNode (nNodesOnArc, ptsA, Behaviour.Transparent);
    nodes [nNodesOnArc + 1] = new CoverNode (nNodesOnArc + 1, ptsB,
                                             Behaviour .Transparent);
    nodes [nNodesOnArc + 2] = new CoverNode (nNodesOnArc + 2, center, radius,
                                             Cursors .SizeAll);
    cover = new Cover (nodes);
    cover .SetClearance (false);
}
```

The `MoveNode()` method of this class has to check the number of the caught node, because nearly all the nodes are responsible for resizing, while the last one (of non-transparent nodes!) provides the moving of the whole sector.

```
public override bool MoveNode (int i, int dx, int dy, Point ptM, MouseButtons btn)
{
    bool bRet = false;
    if (btn == MouseButtons .Left)
    {
        if (i < nNodesOnArc)
        {
            int nRadNew = Convert .ToInt32 (Auxi_Geometry .Distance (center, ptM));
            if (nRadNew != radius && nRadNew >= nMinRadius)
            {
```

```
                radius = nRadNew;
                bRet = true;
            }
        }
        else
        {
            Move (dx, dy);
        }
    }
```

## Sectors with one movable side

File: **Form_CircleSector_OneMoveableSide.cs**
Menu position: *Graphical objects – Basic elements – Sector of a circle – One moveable side*

Objects of the `CircleSector_OneMoveableSide` class have one addition to the previous example: one side is movable, so the angle between two sides (`angleSweep`) can be changed at any moment. The range for this angle is [0, 180] degrees; moving this side one way and another looks like opening and closing a fan. When the covers are visualized then the movable side of a sector is marked with the strip node above it (**figure 8.8**),. If the covers are not shown, the movable side is still obvious, as it is painted by a wider line. This wider line is also very useful, when the angle is decreased to zero. Without such a line, the closed sector would disappear from the screen without any traces; with such a line it is still obvious, from where you can start to open the sector again.

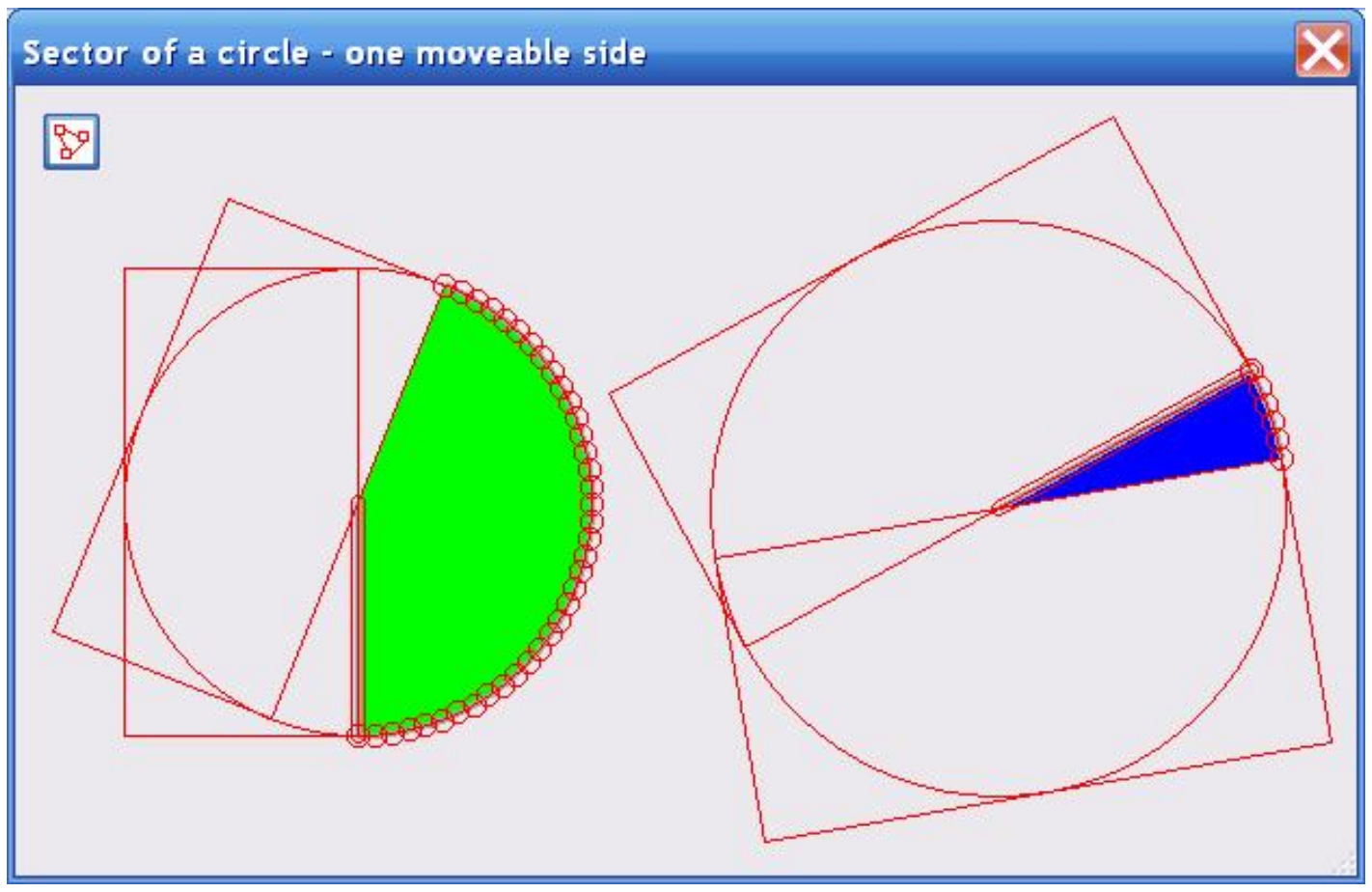

**Fig.8.8** Sectors with one movable side

The cover for such sector differs from the previous case by one strip node, which takes its place after all the circular nodes on the arc. The first side of a sector is unmovable and is determined by the `angleStart`. The second side of a sector is movable; its angle is `angleStart + angleSweep`. Thus we have an additional line in the `DefineCover()` method for this new strip node.

```
public override void DefineCover ()
{
    … …
    nodes [nNodesOnArc] = new CoverNode (nNodesOnArc, center,
           Auxi_Geometry .PointToPoint (center, angleStart + angleSweep, radius));
```

A sector can be easily rotated around, so at the starting moment of changing the sector its unmovable side may have any angle (`angleStart`). The movable side can change the `angleSweep` only inside the [0, 180] degrees range. If the strip node on the movable side of a sector is caught, then the `StartResectoring()` method must be called to calculate the range of angles in which this side can be turned around.

```
    private void OnMouseDown (object sender, MouseEventArgs e)
    {
        if (mover .Catch (e .Location, e .Button))
        {
            GraphicalObject grobj = mover .CaughtSource;
            if (grobj is CircleSector_OneMoveableSide)
            {
                CircleSector_OneMoveableSide sector =
                                                     grobj as CircleSector_OneMoveableSide;
                if (e .Button == MouseButtons .Left &&
                    mover .CaughtNodeShape == NodeShape .Strip)
                {
                    sector .StartResectoring (e .Location);
                }
```

```
                else if (e .Button == MouseButtons .Right)
                {
                        sector .StartRotation (e .Location);
                    }
                }
            }
        }
```

The angles, between which the movable side can go, are called `minAngle` and `maxAngle`. These two angles are determined by the angle of the unmovable side (`angleStart`) and the direction of the allowed movement from this unmovable side, which depends on the `angleSweep` at the moment of the initialization of sector.

```
            angleFull = (angleSweep >= 0) ? Math .PI : -Math .PI;
```

The special compensation for changing the angle of sector (`compensationResectoring`), which you see in the `StartResectoring()` method is used for more accurate turn of the movable side. This is an angle between the mouse position at the starting moment and the angle of the movable side. Though this is a small angle, I cannot ignore it; otherwise it can result in the crash of the program. The cause of this angle appearance is the finite width of the strip node over the movable side. If I would allow to catch the movable side only by the line of this side (one pixel width!), then there would be no need in such compensation, but no user would like such a decision. Catching the side by any point in its vicinity (a strip of 6 pixels width) is much easier, but then consider such a situation. You have a sector opened on the maximum allowed 180 degrees and you catch the movable side one or two pixels outside the sector. If this small difference is ignored and the mouse position is looked at as the new position of the side, then it would be a violation of the rule that the sector cannot exceed 180 degree. For this reason I take into consideration this small `compensationResectoring` angle.

```
    public void StartResectoring (Point ptMouse)
    {
        double angleMouse = Auxi_Geometry .Line_Angle (center, ptMouse);
        double angleSlider = angleStart + angleSweep;
        compensationResectoring =
                              Auxi_Common .LimitedRadian (angleMouse - angleSlider);
        if (angleFull > 0)
        {
            minAngle = angleStart;
            maxAngle = angleStart + angleFull;
        }
        else
        {
            maxAngle = angleStart;
            minAngle = angleStart + angleFull;
        }
    }
```

Whenever the side is moved by the mouse, this compensation is used to calculate the exact angle of the side.

```
public override bool MoveNode (int i, int dx, int dy, Point ptM, MouseButtons btn)
{
    if (btn == MouseButtons .Left)
    {
        … …
        else if (i == nNodesOnArc)      // sliding side
        {
            double angleMouse = Auxi_Geometry .Line_Angle (center, ptM);
            double angleSlider = angleMouse - compensationResectoring;
            … …
```

## Fully resizable sectors

File: **Form_CircleSector_FullyResizable.cs**
Menu position: *Graphical objects – Basic elements – Sector of a circle – Fully resizable*

Though the sectors at **figure 8.9** are fully resizable and both sides can be moved, but only one side of these objects is marked by a wider line. This was done not only to prevent the disappearance of the sectors with the zero angle between the sides, but also for easier explanation. While initializing an object of the `CircleSector_FullyResizable` class, two angles are passed as the parameters: `angleStart` determines the initial position of the first side; `angleSweep` determines the angle between two sides. The sum of these angles determines the position of the second side, which is painted with a wider line. It also simplifies the understanding of the code: the strip nodes on two sides get numbers 0 and 1; the side with the wide line is marked as 0.

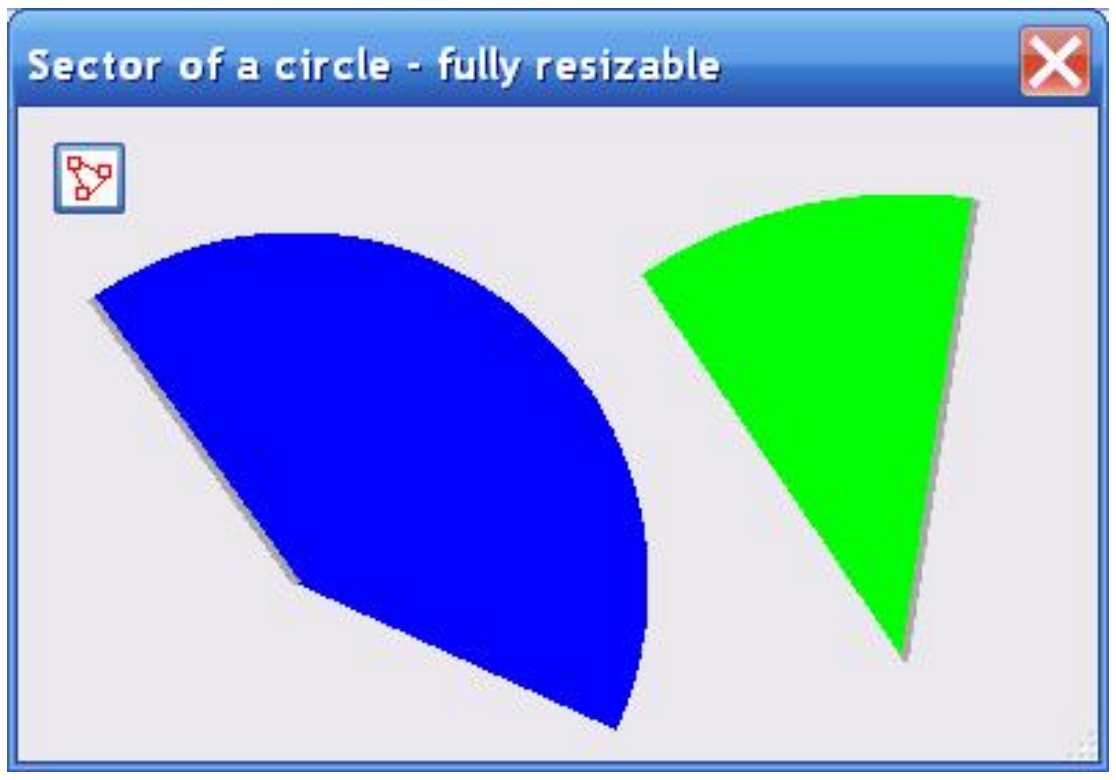

**Fig.8.9** Fully resizable sectors

As the two sides of a sector got numbers 0 and 1, I changed the order of nodes in the cover so that the nodes on those sides also have numbers 0 and 1; this makes the code easier to understand. This is the order of nodes in the cover.

- Strip node on that side, which is shown by a wider line.
- Strip node on another side.
- Circular nodes along the arc.
- Two transparent polygonal nodes.
- Big circular node for the whole circle.

Here is the part of the `DefineCover()` method with the definition of two strip nodes on the movable sides.

```
public override void DefineCover ()
{
    … …
    CoverNode [] nodes = new CoverNode [2 + nNodesOnArc + 2 + 1];
    nodes [0] = new CoverNode (0, center, Auxi_Geometry .PointToPoint (center,
                                              angleStart + angleSweep, radius));
    nodes [1] = new CoverNode (1, center, Auxi_Geometry .PointToPoint (center,
                                                          angleStart, radius));
```

In the case of one movable side (previous example), the range of possible movement of that side `[minAngle, maxAngle]` was calculated at the starting moment, when the side was caught for moving. For two movable sides, this technique works not so well, so I use another algorithm.

The main problem here is to guarantee that the movable side is always going to be not farther than 180 degrees from the unmovable side and never moves into another half of the circle making a sector of more than 180 degrees. For this purpose, at the starting moment of movement one additional point (`ptIn`) is determined in that half of a circle, through which the caught side is allowed to move. The location of this point is determined by the initial sign of the `angleSweep` and also by the side, which is caught for moving.

```
public void StartResectoring (int iNode)
{
    angleStart = Auxi_Common .LimitedRadian (angleStart);
    if (iNode == 0)          // end line
    {
        ptSeg_0 = Auxi_Geometry .PointToPoint (center, angleStart, radius);
        ptSeg_1 = Auxi_Geometry .PointToPoint (center, angleStart+Math.PI, radius);
        if (angleInit >= 0)
        {
            angleToIn = angleStart + Math .PI / 2;
        }
        else
```

```
        {
            angleToIn = angleStart - Math .PI / 2;
        }
    }
    else         // start line
    {
        angleEnd = Auxi_Common .LimitedRadian (angleStart + angleSweep);
        ptSeg_0 = Auxi_Geometry .PointToPoint (center, angleEnd, radius);
        ptSeg_1 = Auxi_Geometry .PointToPoint (center, angleEnd + Math.PI, radius);
        if (angleInit >= 0)
        {
            angleToIn = angleEnd - Math .PI / 2;
        }
        else
        {
            angleToIn = angleEnd + Math .PI / 2;
        }
    }
    ptIn = Auxi_Geometry .PointToPoint (center, angleToIn, radius);
}
```

`ptSeg_0` and `ptSeg_1` are two ends of the diameter, which goes along the side that is currently not moved. The caught side can move only through one half, organized by this diameter; `ptIn` belongs to that side, which is allowed for movement. So the check for the side in the `MoveNode()` method is based on the fact that this auxiliary point and the current mouse position cannot be on different sides of this diameter.

```
public override bool MoveNode (int i, int dx, int dy, Point ptM, MouseButtons btn)
{
    bool bRet = false;
    if (btn == MouseButtons .Left)
    {
        double angleMouse = Auxi_Geometry .Line_Angle (center, ptM);
        if (i == 0)
        {
            if (!Auxi_Geometry .OppositeSideOfLine (ptSeg_0, ptSeg_1, ptM, ptIn))
            {
                angleSweep = Auxi_Common .LimitedRadian (angleMouse - angleStart);
                bRet = true;
            }
        }
        else if (i == 1)
        {
            if (!Auxi_Geometry .OppositeSideOfLine (ptSeg_0, ptSeg_1, ptM, ptIn))
            {
                angleSweep = Auxi_Common .LimitedRadian (angleEnd - angleMouse);
                angleStart = Auxi_Common .LimitedRadian (angleEnd - angleSweep);
                bRet = true;
            }
        }
        … …
```

The same technique could be used in the previous example, but I decided to demonstrate different possibilities.

## Fill the holes

File: **Form_FillTheHoles.cs**
Menu position: *Applications – Fill the holes*

It happened so that all the previous examples in this chapter demonstrated the objects with either one or two transparent nodes in their covers. But there is no limit on the number of transparent nodes in a cover; these are the standard nodes and they can be used in the same way as other nodes with different types of behaviour. Here is an example of the objects, which have (N + 1) nodes in their covers; of those nodes all but one are transparent.

This form differs very much from all the previous examples. It is not a small form to demonstrate one or another special feature with an example of one or few simple objects. On the contrary, it is organized like a real application with tuning of parameters and using context menus with a lot of commands. This form can be easily transformed into a stand alone application; this is the first form in discussion, which starts by one of the commands from the *Applications* position of the main menu. The **Form_FillTheHoles.cs** is designed as a normal user-driven application, but it is too early to discuss the features of such programs, so I will only mention briefly what can be done here. After it I will return to the special features of the `AreaWithHoles` class.

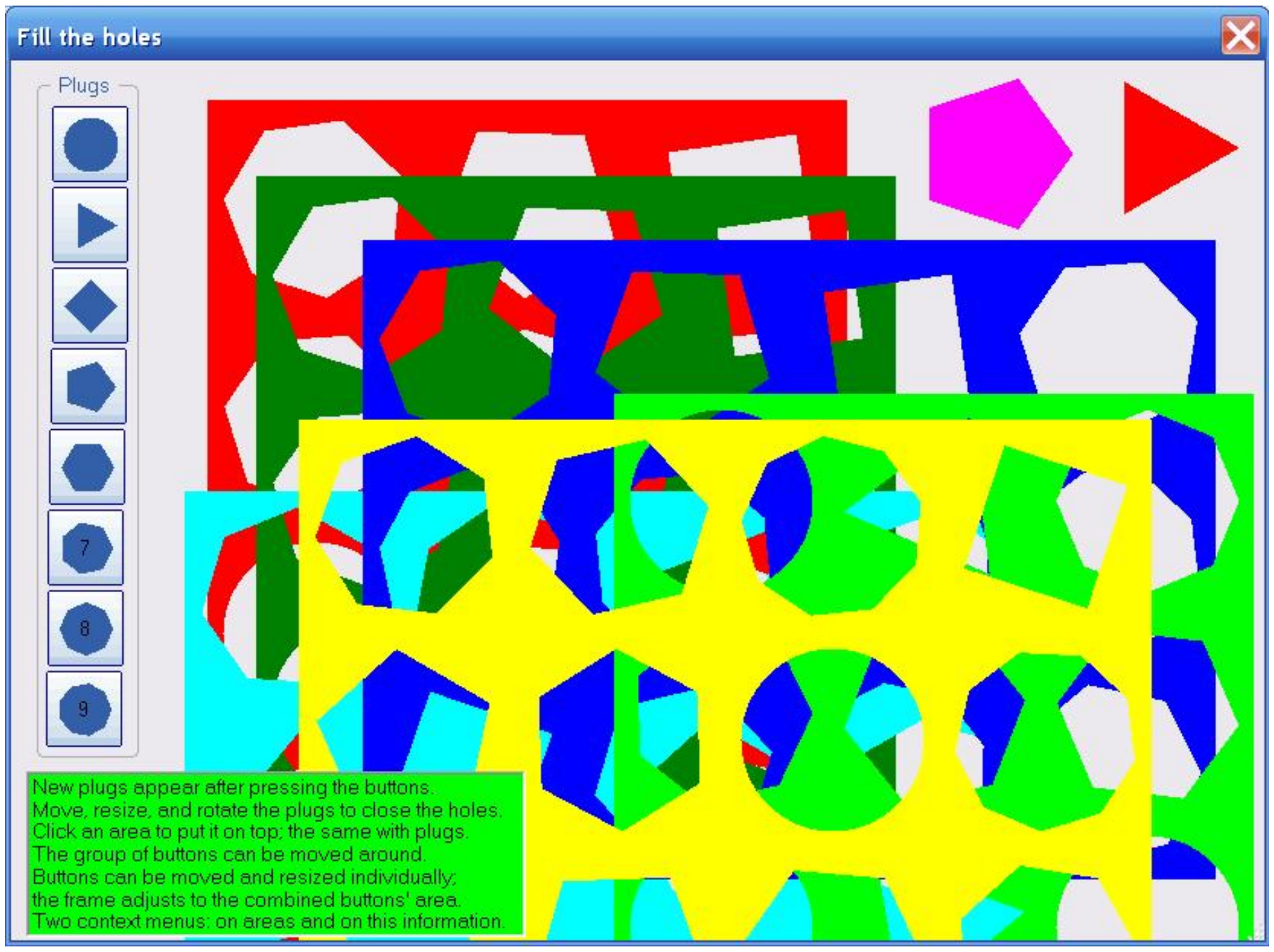

**Fig.8.10** The cover for each rectangular area includes a set of transparent nodes

The initial view of the **Form_FillTheHoles.cs** is going to be slightly different from **figure 8.10**, but not too much. Big rectangular boards with a lot of holes in them are the objects of the `AreaWithHoles` class; the covers of these objects mostly consist of the transparent nodes.

The *Plugs* group contains eight buttons. Each button can be moved individually by its border. On any movement of the inner elements the group adjusts the frame, so it is always shown around all the buttons. The group can be moved by any inner point and by all the points of its frame. The title of the group can be moved left and right between the borders. Group belongs to the `ElasticGroup` class; this class is discussed in details in the chapter *Groups of elements*.

On clicking any button of the group, you receive on the screen a new graphical object of the appropriate shape. These objects can be either a circle or a regular polygon with the number of vertices between 3 and 9. Those objects belong to the `Plug` class; two such objects can be seen in the top right corner of **figure 8.10**.

The green panel with the information in the bottom left corner is the object of the `TextM` class; this class was already discussed in the chapter *Texts*. A context menu can be called on this panel and all its visibility parameters can be tuned via the commands of this menu.

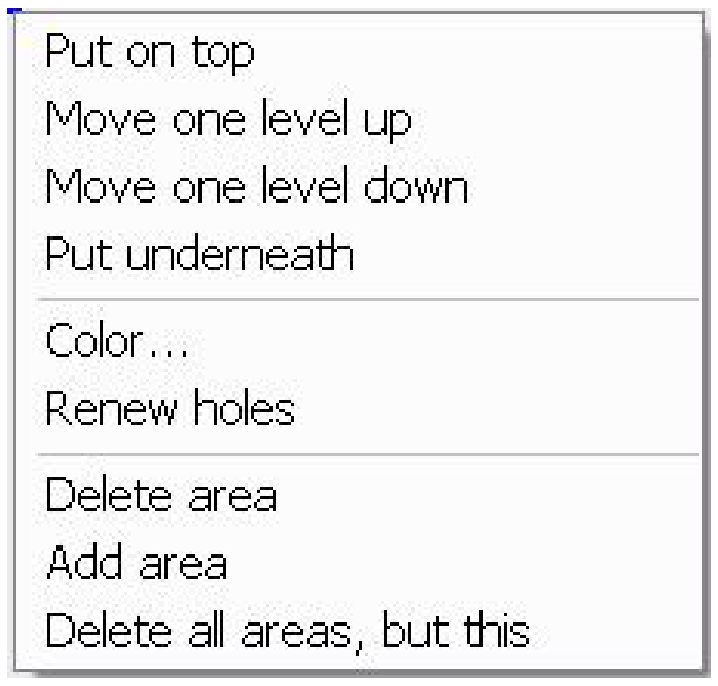

**Fig.8.11** Menu for the colored areas

Another menu can be called on any colored area (board); commands of this menu are divided into several groups (**figure 8.11**). Commands of the first group allow to change the order of areas on the screen. Each area occupies its own level; the number of areas (levels) is unlimited. There are four standard commands, which are used in many examples further on: an object (in this case it is a colored area) can be moved one level up or down and the object can be moved on top of all others or to the bottom.

Commands for tuning the areas are very limited, because I could not think out anything more; you can change the color of any area and you can renew the set of holes in it.

The particular type of each hole is defined at random.

The third group of commands allows to change the number of areas. You can add a new area, delete the touched area, or delete all other areas except the touched one. There is no way to delete the last area in view; the program checks the number of areas and disables the appropriate menu command, if there is only one area left.

I mentioned that the **Form_FillTheHoles.cs** is designed as a normal user-driven application. The time has not come to discuss the rules of design for such applications, but I have to admit that there is definitely one serious flaw in this form: the tuned parameters and some changes are not saved on closing the form and, as a result, are not restored the next time you open the form. When I look at this problem as a programmer, I want to fix this problem; it will take around 15 minutes. Then I remind myself that this is an example in a book and I want to leave it unchanged as a possible exercise. After going through the second part of the book and looking at similar procedures of saving and restoring the views of many further examples, you can do it really quickly. The only things that need saving / restoring here are the sizes of the form, the group, and the panel with information. Both classes `ElasticGroup` and `TextM` have their `IntoRegistry()` and `FromRegistry()` methods to deal with the `Registry`; the use of these methods is demonstrated in many examples of the accompanying Demo application.

The main idea of this form is to close the holes in the areas with the appropriate plugs. Plugs can be moved, rotated, and resized to organize the "well enough fit" with the holes. Now let us see how all these things work.

When the form is opened, three areas with holes are constructed, the *Plugs* group, and the information.

```
private void OnLoad (object sender, EventArgs e)
{
    areas .Add (new AreaWithHoles (new RectangleF (200, 100, 420, 420), 3, 3,
                                   Color .Green));
    areas .Add (new AreaWithHoles (new RectangleF (250, 150, 560, 420), 3, 4,
                                   Color .Blue));
    areas .Add (new AreaWithHoles (new RectangleF (300, 50, 420, 560), 4, 3,
                                   Color .Red));
    groupButtons = new ElasticGroup (this, new Control [] { btn_Circle, btn_N3,
                        btn_N4, btn_N5, btn_N6, btn_N7, btn_N8, btn_N9 }, "Plugs");
    info = new TextM (this, new Point (40, 450), infotext);
    info .BackColor = Color .Lime;
    RenewMover ();
}
```

Plugs and colored areas are organized into two special `Lists`.

```
List<AreaWithHoles> areas = new List<AreaWithHoles> ();
List<Plug> plugs = new List<Plug> ();
```

All the objects must be registered in the mover's queue. Controls have to take the leading positions in this queue, so the group must be at its head. Graphical objects can be included in different order, but I want information to be shown above all other graphical objects. And I want plugs to move atop the areas, as this would be much better for viewing their fit to the holes. Thus we receive such an order of objects in the mover's queue: group, information, plugs, and areas.

```
public void RenewMover ()
{
    mover .Clear ();
    mover .Add (info);
    for (int i = 0; i < plugs .Count; i++)
    {
        mover .Add (plugs [i]);
    }
    for (int i = 0; i < areas .Count; i++)
    {
        mover .Add (areas [i]);
    }
    groupButtons .IntoMover (mover, 0);
}
```

Plugs – circles and regular polygons – are the objects, which were discussed in details in the previous chapters. Resizable circles have the border covered by a set of small overlapping circular nodes; one bigger circular node is used to cover the whole area of an object. The border of any regular polygon is covered by a set of strip nodes between each pair of consecutive vertices; the area of object is covered by a polygonal node.

```
public override void DefineCover ()
{
    CoverNode [] nodes;
    if (shape == Shape .Circle)
    {
        nodes = new CoverNode [nNodesOnCircle + 1];
        for (int i = 0; i < nNodesOnCircle; i++)
        {
            nodes [i] = new CoverNode (i, Auxi_Geometry .PointToPoint (center,
                                 2 * Math .PI * i / nNodesOnCircle, radius), nrSmall);
        }
        nodes [nNodesOnCircle] = new CoverNode (nNodesOnCircle, center,
                                        Convert.ToInt32 (radius), Cursors.SizeAll);
    }
    else
    {
        // first the strips along the sides; the last one is the convex polygon
        nodes = new CoverNode [nVertices + 1];
        PointF [] pts = Vertices;
        for (int i = 0; i < nVertices; i++)
        {
            nodes [i] = new CoverNode (i, pts [i], pts [(i + 1) % nVertices], 5);
        }
        nodes [nVertices] = new CoverNode (nVertices, pts);
    }
    cover = new Cover (nodes);
}
```

The resizing and rotation of circles and regular polygons was already discussed in details in the previous chapters.

The most interesting in the **Form_FillTheHoles.cs** is the `AreaWithHoles` class. There are no limitations on the size of the rectangular area, but there is a minimum allowed distance between the holes, so that they never overlap. Holes can be either circular or in a form of a regular polygon with the number of vertices between 3 and 9. The particular shape of each hole is defined at random. The initial number of holes is set at the moment of initialization by declaring the number of rows and columns.

```
public AreaWithHoles (RectangleF rect, int rows, int columns, Color clr)
{
    nRow = Math .Max (rows, 1);
    nCol = Math .Max (columns, 1);
    rc = new RectangleF (rect .Left, rect .Top,
       Math.Max (rect.Width, nCol * 2 * (spaceAroundHole + Hole .MinimumRadius)),
       Math.Max (rect.Height, nRow * 2 * (spaceAroundHole + Hole .MinimumRadius)));
    DefineHoles ();
    brush = new SolidBrush (clr);
}
```

Holes can be of the same shapes as plugs: either a circle or a regular polygon with the number of vertices between 3 and 9. I have similar programs in which holes are also resizable; this feature adds nothing interesting to the questions under discussion, but makes a code longer. In this application I decided to organize areas with the non-resizable holes, so from the moment of initialization the holes are fixed.

Holes are used as figures to which plugs have to fit. Fitness is estimated by the closeness of two figures: hole and plug. The estimation is based on the comparison of their parameters. For a circle it is enough to have central point and radius. For a regular polygon you need central point, radius of vertices, number of vertices, and an angle.

```
    public class Hole
    {
        Shape shape;
        PointF center;
        float radius;
```

```
        int nVertices;
        double angle;
```

After a set of holes in the area is determined, it is time for the cover design. With the use of transparent nodes, the design of cover is elegant and easy. Each hole is either a circle or a regular polygon. In the first case, the corresponding node gets the circular shape; in the second case, it is a polygonal node, defined by the points for vertices. For any rectangular area and any set of nodes, the cover consists of easily defined nodes on holes (one per each), plus a rectangular node for the whole area. This is the only node in the cover, which is not transparent. Any transparent node must precede in the cover that bigger node, from which it burns out some part; thus the node for the whole area must be the last one in the cover.

```
public override void DefineCover ()
{
    CoverNode [] nodes = new CoverNode [holes .Count + 1];
    for (int i = 0; i < holes .Count; i++)
    {
        if (holes [i] .VerticesNumber == 0)
        {
            nodes [i] = new CoverNode (i, holes [i] .Center, holes [i] .Radius,
                                              Behaviour .Transparent);
        }
        else
        {
            nodes [i] = new CoverNode (i, holes [i] .Vertices,
                                              Behaviour .Transparent);
        }
    }
    nodes [holes .Count] = new CoverNode (holes .Count, rc);
    cover = new Cover (nodes);
}
```

It is funny to mention, but such an interesting object with a lot of holes of different shape has extremely simple both `Move()` and `MoveNode()` methods. They are as simple, as you can find only in the first example of the book with the primitive objects, whose covers consist of a single node. The `AreaWithHoles` class is very close to those primitive objects, because it also has only one node for real movement. The cover may consist of many nodes, but all of them except one give mover only one command: “Forget about this object immediately!”

If the cover has only one node for moving, there is no need in any checking in the `MoveNode()` method; the only needed thing is to call the `Move()` method and that is all.

```
public override bool MoveNode (int i, int dx, int dy, Point ptM, MouseButtons btn)
{
    bool bRet = false;
    if (btn == MouseButtons .Left)
    {
        Move (dx, dy);
        bRet = true;
    }
    return (bRet);
}
```

The `Move()` method is also very simple. For the area itself it would be enough to change one point – the location of the area, but there are holes, which have to be moved synchronously.

```
        public override void Move (int dx, int dy)
        {
            rc .X += dx;
            rc .Y += dy;
            SizeF size = new SizeF (dx, dy);
            for (int i = 0; i < holes .Count; i++)
            {
                holes [i] .Center += size;
            }
        }
```

If the `AreaWithHoles` class turns out to be very simple, then something must be done in the `OnMouseDown()` and `OnMouseUp()` methods of the form.

Two things, which are associated with the plugs, must be started in the `OnMouseDown()` method if any plug is caught.

- If a strip node is caught, it can be only the border of a polygonal plug and then the `Plug.StartScaling()` method must be called.
- If a plug is caught by the Right button, then the `Plug.StartRotation()` method must be called.

In nearly all the examples the code of the `OnMouseDown()` method is divided between the reactions on catching some objects either by the left or right button; each of these branches is divided further on depending on the class of the caught object. In the **Form_FillTheHoles.cs** any interesting situation happens only after catching a `Plug` object, so the consequence of **if** statements is a bit unusual.

```
private void OnMouseDown (object sender, MouseEventArgs e)
{
    ptMouse_Down = e .Location;
    if (mover .Catch (e .Location, e .Button))
    {
        if (mover .CaughtSource is Plug)
        {
            mover .Clipping = Clipping .Unsafe;
            Plug plug = mover .CaughtSource as Plug;
            if (plug .Shape == Shape .Polygon)
            {
                if (e .Button == MouseButtons .Left)
                {
                    if (mover .CaughtNodeShape == NodeShape .Strip)
                    {
                        plug .StartScaling (e .Location);
                    }
                }
                else if (e .Button == MouseButtons .Right)
                {
                    plug .StartRotation (e .Location);
                }
            }
        }
    }
    ContextMenuStrip = null;
}
```

And there is an interesting case of changing the mover's `Clipping` property throughout the work of a form. By default this property is set to `Visual`, so you can move all the objects around the screen, but without losing them from view. The areas, if you do not need them, can be deleted via the menu command (**figure 8.11**). What to do with the unneeded plugs? Are they going to stay somewhere around distracting the attention? There is no menu on the plugs with a command to delete them, though it is not a problem to add such a menu. The change of the `Clipping` property to `Clipping.Unsafe`, when a plug is caught for moving, allows to drop any plug beyond the borders and forget about it. Certainly, an original mover's `Visual` clipping is restored in the `OnMouseUp()` method; when later any other object is caught for moving, it is not allowed to be moved across the borders.

Several actions are started from inside the `OnMouseUp()` method.

The colored area with the holes can be called on top of all others by a simple click.

```
private void OnMouseUp (object sender, MouseEventArgs e)
{
    ptMouse_Up = e .Location;
    double dist = Auxi_Geometry .Distance (ptMouse_Down, e .Location);
    if (mover .Release ())
    {
        GraphicalObject grobj = mover .WasCaughtSource;
```

```
            long id = grobj .ID;
            if (e .Button == MouseButtons .Left)
            {
                if (grobj is AreaWithHoles)
                {
                    if (dist <= 3)
                    {
                        PopupArea (id);
                    }
```

The big area can be either moved around the screen or called to the proscenium. With the smaller plugs there are more options and the consequences of some of them are much more interesting.

```
    private void OnMouseUp (object sender, MouseEventArgs e)
    {
            … …
            if (e .Button == MouseButtons .Left)
            {
                … …
                else if (grobj is Plug)
                {
                    if (!PlugDisappeared (id, ClientRectangle))
                    {
                        if (dist <= 3)
                        {
                            PopupPlug (id);
                        }
                        CheckMatch ((grobj as Plug));
                    }
                }
            }
            else if (e .Button == MouseButtons .Right)
            {
                if (grobj is Plug)
                {
                    CheckMatch ((grobj as Plug));
                }
```

First, the fate of any plug that could be moved across a border and dropped there. It is not in view, it is not distracting your attention any more, but it is still an object in the `List` of plugs. It is not visible, but it exists. The garbage, which has to be deleted. That is the work for the `PlugDisappeared()` method: to delete any plug, which is released outside the `ClientRectangle` area.

Second thing to be done: any plug can be called on top of other plugs, if it was simply clicked (moved for a zero or not more than a couple of pixels).

The last option is the most interesting: check the fitness of the plug to any hole. This action is required both for forward movement and for rotation of any plug.

Part of the checks is easy to do as there is no match between a circle and any polygon or between the polygons with different number of vertices. But other checks depend on your own definition for "good enough fitness". It would be ridiculous to demand that the central points of a plug and a hole have to be at exactly the same point; a small discrepancy must be allowed. Other small discrepancies must be allowed between the radii and the angles.

If all the checks signal about "good enough fitness", then the plug and the hole both annihilate. The after steps can be different. If it was not the last unclosed hole in the area, then nothing else has to be done; take another plug for another hole. If no more holes are left in the area, then they must be renewed in full number. Life goes on. In each of these cases the cover of the area must be renewed, because the number of holes has changed.

# Sliding partitions

Several examples in the previous chapters demonstrate the multicolored objects. Those objects are resizable by borders, but the ratio between their inner parts is fixed at the moment of design and does not change throughout the time of their life. In this chapter I show the similar objects with the movable partitions between the inner parts.

## *Rectangles with sliding partitions*

File: **Form_Rectangles_SlidingPartitions.cs**
Menu position: *Graphical objects – Basic elements – Rectangles – Sliding partitions*

The first example in this chapter shows one more type of resizable rectangles. As you can see, these rectangles are multicolored (**figure 9.1**). In addition to standard resizing by borders and moving by any inner point, the partitions between different colors are movable.

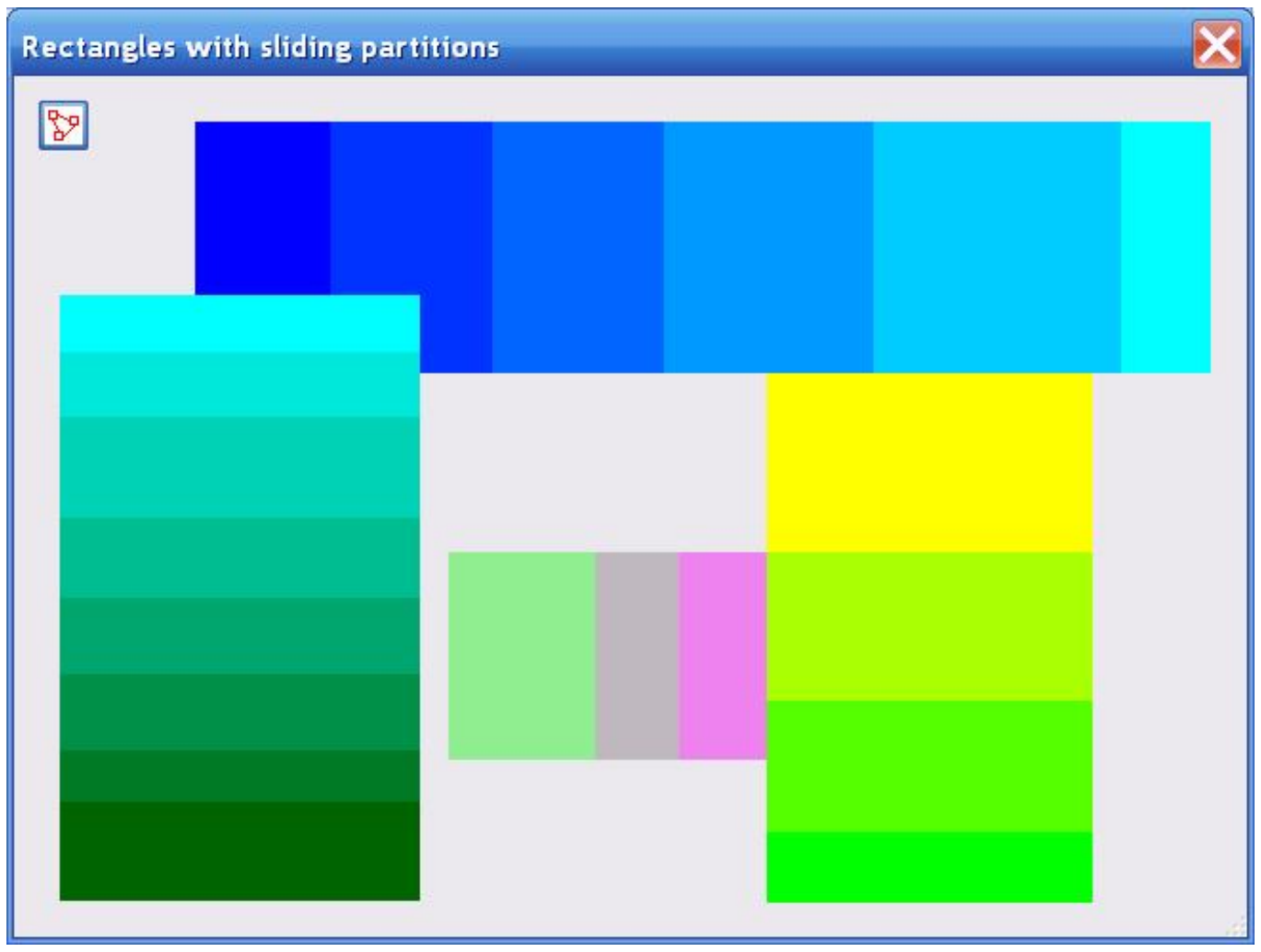

**Fig.9.1** Rectangles with the sliding partitions

Objects of the `Rectangle_SlidingPartitions` class have two restrictions to prevent them from total disappearance on squeezing (`minSize = 10`) and to prevent the disappearance of any of the colored segments (`minSegment = 4`). All the borders – outer and inner – are covered by the thin rectangular nodes (**figure 9.2**), which provide different types of resizing. The order of nodes in the cover is determined by the priority of the allowed movements (as I estimate them) and partly by the width of the sensitive nodes, covering the borders.

The very first example of rectangles in this book was shown with the possibility of changing the width of the sensitive nodes, responsible for moving the borders (**Form_Rectangles_StandardCase.cs**, **fig.3.1**). I found out throughout the years of work with the movable objects that a strip of six pixels width over a border (three pixels on each side of the border line) is just wide enough to find such a node easily with a mouse (in real applications the nodes are never visualized) and at the same time such node covers from other actions only a tiny part of an object. So in this and further examples the width of such nodes is standard and set to six pixels, though you can change them at any moment, if you want.

The nodes on the borders of colored strips in these rectangles play different roles:

- A move of any border between two strips changes only the widths of these two strips and does not affect anything else.
- A move of the outer border changes the size of the whole rectangle. At the same time the ratio between the widths of all the segments (parts of rectangle) does not change, so the sizes of all the segments have to change in order to keep the ratio. Thus, a move of the outer border affects all the segments simultaneously.

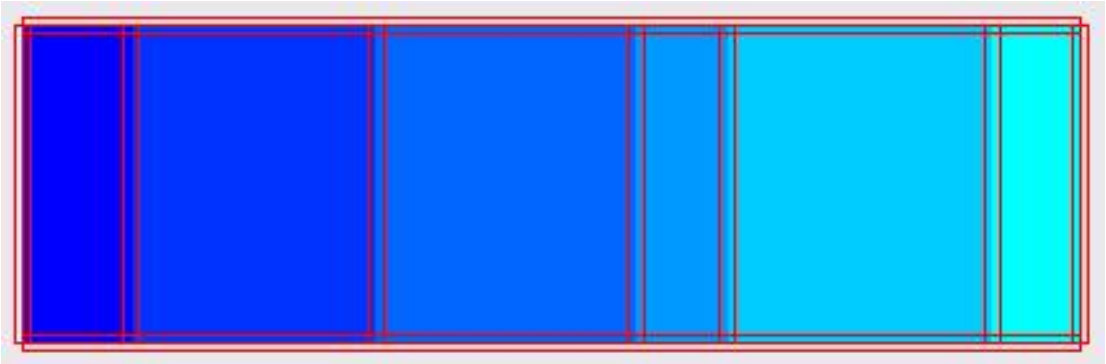

**Fig.9.2** Rectangle with the sliding partitions and its cover

Different reactions on moving different nodes make the order of nodes in the cover one of the most important things. The width of the nodes on all the borders is set to six pixels. At the same time, the minimum size of the colored segments in these rectangles is set to four pixels; I decided that that would be enough to see them. Thus, if you move two inner borders close to each other, you will still see the narrow strip, but its full area is going to be covered by two nodes, which are responsible for resizing, but not for moving the whole object. If you put the nodes on the inner borders at the head in the cover and then squeeze the whole object, you can turn it into unmovable and non-resizable. The inner nodes will cover the whole area of an object; at the same time the inner borders will stay so close to each other that there is no place to move them. Such an object will turn into analogue of local “black hole”, though it is still visible.

To avoid such a ridiculous situation, the nodes of the cover have such an order.

1. Two nodes on those two borders of rectangle, which are parallel to the sliding partitions. These nodes provide the resizing of the whole rectangle without changing the ratio between the segments.

2. A set of nodes on the inner partitions; these nodes allow to change the ratio between the segments.
3. Two nodes on two other borders of rectangle, which are orthogonal to the sliding partitions. These nodes provide the resizing of the whole rectangle.
4. One big node covering the whole rectangle; this node provides the movement of the whole object.

To provide the resizing of an object without changing the ratio between the segments, the whole set of coefficients must be stored at the beginning of such movement. As usual, when any coefficients have to stay unchanged throughout some type of movement between the **MouseDown** and **MouseUp** events, these coefficients must be calculated at the start of this movement by calling the special method from inside the `OnMouseDown()` method. This was demonstrated with rotation or zooming of different objects; exactly the same thing is done by calling the `Rectangle_SlidingPartitions.Distribution()` method. All the nodes of a cover for such rectangles are polygonal (they are all rectangles), so the decision about calling this method must be based on the number of the caught node. This check is easy to organize, as only the catching of the first two nodes requires the call of the `Distribution()` method.

```
private void OnMouseDown (object sender, MouseEventArgs e)
{
    ptMouse_Down = e .Location;
    if (mover .Catch (e .Location, e .Button))
    {
        if (e .Button == MouseButtons .Left &&
            mover .CaughtSource is Rectangle_SlidingPartitions &&
            mover .CaughtNode < 2)
        {
            (mover .CaughtSource as Rectangle_SlidingPartitions).Distribution ();
        }
    }
}
```

This `Distribution ()` method not only calculates the needed ratio coefficients, but also saves the number of the narrowest segment. These two things are needed later for two different types of resizing. The parts of code for horizontal and vertical partitions are similar; the code below is for the case of the vertical partitions (**figure 9.2**).

Before zooming the whole rectangle, it is enough to check only the proposed new size of the narrowest segment against the allowed minimum size of segment. If zooming is allowed, then the new sizes of all the segments are calculated with the set of previously saved coefficients.

```
public override bool MoveNode (int i, int dx, int dy, Point ptM, MouseButtons btn)
{
    if (btn == MouseButtons .Left)
    {
        … …
        if (i == 0)                    // left border
        {
            widthNew = rc .Width - dx;
            if (widthNew * fPart [iMinSegment] >= minSegment)
            {
                rc .X += dx;
                rc .Width -= dx;
                for (int j = 0; j < segmentSize .Length; j++)
                {
                    segmentSize [j] = Convert .ToSingle (rc .Width * fPart [j]);
                }
                bRet = true;
```

Before moving of any inner partition, the proposed new sizes of segments on both sides of the border under move must be checked against the allowed minimum size of segment.

```
public override bool MoveNode (int i, int dx, int dy, Point ptM, MouseButtons btn)
{
    if (btn == MouseButtons .Left)
    {
```

```
… …
else if (i <= segmentSize .Length)      // sliding partitions
{
    wOnLeft = segmentSize [i - 2] + dx;
    wOnRight = segmentSize [i - 1] - dx;
    if (wOnLeft >= minSegment && wOnRight >= minSegment)
    {
        segmentSize [i - 2] += dx;
        segmentSize [i - 1] -= dx;
        bRet = true;
    }
```

## *Circles with sliding partitions*

File: **Form_Circles_SlidingPartitions.cs**
Menu position: *Graphical objects – Basic elements – Circles – Sliding partitions*

Next example demonstrates the use of the sliding partitions in case of the multicolored circles (**figure 9.3**). The resizable multicolored circle of the `CircleNR` class was the first object on which I explained the N-node covers (**Form_NnodeCovers.cs**, **figure 7.2**), but for that class the angle of each segment was determined at the moment of initialization and did not change throughout the movement of a circle or its resizing. Objects of the `Circle_SlidingPartitions` class are different, as the ratio between the sectors can be easily changed by moving the sliding partitions. This class has two restrictions:

- Minimum allowed radius (`nMinRadius = 15`) to prevent a circle from total disappearance.
- Minimum allowed angle for any sector to prevent its disappearance. This value is set to 0.05 radians, so it is approximately three degrees.

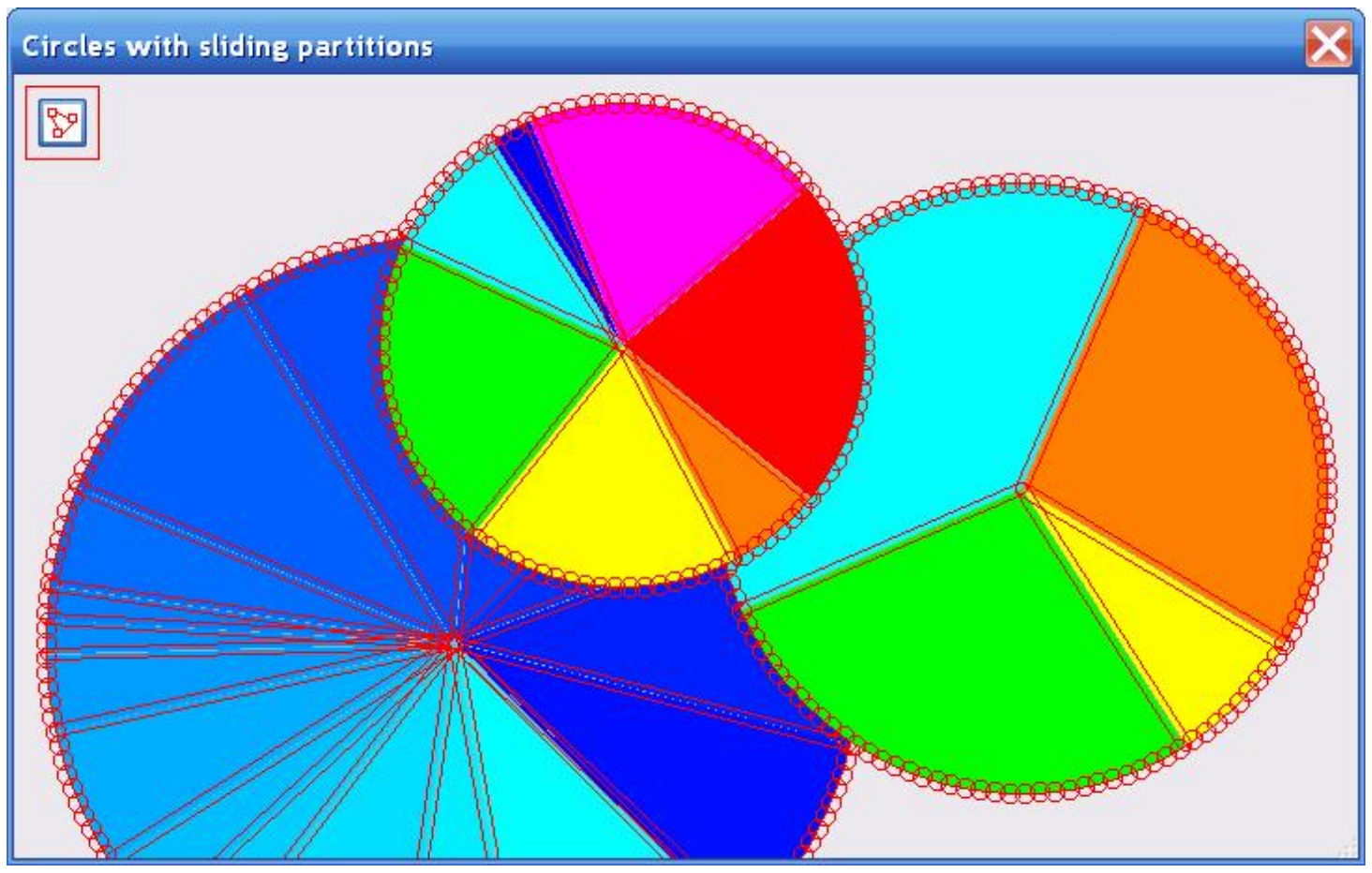

**Fig.9.3** Circles with the sliding partitions

The cover of the new class is similar to the old one, but in addition the partitions between the sectors are covered by the strip nodes. The number of nodes on the border of a circle (`nNodesOnCircle`) is defined by the radius of a circle and the distance between the neighbouring small nodes. The number of strip nodes is equal to the number of colored segments. This is a classical type of cover, in which the smaller nodes precede all the bigger ones, so the order of nodes in this class is: circular nodes on the outer border, strip nodes on the inner partitions, and the big node to cover the whole area of an object.

```
public override void DefineCover ()
{
    CoverNode [] nodes;
    nodes = new CoverNode [nNodesOnCircle + vals .Length + 1];
    for (int i = 0; i < nNodesOnCircle; i++)
    {
        nodes [i] = new CoverNode (i, Auxi_Geometry .PointToPoint (center,
                            2 * Math .PI * i / nNodesOnCircle, radius), nrSmall);
    }
    double angleLine = angle;
    for (int i = 0; i < vals .Length; i++)     // nodes on borders between sectors
    {
        nodes [nNodesOnCircle + i] = new CoverNode (nNodesOnCircle + i, center,
                        Auxi_Geometry .PointToPoint (center, angleLine, radius));
        angleLine += sweep [i];
    }
    nodes [nodes .Length - 1] = new CoverNode (nodes .Length - 1, center,
                                    Convert .ToInt32 (radius), Cursors .SizeAll);
```

```
    cover = new Cover (nodes);
    cover .SetClearance (false);
}
```

In several of the previous examples I have already explained the technique of calculating some needed coefficients at the starting moment of the movement; then these parameters are used throughout the whole time of movement until the object is released. This technique is used both for resizing and rotation; the same technique is used for moving the sliding partitions in the case of the `Circle_SlidingPartitions` class.

When any sliding partition is pressed for moving (well, in reality it is the strip node above this partition, which is really caught), then the range for moving this partition in one direction or another must be determined. This is done by calling the `StartResectoring()` method; the only parameter of this method is the number of the caught node.

```
    private void OnMouseDown (object sender, MouseEventArgs e)
    {
        ptMouse_Down = e .Location;
        if (mover .Catch (e .Location, e .Button))
        {
            GraphicalObject grobj = mover .CaughtSource;
            if (grobj is Circle_SlidingPartitions)
            {
                if (e .Button == MouseButtons .Left)
                {
                    if (mover .CaughtNodeShape == NodeShape .Strip)
                    {
                        (grobj as Circle_SlidingPartitions) .StartResectoring
                                                                  (mover .CaughtNode);
                    }
```

The `StartResectoring()` method looks simple, but I think that some explanations on its code are needed.

```
public void StartResectoring (int iNode)
{
    iBorderToMove = iNode - nNodesOnCircle;
    double angleCaughtBorder = angle;
    for (int i = 0; i < iBorderToMove; i++)
    {
        angleCaughtBorder += sweep [i];
    }
    if (dirDrawing == Rotation .Clockwise)
    {
        iSector_Clockwise = iBorderToMove;
        min_angle_Resectoring = angleCaughtBorder + sweep [iSector_Clockwise];
        iSector_Counterclock = (iSector_Clockwise == 0) ? (vals .Length - 1)
                                                        : (iSector_Clockwise - 1);
        max_angle_Resectoring = angleCaughtBorder - sweep [iSector_Counterclock];
    }
```

Any `Circle_SlidingPartitions` object is described by the starting angle for the first segment (`angle`) and the set of angles for segments (`sweep[]`), so the first thing to do is to calculate the angle for the caught partition (`angleCaughtBorder`). Then the range of moving this partition is calculated, but for this calculation two things must be taken into consideration.

1. The direction of drawing the circle: it can be clockwise or counterclockwise.
2. The finite number of real values, which are behind the segments, these numbers are turned here into the infinitive loop of segments, so the number of the caught partition is used once more to get the angles for the two neighbouring segments on the sides of it. Thus not only the limits for moving this partition are calculated, but the sum of angles for two segments. With the moving of any partition, the ratio between two segments is going to change, but not their sum.

Moving of any node is described in the `MoveNode()` method of the class. For moving the partition, the real allowed range is going to be slightly less than calculated in the `StartResectoring()` method, as this class of circles does not allow the complete disappearance of any segment.

```
public override bool MoveNode (int i, int dx, int dy, Point ptM, MouseButtons btn)
{
    if (btn == MouseButtons .Left)
    {
        … …
        else          // border between two sectors
        {
        … …
            if (min_angle_Resectoring + minSector < angleMouse &&
                angleMouse < max_angle_Resectoring - minSector)
            {
                double part_Counterclock = (max_angle_Resectoring - angleMouse) /
                                     (max_angle_Resectoring - min_angle_Resectoring);
                if (iBorderToMove == 0)
                {
                    angle = angleMouse;
                }
                vals [iSector_Counterclock] =
                                      two_sectors_sum_values * part_Counterclock;
                vals [iSector_Clockwise] =
                                 two_sectors_sum_values –vals [iSector_Counterclock];
                SweepAngles ();
            }
```

## *Set of objects*

File: **Form_SetOfObjects.cs**
Menu position: *Graphical objects – Basic elements – Set of objects*

Examples in several previous chapters demonstrate the objects of different shapes and explain the design of their covers. In the **Form_SetOfObjects.cs** (**figure 9.4**) you can see the same familiar shapes. Here the objects of many different classes are put together in order to analyse their coexistence and to find out, if there are going to be any problems or something special in dealing with an unlimited number of elements of such variety.

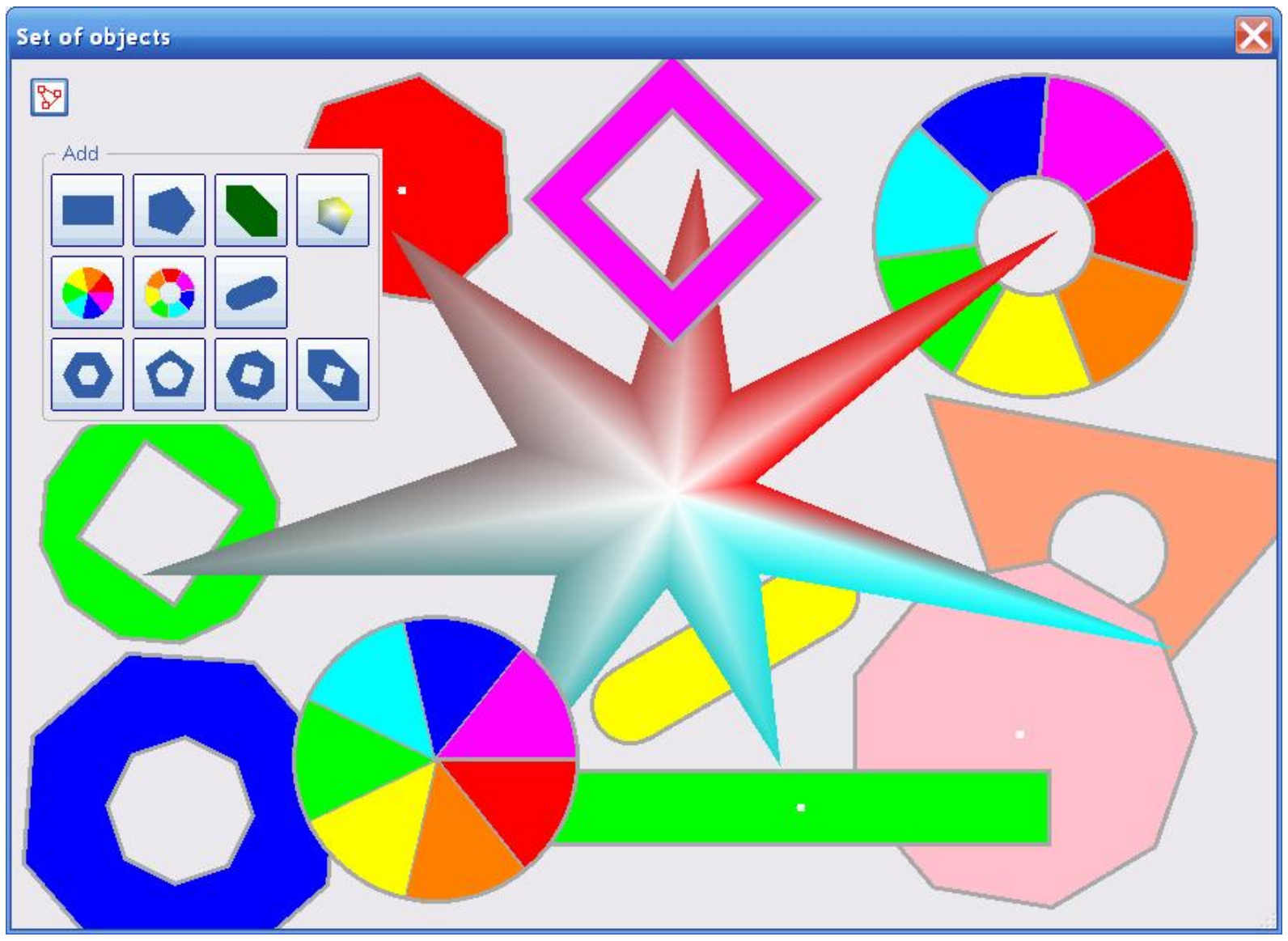

**Fig.9.4** Representatives of objects that are used in this form

There are graphical elements of 11 different classes; all of them are derived from the abstract class `ElementOfSet`. All objects can be moved and rotated by any inner point; all of them are resizable by any border point; circles and rings have the sliding partitions between their sectors. Some objects of different classes might look like they belong to the same class, but only at the beginning at the moment of initialization; later on they can be transformed in different ways.

Here are the types of objects that are used in this form and the short descriptions of their resizing.

| | |
|---|---|
| Rectangle | Resizable by any side and any corner. |
| Circle | Resizable by the border; movable partitions between the sectors. |
| Ring | Resizable by the borders; movable partitions between the sectors. |
| Strip | The curved parts of the border are used to change the length; the straight parts of the border allow to change the width of a strip. |
| Regular polygon | Resizable by the border. |
| Perforated polygon | Resizable by both borders. The two borders are the identical regular polygons with the same number of vertices and the same angle. |
| Chatoyant polygon | Organized as a number of vertices, connected into an infinitive loop, plus a central point. All these N + 1 points can be moved arbitrary around the screen thus causing a reconfiguration. Each pair of consecutive vertices is connected by a strip node; these nodes are used for zooming of an object. |
| Convex polygon | Initialized as a regular polygon; after it all vertices can be moved individually around the screen until the convexity of the polygon is not broken. Each pair of consecutive vertices is connected by a strip node; these nodes are used for zooming of an object. |
| Regular polygon with a regular polygonal hole | Resizable by both borders. The inner and outer borders are both the regular polygons, but they may have different number of vertices and different angles. |
| Regular polygon with a circular hole | Resizable by both borders. |
| Convex polygon with a regular polygonal hole | On initialization both borders, inner and outer, look like the regular polygons. Later the inner border is only resizable, but continues to be regular polygon. The outer border can be reconfigured, but has to be a convex polygon. |

All these objects were already demonstrated in the previous examples and their special features were discussed, but there is one important feature, which was added to all the classes in the **Form_SetOfObjects.cs**: their ability to be resized or rotated can be switched at any moment for any object individually or simultaneously for all of them. (I did not include the switch of movability for objects into this example, though it can be done in the same easy way.)

You can see at **figure 9.4** that the borders of nearly all the figures, except the chatoyant polygon, are marked with the wide grey lines. The wide lines on the borders are used in the **Form_SetOfObjects.cs** for visualization of the ability of objects to be resized. In this form it can be a good indication, but I never use this thing in real applications, and I would not recommend the use of it. The visualization of such feature looks like a good idea, but only when you start working with the user-driven application. After some time, and it happens quickly enough, when you get used to the idea of every object being resized by any border point, you do not need any indication or reminder.

At the same **figure 9.4** several objects are shown with a small white circle in their center. This is also an artificial indication of whether an object can be rotated at this particular moment or not; this indication is used only in this particular form and I would not recommend using it in real applications.

When any object is caught by mover, it can be a starting point for some resizing or rotation, which is decided inside the `OnMouseDown()` method on the basis of the pressed button.

```
private void OnMouseDown (object sender, MouseEventArgs e)
{
    ptMouse_Down = e .Location;
    if (mover .Catch (e .Location, e .Button))
    {
        if (e .Button == MouseButtons .Left)
        {
            if (mover .CaughtSource is ElementOfSet)
            {
                StartResizing (e .Location);
            }
        }
        else if (e .Button == MouseButtons .Right)
        {
            StartRotation (e .Location);
        }
    }
    ContextMenuStrip = null;
}
```

In this piece of code the `StartResizing` is only the generic name for a whole set of methods that are slightly different for each class of objects. For many of them it is a zooming operation, which is applied to all parts of an object; for others it can be a change in one direction only. To transform the general `StartResizing()` call into the specific method of any particular class, first this class must be determined. Also the methods of different classes depend on some additional parameters like the number of the node or its shape; the use of those additional parameters is determined by the design of cover for each class.

```
private void StartResizing (Point pt)
{
    GraphicalObject grobj = mover .CaughtSource;
    if (grobj is RegPoly_EOS)
    {
        (grobj as RegPoly_EOS) .StartScaling (pt);
    }
    else if (grobj is ConvexPoly_EOS)
    {
        (grobj as ConvexPoly_EOS) .StartScaling (pt, mover .CaughtNode);
    }
    else if (grobj is ChatPoly_EOS)
    {
        (grobj as ChatPoly_EOS) .StartScaling (pt, mover .CaughtNodeShape);
    }
    else if (grobj is Circle_EOS)
    {
        if (mover .CaughtNodeShape == NodeShape .Strip)
        {
            (grobj as Circle_EOS) .StartResectoring (mover .CaughtNode);
        }
    }
```

Usually the resizing of objects of any class is started not by any node of its cover, but only by some of them. The decision is often based on the shape of the caught node, which is one of the parameters that are received from mover. The checking of the needed conditions can be done in the form or in the particular class; in the last case this parameter must be passed into the method of the class. I do not see any advantages or disadvantages in either case; both of them work fine; you can see both ways in the lines of code above.

The start of rotation looks more straightforward: the abstract method from the base `ElementOfSet` class is overridden in each particular class and has the same parameter – point of the mouse press – for all of them.

The `OnMouseMove()` method is usually very simple; there must be a single call to one of the mover's methods, and this would be enough for all movements of all the classes as the details are in the `MoveNode()` methods of those classes. However, you can see the call to the `Update()` method in this form, if objects of two classes are under move.

```
private void OnMouseMove (object sender, MouseEventArgs e)
{
    ptMouse_Move = e .Location;
    if (mover .Move (e .Location))
    {
        GraphicalObject grobj = mover .CaughtSource;
        if (grobj is ElasticGroup || grobj is SolitaryControl)
        {
            Update ();
        }
        Invalidate ();
    }
}
```

When you deal only with the graphical objects, then the use of the standard double buffering solves the problem of the screen flickering. When you start to move controls or groups of controls, then the system makes the decision about the proper moment for their redrawing, and it is always done with some delay. Adding a couple of lines into the `OnMouseMove()` method solves the problem by enforcing the immediate update of the screen.

When any object is finally released by mover, it is time to decide about some actions and it is not always very easy to do. I have already written at the beginning that after releasing an object by the left button, there is a choice to interpret the command as a finished movement or an order to bring an object on top of others. On releasing an object by the right button, there is a similar choice between an ordinary rotation and a call for context menu. The **Form_SetOfObjects.cs** demonstrates all these variants; as usual, the choice is made on the basis of a distance between two points, where the mouse was pressed and released.

```
private void OnMouseUp (object sender, MouseEventArgs e)
{
    ptMouse_Up = e .Location;
    double dist = Auxi_Geometry .Distance (ptMouse_Down, e .Location);
    int iWasCaught, iNode;
    NodeShape shape;
    if (mover .Release (out iWasCaught, out iNode, out shape))
    {
        GraphicalObject grobj = mover [iWasCaught] .Source;
        if (e .Button == MouseButtons .Left)
        {
            if (grobj is ElementOfSet)
            {
                (grobj as ElementOfSet) .RedefineCover ();
                if (dist <= 3)
                {
                    CheckPopup (iWasCaught, iNode, shape);
                }
                Invalidate ();
            }
        }
        else if (e .Button == MouseButtons .Right)
        {
            if (dist <= 3 && grobj is ElementOfSet)
            {
                elemPressed = grobj as ElementOfSet;
                ContextMenuStrip = menuOnFigures;
            }
        }
    else
    {
        if (e .Button == MouseButtons.Right && dist <= 3 && figures.Count > 0)
        {
            ContextMenuStrip = menuOnEmpty;
        }
    }
}
```

You can be surprised to see the call for the `RedefineCover()` method on release of any `ElementOfSet` object by the left button. (In reality it will be any object, derived from this class.) I have already explained earlier that the cover must be redefined only when the change of sizes requires the new number of nodes in the cover. This happens only in the case of N-node covers, and even then this requirement is not always mandatory, but may depend on the construction of cover and the released node (a strip is the best demonstration of all these **if**). Of the 11 classes of graphical objects, shown at **figure 9.4**, only four have such type of covers, so the use of the `RedefineCover()` call in the `OnMouseUp()` method is a simplification of code. I prefer to add empty `RedefineCover()` methods into the classes of other figures instead of adding more **if** statements into the method above, but you can decide differently.

There can be any number of different objects in this form; all of them can be moved around, so there can be many overlapped figures. If the distance between the points of pressing and releasing the mouse is really short (not greater than three pixels) then the pressed object is moved on top of all others.

```
if (dist <= 3)
{
    CheckPopup (iWasCaught, iNode, shape);
}
```

When the number of object in the mover's queue is known, it is easy to make several steps:

1. Get the **id** of this object.

2. Find this object in the `List` of all the objects.
3. Move the object to the first position in the `List`.
4. Renew the mover's queue according to the new order of objects.
5. Redraw the screen.

I used the same technique in many of the previous examples, but in all of them it was enough to pass a single parameter to such method – the number of the pressed object. Then why are there three parameters in the `CheckPopup()` method in this form? From my point of view, this form of the method is correct and in all the previous examples the simplified version was used in order not to distract attention from more important features.

The covers for nearly all the figures in the **Form_SetOfObjects.cs** have one common feature: a single node to cover the whole area and several or many nodes to cover the border(s). From my observations, a tiny move of a border to adjust the view by one or two pixels is used much more often than the tiny move of the whole object. So, if there is a tiny move of any node on a border, then it is more likely to be the border adjustment than an attempt to bring this object atop. With the small move of the main area of an object the probability is reverse and it is more likely to be an attempt to put this object in full view on top of others. For these reasons I added the analysis of the additional parameters into the `CheckPopup()` method. For eight classes of 11 there is a checking of the shape of the node, as there is only one polygonal node, and this is the node that covers the whole area of an object. For three remaining classes I am checking if it is the last node in the cover. (Usually the area of an object is covered by the biggest node; all smaller nodes on borders are included into the cover ahead of that big one.)

```
private void CheckPopup (int iInMover, int iNode, NodeShape shape)
{
    bool bPopup = false;
    Figure figure = (mover [iInMover] .Source as ElementOfSet) .Figure;
    switch (figure)
    {
        case Figure .Rectangle:
        case Figure .RegularPolygon:
        case Figure .ConvexPolygon:
        case Figure .ChatoyantPolygon:
        case Figure .PerforatedPolygon:
        case Figure .RegPoly_CircleHole:
        case Figure .RegPoly_RegPolyHole:
        case Figure .ConvexPoly_RegPolyHole:
            if (shape == NodeShape .Polygon)
            {
                bPopup = true;
            }
            break;
        case Figure .Circle:
        case Figure .Ring:
        case Figure .Strip:
            if (iNode == (mover [iInMover].Source as ElementOfSet) .NodesCount - 1)
            {
                bPopup = true;
            }
            break;
    }
    if (bPopup)
    {
        PopupFigure (mover [iInMover] .Source .ID);
    }
}
```

What other features can be found in the **Form_SetOfObjects.cs**? The use of context menus. In the really complex applications, which are discussed later, there can be many different context menus in a single form. The standard practice is to have a personal menu for each class of objects, so there can be up to 10 different menus or more. In the current form, there are many different classes, but all the figures are derived from the `ElementOfSet` class, so one menu is enough for all of them; another menu can be opened at any empty spot.

**Figure 9.5** shows the menu opened on one of the figures. Several commands of the upper group allow to change the order of objects. These four commands to move an object one position up or down and to move it to one or another end of the `List`, these are the standard commands that are used throughout all of my applications. I found this set of commands enough even for the most complicated cases with a lot of objects on screen.

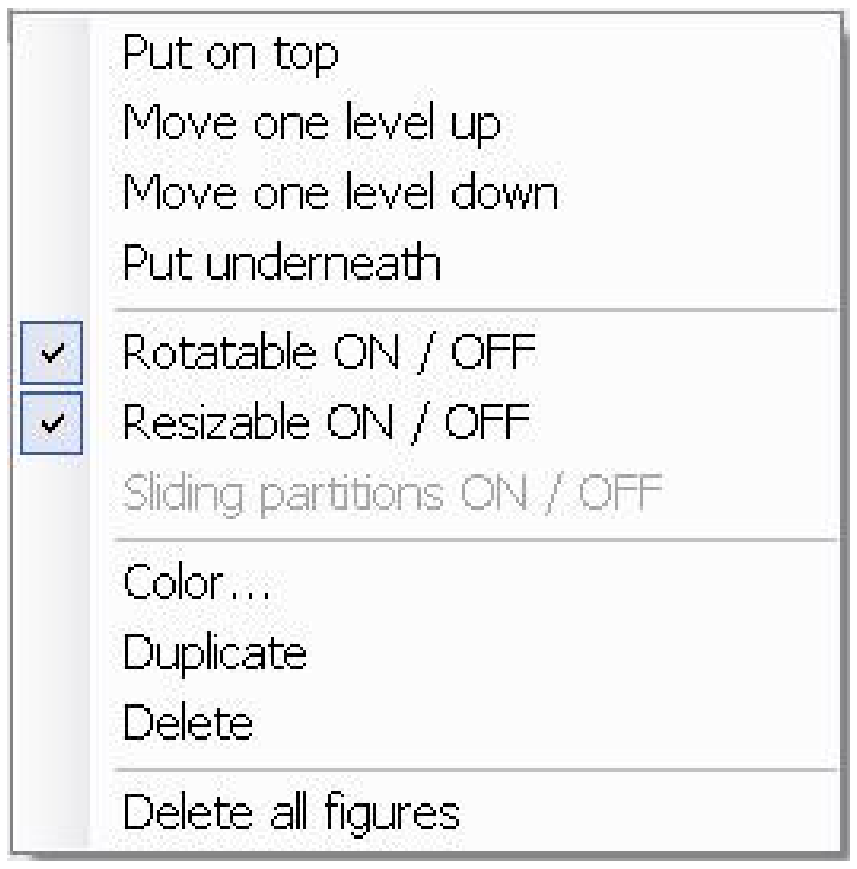

**Fig.9.5** Context menu on one of the figures

The second group of commands deals with the parameters of movability. Two first commands allow to switch between rotatable – non-rotatable and resizable – non-resizable. The third command allows to fix – unfix the internal partitions, so this command can be applied only to circles and rings; for all other objects this line in menu is disabled.

**Figure 9.6** demonstrates the menu, which can be called at any empty place of the form. The commands of this menu regulate the rotatability and resizability (outer and inner) of the objects in view, only these commands are applied to all the objects simultaneously.

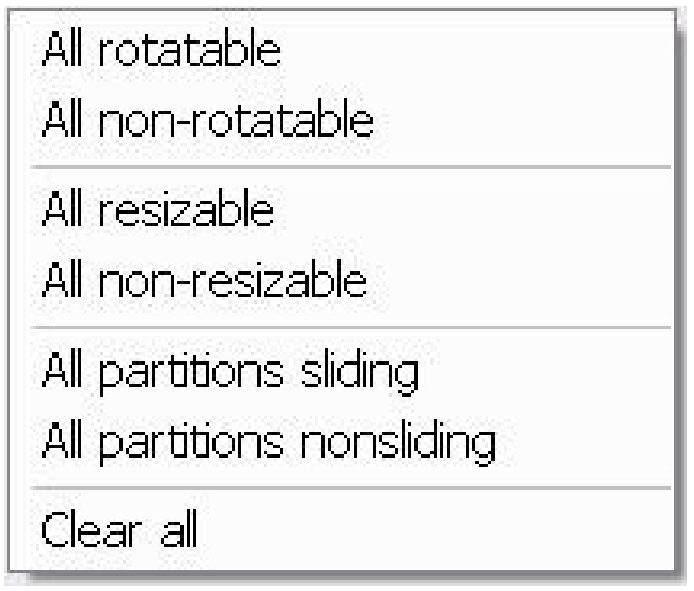

**Fig.9.6** Menu at empty places

The regulation of the object's involvement in rotation is a small addition to its ability to be rotated. The rotation process was described in details in the chapter *Rotation*. When an object is caught by the right button, this is the starting moment of rotation and the object's `StartRotation()` method must be called. All classes derived from the `ElementOfSet` class have such functions, but they differ in one thing. The only parameter for such a call is the current mouse position, which is used to calculate either a single compensation angle or a whole array of compensations. For some of the objects from the **Form_SetOfObjects.cs** this single compensation angle is a difference between the angle to the mouse and the object's angle; objects of the `Rectangle_EOS` class are of such a type. Other objects may have several independent basic points; for each of them the angle between the mouse position and this particular basic point must be calculated; objects of the `ChatPoly_EOS` class belong to this group. To rotate an object, this single compensation or the whole array of compensations is used to calculate the position of all the basic points in that part of the `MoveNode()` method, which is used, when an object is moved by the right button.

The regulation of the object's involvement in rotation is organized in identical way for all the objects. The base class `ElementOfSet` has a field `bRotate`. On initialization any graphical object in this form is rotatable, because in the constructor of the base class this field is assigned the **`true`** value.

```
public ElementOfSet ()
{
    bResize = true;
    bRotate = true;
    angle = 0;
    penBorder = new Pen (Color .DarkGray, 3);
    brushAnchor = new SolidBrush (Color .White);
}
```

The command from menu (**figure 9.5**) allows to switch this value for the pressed object.

```
private void Click_miRotatable (object sender, EventArgs e)
{
    elemPressed .Rotatable = !elemPressed .Rotatable;
    Invalidate ();
}
```

The call to redraw the form is included here into the `Click_miRotatable()` method, because some objects in this form can inform about their possible involvement in rotation by a small white circle in rotation center; in the real applications such thing usually do not exist, so the same thing goes without any repainting. But the change of the `bRotate` value is very important as the check of this value is included into that part of the `MoveNode()` method, which is associated with the move by the right button and which provides the rotation of an object. You can find this line in the methods for all the 11 classes.

```
public override bool MoveNode (int i, int dx, int dy, Point ptM, MouseButtons btn)
{
    bool bRet = false;
    if (btn == MouseButtons .Left)
    {
        … …
    else if (btn == MouseButtons .Right  &&  bRotate)
    {
        … …
    }
```

The switch between the resizable – non-resizable states of the same object is organized in absolutely different way. For the simple objects, which do not have independently moving parts (all figures in this form are of such a type), the easiest way is to change the cover. In the `DefineCover()` methods of all these figures you can see two branches of the code. The second branch is always shorter and very simple, as it does not include the nodes on the borders.

```
public override void DefineCover ()
{
    CoverNode [] nodes;
    if (bResize)
    {
        … …
    }
    else
    {
        … …
    }
    cover = new Cover (nodes);
}
```

When an object is resizable, the smaller nodes on the border nearly always precede in the cover the bigger nodes responsible for moving an object. Though these bigger nodes are the same in both cases, but they have different numbers for resizable and non-resizable objects of the same class, so the part of code in the `MoveNode()` method, which deals with forward movement of objects, has to be divided into two branches also. And the only thing that is needed in the `MoveNode()` method for non-resizable objects is to call the `Move()` method.

```
public override bool MoveNode (int i, int dx, int dy, Point ptM, MouseButtons btn)
{
    if (btn == MouseButtons .Left)
    {
        if (bResize)
        {
            … …
        }
        else
        {
            Move (dx, dy);
        }
```

The fixing – unfixing of the inner partitions (this is applied only to circles and rings) can be regulated by the special parameter **bFixSectors** (this is how I organize it in the **Form_SetOfObjects.cs**), or can be regulated by the same `bResize` parameter. It depends on how much flexibility you are ready to give to the users of your applications.

In the **Form_SetOfObjects.cs** the fixing / unfixing of the partitions is organized in such a way. When the corresponding command of the menu is called, the `SwitchSectorsFixing()` method of the pressed element is called. This menu command is enabled only for elements of two classes, so only two derived classes have this method. The redrawing of the form is needed in this case, because there is a possibility that covers are shown.

```
private void Click_miSlidingPartitions (object sender, EventArgs e)
{
    if (elemPressed .Figure == Figure .Circle)
    {
        (elemPressed as Circle_EOS) .SwitchSectorsFixing ();
```

```
            }
            else                  // elemPressed .Figure == Figure .Ring)
            {
                (elemPressed as Ring_EOS) .SwitchSectorsFixing ();
            }
            Invalidate ();
        }
```

The whole process for circles and rings is identical, so let us look at the `Circle_EOS` class. The called method switches the value of the `bFixSectors` and calls the redefinition of the cover.

```
        public void SwitchSectorsFixing ()
        {
            bFixSectors = !bFixSectors;
            RedefineCover ();
        }
```

The `RedefineCover()` method simply calls the `DefineCover()` method, but with the preliminary calculation of the needed nodes.

```
        public override void RedefineCover ()
        {
            NodesOnBorders ();
            DefineCover ();
        }
```

The `NodesOnBorders()` method has to define two different numbers, which can vary. If the circle is resizable, then it has to calculate the number of the small circular nodes (`nNodesOnCircle`) that will cover the border of the circle; otherwise this number is zero. The number of the nodes on the borders between sectors (`nNodesOnSides`) is either equal to the number of sectors or zero.

```
    private void NodesOnBorders ()
    {
        if (bResize)
        {
            nNodesOnCircle =
                  Convert .ToInt32 ((2 * Math .PI * radius) / distanceNeighbours);
        }
        else
        {
            nNodesOnCircle = 0;
        }
        nNodesOnSides = bFixSectors ? 0 : vals .Length;
    }
```

The `NodesOnBorders()` method prepares two values; each value can be zero or positive, so there are four different situations for any `Circle_EOS` object. The `Circle_EOS.DefineCover()` method simply uses these values.

```
    public override void DefineCover ()
    {
        CoverNode [] nodes;
        nodes = new CoverNode [nNodesOnCircle + nNodesOnSides + 1];
        for (int i = 0; i < nNodesOnCircle; i++)
        {
            nodes [i] = new CoverNode (i, Auxi_Geometry .PointToPoint (center,
                                  2 * Math .PI * i / nNodesOnCircle, radius), nrSmall);
        }
        double angleForLine = angle;
        for (int i = 0; i < nNodesOnSides; i++) // nodes on borders between sectors
        {
            nodes [nNodesOnCircle + i] = new CoverNode (nNodesOnCircle + i, center,
                         Auxi_Geometry .PointToPoint (center, angleForLine, radius));
            angleForLine += sweep [i];
```

```
        }
        nodes [nodes .Length - 1] = new CoverNode (nodes .Length - 1, center,
                                                   Convert .ToInt32 (radius), Cursors .SizeAll);
        cover = new Cover (nodes);
        cover .SetClearance (false);
    }
```

There are no checks of the Boolean values in this method; everything is defined by the two prepared values. If the circle is not resizable, then the number of circular nodes on the border is zero and the first loop is skipped. If the borders between the sectors are declared fixed, then the number of nodes on these borders is zero and the second loop is skipped. The only node, which always exists in the cover of his class, is the big circular node to move an object. I have written about changing the object's movability in the *Texts* chapter; there will be more about it in the examples with more complex objects further on.

When any of the parameters regulating the ability to resize or rotate the objects are changed, the cover has to be renewed by calling the `DefineCover()` method, but the call for `RenewMover()` method is not needed, as the number of objects registered with the mover is not changed. This is true for all simple objects. In the next chapter you will see that the situation with the complex objects is different.

Small addition at the last moment. The example, demonstrated in the **Form_SetOfObjects.cs**, is very similar to one of the examples from the article "*On the theory of moveable objects*" (see *Programs and documents*). Only at the last moment checking, when the book was finished and I was looking for some possible discrepancies between the book and the previous articles, I found out that I forgot to include into the **Form_SetOfObjects.cs** the uniting of an arbitrary set of objects into the groups. I decided not to change this example at the last moment; you can look for it in the mentioned article and its accompanying program. Also, the similar example of uniting different graphical objects into the groups is demonstrated further on in the chapter *An exercise in painting*; that example **Form_Village.cs** shows the better implementation, than was shown in the article nearly a year ago.

# Complex objects

In one of the previous chapters (*Texts*) I have demonstrated an example with some related movements. The objects in that example were independent, but movement of one of them could affect other objects. More common is the situation, when the related movements exist between the parts of some complex objects; the parts of such objects can be involved both in individual and related movements. This chapter looks into the details of such situations.

## *Rectangles with comments*

File: **Form_RectanglesWithComments.cs**
Menu position: *Graphical objects – Complex objects – Rectangles with comments*

In the **Form_RectanglesWithComments.cs** (**figure 10.1**) you can organize any number of rectangles (class `RectangleWithComments`); each rectangle may have an arbitrary number of comments (class `CmntToRectangle`). As usual, there are no limitations on anything user would like to do with these objects: add, delete, remove, change the order, or change the parameters at any moment. There is a single button to add the new rectangles; everything else is done with a mouse or via a couple of context menus (one for rectangles, another for comments). The individual and related movements of these objects are organized in such a way.

- When any rectangle is moved, then all its comments move synchronously with the rectangle.
- When any rectangle is resized, then all its comments move, but each one keeps the same relative position to the rectangle as it was before. The new position of comment depends on whether the particular comment was originally inside or outside the parental rectangle. (Position of comment is described by its central point.) When a comment is outside the rectangle, then its distance from the rectangle is kept constant regardless of the change of rectangle. If a comment is inside, then its relative position inside the rectangle is kept unchanged.
- Any comment can be moved and rotated freely; such movements have no effect on anyone else.

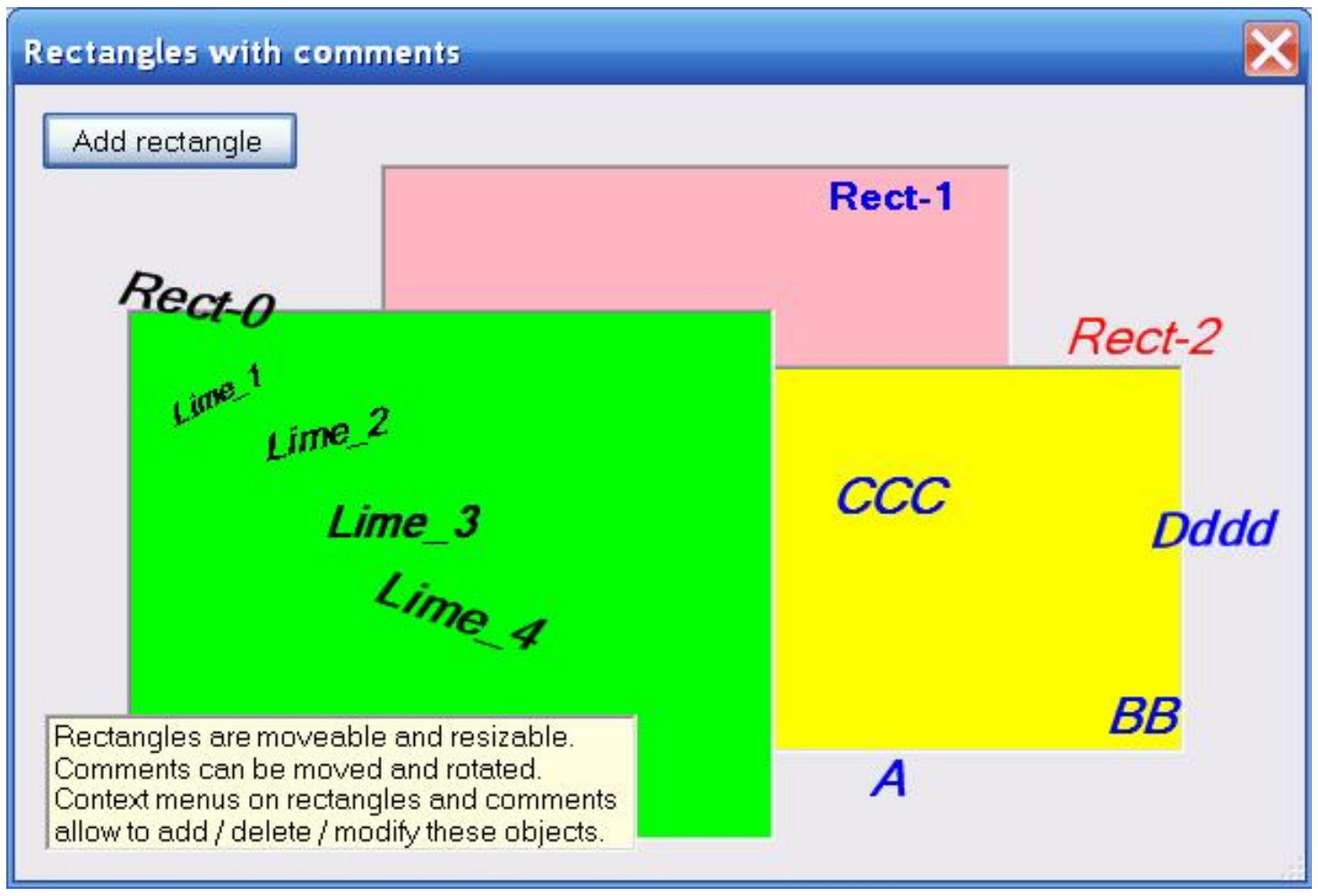

**Fig.10.1** Rectangles with comments

First, let us look into the `CmntToRectangle` class to understand, what has to be done for its participation in such movements.

```
public class CmntToRectangle : TextMR
{
    RectangleF rcParent;
    double xCoef;
    double yCoef;
```

The `CmntToRectangle` class is derived from the `TextMR` class, so the forward movement and rotation work automatically without any mentioning of an object in the code. But this is not enough for our purpose, so there are three new fields in this class of comments: the first one is the "parent" rectangle; two others are the coefficients to describe the position of comment in relation to this rectangle. A `CmntToRectangle` object can be initialized either by assigning these coefficients or the location of this comment. In the last case the coefficients are obtained with one standard method from the **MoveGraphLibrary.dll**.

```
Auxi_Geometry .CoefficientsByLocation (rcParent, pt, out xCoef, out yCoef);
```

This is how the coefficient along the horizontal scale is calculated.

- If the point is to the left of rectangle, then it is the distance from the point to the left border with the negative sign.

- If the point is to the right of rectangle, then it is the distance from the point to the right border with the positive sign.
- If the point is between the left and right border of rectangle, then the coefficient belongs to the [0, 1] range with 0 on the left border and 1 – on the right.

The coefficient along the vertical scale is calculated in the similar way with 0 on the upper border and 1 – on the lower border of rectangle.

Any comment has two instruments to set its position in relation to the parent: the area of parent (rectangle) and the coefficients to describe the position in relation to this rectangle. On any movement of either a rectangle or a comment only one of these parameters is changed; then the second one is used to recalculate the position.

- If a rectangle is moved or resized, then the fixed coefficients are used to calculate the new position of comment according to the changed area.

```
public RectangleF ParentRect
{
    get { return (rcParent); }
    set
    {
        rcParent = value;
        Location = Point .Round (Auxi_Geometry .LocationByCoefficients
                                                 (rcParent, xCoef, yCoef));
    }
}
```

- If a comment is moved, then its new point, received from its `Location` property, is used to calculate the new coefficients in relation to the unchanged rectangle.

```
public override void Move (int dx, int dy)
{
    base .Move (dx, dy);
    Auxi_Geometry .CoefficientsByLocation (rcParent, Location,
                                           out xCoef, out yCoef);
}
```

This is a pretty simple mechanism, which provides the correct involvement of comments in individual and related movements; now let us look at the code of the `RectangleWithComments` class. The class is designed as simple, as possible: one colored rectangle and a `List` of related comments.

```
public class RectangleWithComments : GraphicalObject
{
    RectangleF rc;
    SolidBrush brush;
    List<CmntToRectangle> comments = new List<CmntToRectangle> ();
```

Though the "parent" rectangle may have an arbitrary number of related comments and all these comments are involved in some synchronous or related movements with their "parent", but these facts do no affect the cover of rectangle in any way. The rectangle has its own cover; the comments have their covers, and the design of these covers have nothing to do with any kind of related movements, which are organized in other ways. A rectangle has to be moved by any inner point and resized by any border point, so the needed type of cover is standard; it was already demonstrated with the very first example of rectangles in this book (**Form_Rectangles_StandardCase.cs**, **figure 3.1**).

```
public override void DefineCover ()
{
    cover = new Cover (rc, Resizing .Any);
}
```

The `Move()` and `MoveNode()` methods are simple, but there is one small addition: whenever there is any change in position or size of the rectangle, all its associated comments must be informed about the change.

```
public override void Move (int dx, int dy)
{
    rc .X += dx;
    rc .Y += dy;
```

```
            InformRelatedElements ();
        }
```

This passing of the information from “parent” to all the linked comments is crucial for organizing the related movements, but in reality it is primitive.

```
        private void InformRelatedElements ()
        {
            foreach (CmntToRectangle comment in comments)
            {
                comment .ParentRect = rc;
            }
        }
```

I have already shown what this `CmntToRectangle.ParentRect` is going to do: the location of comment is adjusted to the new rectangle by using the existing coefficients of comment.

Now, when the details of the two involved classes – `RectangleWithComments` and `CmntToRectangle` – are explained, it is time to look at how it all works in the form, where the objects of these two classes are involved in individual and related movements..

The first and the most important thing in organizing such movements is the correct registering of the complex `RectangleWithComments` objects in the mover’s queue. Here are several statements that must be considered.

- A rectangle might have an arbitrary number of comments.
- The comments and rectangles can move individually; each of them has its own cover, so each one must be registered in the mover’s queue individually.
- The comments can be placed anywhere, but they are always shown atop their “parent” rectangle. To be shown above the rectangle, the comments must be painted after this rectangle. When any comment is placed above its “parent” rectangle and the mouse is pressed on the comment, then the expected object to be caught is a comment and not a rectangle. Thus all the comments must be registered before their parent in the mover’s queue.
- The comments can be added and deleted at an arbitrary moment; the request for such actions can come from different places in the code, for example, as a reaction on selection of different menu commands.

Combine these statements together and it will be obvious that it is unreliable to change the queue of the movable objects manually on any changes of the situation with comments. Such complex objects with individually movable “children” always have to use a single method, which guarantees the correct registration regardless of the number of components. Here is such `IntoMover()` method for the `RectangleWithComments` class.

```
        public void IntoMover (Mover mover, int iPos)
        {
            mover .Insert (iPos, this);
            for (int i = comments .Count - 1; i >= 0; i--)
            {
                mover .Insert (iPos, comments [i]);
            }
        }
```

While registering the `RectangleWithComments` object, first the cover of rectangle is inserted into the mover’s queue at the required position; then all the related comments are inserted at the same position ahead of their “parent” rectangle. Thus a rectangle will be correctly registered together with all its comments regardless of their number.

The change in the number of movable objects in the form might happen in different cases: add a new rectangle, delete a rectangle with all its comments, and add or delete any comment. In each case the `RenewMover()` method is called.

```
        void RenewMover ()
        {
            mover .Clear ();
            for (int i = rects .Count - 1; i >= 0; i -- )
            {
                rects [i] .IntoMover (mover, 0);
            }
            mover .Insert (0, info);
```

```
        mover .Insert (0, btnAddRectangle);
    }
```

These two methods of the form and of the class work in pair:

- The RectangleWithComments.IntoMover() method guarantees that any rectangle is registered correctly regardless of the number of its comments.
- The RenewMover() method guarantees that all the objects of the form are registered fully and correctly. Pay attention that the comments are not even mentioned in this method, because their correct registration is hidden inside the RectangleWithComments.IntoMover() method.

The IntoMover() method is designed for each complex class with the individual and related movements of the parts. The RenewMover() method is developed for each form with the changing number of movable / resizable objects.

Is there any difference in mover's dealing with the simple objects and complex objects? Not a bit! Mover does not know anything about the real objects but deals only with their covers. If any node is caught for moving, this is translated into the MoveNode() method of the corresponding object. Mover does not know anything about whether it is going to be an individual movement or a synchronous one; only the correct method of the caught object is invoked, so if the related movements must be started, then the request for it must be somewhere inside the Move() or MoveNode() methods of the complex object. In the case of the RectangleWithComments class the synchronous or related movement of all the comments is started by the InformRelatedElements() method.

This is the procedure to organize the individual and related movements for rectangles with comments, but exactly the same technique is used for other types of complex objects. The calculation of the specific coefficients can be different and depends on the shape of the "parent" object, but the idea is exactly the same. It also does not matter, how many levels of related objects are included into the chain of linked objects. This example has only two levels (rectangle – comments); one of the further examples will have more levels of related objects.

**Note**. The CmntToRectangle class is especially designed for demonstration of this example. The **MoveGraphLibrary.dll** includes an absolutely identical in behaviour class CommentToRect. This **CommentToRect** class is used in further examples and with other classes in DLL whenever the comment to any movable / resizable rectangular area is needed.

## *Regular polygon with comments*

File: **Form_RegPolyWithComments.cs**
Menu position: *Graphical objects – Complex objects – Regular polygon with comments*

**Form_RegPolyWithComments.cs** (**fig.10.2**) has only one regular polygon, to which an arbitrary number of comments can be added. Polygon belongs to the RegPolyWithComments class. As any regular polygon, it is defined by the central point, radius of vertices, number of vertices, and an angle of the first vertex.

```
public class RegPolyWithComments : GraphicalObject
{
    Point ptCenter;
    int nRadius;
    int nVertices;
    double angle;
    SolidBrush brush;
    List<CmntToRegPoly> comments = new List<CmntToRegPoly> ();
```

Comments belong to the CmntToRegPoly class, which is derived from the TextMR class, so they are moved and rotated automatically. Though these comments can move individually, they are not absolutely independent, but have some relation with the polygon. Because of this relation, the position of comments must be linked by some parameters with the position and size of a polygon.

- The first of these parameters is an angle. But this is not the angle from the center of a polygon to the center of comment; this would be not enough. When the polygon is rotated, the comments must be also rotated with the same angular velocity, so this parameter is the difference between the angle to the center of comment and the angle of polygon, which is the angle to the first vertex.
- The second parameter is the special coefficient, describing the distance between two centers. If the comment is inside the radius of vertices, then this coefficient is equal to the ratio between the distance to comment from the center of polygon and the radius of vertices; thus for such a case the coefficient is from the [0, 1] range. When a comment is outside the radius of vertices, then the coefficient is equal to the distance from comment to the circle of vertices.

```
public class CmntToRegPoly : TextMR
{
    Point ptCenterPoly;
    float radiusVertices;
    double anglePoly;
    double coef;              // special coefficient to the circle of vertices
    double angleToPolyStart;
```

The polygons can be moved by any inner point and resized by any border point. What is interesting to look at, is the related movement of all the comments, when a polygon is either moved or resized. On all the possible changes in position, size or angle of a polygon, the comments must react in such a way.

- When the polygon is moved forward, the comments have to make a synchronous movement.
- When the polygon is resized, the comments retain their coefficients, so they simply move closer to the center of polygon or farther from it, depending on which way the polygon is changed.
- When the polygon is rotated, all the comments retain their angles in relation to the polygon, so they have to rotate synchronously.

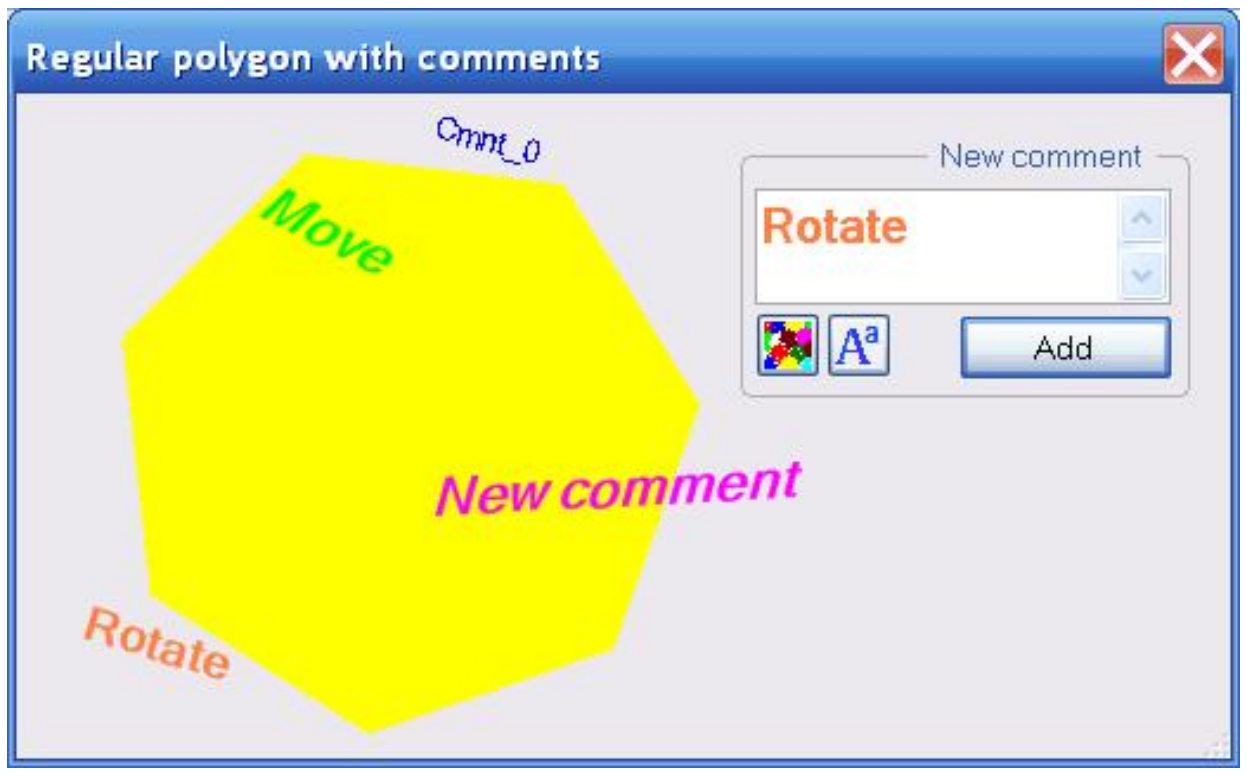

**Fig.10.2** Regular polygon with comments

The coding of the related movements in this example is organized in the same way, as in the previous: the `Move()` and `MoveNode()` methods of the `RegPolyWithComments` class must inform all the related comments about any possible changes in the size, position or angle of a polygon. Exactly these three parameters are passed to the comments on any change of a "parent".

```
private void InformRelatedElements ()
{
    foreach (CmntToRegPoly comment in comments)
    {
        comment .NewParentParams (ptCenter, nRadius, angle);
    }
}
```

This method has to calculate the new position of comment according to any change in the parameters of polygon.

```
public void NewParentParams (Point ptPolyCenter, int fVerticesRadius,
                             double angleParent)
{
    ptCenterPoly = ptPolyCenter;
    radiusVertices = fVerticesRadius;
    anglePoly = angleParent;
    Location = Auxi_Geometry .CirclePointBySpecialCoefficient (ptCenterPoly,
           Convert .ToInt32 (radiusVertices), anglePoly + angleToPolyStart, coef);
}
```

Certainly, there can be another way of doing the same thing. Only one of the parameters of polygon is changed at any particular moment; this depends on the type of movement in which the polygon is involved.

- By the forward movement only the central point is changed.
- By the resizing only the radius of vertices is changed.
- By rotation only the angle of the first vertex is changed.

Thus, on any kind of movement it is enough to pass to the comments the new value of a single parameter, but in this case three different methods must be used. Which way you prefer to write the code, is up to you; both ways will work.

## Identification

Before starting with the next example, I would like to go into the details of the identification of objects. I have already used it in some of the previous examples, but with the increase of objects in the forms and with the growing complexity of the objects you will see the use of the identification numbers more and more often.

Usually you have not one, but few or many graphical objects in your form, so any actions with these objects like deleting, changing the order or modifying start with the identification of the selected or touched object. There is no need in thinking out the identification mechanism for any new class of objects, as there is already one, which works for all of them.

Any movable / resizable object gets a unique identification number at the moment of construction; the number is generated automatically in the constructor of the base class (`GraphicalObject`).

```
id = Auxi_Common .UniqueID;
```

This number can be set or get with the ID property of the same base class. As any movable / resizable object is derived from the `GraphicalObject` class, then getting the **id** number for any object is trivial. There can be situations, when you need to set the **id** number after the construction of an object. For noncomplex objects it is also trivial and can be done with the same ID property; for complex object there is a big chance that you will have to override this property; this is explained a bit further on.

There are three situations, in which the identification is at high demand:

- When an object is caught, it is good to know what was really caught.
- When an object is moved.
- When an object is released.

From the point of using mover's methods, the first two situations are identical, as in both of them some object is caught at the moment of identification.

Usually there exists some collection of objects, which are currently used in the form; this collection can be organized as a `List`, or an array, or in some other way. When there are a lot of objects of different classes, you will often have several lists or arrays. When an object is caught by the mover, the first task is to identify this object as some element in the collection. The caught object is derived from the `GraphicalObject`; mover's methods allow to get the base `GraphicalObject` behind the caught one.

```
GraphicalObject grobj = mover .CaughtSource;
```

or

```
GraphicalObject grobj = mover [mover .CaughtObject] .Source
```

The identification number of this object is easily obtained.

```
long id = grobj .ID;
```

With the known **id**, you can look through all the elements of the form until the needed element is found, but in such a way the search is organized only in simple cases, when you have the objects of only one class. If there are several `Lists` for different classes of objects, then you first determine the exact class of the caught object and then look through the corresponding `List`. This is how it is done in the **Form_FillTheHoles.cs** (**figure 8.10**)

```
if (grobj is AreaWithHoles)
{
    … …
}
else if (grobj is Plug)
{
    … …
}
```

When an object was just released, then the identification of the base `GraphicalObject` behind it is slightly different.

```
GraphicalObject grobj = mover .WasCaughtSource;
```

or

```
GraphicalObject grobj = mover [mover .WasCaughtObject] .Source
```

With the complex objects, there is one more aspect of using the identification numbers. The parts of complex objects are often involved both in synchronous movements with other parts and in individual movements. This means that not only the "parent" object is registered in the mover's queue, but all these parts receive their personal places in the same queue; certainly, all the parts have their own **id** numbers. When any element is caught by mover, the search through all the parts of all the objects in the form can be organized, but this search can be significantly simplified, if any part of a complex object would keep the ID of its parent. It would be enough to contain the ID of the direct parent (one level up), as in the same way you can track the relations from level to level and decipher the whole chain of relations between the linked objects.

An object of the `GraphicalObject` class has a field, which is used for parent's identification. When a constructor of this base class is called, this field is assigned a zero value.

```
long idParent = 0;
```

This field can be get / set by the `GraphicalObject.ParentID` property. When you work with the simple objects, this property is not needed, but for the complex objects it becomes very useful and important. Usually, but not always. In the previous example of the regular polygon with comments, there is only one polygon in the **Form_RegPolyWithComments.cs** (**figure 10.2**), so you cannot make a mistake about the parent of any new comment. As a result, there is no need in the `ParentID` property in that form.

Absolutely different situation is in the example of rectangles with comments from the **Form_RectanglesWithComments.cs** (**figure 10.1**). In that form you can have on the screen an arbitrary number of rectangles and each of them may have any number of comments. You can call a context menu on one of the comments and select the command to delete this comment. Do you need to look through all the comments of all the rectangles to identify the one to be deleted, or is there any better way of identification? To find everything about this process, we have to start with the adding of a new comment to any rectangle, or maybe even with the construction of the new rectangle.

When a new rectangle is ordered on the screen, it appears with one comment.

```
private void Click_btnAddRectangle (object sender, EventArgs e)
{
    RectangleWithComments rwc = new RectangleWithComments (RandomPoint, 160,
                              100, Auxi_Colours .ColorPredefined (rand .Next (7)));
    rwc .AddComment (new CmntToRectangle (this, rwc .Rectangle, 0.4, 0.4,
                                         "Rect-" + iNewRect));
    rects .Insert (0, rwc);
    RenewMover ();
    Invalidate ();
    iNewRect++;
}
```

You can also add any number of comments to any rectangle by using a command from the context menu, called on rectangle.

```
private void Click_miAddCommentQuick (object sender, EventArgs e)
{
    rectPressed .AddComment (this, ptMouse_Up, "New comment");
    RenewMover ();
    Invalidate ();
}
```

What is interesting for our discussion is that in both cases the `RectangleWithComments.AddComment()` method is called; different versions of this method vary only in defining the position of the new comment in relation to rectangle.

```
public void AddComment (Form form, PointF pt, string txt)
{
    CmntToRectangle cmnt = new CmntToRectangle (form, rc, pt, txt);
    cmnt .ParentID = ID;
    comments .Add (cmnt);
}
```

The new comment gets the unique **id** on initialization and its **idParent** gets zero value, but in the next line this field gets the **id** of the rectangle, to which the new comment is added. Thus each comment gets the identification number of its parent. If you decide to delete any comment, you can do it through the context menu, called on this comment (`menuOnCmnt`).

```
private void OnMouseUp (object sender, MouseEventArgs e)
{
    ptMouse_Up = e .Location;
```

```
        if (mover .Release ())
        {
            GraphicalObject grobj = mover .WasCaughtSource;
            double dist = Auxi_Geometry .Distance (ptMouse_Down, e .Location);
            if (e .Button == MouseButtons .Left)
            … …
            else if (e .Button == MouseButtons .Right)
            {
                if (dist <= 3)
                {
                    if (grobj is RectangleWithComments)
                    … …
                    else if (grobj is CmntToRectangle)
                    {
                        cmntPressed = grobj as CmntToRectangle;
                        CommentIdentification ();
                        ContextMenuStrip = menuOnCmnt;
                    }
```

Calling of the needed menu is a standard procedure, in which the class of the released object is checked and the distance between the mouse positions for **MouseDown** and **MouseUp** events must be small enough, but there is also a call for the `CommentIdentification()` method before the menu is really opened. The name of this method can be a bit confusing (it has to be called "full comment identification"), because it identifies not the comment, but the rectangle, to which the pressed comment belongs.

```
        private void CommentIdentification ()
        {
            long idRect = cmntPressed .ParentID;
            for (int i = rects .Count - 1; i >= 0; i--)
            {
                if (idRect == rects [i] .ID)
                {
                    rectPressed = rects [i];
                    break;
                }
            }
        }
```

This method finds the parent rectangle by looking through the list of rectangles and comparing their **id** with the **idParent** of the pressed comment. The search for the parent is quick and easy. Now, when the menu on comment is opened, the parent rectangle for this comment is already known. If you order to delete the comment, it is enough to look through the list of comments for this particular rectangle and find the comment to be deleted by its **id**.

```
        private void Click_miDeleteCmnt (object sender, EventArgs e)
        {
            long id = cmntPressed .ID;
            for (int i = rectPressed .Comments .Count - 1; i >= 0; i--)
            {
                if (id == rectPressed .Comments [i] .ID)
                {
                    rectPressed .Comments .RemoveAt (i);
                    RenewMover ();
                    Invalidate ();
                    break;
                }
            }
        }
```

Each step looks simple and all of them are really simple. But exactly the same simple system of identification allows to organize the systems of different objects in much more complicated cases of scientific / engineering applications or in the programs for financial analysis, where a lot of different objects with different relations are involved. Those examples are demonstrated and discussed further on in the second part of the book.

## *Plot analogue*

File: **Form_PlotAnalogue.cs**
Menu position: *Graphical objects – Complex objects – Plot analogue*

Plotting, which is used in scientific and engineering applications, is going to be the theme of discussion in this book, but a bit later. Such plots have a lot of different parameters, which are very important and crucial for their use in different applications. In this section I want to exclude from discussion a lot of details and look only at the questions of design of plots as the complex objects.

A real plot consists of a single rectangular plotting area, an arbitrary number of horizontal and vertical scales, and an arbitrary number of textual information, associated with the plotting area and scales. Not only the number of all the additional parts is arbitrary, but their positions also. Scales can be placed at any side of the plotting area and atop the area itself. Comments can be placed anywhere in relation to the associated plotting area or scale; any comment can be also turned on any angle.

The real scales consist of the main line, ticks, and numbers (or words). These parts together occupy some rectangular area. In my scale analogues (`HorScaleAnalogue` and `VerScaleAnalogue` classes), the combinations of those parts are shown as rectangular areas (**figure 10.3**). This is the biggest simplification of the demonstrated model; all other things work as they do in the real plots.

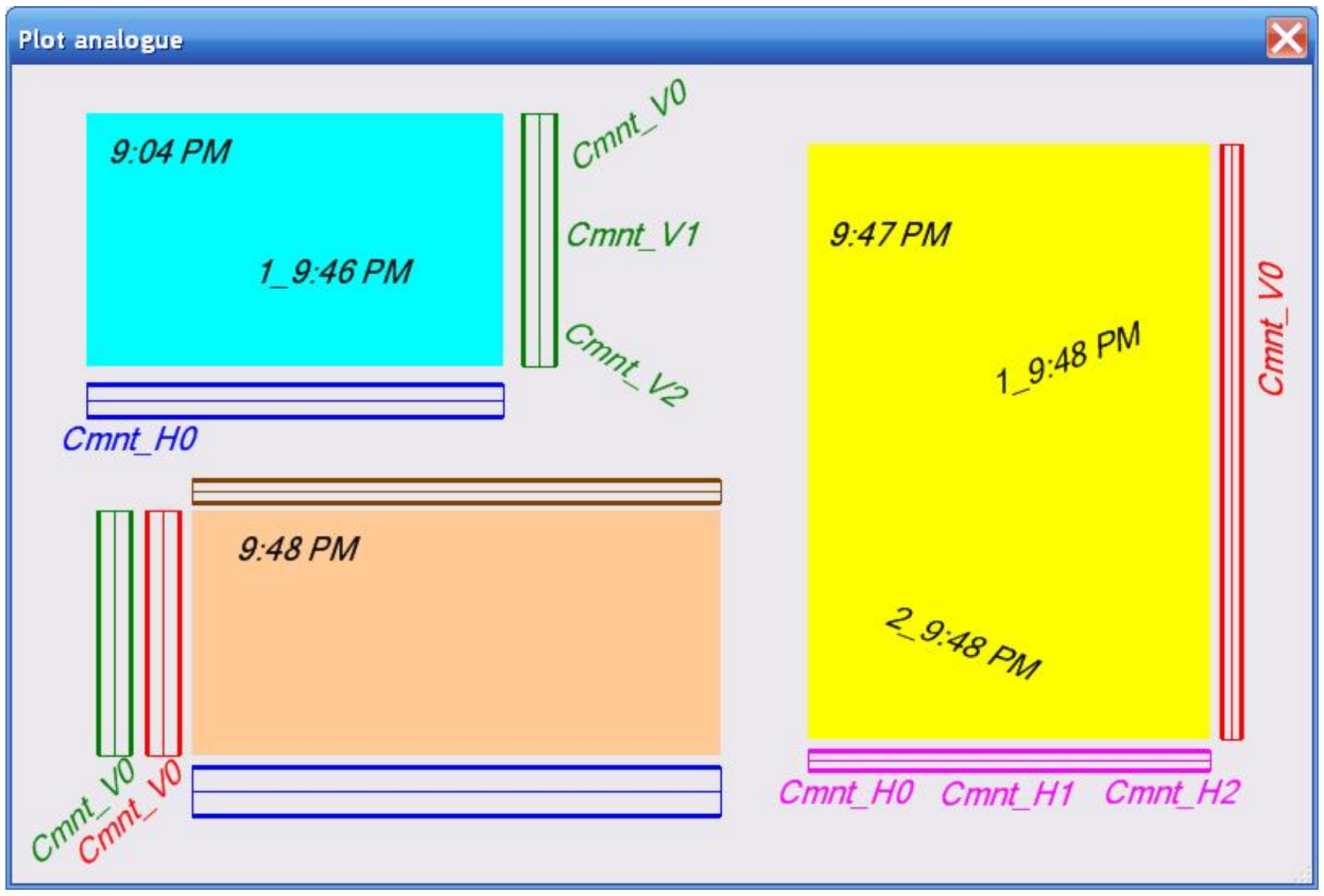

**Fig.10.3** These objects work as plot analogues with plotting areas, scales, and comments

The <u>comments</u> that are used in this example both for the plotting areas and for the scales are of the `CommentToRect` class. The class is derived from the `TextMR` class, so all these comments can be moved and rotated at any moment. Any comment is placed by its central point; the rotation goes around the same point. Positioning of any comment in relation to its "parent" rectangle is defined by two coefficients: one of them is for horizontal positioning, another – for vertical. This is how the coefficient for horizontal positioning is calculated.

- If the central point of comment is to the left of rectangle, then it is the distance from the point to the left border; coefficient is negative.
- If the central point is to the right of rectangle, then it is the distance from the point to the right border; coefficient is positive.
- If the point is between the left and right border of rectangle, then the coefficient belongs to the [0, 1] range with 0 on the left border and 1 – on the right.

The coefficient along the vertical scale is calculated in the similar way with 0 on the upper border and 1 – on the lower border of rectangle.

The size of any comment depends on the text and the used font. In real applications, the font for each comment can be set independently. To simplify the code of this example, I included the possibility of changing the font of comments, but only simultaneously for all of them. Anyway, it makes the code shorter, but does not affect the main topic of this example.

Comments can be related either to a plotting area or a scale. Comment does not know about the class of its parent, and there is no need for such knowledge. In any way the "parent" has a rectangular area; the knowledge of this area in combination with two positioning coefficients is enough for correct involvement of any comment in individual movements and related movements with its "parent". The use of `rcParent` and two coefficients was explained in the example of rectangles with comments (**Form_RectanglesWithComments.cs**, **fig.10.1**); the `CmntToRectangle` class from that example uses exactly the same technique.

The scales can be either horizontal (`HorScaleAnalogue` class) or vertical (`VerScaleAnalogue` class). Two of these classes differ only by the direction of the main line, but otherwise they are identical, so I will write further on only about the horizontal scales.

```
public class HorScaleAnalogue : GraphicalObject
{
    Rectangle rcParent;          // main plotting area
    Point pt_LT, pt_RB;          // end points of the central line
    int hHalf;                   // half of the height of the scale's rectangle
    double posCoef;              // positioning coefficient to the main plotting area
    Pen penThin, penThick;
    List<CommentToRect> comments = new List<CommentToRect> ();
```

The scale is shown as a central line, described by two end points (**`pt_LT`** and **`pt_RB`**), and the sensitive rectangle around it, marked with two wider lines. The size of this sensitive rectangle to each side from the central line is described by the **`hHalf`** parameter; thus the height of the rectangle for scale is `2 * hHalf`. Scale can be positioned anywhere in relation to the associated plotting area. The position of the scale is described by the **`posCoef`** parameter; this coefficient is determined in exactly the same way, as the positioning coefficients of comments in relation to rectangle. When we talk about the positioning of the horizontal scale in relation to the main plotting area, this coefficient describes the vertical position of the main line of this scale. The left and right borders of such scale are exactly the same as of the plotting area.

Scale can be moved to any place at the side of the plotting area or even atop of it; scale can also change its size, so, when the scale is movable and resizable, its cover consists of three nodes. It is possible to turn a scale into unmovable; in this case the cover consists of one node with the behaviour changed to `Behaviour.Frozen.` When it happens, the scale becomes unmovable by itself, but it is still moved around, whenever the "parent" plot is moved. The `Behaviour.Frozen` means that the mover cannot grab an object for relocation, but still recognizes the node, so, for example, a context menu can be still called on such an object.

```
public override void DefineCover ()
{
    int cxL = pt_LT .X;
    int cxR = pt_RB .X;
    int cyT = pt_LT .Y - hHalf;
    int cyB = pt_LT .Y + hHalf;
    Rectangle rect = Rectangle .FromLTRB (cxL, cyT, cxR, cyB);
    CoverNode [] nodes;
    if (Movable)
    {
        nodes = new CoverNode [] {
                      new CoverNode (0, new Point (cxL, cyT), new Point (cxR, cyT)),
                      new CoverNode (1, new Point (cxL, cyB), new Point (cxR, cyB)),
                      new CoverNode (2, rect, Cursors .SizeNS) };
    }
    else
    {
        nodes = new CoverNode [] {
                         new CoverNode (0, rect, Behaviour .Frozen, Cursors .Hand) };
    }
    cover = new Cover (nodes);
    cover .SetClearance (false);
}
```

A scale has two fields to adjust or retain its position along the plotting area whenever this area is changed or the scale is moved. The first one is the plotting area `rcParent`; the second one is the already mentioned coefficient `posCoef`. On any movement of either plot or scale only one of these parameters is changed.

- If the plotting area is moved or resized, then the fixed coefficient is used to calculate the position of scale according to the changed area.

```
public Rectangle ParentRect
{
    get { return (rcParent); }
```

```
        set
        {
            rcParent = value;
            int cy = Convert .ToInt32 (Auxi_Geometry .CoorByCoefficient
                                  (rcParent .Top, rcParent .Bottom, posCoef));
            pt_LT = new Point (rcParent .Left, cy);
            pt_RB = new Point (rcParent .Right, cy);
            CommentsNotification ();
            DefineCover ();
        }
    }
```

- If a scale is moved, then its new coordinate (location of one of the end points of its main line) is used to calculate the new coefficient in relation to the unchanged rectangle.

```
    public override void Move (int dx, int dy)
    {
        pt_LT .Y += dy;
        pt_RB .Y += dy;
        posCoef = Auxi_Geometry .CoefficientByCoor (rcParent .Top,
                                               rcParent .Bottom, pt_LT .Y);
        CommentsNotification ();
    }
```

Regardless of whether it was an enforced movement of scale as a result of the plot's move or the scale itself was moved by the mouse, in any case the comments, associated with the scale, have to be informed about the new position and size of the scale. This is done via the `CommentsNotification()` method.

```
    private void CommentsNotification ()
    {
        Rectangle rect = Rectangle;
        foreach (CommentToRect comment in comments)
        {
            comment .ParentRect = rect;
        }
    }
```

What happens, when the comment gets the new rectangular area of its "parent", was already explained.

The complex object representing the plot analogue has very limited number of parameters, of which only four are interesting for the current discussion: the rectangular area for plotting and three lists of associated elements.

```
public class PlotAnalogue : GraphicalObject
{
    Rectangle rc;
    List<HorScaleAnalogue> horScales = new List<HorScaleAnalogue> ();
    List<VerScaleAnalogue> verScales = new List<VerScaleAnalogue> ();
    List<CommentToRect> comments = new List<CommentToRect> ();
```

The cover for the `PlotAnalogue` objects is very simple: there is only a single rectangle, which can be resized by any side or corner, so the standard cover can be used.

```
public override void DefineCover ()
{
    cover = new Cover (rc, Resizing .Any, 6, 3);
}
```

The `Move()` method is primitive; in it only the location of rectangle must be changed.

```
public override void Move (int dx, int dy)
{
    rc .Location += new Size (dx, dy);
    InformRelatedElements ();
}
```

The MoveNode() method is definitely longer, as there are nine nodes in the cover, but the code is simple. Whenever the plotting area is moved or resized, all the related parts must be informed about the change; this is done with the InformRelatedElements() method. The ParentRect property of scales and comments are called to adjust their positions; these properties were already discussed a bit earlier.

```
private void InformRelatedElements ()
{
    foreach (CommentToRect comment in comments)
    {
        comment .ParentRect = rc;
    }
    foreach (HorScaleAnalogue scale in horScales)
    {
        scale .ParentRect = rc;
    }
    foreach (VerScaleAnalogue scale in verScales)
    {
        scale .ParentRect = rc;
    }
}
```

When the ParentRect property of the scale gets the new value of the plotting area, then the new position of the scale is calculated and the CommentsNotification() method of the scale is called. Through this method, all the comments of the scale get the new area of the scale and adjust their positions to it. Thus the adjusting of positions is done as a chain reaction from the top (PlotAnalogue class) to the bottom (CommentToRect class). Whether there are two elements in the chain (plot – scale) or three (plot – scale – comment), it does not matter; all the elements in the chain adjust their positions.

I have already mentioned that the complex objects cannot be registered with the mover by the mover.Add() or mover.Insert() methods. Instead, the IntoMover() method must be designed for each class of complex objects. There are too many places, where the set of elements of the complex classes can be changed; on each such change the mover's queue must be renewed, and it cannot be done manually. The class itself has to guarantee that any member of this class is going to be registered correctly regardless of the particular set of its parts.

```
new public void IntoMover (Mover mover, int iPos)
{
    mover .Insert (iPos, this);
    for (int i = comments .Count - 1; i >= 0; i--)
    {
        mover .Insert (iPos, comments [i]);
    }
    for (int i = horScales .Count - 1; i >= 0; i--)
    {
        horScales [i] .IntoMover (mover, iPos);
    }
    for (int i = verScales .Count - 1; i >= 0; i--)
    {
        verScales [i] .IntoMover (mover, iPos);
    }
}
```

This is the PlotAnalogue.IntoMover() method. When it is over, you have all parts of the plot in the mover's queue in such an order.

1. All the vertical scales.
2. All the horizontal scales.
3. All the comments of the main plotting area.
4. The PlotAnalogue object itself.

The scales themselves are also the complex objects and the same rule of mandatory use of their IntoMover() method must be applied. You can see that this is the way the scales are registered as parts of the PlotAnalogue object; here is the code of the IntoMover() method for scales. Absolutely the same main idea of design: first include all the parts of the complex object, then the object itself. Both classes of scales – HorScaleAnalogue and VerScaleAnalogue – have identical IntoMover() methods.

```
new public void IntoMover (Mover mover, int iPos)
{
    mover .Insert (iPos, this);
    for (int i = comments .Count - 1; i >= 0; i--)
    {
        mover .Insert (iPos, comments [i]);
    }
}
```

The PlotAnalogue class represents the analogue of really complex object, which was only simplified to demonstrate without any unessential details the use of such objects in many applications. The **Form_PlotAnalogue.cs**, in which this example works, includes also a lot of features, which are standard for all the user-driven applications. Among all the examples, used to this moment, this is the first form with such features, so I would like to mention some of them.

The user-driven applications are designed according to four main rules.

**Rule 1**. All the elements are movable.

**Rule 2**. All the visibility parameters must be easily controlled by the users.

**Rule 3**. The users' commands on moving / resizing of objects or on changing the parameters of visualization must be implemented exactly as they are; no additions or expanded interpretation by developer are allowed.

**Rule 4**. All the parameters must be saved and restored.

The time for detailed discussion of these rules will come later (the second half of the book is devoted to this); here I want only to mention, how these rules are transformed into the code of this particular **Form_PlotAnalogue.cs**.

Rule 1. There are no unmovable elements in this form. Some elements (scales) can be turned into unmovable, but only by the user. The need for such change of movability was announced by the users of the very complicated applications; here I only demonstrate the possibility and its realization. Changing of the movability of scale is done via the context menu, which can be called on any scale. Usually the change of movability does not affect the view of the objects, but in this example I even made this feature obvious by changing the style of the lines, with which the scales are drawn: movable sales are shown by solid lines, fixed scales – by the dotted lines.

Rule 2 is demonstrated in this form in several ways, but, as it is usually done, the control of the visual parameters is organized via the context menus. There are three types of objects in this form (areas, scales, and comments); each one has its own menu. For areas and scales the color can be changed by clicking the corresponding line in context menu; for comments there is no such line, but that was done on purpose. This is a form to demonstrate the complex objects and the relations between their parts. To make the whole thing more obvious, the comments of the scales always have the color of their "parent".

Changing the set of objects on the screen and changing their order are also among the parameters of visualization; all these things are organized via the context menus.

Rule 3. All the moving / resizing operations are implemented according to the users' commands without any addition on my side, as a developer. Areas, scales, and comments can be moved anywhere; the resizing of areas is unlimited.

Rule 4. An arbitrary number of areas, scales, and comments can be put on the screen. The order of areas and visual parameters of all the elements are also decided by the user. Whatever view is organized for the form, will be saved on closing and restored the next time, when the form is opened again. Saving / restoring are organized via the Registry. As this is the first example, in which the saving and restoring are organized in the way they always have to be, let us look into some details of this process.

All classes, which are used in the form, have to have the methods to save and restore their parameters via Registry and binary files. These two pairs of methods differ in details of writing and reading, but they are used to save exactly the same data, so it is enough to look into one of them; the **Form_PlotAnalogue.cs** uses the Registry to preserve the data. Of the four different classes used in this form (CommentToRect, HorScaleAnalogue, VerScaleAnalogue, and PlotAnalogue), the first one comes from the **MoveGraphLibrary.dll**, so it has all the needed methods. The needed methods for other three classes can be found in the code of these classes. Methods of the complex classes have to save / restore the data for all its parts, so they use the corresponding methods of these parts.

The data describing all the plots, scales, and comments on the screen is saved, when the form is going to be closed.

```
private void OnFormClosing (object sender, FormClosingEventArgs e)
{
    SaveInfoToRegistry ();
}
```

At the next opening of the form the `Registry` is checked for having any information about this form.

```
private void OnLoad (object sender, EventArgs e)
{
    RestoreFromRegistry ();
    if (bRestore == false)
    {
        PlotAnalogue area = new PlotAnalogue (this,
                                      new Rectangle (100, 100, 300, 200), Color .Cyan,
                                 fntCmnts, Side .S, Color .Blue, Side .E, Color .Green);
        areas .Add (area);
    }
    RenewMover ();
}
```

If there is any information about the previous work with this form and the information is restored correctly, then it provides all the needed data about the size of the form and all the `PlotAnalogue` objects with all their parts that have to be shown in the form. If there were no data or some mistakes occur on its reading, then one new area is organized.

The **Form_PlotAnalogue.cs** demonstrates the work of the analogues of the plots. A lot of features of the real plots, scales, and comments are excluded to make the explanation more clear. At the same time the work of this form is made as close to the real applications, as possible; you can see here nearly the same organization of the work of a form as you will see further on in the examples of the real scientific applications. For example, all the actions inside the **`MouseUp`** event are exactly the same as in the most complex applications.

```
private void OnMouseUp (object sender, MouseEventArgs e)
{
    bFormInMove = false;
    ptMouse_Up = e .Location;
    double dist = Auxi_Geometry .Distance (ptMouse_Down, e .Location);
    if (mover .Release ())
    {
        GraphicalObject grobj = mover .WasCaughtSource;
        if (e .Button == MouseButtons .Left)
        {
            if (grobj is PlotAnalogue && dist <= 3)
            {
                Identification (mover .WasCaughtObject);
                if (iAreaPressed > 0)
                {
                    AreaOnTop ();
                }
            }
        }
        else if (e .Button == MouseButtons .Right && dist <= 3)
        {
            Identification (mover .WasCaughtObject);
            MenuSelection (grobj);
        }
    }
    else
    {
        if (e .Button == MouseButtons .Right && dist <= 3)
        {
            ContextMenuStrip = menuOnEmpty;
        }
```

```
            }
        }
```

If any object is released, then it all depends on the button by which it was previously caught. If it was the left button, then it is either some movement or the change of the object's order. The last one can be done not for any object, but only for the most important objects of the form; in this case it is the `PlotAnalogue` class. The related objects (comments and scales) are shown only together with their "parents"; there are strict rules of showing all "children" in the correct order, so the voluntary change of the order for "children" is not allowed. If any object was released by the right button, then the menu selection is organized. In both cases any action is preceded by the full object's identification. The full identification includes not only the class of the released object, but, if needed, the whole chain of related objects. For example, if you call the menu on some comment of one of the scales, then the `Identification()` method has to provide the order of plot, with which this scale is associated, the order of scale in one or another `List` of scales (horizontal or vertical) of this plot, and the order of comment in the `List` of comments of this scale.

If no object is released, but the Right mouse button was released, then it can be a request for the context menu at an empty spot.

There is one more interesting feature in the **Form_PlotAnalogue.cs** – this form can be moved by any inner point. Certainly, by any empty point of the form; by any point, where no object can be caught. The mechanism is very simple and will be explained further on in the second part of the book. Here I want only to inform about such a possibility and to mention that it was born as a consequence of the rules of user-driven applications. If any object can be moved by any inner point, then why the form itself cannot be moved by any inner point? Certainly it can and this is very easy to implement.

## Track bars

File: **Form_Trackbars.cs**
Menu position: *Graphical objects – Complex objects – Track bars*

Track bars are the well known elements of the interface design. The idea of an object in which a small bar slides along another bar and allows to select any value from some range can be very useful in many situations. Unfortunately, the implementation of this control in VisualStudio is so clumsy that after each attempt of using the standard `TrackBar` control I had to reject it and turn to something else. But the main idea of the track bar is perfect and a good element of such a type is so desirable from time to time that it has to be used. So, I designed the graphical `Trackbar` class which fits perfectly with all other elements of the user-driven applications and with the rules of such applications.

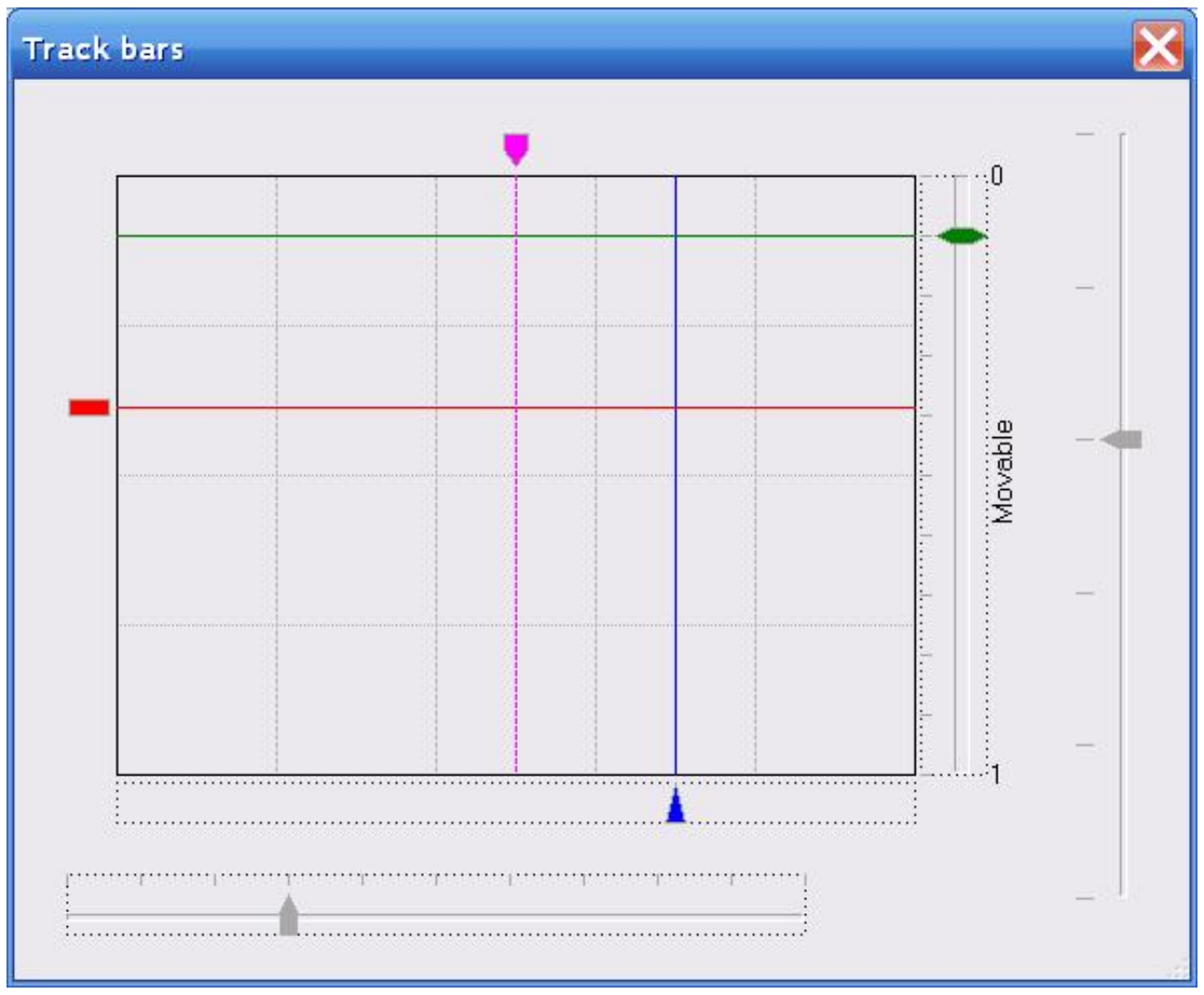

**Fig.10.4** Track bars can be used as independent objects or together with any other rectangular objects.

Objects of the `Trackbar` class can be used in two different ways: it can be a stand alone element, or it can be associated with some rectangular area. The difference in object's features for these two cases is so significant that it required the implementation of two different covers for those two cases. Both cases of using the `Trackbar` objects are demonstrated in the **Form_Trackbars.cs** (**figure 10.4**).*

The small sliding bar belongs to the `TrackbarSlider` class. There are four different variations of the shape of sliders defined by such enumeration.

```
enum SliderShape {Triangle, Rectangle, Pentagon, Sexangle };
```

As seen from **figure 10.4**, sliders of different shapes have different number of vertices, but they are all polygons, so the cover for any slider consists of a single polygonal node. Four parameters describe the sizes of any slider:

- Length of the head part.

* The **Form_Trackbars.cs** contains and uses the `TrackbarC` class, which is the exact copy of the `Trackbar` class from the **MoveGraphLibrary.dll**.

- Length of the straight part.
- Length of the tail part.
- Half of the width.

Not all four parameters are used for each type of a slider. For example, the triangular and rectangular sliders use only one parameter of the first three.

Other important parameters of the `TrackbarSlider` class include the side (or the direction) of the spike and two points, between which the spike of a slider can move. For rectangular slider which has no spike, these are the end points for the movement of the middle point of that side.

All these parameters are set at the moment of initialization.

```
public TrackbarSlider (Side side_head, Point ptLT, Point ptRB, double poscoef,
                       SliderShape shape, int h_headslope, int h_straight,
                       int h_backslope, int w_half)
```

If the shorter version of the `TrackbarSlider` constructor is used, then the slider gets the pentagonal shape; the track bar at the bottom of **figure 10.4** demonstrates the default view of the slider.

Positioning coefficient of a slider is always from the `[0, 1]` range and determines the placement of the spike between the two end points `ptLT` and `ptRB`. Thus regardless of the direction of spike, the 0 value corresponds to the left (or upper) point of the spike's movement.

The **Form_Trackbars.cs** includes six different objects of the `TrackbarC` class. These objects have to demonstrate both the dependent and independent track bars, to show the sliders of all four different shapes, and to demonstrate some other interesting features.

```
private void OnLoad (object sender, EventArgs e)
{
    if (!bRestore)
    {
        plot = new Plot (this, new Rectangle (100, 100, ClientSize .Width / 2,
                                               ClientSize .Height / 2));
        plot .HorScales [0] .Visible = false;
        plot .VerScales [0] .Visible = false;
        Rectangle rc = plot .PlottingArea;

        trackbar_Btm = new TrackbarC (this, rc, 4, Side.N, 0.7, false,0,0,0, null);
        trackbar_Btm .Slider .InnerColor = Color .Blue;
        trackbar_Btm .Slider .Shape = SliderShape .Triangle;
        trackbar_Btm .Slider .HalfWidth += 1;

        trackbar_Left = new TrackbarC (this, rc, -4, Side .E, 0.15, false,
                                       0, 0, 0, null);
        trackbar_Left .Slider .InnerColor = Color .Red;
        trackbar_Left .ShowAreaBorder = false;
        trackbar_Left .Slider .Shape = SliderShape .Rectangle;
        trackbar_Left .Movable = false;

        trackbar_Right = new TrackbarC (this, rc, 10, Side .W, 0.1,
               SliderShape .Sexangle, 8, 10, 8, 4, true, 2, 5, 11, Draw_TrackbarR);
        trackbar_Right .Slider .InnerColor = Color .Green;
        trackbar_Right .AddComment (6, 0, "0", Font, 0, ForeColor, false);
        trackbar_Right .AddComment (6, 1, "1", Font, 0, ForeColor, false);
        trackbar_Right .AddComment (8, 0.5, "Movable", Font, 90, ForeColor, true);

        trackbar_Top = new TrackbarC (this, rc, -5, Side .S, 0.5,
                   SliderShape .Pentagon, 8, 8, 0, 6, false, 0, 0, 0, null);
        trackbar_Top .Slider .InnerColor = Color .Magenta;
        trackbar_Top .ShowAreaBorder = false;
        trackbar_Top .Movable = false;
        trackbar_Top .AddComment (trackbar_Top .Slider .PositionCoefficient, -8,
                trackbar_Top .Slider .PositionCoefficient .ToString ("F2"),
```

```
                                              Font, 0, Color .Blue, false);
        int cy = plot .Underlayer .Area .Bottom + 90;
        TrackbarSlider slider = new TrackbarSlider (Side .N, new Point (60, cy),
                                      new Point (ClientSize .Width * 2 / 3, cy), 0.3);
        trackbarHor = new TrackbarC (this, slider, true,5,5,11, Draw_Trackbar_Hor);

        int cx = plot .Underlayer .Area .Right + 120;
        slider = new TrackbarSlider (Side .W, new Point (cx, 60),
                                    new Point (cx, ClientSize .Height * 3 / 4), 0.4);
        trackbarVer = new TrackbarC (this, slider, true,4,8, 6, Draw_Trackbar_Ver);
        trackbarVer .ShowAreaBorder = false;
    }
    penSpec = new Pen (trackbar_Top .Slider .InnerColor);
    penSpec .DashStyle = DashStyle .Dash;
```

The names of four dependent track bars include the tip of their initial position in relation to the main area of the plot. Two of these track bars are movable, so their names can become confusing, if I continue to use them, while discussing some features. Usually the parameters of the track bars can be changed via the commands of the context menu, but I just found out that there is no menu in this form (looks like I forgot about it…), so in the further explanation I can specify any of the dependent track bars by the color of its slider.

Objects of the `Trackbar` class can be used either as independent or dependent elements. The exact type of use is determined at the moment of initialization. In case of dependency, the list of parameters must include the rectangle, with which the track bar is associated, and the positioning coefficient in relation to this rectangle.

```
public Trackbar (Form form, Rectangle rc, double trackbarcoef,
                 Side side_head, double slidercoef, bool showticks,
              int spaceToTicks, int ticklength, int ticknum, Delegate_Draw onDraw)
```

The `trackbarcoef` coefficient describes the position of the track bar in relation to the "parent" rectangle. This coefficient determines the positioning of the line along which the spike of a slider is moving. For example, look at the horizontal track bar with a blue slider. When the spike is below the rectangle, then this coefficient is positive with the value equal to the distance between the spike and the bottom of the rectangle. If the track bar is moved up with the spike somewhere inside the rectangle, then this coefficient will be from the `[0, 1]` range (0 corresponds to the upper border, 1 – to the lower border of rectangle). If the spike is above the rectangle, then the coefficient is equal to the distance from the spike to the upper border of rectangle taken with the negative sign.

The dependent `Trackbar` object can be moved only along one axis across the associated rectangle. There are four dependent track bars at **figure 10.4**; those with the blue and violet sliders can be moved only up or down; the track bars with the red and green sliders can be moved only left or right. A dependent track bar cannot be resized by itself, so the cover of such track bar consists of a single polygonal (rectangular) node. The cursor above this node informs about the possible movement of the track bar; the cursor depends on whether the track bar is horizontal or vertical.

```
public override void DefineCover ()
{
    if (Movable)
    {
        if (bDependent)
        {
            cover = new Cover (SliderTicksArea, Resizing .None);     // 1 node
            if (m_slider .HeadSide == Side .N || m_slider .HeadSide == Side .S)
            {
                cover .SetCursor (Cursors .SizeNS);
            }
            else
            {
                cover .SetCursor (Cursors .SizeWE);
            }
        }
        else {
            … …
        }
    }
```

```
    else
    {
        cover = new Cover (SliderTicksArea, Resizing .None);     // 1 node
        cover [0] .SetBehaviourCursor (Behaviour .Frozen, Cursors .Default);
    }
}
```

You can see from this code that there is a variant of a unmovable track bar. What is it for in the world of all movable objects? This will be discussed in the second part of the book, when I turn to ideas and design of user-driven applications, but here is only a short remark.

Suppose that you organized some plotting area with a track bar at the side of this area. You moved the track bar in such a way that it is placed exactly as you like it to be. The plotting area is movable and resizable, so it can be positioned anywhere around the screen. In reaction to all possible moving / resizing of the plotting area, the track bar is going to retain its relative position to the plotting area. This area can be moved by any inner point and resized by any border point. A track bar is often positioned next to the rectangle, with which it is associated, so the nodes to move a track bar and to move / resize the plotting area often stay next to each other or even overlap. In such situation, it can easily happen that instead of moving or resizing the plotting area, the track bar can be moved. To avoid such accidental movement of the track bar, it can be declared temporarily unmovable, for example, via a context menu. An unmovable track bar continues to react correctly on all the moving / resizing of the plotting area; it is also recognized by the mover, as it is simply “frozen”, so the context menu on the unmovable track bar can be called in exactly the same way as on a movable one.

The switch to the unmovable track bar does not mean that its slider cannot move! The slider continues to work regardless of the movability of the “parent” track bar. For example, two dependent track bars in the **Form_Trackbars.cs** are declared unmovable on initialization (red and violet), but all four sliders work.

The `MoveNode()` method of an object with a cover consisting of a single node must be a simple one: it has to call the `Move()` method, but the parameters passed to this method depend on whether it is a horizontal or a vertical track bar. The coefficient, which describes the position of the track bar in relation to the rectangular area, must be also recalculated.

```
public override bool MoveNode (int i, int dx, int dy, Point ptM, MouseButtons btn)
{
    bool bRet = false;
    if (btn == MouseButtons .Left)
    {
        if (bDependent)
        {
            if (m_slider .HeadSide == Side .N || m_slider .HeadSide == Side .S)
            {
                Move (0, dy);
                coef = Auxi_Geometry .CoefficientByCoor (rcParent .Top,
                                        rcParent .Bottom, SliderEndPoints [0] .Y);
            }
            else
            {
                Move (dx, 0);
                coef = Auxi_Geometry .CoefficientByCoor (rcParent .Left,
                                         rcParent .Right, SliderEndPoints [0] .X);
            }
```

As always happens with the complex objects, when the “parent” is moved or resized, all the “children” must be notified about the new position and dimensions of the “parent”. In the case of the **Form_Trackbars.cs**, four different `Trackbar` objects are associated with a single `Plot` object, so on any move of this plotting area those four track bars must be informed by providing them with the value of the new rectangle.*

* The `Plot` class is discussed further on in the chapter *Applications for science and engineering*, but here I want to attract your attention to one feature of using this class. Any plot is a complex object and the `Plot` class includes a part, which belongs to the `RectCorners` class; this is a special addition to improve the resizing of the main plotting area. Thus the resizing of the main plotting area can occur, when either `Plot` or `RectCorners` object is caught by a mover, so both classes must be mentioned as a possible cause to inform all the related track bars about the change of the plotting area.

```
    private void OnMouseMove (object sender, MouseEventArgs e)
    {
        if (mover .Move (e .Location))
        {
            GraphicalObject grobj = mover .CaughtSource;
            … …
            else if (grobj is Plot || grobj is RectCorners)
            {
                Rectangle rc = plot .PlottingArea;
                trackbar_Btm .ParentRect = rc;
                trackbar_Left .ParentRect = rc;
                trackbar_Right .ParentRect = rc;
                trackbar_Top .ParentRect = rc;
            }
            Invalidate ();
```

With the new associated rectangle passed as a parameter and the positioning coefficient stored in each track bar, the new end points for movement of slider are calculated. The points depend on the orientation of a particular track bar.

```
public Rectangle ParentRect
{
    get { return (rcParent); }
    set
    {
        if (bDependent)
        {
            rcParent = value;
            int cx, cy;
            Point ptLT, ptRB;
            switch (m_slider .HeadSide)
            {
                case Side .N:
                case Side .S:
                default:
                    cy = Auxi_Geometry .CoorByCoefficient (rcParent .Top,
                                                           rcParent .Bottom, coef);
                    ptLT = new Point (rcParent .Left, cy);
                    ptRB = new Point (rcParent .Right, cy);
                    break;
                case Side .W:
                case Side .E:
                    cx = Auxi_Geometry .CoorByCoefficient (rcParent .Left,
                                                           rcParent .Right, coef);
                    ptLT = new Point (cx, rcParent .Top);
                    ptRB = new Point (cx, rcParent .Bottom);
                    break;
            }
            m_slider .SliderEndPoints = new Point [] { ptLT, ptRB };
            CalcAreas ();
            foreach (CommentToRect cmnt in m_comments)
            {
                cmnt .ParentRect = rcArea;
            }
        }
    }
}
```

There is also one more level of related movements as any `Trackbar` object can be associated with a set of comments. (At **figure 10.4**, you can see three comments associated with the track bar with the green slider.) These comments are of the `CommentToRect` class. The comments are positioned in relation to the rectangular area, which unites the area of the possible movement for slider and the area of ticks, if the ticks are shown. After the end points of the slider's movement are determined, this united area is calculated by the `CalcAreas()` method; the new "parent" rectangle is sent to all the associated comments via the `CommentToRect.ParentRect` property.

When the `Trackbar` object is used as an independent element, then it has a different cover. Such object can be moved around the screen, so its whole area is covered by a rectangular node, but in this case the cursor over this node is different. In addition, the length of such track bar can be changed, so this is a classical three nodes cover of a rectangle which can be resized along one axis only.

```
public override void DefineCover ()
{
    if (Movable)
    {
        if (bDependent)
        {
            … …
        else
        {
            if (m_slider .HeadSide == Side .N || m_slider .HeadSide == Side .S)
            {
                cover = new Cover (SliderTicksArea, Resizing .WE);     // 3 nodes
            }
            else
            {
                cover = new Cover (SliderTicksArea, Resizing .NS);     // 3 nodes
            }
        }
```

The `MoveNode()` method in such a case is definitely longer as there are three nodes in a cover, but the code is simple. The only thing to be checked is the new proposed length of the track bar against the minimum allowed length.

Regardless of whether a `Trackbar` object is dependent or independent, it may have an arbitrary number of associated comments. Each of those comments can be individually movable or not. If movable, it can be also rotated, as the `CommentToRect` class is derived from the `TextMR` class; objects of the last class are rotatable by default. The track bar with the green slider has three associated comments, of which two are unmovable and the third one is movable.

The track bar with the violet slider also has a comment, but you do not see it at the picture. Any object has a visibility parameter; in further examples you will see different cases, when such parameter can be switched ON / OFF by a user of a program thus changing the view at any moment. The switch of visibility is often implemented in the complex applications with a lot of information on the screen; users can hide some information (or even bigger objects) and visualize them again only when they are really needed. In the case of the track bar with the violet slider, I demonstrate the semi-automatic switch of the visibility.

The track bars along the plotting areas are often combined with the sliding line across the area (see **figure 10.4**). It is very informative to show somewhere the parameter, associated with the current position of the slider. It is easy to show this information all the time by simply changing the text of comment, but in this case there is even an addition to this: the parameter is shown, but only if the violet slider is moved along its "parent" track bar or when the mouse cursor is simply moved across this slider even without grabbing it for movement.

```
    private void OnMouseMove (object sender, MouseEventArgs e)
    {
        if (mover .Move (e .Location))
        {
            GraphicalObject grobj = mover .CaughtSource;
            if (grobj is TrackbarSlider && grobj .ID == trackbar_Top .Slider .ID)
            {
                UpdateTrackbarComment ();
            }
            … …
        }
        else
        {
            MoverPointInfo info = mover .PointInfoUpper (e .Location);
            if (info .ObjectNum >= 0 &&
```

```
                mover [info .ObjectNum] .Source is TrackbarSlider &&
                mover [info .ObjectNum] .Source .ID == trackbar_Top .Slider .ID)
            {
                if (trackbar_Top .Comments [0] .Visible == false)
                {
                    UpdateTrackbarComment ();
                    Invalidate ();
                }
            }
            else if (trackbar_Top .Comments [0] .Visible == true)
            {
                trackbar_Top .Comments [0] .Visible = false;
                Invalidate ();
            }
        }
```

The method to change the view of the comment is called `UpdateTrackbarComment()`; you can see two calls to this method in the above shown piece of code. Let us begin with the second part of this code, which works when the mouse is moved around and no object is caught. In the very first chapter of this book in the subsection *From algorithm to working programs* I mentioned that mover can give information about objects at any point. This information is given in the form of a `MoverPointInfo` object. It can be an information about the upper element at the spot, or it can be an information about all the objects that overlap at the specified location. Just now I am especially interested in situation, when the mouse cursor is above the particular slider, so I am interested in information about the upper object under the cursor.

```
MoverPointInfo info = mover .PointInfoUpper (e .Location);
```

The `MoverPointInfo` gives me all the needed information about the existence of object under cursor and about the parameters of the object, if there is any. If there is any object under the mouse, then its number in the mover's queue is either zero or bigger.

```
if (info .ObjectNum >= 0 &&
```

Knowing this number, I can check the class of object under cursor.

```
mover [info .ObjectNum] .Source is TrackbarSlider &&
```

If it is a `TrackbarSlider` object, then I can check if it is the violet slider; I check not the color, but the slider itself by checking its identification number.

```
mover [info .ObjectNum] .Source .ID == trackbar_Top .Slider .ID)
```

If it is really the needed slider, then I call the `UpdateTrackbarComment()` method, which fills the comment of the track bar with the needed information and makes this comment visible.

```
if (trackbar_Top .Comments [0] .Visible == false)
{
    UpdateTrackbarComment ();
    Invalidate ();
}
```

The information, which is shown above the violet slider, is only the coefficient of this slider, which is from the [0, 1] range. In the real application I would use this coefficient to calculate and show the real value from the X range of the plotting area.

```
private void UpdateTrackbarComment ()
{
    trackbar_Top .Comments [0] .NewLocationByCoefficients (
                                    trackbar_Top .Slider .PositionCoefficient,
                                    trackbar_Top .Comments [0] .YCoefficient);
    trackbar_Top .Comments [0] .Text =
                        trackbar_Top .Slider .PositionCoefficient .ToString ("F2");
    trackbar_Top .Comments [0] .Visible = true;
}
```

The comment, associated with that track bar, becomes visible, when the cursor is moved into its area. You press the violet slider and move it left or right; then the first part of the `OnMouseMove()` method works. This part also includes the call to the `UpdateTrackbarComment()` method, so:

- The position of comment changes so as to be always above the slider.
- The text of comment is changed according to the current coefficient of slider.
- The comment is visible.

When the violet slider is released, the mouse cursor is still above this slider, so the information is still shown. But the moment you move the cursor outside this slider, another part of the `OnMouseMove()` method is called. This is the part when no object is called and the violet slider is not under cursor any more.

```
else if (trackbar_Top .Comments [0] .Visible == true)
{
    trackbar_Top .Comments [0] .Visible = false;
    Invalidate ();
}
```

The visibility of comment is switched OFF and the comment becomes invisible until the next moment when the mouse again arrives over this slider.

# Movement restrictions

From the very first example and up till now I was trying to explain how to organize one type of movement or another. A majority of these movements were absolutely unrestricted, or if there were some limitations on movements, they were not the main theme of the discussion. However, there are situations when the movement restrictions become very important. These situations can be of different types; some of them are discussed in this chapter.

The term *movement restrictions* can be applied to different situations, so it is better to begin with some definitions, as a solution for each case depends on the clear understanding of what is really meant by the term. Here is the list of situations, which are discussed in this chapter.

- General restrictions on moving all the objects around the screen.
- Personal restrictions on the object's sizes, which are not related to any other objects.
- Restrictions caused by the coexistence of objects on the screen. Prevention of overlapping.

## *General restrictions on moving all the objects*

Suppose that you are going to develop your application on the basis of the movable objects. For this, you have to design all the needed objects according to some basic rules, which mean writing `DefineCover()`, `Move()`, and `MoveNode()` methods for their classes. This is a necessary, but not sufficient thing to make a screen object mobile. To make an object really movable, it has to be registered with a mover. Mover is an object that organizes and supervises the whole moving process. You can declare and initiate a mover as any other object.

```
Mover mover = new Mover ();
```

You can try this statement in nearly any of the previous examples, for example, in any form dealing with rectangles or polygons. Then you can grab any screen object with a mouse as you were doing it before, move the caught object across the border of the form, and release it there. Do you know what will happen then? Well, it depends on the side of the form, across which you have smuggled out this object.

- If this is either the right or bottom side of the form and the form is resizable, then you can enlarge the form until you see that object again; after it you can move it back.
- If this is either the left or top side of the form, then there is no way to see this object again. It is gone. It still exists, but it is inaccessible.

I do not think that any client would like the applications with the voluntarily disappearing objects. I also do not think that you can give anyone an application with an additional warning like "Do not move any object across this and that border"; that is why you will not find the shown type of mover's initialization anywhere in the program. Instead, in all the forms you see

```
mover = new Mover (this);
```

After such initialization, the mouse with the caught object can be moved only to the borders of the form, but not across.

These two variants of mover's initialization demonstrate two of three available levels of mover's clipping

```
public enum Clipping { Visual, Safe, Unsafe };
```

- `Visual` – elements can be moved only inside the visible area.
- `Safe` – elements can be moved from view only across the right and bottom borders of the form.
- `Unsafe` – elements can be moved from view across any border.

When I write *clipping* or *mover's clipping*, I mean the restrictions of the mouse movement, <u>when any object is caught by the mouse (by the mover!)</u>. This clipping is set at the moment, when mover grabs an object, and works until the moment of its release. At the very same moment, when an object is released, all the restrictions are eliminated, and the mouse can be moved anywhere.

When the mover is initiated without any parameter, then the `Unsafe` level of clipping is set; when the initiation is done with a parameter – the `Visual` level is set. Mover can be used not only in the form, but, for example, on a panel; in such case the panel is used as a parameter and the same clipping ideas are applied to the area of a panel.

The clipping level is set at the moment of mover's initiation; it can be also changed at any moment later by the `mover.Clipping` property, but there are some peculiarities, depending on the situation.

- If no object is caught by the mover at the moment of calling this property, then the `mover.Clipping` can be used to set any needed level. For example, in the **Form_SetOfObjects.cs** (**figure 9.4**) the clipping level is changed at the beginning to allow the move of any object across the right and bottom borders of the form. In this way you can move any object out of view across these borders, but you can always find them later by enlarging the form. Such type of temporarily hiding the objects is a standard and widely used feature of the user-driven applications.

```
public Form_SetOfObjects ()
{
    InitializeComponent ();
    mover = new Mover (this);
    mover .Clipping = Clipping .Safe;
```

- If an object is already caught by the mover, then the same `mover.Clipping` property can be used, but only if the new level widens the area of clipping. Thus, the `Visual` level can be changed to any other; the `Safe` level can be switched to `Unsafe`, but not to `Visual`.

Let us look once more at the **Form_FillTheHoles.cs** (**figure 11.1**).

```
public Form_FillTheHoles ()
{
    InitializeComponent ();
    mover = new Mover (this);
}
```

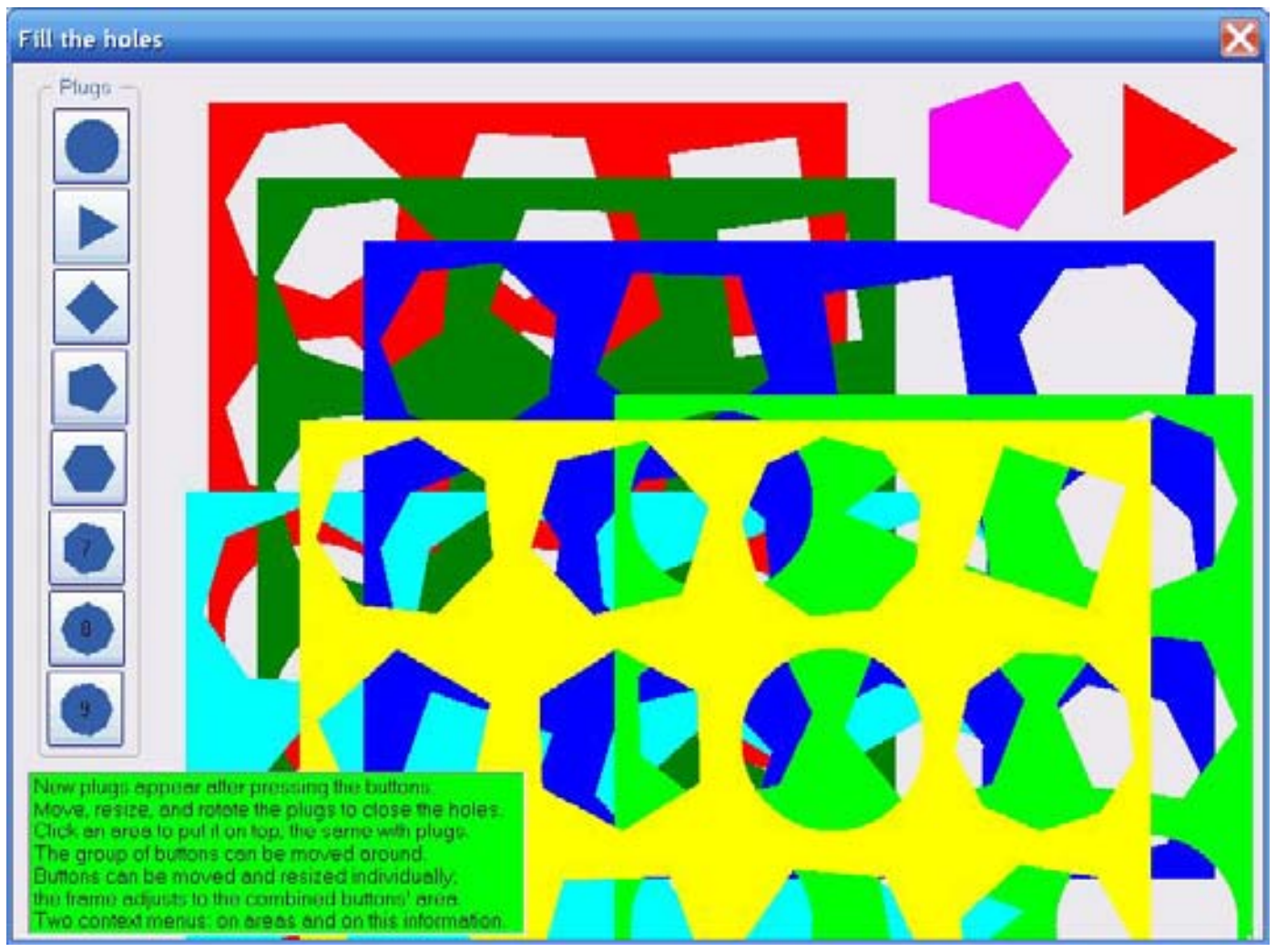

**Fig.11.1** Only the "plugs" can be moved across the right and bottom borders; other objects are not allowed across any border

When the form is constructed, the mover is initialized in the standard way with the form, as a parameter, so the clipping level is automatically set to `Visual` and any caught object is not supposed to move across any border. The real situation is more interesting. The green rectangular area with some explanation for this form, the group with the buttons, and colored boards with the holes cannot be moved across the borders and are always visible. At least partly! The boards can be deleted by the command of its context menu, so there is a way to get rid of some boards. But the "plugs" (circles and regular polygons, which are used to close the holes) have no context menu, so there is no way to get rid of the unneeded "plugs"… except throwing them across any border. Why and how it became possible?

In the `OnMouseDown()` method of the form, the mover's clipping is changed to `Unsafe`, if the "plug" was caught.

```
private void OnMouseDown (object sender, MouseEventArgs e)
{
    ptMouse_Down = e .Location;
    if (mover .Catch (e .Location, e .Button))
    {
        if (mover .CaughtSource is Plug)
        {
            mover .Clipping = Clipping .Unsafe;
```

This change of clipping level is possible, because it widens the clipping area (from `Visual` to `Unsafe`). The `Unsafe` clipping allows to move the caught object across the borders, but that level of clipping is achieved only when a "plug" was caught, so the analysis of the situation with any object beyond the borders is limited to the situation of a "plug being out of view".

```
private void OnMouseUp (object sender, MouseEventArgs e)
{
    … …
    if (mover .Release (out iWasCaught))
```

```
        {
            GraphicalObject grobj = mover .WasCaughtSource;
            long id = grobj .ID;
            … …
                else if (grobj is Plug)
                {
                    if (!PlugDisappeared (id, ClientRectangle))
                    {
                        … …
        }
        mover .Clipping = Clipping .Visual;
    }
```

If any object is released by the mover, and this object happened to be a "plug", then the `PlugDisappeared()` method is used. It identifies the "plug" via the **id** of the caught object, gets the rectangular area around this "plug", and if this area is out of view, then the "plug" is deleted, and the mover's queue is renewed.

```
private bool PlugDisappeared (long id, Rectangle rcClient)
{
    bool bRet = false;
    for (int i = 0; i < plugs .Count; i++)
    {
        if (plugs [i] .ID == id)
        {
            Rectangle rcPlug = Rectangle .Round (plugs [i] .RectAround);
            if (Rectangle .Empty == Rectangle .Intersect (rcPlug, rcClient))
            {
                plugs .RemoveAt (i);
                RenewMover ();
                bRet = true;
            }
            break;
        }
    }
    return (bRet);
}
```

If an object was moved across the right or lower border, the total disappearance of this object can be checked by enlarging the form. To avoid further possibility of moving any object across the borders, the original `Visual` level of clipping has to be restored. This is done at the end of the `OnMouseUp()` method. At this moment, it is possible to switch back from the wider to narrower area of clipping (from `Unsafe` to `Visual`), because an object was already released before by the `mover.Release()` method. Without any object being caught at the moment, the mover's clipping can be set at any level.

## Personal restrictions on the sizes of objects

The personal restrictions on the sizes of objects are always checked inside the object's `MoveNode()` method. The variety of possibilities is wide and depends on the shape of objects and the way they are supposed to be used. You can find a lot of cases among the examples that were already discussed.

The most common is the restriction on minimum size, so that an object cannot disappear from view as a result of shrinking. For example, this is the case of the `RegPoly_Variants` class in the **Form_RegPoly_Variants.cs** (see **figure 6.1**). When regular polygons are supposed to be resized by vertices, then there is a small circular node atop of each vertex. Any vertex can be moved by mouse; the proposed radius for all the vertices is calculated as the distance between the mouse and the central point of the polygon; this distance is checked against the minimum allowed radius.

```
public override bool MoveNode (int i, int dx, int dy, Point ptM, MouseButtons btn)
{
    double distance;
    if (btn == MouseButtons .Left)
        … …
        case PolygonType .ZoomByVertices:
```

```
                if (i < nVertices)
                {
                    distance = Auxi_Geometry .Distance (ptC, ptM);
                    if (distance >= minR)
                    {
                        radius = Convert .ToInt32 (distance);
                    }
```

The **Form_Polygons_Convex.cs** (see **figure 6.4**) deals with the convex polygons of the `Polygon_Convex` class. There is one size restriction for these polygons: the minimum allowed distance between any pair of consecutive vertices (`minSide`).

```
public override bool MoveNode (int i, int dx, int dy, Point ptM, MouseButtons btn)
{
    if (btn == MouseButtons .Left)
    {
        if (i < pts .Length)
        {
            PointF ptNew = new PointF (pts [i] .X + dx, pts [i] .Y + dy);
            int nVertices = pts .Length;
            int jNext = (i + 1) % nVertices;
            int jPrev = (i + nVertices - 1) % nVertices;
            if (Auxi_Geometry .Distance (pts [jPrev], ptNew) > minSide &&
                Auxi_Geometry .Distance (pts [jNext], ptNew) > minSide &&
```

The rectangles of the `RectStandard` class, which are used in the **Form_Rectangles_StandardCase.cs** (see **figure 3.1**), may have both the minimum and maximum size restrictions. In the shown piece of code, the checking is done for the move of the top left corner of the rectangle. The expected new height of the rectangle is calculated and checked against the allowed height range `[hMin, hMax]`; there is also a similar checking of the proposed width against the allowed width range `[wMin, wMax]`.

```
public override bool MoveNode (int i, int dx, int dy, Point ptM, MouseButtons btn)
{
    if (btn == MouseButtons .Left)
    {
        int wNew, hNew;
        switch (resize)
        {
            case Resizing .Any:
                …
                else if (i == 0)        //LT corner
                {
                    hNew = rc .Height - dy;
                    if (hMin <= hNew && hNew <= hMax)
                        MoveBorder_Top (dy);
                        bRet = true;
                    }
                    wNew = rc .Width - dx;
                    if (wMin <= wNew && wNew <= wMax)
```

Rings in the **Form_Rings.cs** (see **figure 8.1**) belong to the `RingT` class. The inner border of a ring can be moved, if the new inner radius is going to be not less than the allowed minimum radius, and the new width of a ring cannot become less than the minimum allowed width.

```
public override bool MoveNode (int i, int dx, int dy, Point ptM, MouseButtons btn)
{
    if (btn == MouseButtons .Left)
    {
        … …
        else if (i >= nNodesOnOuter)
        {
            int newInner = Convert.ToInt32 (Auxi_Geometry.Distance (center, ptM));
            if (minInner <= newInner && newInner <= rOuter - minWidth)
```

There is one common feature for all these examples: based on the parameters of the `MoveNode()` method, the proposed sizes of an object are calculated and compared with the allowed. If the new sizes are allowed, then this method returns `true`; only in this case the movement of a caught node will be done and the movement of an object will be fulfilled. If any checking failed, then the caught node and the associated object are not moved. At the same time there is no restriction on moving of the mouse cursor, so it continues to move, living the caught node (and the object) behind. I do not see any problem with this and never add in my applications any special code to stop the mouse in such a case. However, some people may want to anchor the cursor at that point of an object, where it was initially caught, and not allow the cursor to move on, when an object was stopped by some restrictions. There is a way to organize such a thing; I will demonstrate this technique a bit later, but I want to underline beforehand that this has nothing to do with clipping.

## *Restrictions caused by other objects*

There are going to be two series of examples in this section. The first one includes the sliders in the rectangular areas; this type of objects is often used inside the plotting areas in scientific and engineering programs. Another group of examples is about the circles in rectangular area. First of those examples use the same technique as sliders, but further work in this direction led me to another technique, which is discussed in the next section.

### Sliders in resizable rectangle

Let us consider a plotting area with one or several graphs in it. It is not a rare situation, when you would like to add some sliders into this area. In the chapter *Complex objects* I have demonstrated the track bars; some of them were used with rectangles and were associated with the sliders; those colored lines were moved across the rectangles synchronously with the movements of the track bars. Not all the plots require track bars; you can exclude them from the set of movable objects, but add the sliders, which can be moved inside the same rectangles. In some cases, such sliding lines are useful to associate special points of the graph with the numbers on the scale. The scales are usually positioned at the sides of the plotting area; if the plotting area is big, then the special points of interest (maximum of the graph, minimum, or something else) can be far away from the scales; the sliders make the estimation of the values in these points much easier. The best way would be to demonstrate those sliders not on some rectangles, but with the class, which is used in my applications for plotting in all the scientific and engineering applications. Only I do not want to use the `Plot` class in any example before starting the discussion of that class, so for now I am going to use more primitive `ResizableRectangle` class as the background for the sliders.

The `ResizableRectangle` class is included into the **MoveGraphLibrary.dll** and described in the **MoveGraphLibrary_Classes.doc**. It is a simple resizable rectangle, for which different types of resizing can be organized. There is a limit on the minimum size of rectangle, which prevents it from accidental disappearance. The possibilities of resizing are defined by some of parameters on initialization. For all three examples in this subsection, the rectangles can be resized in any way by borders and by corners.

The sliders, used in the examples of this subsection, belong to the `SliderInRectangle` class.

```
public class SliderInRectangle : GraphicalObject
{
    RectangleF rc;
    Pen pen;
    LineDir dir;
    PointF ptLT, ptRB;
```

A slider can be either horizontal or vertical. Visually it is a straight line from one border of rectangle to the opposite. A slider can be moved between two sides of rectangle. A slider can be moved by any point; only the very short parts of a slider on both of its ends are not sensitive and cannot be used to move a slider. All sliders have to be painted over the rectangle, on which they reside; certainly, they have to stay in the mover's queue ahead of this rectangle. But I want the rectangular area to be resizable by any border point regardless of the positions of sliders; for this purpose the cover for a slider does not close the ends of the slider and does not block any part of nodes on the borders of rectangle. The cover for slider is primitive and consists of a single rectangular node. The mouse cursor over a slider informs about the possible direction of the slider's movement; this is always orthogonal to the line of slider, so the cursor over the slider depends on the direction of its line.

```
public override void DefineCover ()
{
    RectangleF rcNode;
    CoverNode [] nodes = new CoverNode [1];
    if (dir == LineDir .Hor)
```

```
        {
            rcNode = new RectangleF (rc .Left + halfsense, ptLT .Y - halfsense,
                                     rc .Width - 2 * halfsense, 2 * halfsense);
            nodes [0] = new CoverNode (0, rcNode, Cursors .SizeNS);
        }
        else
        {
            rcNode = new RectangleF (ptLT .X - halfsense, rc .Top + halfsense,
                                     2 * halfsense, rc .Height - 2 * halfsense);
            nodes [0] = new CoverNode (0, rcNode, Cursors .SizeWE);
        }
        cover = new Cover (nodes);
    }
```

The individual movement of any slider is restricted by two values; depending on the orientation of slider, it is either left and right limits or top and bottom limits. If the borders of the associated rectangle would be the only restrictions, then the `MoveNode()` method would be a bit simpler. But I want to use the same `SliderInRectangle` class for the case, when there are several sliders of the same type and the neighbouring sliders can put additional restrictions on their siblings.

The restrictions that can be imposed by the neighbours are called `fLeft` and `fRight` for the vertical sliders and `fTop` and `fBottom` for horizontal sliders. By default the values for all four of these restrictions are set far beyond the borders of any rectangle, so in the case of “no neighbours” they do not have any effect on the movements of slider. When the positions of neighbours have to be taken into consideration, then the narrowest range, based both on the borders of rectangle and the positions of neighbours, defines the limits of the movements for slider.

```
public override bool MoveNode (int i, int dx, int dy, Point ptM, MouseButtons btn)
{
    bool bRet = false;
    if (btn == MouseButtons .Left)
    {
        if (dir == LineDir .Hor)
        {
            float cyNew = ptLT .Y + dy;
            if (Math .Max (fTop, rc .Top) <= cyNew &&
                                          cyNew <= Math .Min (fBottom, rc .Bottom))
            {
                Move (dx, dy);
                bRet = true;
            }
        }
        else
        {
            float cxNew = ptLT .X + dx;
            if (Math .Max (fLeft, rc .Left) <= cxNew &&
                                            cxNew <= Math .Min (fRight, rc .Right))
            {
                Move (dx, dy);
                bRet = true;
            }
        }
    }
    return (bRet);
}
```

Let us begin with the case when there is only one horizontal and one vertical slider in rectangle (**figure 11.2**).

File: **Form_SlidersInRectangle.cs**
Menu position: *Graphical objects – Movement restrictions – Sliders in rectangle*

```
public Form_SlidersInRectangle ()
{
    InitializeComponent ();
    mover = new Mover (this);
```

```
    rr = new ResizableRectangle (new RectangleF (100, 100, 400, 300),
                  new Size (100, 100), new Size (500, 500), DrawRect, ChangeRect);
    sliderHor = new SliderInRectangle (rr .Rectangle, LineDir .Hor, 0.7,
                                       new Pen (Color .Blue));
    sliderVer = new SliderInRectangle (rr .Rectangle, LineDir .Ver, 0.3,
                                       new Pen (Color .Green, 3));
    mover .Add (sliderHor);
    mover .Add (sliderVer);
    mover .Add (rr);
}
```

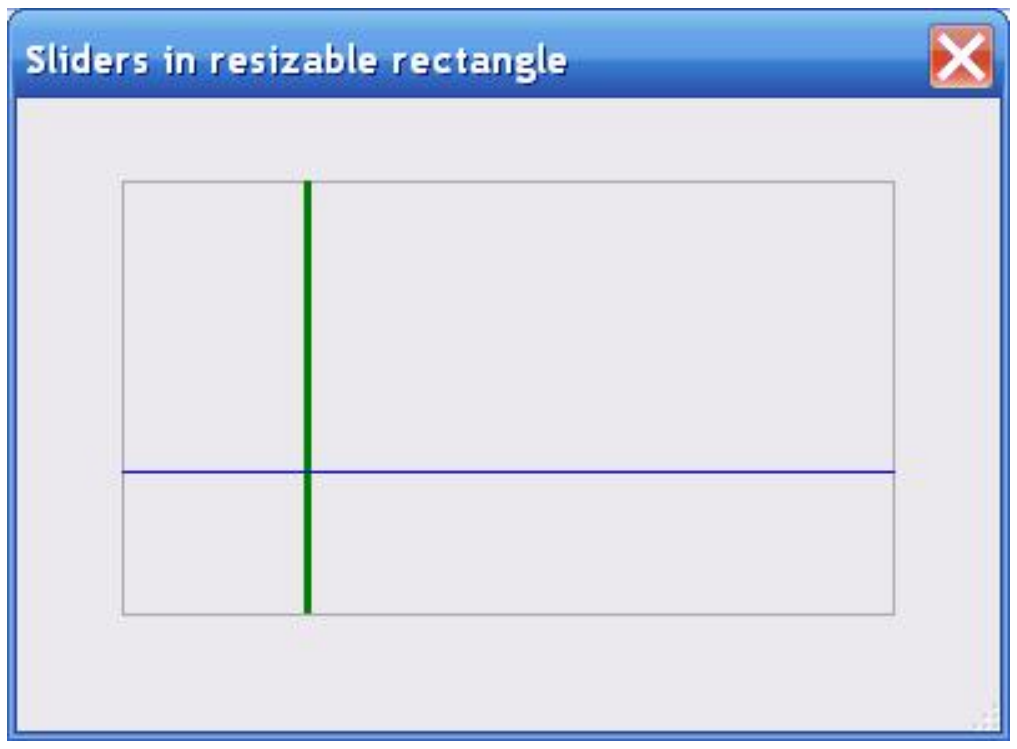

**Fig.11.2** Sliders in rectangle

The resizable rectangle gets the initial size plus its minimum and maximum sizes on initialization. Two sliders are positioned inside the rectangle according to the initial value of the positional coefficient. This coefficient is from the `[0, 1]` range with the 0 value assoziated with the left or top border, while the right and bottom borders are associated with the coefficient 1.

The moving / resizing of the rectangle and the movement of the sliders are so simple that three mouse events, through which all these things are organized, cannot be simpler.

```
    private void OnMouseDown (object sender, MouseEventArgs e)
    {
        mover .Catch (e .Location, e .Button);
    }
    private void OnMouseUp (object sender, MouseEventArgs e)
    {
        mover .Release ();
    }
    private void OnMouseMove (object sender, MouseEventArgs e)
    {
        if (mover .Move (e .Location))
        {
            Invalidate ();
        }
    }
```

The sliders move only inside the rectangle, because they were provided with this rectangle on their initialization. But what happens, when the rectangle is resized or moved? In such a case the sliders have to be informed about the new borders of rectangle. This is done through the method `ChangeRect()`, which is the one of the parameters on initialization of rectangle. This method informs the sliders about any change of rectangle.

```
    private void ChangeRect ()
    {
        sliderHor .Area = rr .Rectangle;
        sliderVer .Area = rr .Rectangle;
    }
```

In this `SliderInRectangle.Area` property the correct order of expressions is the most important thing.

- First, the positioning coefficient of the slider in the current rectangle is calculated.

```
    public RectangleF Area
    {
        set
        {
            float coor;
            double coef;
            if (dir == LineDir .Hor)
            {
                coef = Auxi_Geometry.CoefficientByCoor (rc.Top, rc.Bottom, ptLT.Y);
```

- Then the new value of rectangle is applied.

```
rc = value;
```

- Now the calculated coefficient can be used to determine the position in the changed rectangle and the two end points of the slider can be calculated.

```
coor = Auxi_Geometry.CoorByCoefficient (rc.Top, rc.Bottom, coef);
ptLT = new PointF (rc .Left, coor);
ptRB = new PointF (rc .Right, coor);
```

The above shown expresions are for the case of the horizontal slider; the vertical sliders are moved in similar way.

```
else
{
    coef = Auxi_Geometry.CoefficientByCoor (rc.Left, rc.Right, ptLT.X);
    rc = value;
    coor = Auxi_Geometry.CoorByCoefficient (rc .Left, rc .Right, coef);
    ptLT = new PointF (coor, rc .Top);
    ptRB = new PointF (coor, rc .Bottom);
}
DefineCover ();
```

File: **Form_SlidersChangeableOrder.cs**
Menu position: *Graphical objects – Movement restrictions – Sliders with changeable order*

The next example includes more sliders of each direction (**figure 11.3**), but all of them are still movable inside the whole area of rectangle. This means that these sliders do not pay attention to the positions of other sliders; the only important thing for them is the parental rectangle. So everything is going exactly as in the previous example; only the whole set of sliders must be informed about any change of rectangle.

```
private void ChangeRect ()
{
    for (int i = 0; i < horSliders .Count; i++)
    {
        horSliders [i] .Area = rr .Rectangle;
    }
    for (int i = 0; i < verSliders .Count; i++)
    {
        verSliders [i] .Area = rr .Rectangle;
    }
}
```

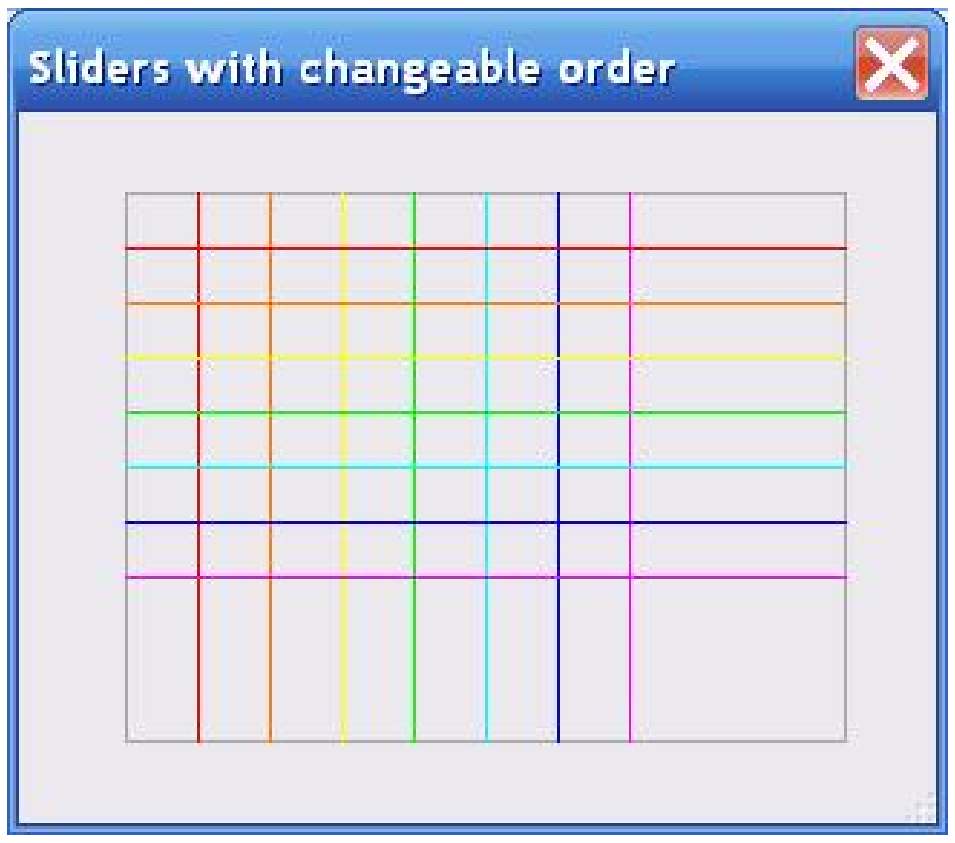

**Fig.11.3** Rectangle with independent sliders

File: **Form_SlidersUnchangeableOrder.cs**
Menu position: *Graphical objects – Movement restrictions – Sliders with unchangeable order*

The situation, when there are multiple sliders in the plotting area and these sliders affect each others positioning, is common enough. This will be the case of the third example in a row. Sliders of each type (direction) make up their own `List`.

```
public partial class Form_SlidersUnchangeableOrder : Form
{
    Mover mover;
    List<SliderInRectangle> horSliders = new List<SliderInRectangle> ();
    List<SliderInRectangle> verSliders = new List<SliderInRectangle> ();
    ResizableRectangle rr;
```

I do not need to demonstrate different figure for this example, as initially the positioning of the sliders is exactly the same, as in the previous one. The main difference is that no slider can move over its neghbour, so each slider has its own limits for individual movement; these limits also depend on whatever movement the neghbours have made before. All the interesting things are going to happen inside the `OnMouseDown()` method, when any slider is caught by the mover.

```
private void OnMouseDown (object sender, MouseEventArgs e)
{
    if (mover .Catch (e .Location, e .Button) &&
```

```
        e .Button == MouseButtons .Left &&
        mover .CaughtSource is SliderInRectangle)
    {
        SliderInRectangle sliderPressed = mover .CaughtSource as SliderInRectangle;
        long id = sliderPressed .ID;
        int iPressed;
        if (sliderPressed .Direction == LineDir .Hor)
        {
```

First, the slider which was caught has to be identified. It has to be identified not only by its **id** (this is easy), but, based on this **id,** the position of slider in the `List` of sliders of the same direction must be determined.

```
            for (iPressed = horSliders .Count - 1; iPressed >= 0; iPressed--)
            {
                if (id == horSliders [iPressed] .ID)
                {
                    break;
                }
            }
```

The order of the caught slider in the `List` is important, because for the first slider the upper boundary of its movement (I am looking here at the case of the horizontal sliders) is determined only by the border of rectangle, while for all others it is determined by the position of the previous slider from the `List`.

```
            if (iPressed > 0)
            {
                sliderPressed .Limit_Top = horSliders [iPressed - 1] .Ends [0] .Y;
            }
            else
            {
                sliderPressed .Limit_Top = rr .Top;
            }
```

The same happens with the lower border of movement for slider: for the last slider in the `List` it is determined by the border of rectangle; for all others it is determined by the position of the next slider.

```
            if (iPressed < horSliders .Count - 1)
            {
                sliderPressed.Limit_Bottom = horSliders [iPressed + 1].Ends [0].Y;
            }
            else
            {
                sliderPressed .Limit_Bottom = rr .Bottom;
            }
```

The call to the `SliderInRectangle.Limit_Top` sets the restriction, imposed by the position of the neghbouring slider.

```
        public float Limit_Top
        {
            set { fTop = value; }
        }
```

There is also another restriction, which is imposed by the rectangle itself. When the slider is moved, the `MoveNode()` method compares two limitations and picks up the more binding.

```
public override bool MoveNode (int i, int dx, int dy, Point ptM, MouseButtons btn)
{
    … …
    if (Math.Max (fTop, rc.Top) <= cyNew && cyNew <= Math.Min (fBottom, rc.Bottom))
    {
```

I prefer to use the system of two restrictions, because in the complex applications the sliders can be added and deleted at any moment and change the whole situation.

## Balls in rectangles

File: **Form_BallsInRectangles.cs**
Menu position: *Graphical objects – Movement restrictions – Balls in rectangles*

When the **Form_BallsInRectangles.cs** (**figure 11.4**) is opened, there is only one board with the balls in view. At any moment the new boards can be added. There can be from two to 15 balls on each board; balls may have different radii, or the same radius for all the balls on board can be set.

- Balls can be moved only inside their "parent" board.
- Boards can be moved around and resized. The squeezing of a board can push the balls closer to each other, so that all of them are kept inside the board.

These two statements about the behaviour of balls show that the restrictions on their movements are based not on their resizing limitations, but on positions of objects from the different class.

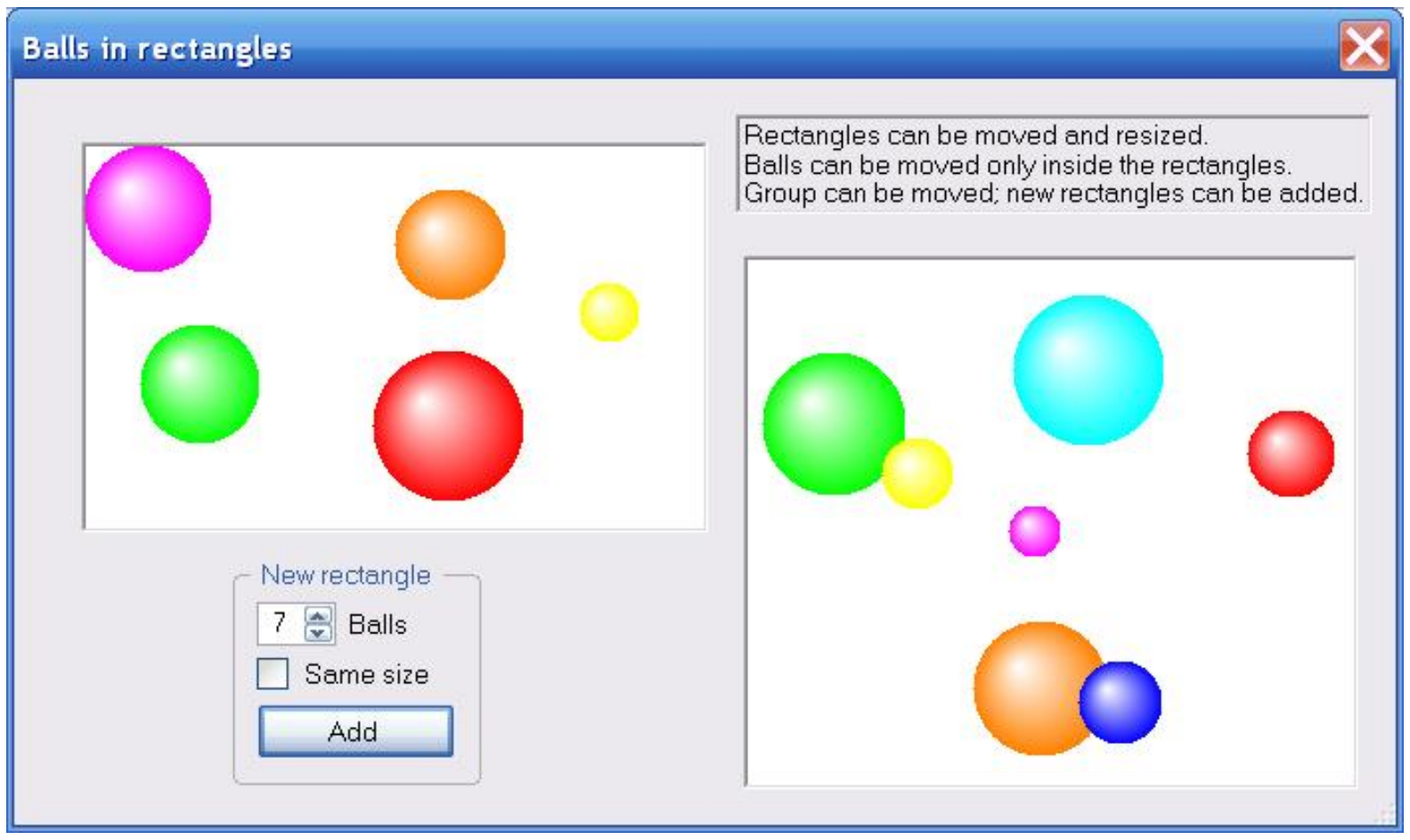

**Fig.11.4** Balls in rectangles

The important parameters of the rectangular board (class `RectWithBalls`) are the area itself and the `List` of balls that live on this board. A board also has minimum and maximum sizes with both width and height allowed to change inside the `[100, 500]` range.

```
public class RectWithBalls : GraphicalObject
{
    Rectangle rc;
    List<Ball> balls = new List<Ball> ();
    RectRange range = new RectRange (100, 500, 100, 500);
```

A board is a simple rectangle with the standard resizing by all the corners and sides, so its cover is standard.

```
public override void DefineCover ()
{
    cover = new Cover (rc, Resizing .Any);
}
```

A board can be also moved around by any inner point. While a board is moved, all its balls must move synchronously.

```
public override void Move (int dx, int dy)
{
    rc .Location += new Size (dx, dy);
    foreach (Ball ball in balls)
    {
        ball .Move (dx, dy);
    }
    SetAreas ();
}
```

A ball can move by itself or as a result of such synchronous movement with its "parent". In each case the `Ball.Move()` method is called, but for the enforced movement it is not enough. Any ball can move only inside the board. When the area has changed, a ball must be informed about this change; the `SetAreas()` method informs all the associated balls about their new areas. Position of a ball is determined by the location of its center, so for each ball the area of its movement is determined by the rectangle of a board but deflated by the radius of a ball on each side.

```
private void SetAreas ()
{
    foreach (Ball ball in balls)
    {
```

```
        ball .SetArea (rc .Left + (ball.Radius + 1), rc .Right - (ball.Radius + 2),
                       rc .Top + (ball.Radius + 1), rc.Bottom - (ball.Radius + 2));
    }
}
```

Resizing of a board is done by its `MoveNode()` method. First, personal restrictions on the moving sizes must be taken into consideration: the sizes of board must always stay inside the allowed range. If the proposed size of a board is inside the range, then the side can be moved.

```
public override bool MoveNode (int i, int dx, int dy, Point ptM, MouseButtons btn)
{
    if (btn == MouseButtons .Left)
    {
        int hNew, wNew;
        if (i == 8)
        {
            Move (dx, dy);
        }
        else if (i == 0)        //LT corner
        {
            hNew = rc .Height - dy;
            if (range .MinHeight <= hNew && hNew <= range .MaxHeight)
            {
                MoveBorder_Top (dy);    // SetAreas() inside
                bRet = true;
            }
            wNew = rc .Width - dx;
            if (range .MinWidth <= wNew && wNew <= range .MaxWidth)
            {
                MoveBorder_Left (dx);
                bRet = true;
            }
        }
```

On moving any side of the board, the `SetAreas()` method must be called to inform all the balls about the change. Here is a simple reaction on moving the upper side of the board.

```
        private void MoveBorder_Top (int dy)
        {
            rc .Y += dy;
            rc .Height -= dy;
            SetAreas ();
        }
```

A ball (`Ball` class) is a very simple object. Its position is determined by the central point, the radius is fixed at the moment of initiation, and four coordinates describe a rectangle in which the center of a ball can move.

```
    public class Ball : GraphicalObject
    {
        Point center;
        int radius;
        int cxLeft, cxRight, cyTop, cyBtm;    // center moves inside these limits
```

As a non-resizable circle, a ball has a very simple standard cover.

```
        public override void DefineCover ()
        {
            cover = new Cover (new PointF [] { center }, radius);
        }
```

The `Move()` method is extremely simple as it requires only the relocation of a central point.

```
        public override void Move (int dx, int dy)
        {
            center += new Size (dx, dy);
```

```
        }
```

The MoveNode() method is the place where the proposed move of the ball's center is checked against the allowed area of its existence.

```
public override bool MoveNode (int i, int dx, int dy, Point ptM, MouseButtons btn)
{
    bool bRet = false;
    if (btn == MouseButtons .Left)
    {
        int cxNew = center .X + dx;
        int cyNew = center .Y + dy;
        if (cxLeft <= cxNew && cxNew <= cxRight)
        {
            Move (dx, 0);
            bRet = true;
        }
        if (cyTop <= cyNew && cyNew <= cyBtm)
        {
            Move (0, dy);
            bRet = true;
        }
    }
    return (bRet);
}
```

The Ball.MoveNode() method is the place where the personal movement of a ball is checked against the area of a board, so here the movement of a Ball object is restricted by the parameters of the RectWithBalls object. But where is the place to enforce the movement of the ball when the board is shrinking? This is inside the already mentioned Ball.SetArea() method.

```
    public void SetArea (int cxL, int cxR, int cyT, int cyB)
    {
        cxLeft = cxL;
        cxRight = cxR;
        cyTop = cyT;
        cyBtm = cyB;
        center = new Point (Math .Min (Math .Max (cxLeft, center .X), cxRight),
                            Math .Min (Math .Max (cyTop, center .Y), cyBtm));
        DefineCover ();
    }
```

As I already mentioned, this method is called whenever a board is moved or resized. The new coordinates of the allowed area for the ball are passed as the parameters; the center of the ball is adjusted to these new limits of its existence. If the center is inside the new borders, then it is not changed and the ball stays in its current place. If the center point happens to be outside the new borders, it is changed in such a way, as to be inside. The ball is pushed inside by the moving border.

Are there any interesting features in this example, which were already mentioned before and are used here? Yes, the registering with the mover. A board with its balls is a complex object; the parts of such object are involved both in individual and related movements, so any RectWithBalls object must be registered by its IntoMover() method.

```
        public void RenewMover ()
        {
            mover .Clear ();
            for (int i = rects .Count - 1; i >= 0; i--)
            {
                rects [i] .IntoMover (mover, 0);
            }
            mover .Insert (0, info);
            mover .Insert (0, lrNew);
        }
```

The balls are shown atop their board, so the balls must precede their parent board in the mover's queue. All the parts of the `RectWithBalls` object must be registered in the mover's queue beginning from the `iPos` position, so it is better to use the `Mover.Insert()` method.

```
new public void IntoMover (Mover mover, int iPos)
{
    mover .Insert (iPos, this);
    for (int i = balls .Count - 1; i >= 0; i--)
    {
        mover .Insert (iPos, balls [i]);
    }
}
```

## No same color overlapping

File: **Form_NoSameColorOverlapping.cs**
Menu position: *Graphical objects – Movement restrictions – No same color overlapping*

The view of the **Form_NoSameColorOverlapping.cs** (**figure 11.5**) is very similar to the previous example. There is a rectangular board with the balls inside. To simplify the code and leave only essential things, I left only one area (though it is very easy to organize adding as many of them, as you want) and even this area is non-resizable.

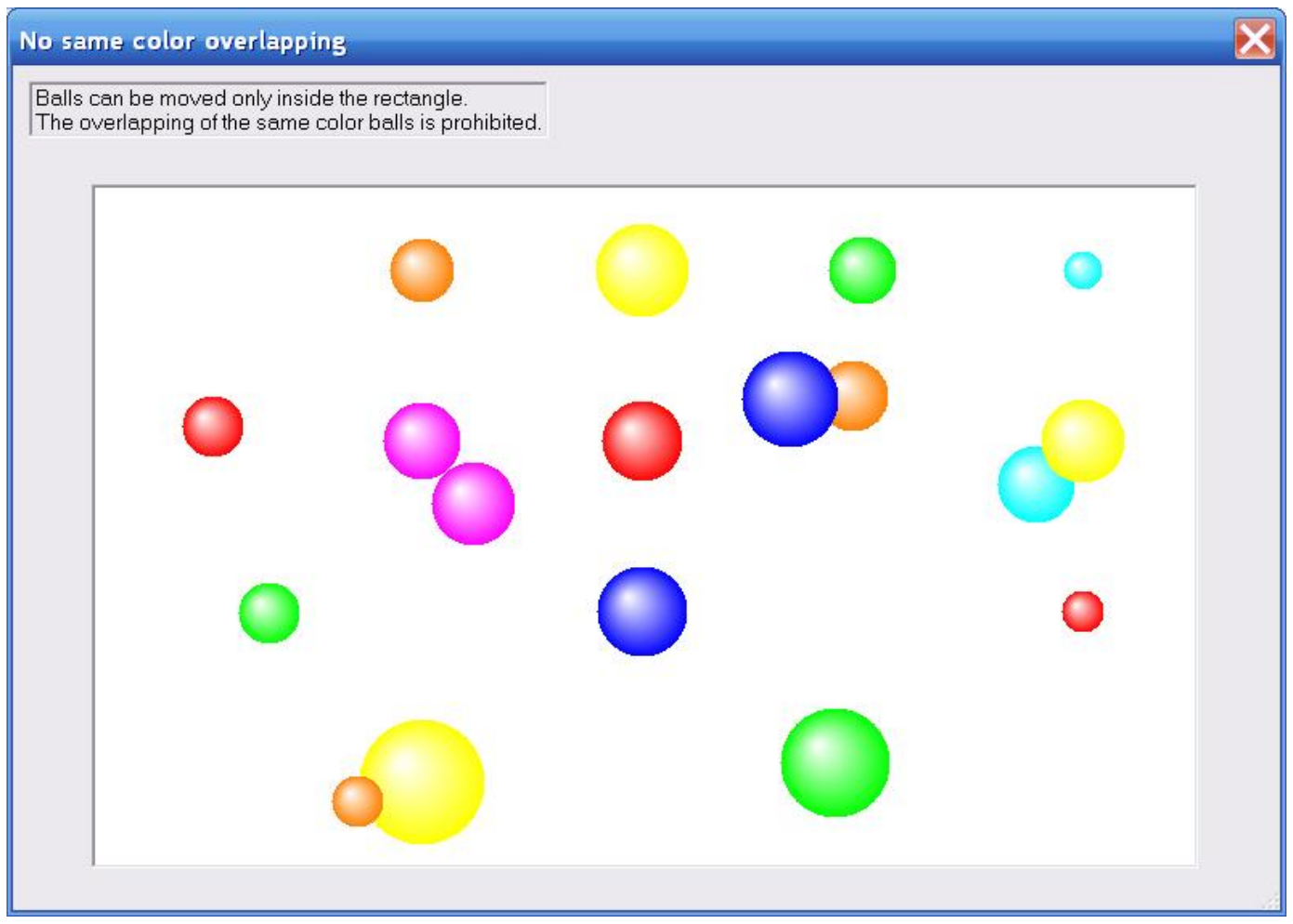

**Fig.11.5** No overlapping of the same color balls is allowed

The balls can move only inside the board, but there is an additional restriction: the balls of the same color cannot overlap.

The boards (class `BoardWithBalls`) and the balls (class `Ball_SCNO`) are very similar to the classes, used in the previous example. The main and the most interesting difference is in the `Ball_SCNO.MoveNode()` method. This is the place to do all the checking for the possibility of movement.

```
public override bool MoveNode (int i, int dx, int dy, Point ptM, MouseButtons btn)
{
    bool bRet = false;
    if (btn == MouseButtons .Left)
    {
        int cxNew = center .X + dx;
        int cyNew = center .Y + dy;
        if (cxLeft <= cxNew && cxNew <= cxRight && cyTop <= cyNew && cyNew <=cyBtm)
        {
            Point ptCenterNext = new Point (cxNew, cyNew);
            foreach (Ball_SCNO ball in balls)
            {
                if (ball .ID != ID  &&  ball .ColorEdge == clrEdge  &&
                    Auxi_Geometry .Distance (ball .Center, ptCenterNext) <=
                                                           ball .Radius + radius)
                {
                    return (false);
                }
            }
            Move (dx, dy);
            bRet = true;
        }
    }
    return (bRet);
}
```

With the known move of the mouse cursor `(dx, dy)`, the proposed coordinates of the ball's center are calculated.

```
int cxNew = center .X + dx;
int cyNew = center .Y + dy;
```

The move is allowed only if the center stays inside the allowed area.

```
if (cxLeft <= cxNew && cxNew <= cxRight && cyTop <= cyNew && cyNew <=cyBtm)
```

But even if the proposed move is allowed by the board, this is not the end of checks; the proposed new position must be checked against all other balls on the board. An overlapping of two balls is easy to find, as in this case the distance between their centers must be less than the sum of their radii. Such check is needed only against the balls of the same color.

```
foreach (Ball_SCNO ball in balls)
{
    if (ball .ID != ID &&
        ball .ColorEdge == clrEdge &&
        Auxi_Geometry .Distance (ball .Center, ptCenterNext) <=
                                                  ball .Radius + radius)
```

The checking works; the balls of the same color cannot overlap, but I am not satisfied with the overall behaviour. Let us see if there is a way to improve it.

## *Overlapping prevention*

If we talk about the different types of restrictions, then the prevention of overlapping is definitely not the new type, but is the part of the restrictions from other objects. The reason to divide them and to unite several further examples into another section is only one: next few examples use another technique. In the previous examples an object could be stopped, but the mouse continued to move. In all the examples of this section the mouse is stopped with the stopped object and stays exactly on that spot of an object, where it first caught it. If this would be organized for moving an object of an arbitrary shape inside the rectangular area, then it would be possible to calculate the rectangular clipping area for a mouse and use the standard clipping procedure. If an object is supposed to move inside the non-rectangular area, then the standard clipping is of no help. For such cases the `Mover` class has something else in its store.

### Adhered mouse

File: **Form_AdheredMouse.cs**
Menu position: *Graphical objects – Movement restrictions – Adhered mouse*

The **Form_AdheredMouse.cs** looks very simple with only few objects in view: a board in the form of a regular polygon, a ball on this board, and one control to change the number of vertices in the board (**figure 11.6**). The board belongs to the class `RegularPolygonWithBall` and can be moved and rotated. It has the form of a regular polygon, so it requires for its initialization a central point, radius of vertices, number of vertices, and the initial angle (angle to the first vertex).

```
public class RegularPolygonWithBall : GraphicalObject
{
    PointF ptC;
    float radius;
    int nVertices;
    double angle;
    SolidBrush brush;
    BallInsideConvexPoly ball;
    int radBall = 20;
```

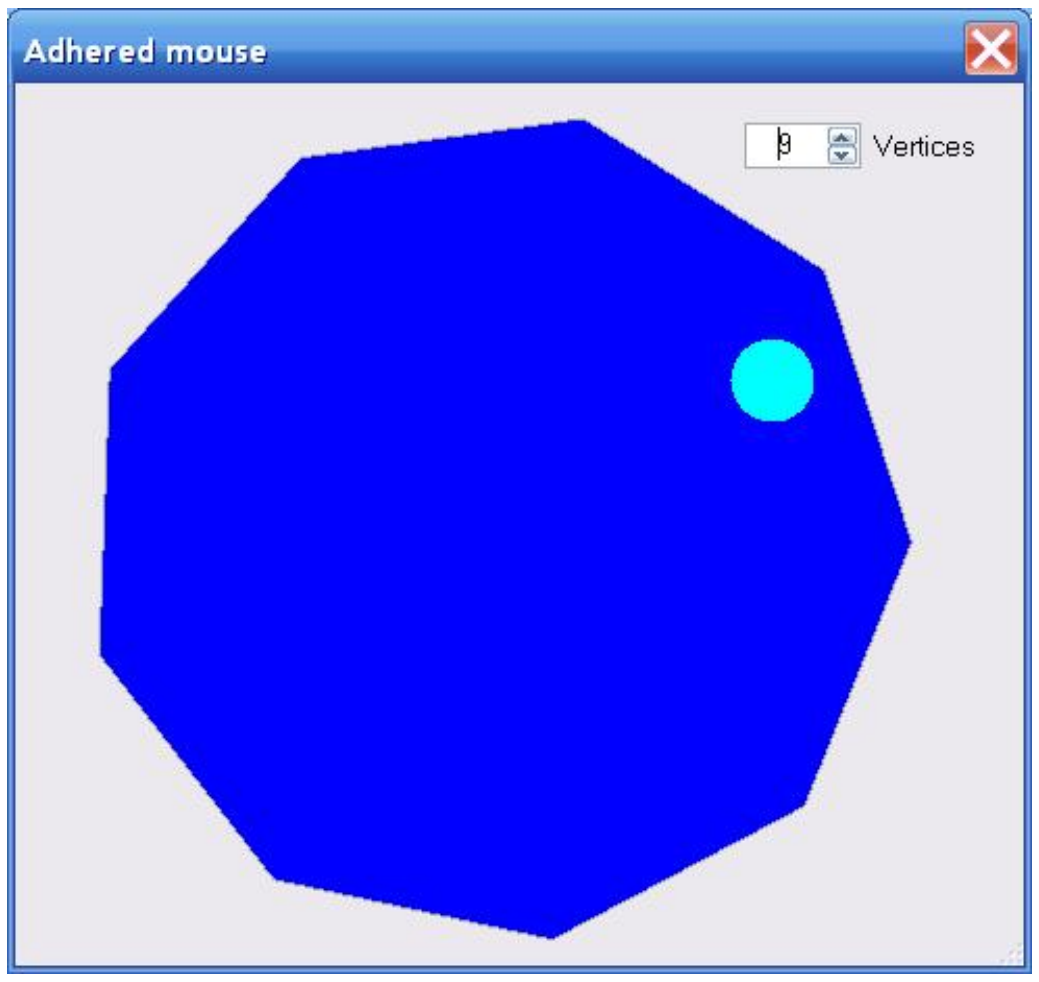

**Fig.11.6** The ball can move only on the board

The ball belongs to the `BallInsideConvexPoly` class. Though the name of the board informs that its form is a regular polygon and the name of the ball's class hint about a convex polygon, this is not a mistake. The ball is designed to move inside a convex polygon without crossing its borders; the use of this ball inside the regular polygon is a private case.

The board is the classical case of a complex object: whenever it is moved, the ball moves with it, but the ball can also move individually. If during the movement the ball runs into the border, it stops and the cursor not goes any farther by itself, but stays adhered to the same place on the ball, by which it originally has caught it.

Among the parameters of the `RegularPolygonWithBall` constructor you can find several expected values to describe a regular polygon, but you can find also two strange parameters, which were never used in constructors of graphical objects of any shape or complexity in all the previous examples. One of them is the form, in which the whole work is done; another is the mover, which supervises the moving process.

```
public RegularPolygonWithBall (Form frm, Mover mvr, PointF center, float rad,
                    int vertices, double angleDegree, Color clrPoly, Color clrObj)
{
    ptC = center;
    radius = Math .Max (Math .Abs (rad), 200);
    nVertices = Math .Min (Math .Max (3, Math .Abs (vertices)), 12);
    angle = Auxi_Convert.DegreeToRadian (Auxi_Common.LimitedDegree (angleDegree));
    brush = new SolidBrush (clrPoly);
    pen = new Pen (Color .DarkGray);
    ball = new BallInsideConvexPoly (frm, mvr, ptC, radBall, clrObj, BallAreal);
}
```

To be absolutely correct, the form is passed as a parameter to some objects, which are based on or use the texts, because the size of a text must be calculated, but if there is no text in an object, then its constructor does not need a form among the parameters. A mover is not needed among the parameters at all, because an object does not need to know whether it is movable or not. Mover supervises the moving / resizing of all the objects from its queue, but the objects cannot regulate a mover or demand anything from their supervisor. At least it was so up till this example, so there is definitely something special about it. As you can see from the code, both of these special parameters are not even saved in the `RegularPolygonWithBall` object, but are redirected into the constructor of the subordinate ball, so the `BallInsideConvexPoly` class is the real recipient of both of these parameters.

```
public BallInsideConvexPoly (Form frm, Mover mvr, PointF pt, float rad,
                             Color clr, PointF [] areal)
{
    form = frm;
    supervisor = mvr;
    center = pt;
    radius = Math .Max (minRadius, Math .Abs (rad));
    ptsAllowed = areal;
    brush = new SolidBrush (clr);
}
```

In addition to these two parameters, the ball gets the areal, in which its center can move. For a board with a shape of a regular polygon this smaller area is easily calculated, when the radius of a ball and the geometry of a board are known, which includes the central point, radius and number of vertices, and the angle of the first vertex. But the points for the perimeter of the area can be calculated in some other way and the ball will continue to move inside these borders; the only restriction is that the array of points must represent a convex polygon.

Moving of all the objects is organized with the standard three mouse events. When any object is pressed by the mouse and caught by the mover, there are two special situations. When the rotation of the board has to start in response to a right mouse press, the reaction is standard and was discussed earlier: the compensation angle must be calculated. When the ball is pressed with a left button, nothing new is expected, because it is a standard movement of a primitive object. But this is the place, where something interesting starts.

```
private void OnMouseDown (object sender, MouseEventArgs e)
{
    if (mover .Catch (e .Location, e .Button))
    {
        GraphicalObject grobj = mover .CaughtSource;
        if (e .Button == MouseButtons .Left)
        {
            if (grobj is BallInsideConvexPoly)
            {
                (grobj as BallInsideConvexPoly) .InitialMouseShift (e .Location);
            }
        }
        else if (e .Button == MouseButtons .Right)
        {
```

```
            if (grobj is RegularPolygonWithBall)
            {
                (grobj as RegularPolygonWithBall) .StartRotation (e .Location);
            }
        }
    }
}
```

A ball is an object of solid sizes. It can be pressed far away from its center, so the difference between the pressed point and the center of the ball must be remembered to organize the accurate moving further on.

```
        public void InitialMouseShift (Point pt)
        {
            dxMouseFromCenter = pt .X - center .X;
            dyMouseFromCenter = pt .Y - center .Y;
        }
```

The movement of any object is described by its `MoveNode()` method. Many previous examples demonstrated that if there exist any kind of restrictions, then they are used inside this method to determine the possibility of moving. Ball is a primitive object with the cover consisting of a single node, so there is no check for the number of the pressed node. But ball has a limitation of its movement in the shape of a convex polygon, inside which the center of a ball can move; the proposed movement of the ball is checked against this area.

```
public override bool MoveNode (int i, int dx, int dy, Point ptM, MouseButtons btn)
{
    bool bRet = false;
    if (btn == MouseButtons .Left)
    {
        PointF centerNew = new PointF (ptM .X - dxMouseFromCenter,
                                       ptM .Y - dyMouseFromCenter);
        if (Auxi_Geometry .InsideConvexPolygon (centerNew, ptsAllowed))
        {
            Center = centerNew;
            bRet = true;
        }
        else
        {
            supervisor .MouseTraced = false;
            Cursor .Position = form.PointToScreen (Point .Round (
                                          new PointF (center.X + dxMouseFromCenter,
                                                      center.Y + dyMouseFromCenter)));
            supervisor .MouseTraced = true;
            bRet = false;
        }
    }
    return (bRet);
}
```

When the proposed position of the ball's center is going to be inside the allowed area, then the ball can be moved.

```
        if (Auxi_Geometry .InsideConvexPolygon (centerNew, ptsAllowed))
        {
            Center = centerNew;
            bRet = true;
        }
```

The real problem is in the case, when the proposed movement is not allowed, because the <u>mouse cursor has already moved</u>! The cursor moved, but the movement of the ball is not allowed, so the cursor must be returned back. Easy? This movement of the cursor there and back again in an instant is absolutely invisible; nobody would see it, but there is a problem. The ball is caught by the mover, so each move of the cursor is transformed into the movement of the ball. If processed in the normal way, this back movement of the cursor is going to be transformed into the synchronous movement of the ball, so the cursor moves back, but the ball synchronously moves from it. The only way to avoid such thing is to cut temporarily the link between the cursor and the caught ball for this back movement of a cursor. This is done by the use of the Mover.MouseTraced property twice. First you cut the link, then you change back the position of the cursor, and then you reinstall the link between the mouse (mover) and the caught object.

```
supervisor .MouseTraced = false;
Cursor .Position = form.PointToScreen (Point .Round (
                                new PointF (center.X + dxMouseFromCenter,
                                            center.Y + dyMouseFromCenter)));
supervisor .MouseTraced = true;
bRet = false;
```

What is important that all the mover's parameters are not affected by this temporarily cut of the link with the mover, so the caught object, the caught node, and everything else are unchanged. Because the move is not allowed, the MoveNode() method immediately returns false value; the caught object (ball) waits for the next movement.

Now you can see why an object, which is glued with the mouse throughout the period of movement, has to get those two additional parameters on initialization. The **form** is needed for the transformation of coordinates from one system to another; the **mover** (supervisor) is needed, because only mover can cut and reinstall the link with an object.

One more important detail. There are two ways to determine the new position of an object by the parameters of the MoveNode() method: either to rely on the pair of shifts along two axis (dx, dy) or on the mouse position (ptM). I mentioned several times that I always use the second choice throughout the rotations, but for normal forward movement, resizing, and reconfiguring I prefer the first option. This is correct anywhere except the cases, when I have to use this Mover.MouseTraced property to go back and force. In all such cases the cursor has to be glued at the same point of an object throughout the whole movement. The shift from some object's basic point to the cursor is calculated at the first moment and has to be unchanged throughout the process. It works much better, if in such a case the proposed position of this basic point is calculated from the changing cursor position with the help of that shift.

```
PointF centerNew = new PointF (ptM .X - dxMouseFromCenter,
                               ptM .Y - dyMouseFromCenter);
```

## Ball in labyrinth

File: **Form_BallInLabyrinth.cs**
Menu position: *Graphical objects – Movement restrictions – Ball in labyrinth*

The next example demonstrates the use of the same technique in some new environment. There is a small ball, which hopes to find its way through the labyrinth (**figure 11.7**). The labyrinth (class Labyrinth) consists of a set of walls (class Wall). The ball (class BallSV) can be moved along the corridors through the labyrinth. If the ball runs into the wall, it does not move farther on and the cursor does not move either. The cursor adheres to that point of the ball, which it initially pressed, and stays at this point of a ball throughout the whole move until the moment of release.

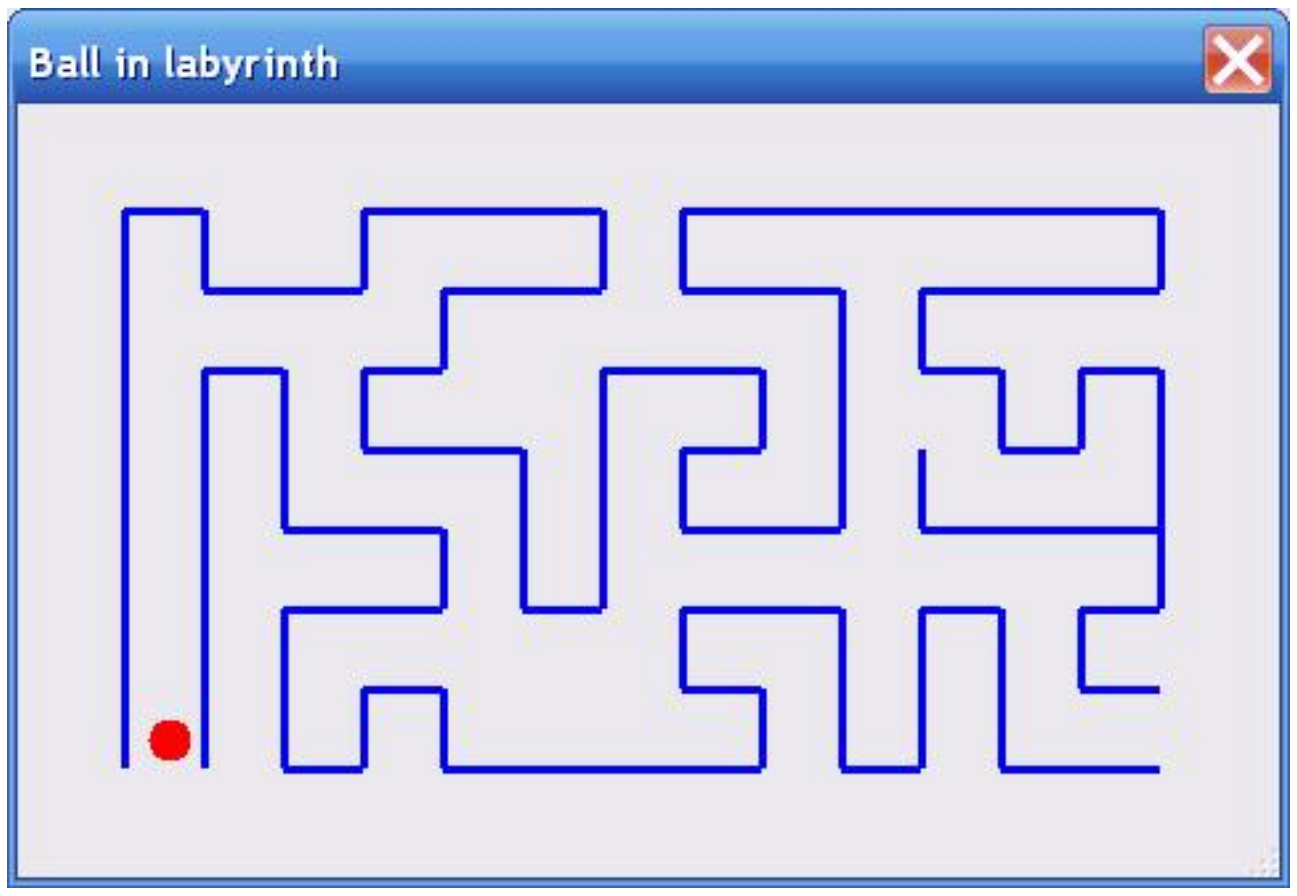

**Fig.11.7** Ball in labyrinth

A ball of the BallSV class is a standard non-resizable circle. To move such an object, it is enough to give it a primitive cover, consisting of a single circular node. To define any simple circle, only a central point and radius are needed.

```
public class BallSV : GraphicalObject
{
    Form form;
    Mover supervisor;
    Point center;
    int radius;
    Labyrinth lab;
```

But this ball is going to move in the area, where its move is restricted by some structure. When the movement is blocked, the ball is supposed to stop and the mouse must be adhered to the same spot on the surface of the ball. Thus three additional parameters must be passed to the ball on initialization: form, mover, and labyrinth.

```
public BallSV (Form frm, Mover mvr, Point pt, int r, SolidBrush brsh, Labyrinth lb)
```

A human eye is a perfect instrument, which can see the discrepancy of one or two pixels. In this case, which requires an absolute accuracy, when the ball is caught by the mover (by the mouse!), it is impossible to simplify the task by assumption that the ball is caught at its central point. The initial shift between the center of a ball and the mouse position must be remembered and used throughout the whole movement. For this purpose, the `InitialMouseShift()` method is called from inside the `OnMouseDown()` method.

```
    private void OnMouseDown (object sender, MouseEventArgs e)
    {
        if (mover .Catch (e .Location, e .Button))
        {
            if (mover .CaughtSource is BallSV)
            {
                (mover .CaughtSource as BallSV) .InitialMouseShift (e .Location);
            }
        }
    }
```

The place, where the possibility of ball's movement is checked, is certainly its `MoveNode()` method.

```
public override bool MoveNode (int i, int dx, int dy, Point ptM, MouseButtons btn)
{
    bool bRet = false;
    if (btn == MouseButtons .Left)
    {
        bRet = true;
        Point ptFrom, ptTo, ptCross;
        Point ptCenterNew = new Point (ptM .X - dxMouseFromCenter,
                                       ptM .Y - dyMouseFromCenter);
        for (int j = 0; j < lab .Walls .Count; j++)
        {
            lab .Segment (j, out ptFrom, out ptTo);
            if (Auxi_Geometry .Distance_PointSegment (ptCenterNew, ptFrom, ptTo)
                                                                         <= radius
                || Auxi_Geometry .Segment_Crossing (ptFrom, ptTo, center,
                                                     ptCenterNew, out ptCross))
            {
                supervisor .MouseTraced = false;
                Cursor .Position = form .PointToScreen (new Point (center .X +
                                dxMouseFromCenter, center .Y + dyMouseFromCenter));
                supervisor .MouseTraced = true;
                return (false);
            }
        }
        Center = ptCenterNew;
        bRet = true;
    }
    return (bRet);
}
```

This is also the case, where the link between the mover and the caught ball must be cut temporarily, so the whole procedure is organized in the way similar to the previous example.

1. The calculation of the proposed position of a ball is based not on the pair of mouse movementss `(dx, dy)` but on exact mouse position `ptM` and the shift from the mouse to the center of a ball, which was estimated at the starting point of this movement.

```
    Point ptCenterNew = new Point (ptM .X - dxMouseFromCenter,
                                   ptM .Y - dyMouseFromCenter);
```

2. There are two checks for the possibility of movement. Both checks are done against each wall of the labyrinth.

```
        lab .Segment (j, out ptFrom, out ptTo);
```

3. The first check estimates the distance between the segment of the wall and the center of a ball; it cannot become less than the radius of a ball.

```
if (Auxi_Geometry.Distance_PointSegment (ptCenterNew, ptFrom, ptTo) <= radius
```

4. For the normal move of a ball it would be enough to have this first check only, but then I found that if the ball is moved against the wall at the speed of light or close to it, then it can go through (I hope you heard something about neutrino…). To return this ball back from becoming a particle into a normal screen object, I had to add another check. One method from the **MoveGraphLibrary.dll** solved the problem easily; the ball is not allowed to cross any segment on the way.

```
|| Auxi_Geometry .Segment_Crossing (ptFrom, ptTo, center,
                                            ptCenterNew, out ptCross))
```

5. If any check failed, the mouse cursor has to be returned back. I want to remind that when the `MoveNode()` method is called, the mouse cursor has already moved! To return the cursor back, the link between the mover and the caught object (ball) must be temporarily cut, the cursor returned back, and then the same link reinstated. And do not forget to return **`false`** from the `MoveNode()` method in this case, because the move is not allowed.

```
supervisor .MouseTraced = false;
Cursor .Position = form .PointToScreen (new Point (center .X +
                       dxMouseFromCenter, center .Y + dyMouseFromCenter));
supervisor .MouseTraced = true;
return (false);
```

## Strip in labyrinth

File: **Form_StripInLabyrinth.cs**
Menu position: *Graphical objects – Movement restrictions – Strip in labyrinth*

One more example of moving an object around a lot of obstacles. The same labyrinth is used in the **Form_StripInLabyrinth.cs** (**figure 11.8**), but the movable object is more interesting. It is a resizable strip with two rounded corners. With its sizes, the strip cannot be simply pressed and moved through the labyrinth; it has to be stopped and turned a bit to pass each turn of a corridor.

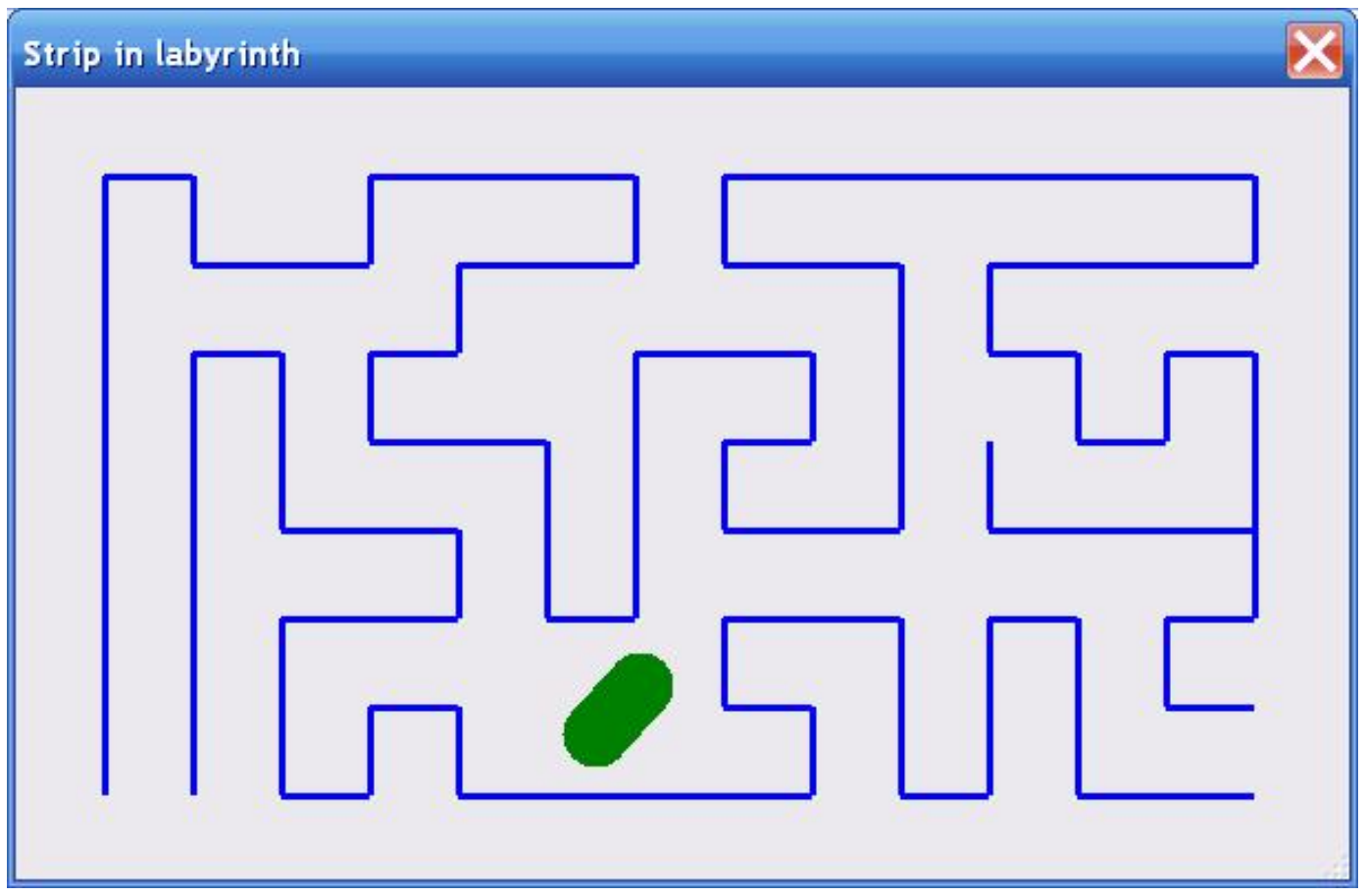

**Fig.11.8** Strip in labyrinth

The strip (class `StripSV`) has a cover that was demonstrated at **figure 7.2**: curved parts of the border are used to change the length of a strip, straight parts of the border – to change the width. Strip can be moved and rotated by any point. The geometry of a strip, the possibility of its rotation, and the changing length require more parameters to describe the object's position, than in case of a circle.

```
public class StripSV : GraphicalObject
{
    Form form;
    Mover supervisor;
    Labyrinth lab;
    PointF ptC0, ptC1;         // central points of the semicircles at the ends
    float radius;              // radius of those semicircles
    double angle;              // angle is calculated from ptc0 to ptc1
    SolidBrush brush;
```

The technique of returning the cursor back, if the proposed movement is not allowed, is similar to what was used in the previous example, so the form, mover, and labyrinth must be also mentioned among the fields of this class.

When any object is grabbed for moving, it usually requires the saving of some parameters, which are not going to change during the initiated movement, but are needed to calculate the object's position throughout this movement. For rotation, it is usually a compensation angle; for zooming it is often some scaling coefficient. In case of a strip we have both, but because of the expectation that on some restricted movements there can be a request to return the cursor back, there is a storing of an additional parameter.

```
private void OnMouseDown (object sender, MouseEventArgs e)
{
    if (mover .Catch (e .Location, e .Button))
    {
        GraphicalObject grobj = mover .CaughtSource;
        if (e .Button == MouseButtons .Left)
        {
            if (grobj is StripSV)
            {
               (grobj as StripSV).StartLengthChange (e.Location, mover.CaughtNode);
            }
        }
        else if (e .Button == MouseButtons .Right)
        {
            if (grobj is StripSV)
            {
                (grobj as StripSV) .StartRotation (e .Location);
            }
        }
    }
}
```

I have already explained the `StartLengthChange()` method for a strip while writing about similar strip before (chapter *Curved borders. N-node covers*). The code for this method is nearly the same, but here you can see an extra call to the `InitialCatch()` method.

```
    public void StartLengthChange (Point ptMouse, int iNode)
    {
        InitialCatch (ptMouse);
        PointF [] pts = CornerPoints ();
        if (iNode < 2)
        {
        }
        else if (iNode < 2 + nNodesOnHalfCircle)
        {
            fStartingDistanceToRect =
                          Auxi_Geometry .DistanceToLine (ptMouse, pts [1], pts [2]);
        }
        else
        {
            fStartingDistanceToRect =
                          Auxi_Geometry .DistanceToLine (ptMouse, pts [0], pts [3]);
        }
    }
```

When an object is pressed by the mouse with the intention to start some movement, it is just the right moment to calculate some parameters, which are fixed for the whole duration of this movement and are used to calculate the new position (or sizes) according to the mouse movement. These parameters are different for different movements and, as a rule, they are calculated in different methods. Up till now we saw three different types of parameters:

- When an object is pressed by the left button for resizing, then a scaling coefficient is calculated.
- When an object is pressed by the right button for rotation, then a compensation angle is calculated; in some cases the distance from the center of rotation is also needed.
- When there is a chance of the future restricted movement, which requires the back move of a cursor, the shift between the pressed point and some object's basic point is calculated.

In the case of the `StripSV` class I decided to combine all these calculations into a single `InitialCatch()` method and call this method on all the occasions. The basic point, from which the shifts are calculated, is also the center of rotation.

```
        private void InitialCatch (PointF pt)
        {
            center = Center;
```

```
            length = Auxi_Geometry .Distance (ptC0, ptC1);
            rMouseFromCenter = Auxi_Geometry .Distance (center, pt);
            double angleMouse = Auxi_Geometry .Line_Angle (center, pt);
            compensation = Auxi_Common .LimitedRadian (angleMouse - angle);
        }
```

If you make a quick search throughout the code of the **Form_StripInLabyrinth.cs** file, you will find out that this method is called not only at the moment of the first mouse press, but also in each part of the `StripSV.MoveNode()` method. This happens because different movements of a strip can be stopped by the walls; all such cases require to return the cursor back.

Any possible move of a strip in labyrinth must be checked against the restrictions. It can be a forward movement, or change of the width, or change of the length, or rotation – any of them can be stopped by the walls. The exact movement depends on the caught node or the pressed button, but all the branches of the strip's `MoveNode()` method are organized in the similar way:

- The cursor has its new location `ptM`.
- The new size of the strip or its new location is calculated on the basis of this mouse position.
- The proposed location of the strip is checked against the walls of the labyrinth by the `Labyrinth.StripAllowedPosition()` method.
- If there are no problems with the proposed position of the strip, then this move or resizing is finalized; if not, then the `ReturnCursor()` method must return the cursor back without disturbing the strip itself.

```
public override bool MoveNode (int i, int dx, int dy, Point ptM, MouseButtons btn)
{
    if (btn == MouseButtons .Left)
    {
        … …
        PointF [] pts = CornerPoints ();
        if (i == 0)
        {
            if (Auxi_Geometry .SameSideOfLine (ptC0, ptC1, ptM, pts [0]))
            {
                fDist = Auxi_Geometry .DistanceToLine (ptM, ptC0, ptC1);
                if (fDist >= minR)
                {
                    radiusNew = Convert .ToSingle (fDist);
                    if (!lab .StripAllowedPosition (ptC0, ptC1, radiusNew))
                    {
                        ReturnCursor ();
                        return (false);
                    }
```

The `Labyrinth.StripAllowedPosition()` method is used with nearly any movement (there is one exception!) and checks, if the proposed position of a strip is not going to be too close to the walls of a labyrinth. This check is based on a calculation of the distance between two segments. One is the segment of the wall; another is a segment between the centers of the strip's semicircles. The distance from this second segment to the border of strip is equal to the radius of semicircles; if the distance between two segments is less than this radius, then such position of a strip is not allowed.

The use of the `StripAllowedPosition()` method in the `MoveNode()` method has one exception: it is not enough for the forward movement of the whole strip. For this case there is a similar `StripAllowedMove()` method, which includes an additional check, which prevents the really quick move of an bject through the wall.

The `ReturnCursor()` method temporarily cuts the link between the movement of the cursor and the movement of an object, returns the cursor back, and then reinstates the link. To return the cursor back, two parameters are used, which were saved by the `InitialCatch()` method: the distance between the mouse and the center of a strip and the compensation angle.

```
public void ReturnCursor ()
{
    center = Center;
```

```
    supervisor .MouseTraced = false;
    Cursor.Position = form.PointToScreen (Point.Round (Auxi_Geometry.PointToPoint (
                                Center, angle + compensation, rMouseFromCenter)));
    supervisor.MouseTraced = true;
}
```

# Individual controls

Controls can be moved individually, in groups, and in some combinations with graphical objects. This chapter is only about the moving / resizing of the solitary controls.

Applications are constructed of the elements of two different types: graphical objects and controls. In reality, all the screen elements are graphical, but from the very beginning of the modern systems the controls were declared special by manufactures and were designed in a special way. I would prefer to design applications exclusively on the basis of graphical objects. Some controls can be easily substituted with the graphical analogues; these changes have a lot of advantages without any visible negative effects. (At least, I do not know any such negative effects.) The `Trackbar` class, demonstrated in one of the previous chapters, is one example of switching from the standard control to its graphical analogue. There are other controls, which require more work for such substitution, but I think that this is the right way of progress in programming and design of applications. But at the current moment a lot of applications are based on controls, so it is important to look into the problems of moving and resizing such elements.

When you design and work with the applications, composed of the fixed controls with the occasionally repainted background, you have no problems with the controls. You simply got used to whatever is provided with them (all their properties and events) and what you can get of them, as a developer. The users also got used to the view of controls and their behaviour throughout the last quarter of a century and demonstrate the classical Pavlov's reflex: if they see the button, they click it; if they see some kind of a list, they scroll it and press the needed line. And it really works in such a way, which makes the reflex only stronger.

What none of the users try to do is to move the buttons, for example, to another place, even if they think that another place would be better. Well, I think that some users tried to do it, when there was no one around to laugh at their attempts, but immediately found out that that was impossible. The reflex was fixed forever: you can click the controls, but not move.

In reality this statement about the controls being not movable is absolutely wrong; they are movable, they are even developed to be movable and resizable, but these features are hidden from the users and are occasionally used by the developers to make an impression. (Like some people in the passed centuries made their living on the ground of knowing, which gradients to through into the flame to produce the colored smoke. Some of those people were burnt for such knowledge, but that is an absolutely different story.) The popular dynamic layout is based on ability of controls to be moved and resized.

Controls can be easily moved and resized by using their properties; the demonstration of a button running away from the approaching mouse is just a funny example of using these properties. (Write the code for one mouse event, consisting of several primitive lines, and the users will be amazed with the behaviour of those crazy controls.) But this is an example of how the developers can use these features, because such a running away control is going to move according to the rules, predefined by the developer. I am writing about the mechanism that allows USERS to move controls in any way they would like to do, so it is absolutely different, though I use exactly the same properties. It is the question of who is managing the movements of controls: whether the algorithm is absolutely predetermined by the developer (the control is moved for predefined number of pixels in predetermined direction, when the mouse arrives over it) or all the movements are determined only by users, when they would decide to move a control one way or another. I think that users and only users must get the full control of moving and resizing the controls.

## *Moving solitary controls*

File: **Form_SolitaryControls.cs**
Menu position: *Individual controls – Solitary controls*

The idea of moving and resizing controls by a mouse is exactly the same, as with all the graphical objects: press and move, or press and resize, but there is a small problem. The standard procedure for the graphical objects is to move them by their inner points and resize by the borders. Unfortunately, the whole area of any control cannot be used for anything new, as controls are designed to use their every inner point for the mouse clicks with already declared purposes. The reflex is so strong that it cannot be changed. Thus the only chance to move and resize the controls is to use the vicinity of their borders. And as we need both the moving and resizing, then the frame of a control is going to be divided between the areas to start moving and resizing.*

* There is some inconsistency in using the border of the graphical objects for resizing only, but the border of the controls for both actions. That is one of the consequences of having controls on the screen. Controls can be redesigned and treated like all other objects, but this would be really a big step, though I think it will happen some time in the future. There is absolutely nothing that demands the existence of those controls on the special level. Any one of them can be replaced with

Certainly, this frame around a control is used as a cover. It looks a bit strange to have a cover outside an object's area, but there are two remarks to this situation. First, the cover often goes outside the real object; in fact, it goes outside every time, when an object is resizable; the covering of the border makes sensitive an area on both sides of the border. Second, this cover, used for moving and resizing of a control, belongs not to the control itself, but to a `SolitaryControl` object, which wraps this control. This object – a frame around the control – is registered in the mover's queue as any other object. The only difference is that the moving and resizing of this frame are directly translated into the moving and resizing of the associated control, making it movable and resizable. The only question in organizing such a frame is the placement of the nodes for resizing, but several applications use the same technique, so the practice is already known. For example, when you are setting the size of the proposed control in Visual Studio, you resize it by the small squares in the corners or in the middle of the sides. So, the best and expected places for the nodes to resize any control are obvious, other details depend on the purpose of the needed resizing.

**Form_SolitaryControls.cs** demonstrates several individually movable controls with their covers (**figure 12.1**). There are also three other controls in this form, which are shown without covers. Those three controls are organized into a special group of the `LinkedRectangles` class; an object of this class can be moved by any inner point; the class will be discussed a bit later in the chapter *Groups of elements*. Now let us return to those controls which can be moved individually and which are shown at the figure with their covers.

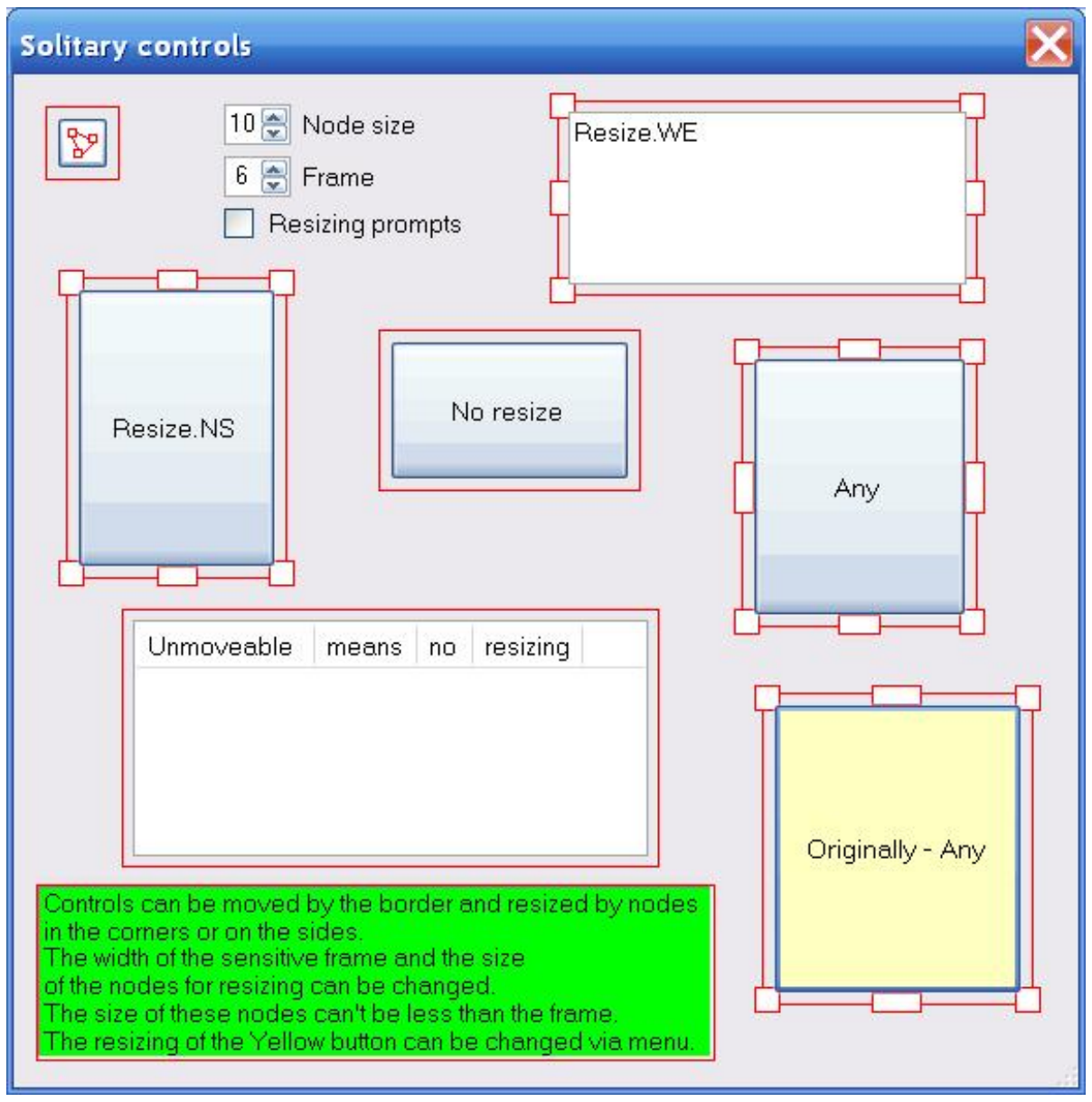

**Fig.12.1** Individually moved controls and their covers

All these seven controls are turned into movable objects in exactly the same way:

- A `SolitaryControl` object is constructed around a control.
- The designed `SolitaryControl` object is registered in the mover's queue.

```
mover .Add (new SolitaryControl (btnCovers, true, nodesize, frame));
mover .Add (new SolitaryControl (textboxResizeWE, true, nodesize, frame));
mover .Add (new SolitaryControl (btnResizeNS, true, nodesize, frame));
mover .Add (new SolitaryControl (btnMoveNoResize, true, nodesize, frame));
mover .Add (new SolitaryControl (btnResizeAny, true, nodesize, frame));
mover .Add (new SolitaryControl (listviewResizeNotMove, false, nodesize, frame));
mover .Add (new SolitaryControl (btnOriginallyAny, true, nodesize, frame));
```

The `SolitaryControl` constructor, which is used here, has four parameters.

```
SolitaryControl (Control ctrl,          // control to become movable
                 bool bMove,            // declares an object to be movable or not
                 int nodesize,          // size of the nodes in the corners
                 int frame)             // the frame's width
```

As you can see, there is no parameter to declare the type of resizing, but the controls in this form can be resized in different ways. The mover determines the type of resizing for a particular control by analysing its sizes and the values of its `MinimumSize` and `MaximumSize` properties.

---

the graphical object that looks exactly the same, behave exactly the same, but does not demand the special status. I designed such objects before, other people were doing similar things, so this is not something absolutely unique and never heard about. It is impossible for the users to distinguish the current day controls from the similarly designed graphical objects; they will not see the difference at all. For programmers, such graphical objects are much better for the forms' design, as they can be of an arbitrary shape. Certainly, this need some work to be done, so that the new graphical "controls" will be as easy in use for design, as our modern day controls, but I think that the word *progress* is a correct one to describe such a work to be done.

- If these MinimumSize and MaximumSize properties are not changed from their default values (0, 0) or their values are the same as the size of a control, then this control is non-resizable. These are the buttons in the top left corner and in the center of the form.
- If these properties provide a range for one direction only (width or height), then this control is resizable in this direction only. The button in the middle of the left side can change only its height; the button in the top right corner can change only its width.
- If the properties provide the ranges for both directions, then this control is fully resizable; this is the case of the button in the middle of the right side.

There is one exception in the last case: if the second parameter in the above shown constructor is false, then this control becomes unmovable and automatically non-resizable regardless of the size range provided by its properties! This is the case of the ListView control in the bottom left corner. According to its properties, it has to be fully resizable, but it is forced to be unmovable and, as a consequence, non-resizable.

The frame parameter of the constructor determines the width of the sensitive frame around the borders of a control; by pressing inside this area, the control can be moved. At **figure 12.1**, the covers are visualized and these sensitive areas around the controls are shown by the big red frames. The width of a frame cannot be less than 2 pixels; by default it is 6 pixels. If the second parameter in the constructor - bMove - is set to false, then such control is unmovable. The unmovable control is automatically non-resizable regardless of those two properties of a control. But it still has a frame around the borders; mover recognizes this area (and thus a control), so, for example, a context menu can be called for the control.

Two big controls at **figure 12.1** demonstrate the same type of cover, consisting of a single red frame around. One of these controls – the button in the middle – is movable; another – the ListView control in the bottom left corner – is unmovable. Users are informed about the difference in their behaviour by different shapes of cursor above those covers.

The process of making any solitary control movable / resizable differs from the same process for graphical objects in several aspects.

1. The absence of DefineCover(), Move(), and MoveNode() methods. Simply forget about these methods, when you deal with the individual controls.
2. The resizing ranges are determined by the MinimumSize and MaximumSize properties of a control and by comparing these values with the sizes of a control.
3. A control can be registered with the mover only indirectly through the SolitaryControl class

```
mover .Add (new SolitaryControl (ctrl, …));
mover .Insert (iPos, new SolitaryControl (ctrl, …));
```

In many cases you can see a shorter form of the registration for control, but this is only a shorter notation and nothing else.

```
mover .Add (ctrl);
mover .Insert (iPos, ctrl);
```

Even in such a case the control is wrapped into the SolitaryControl object, but this is done by the mover itself. Thus prepared SolitaryControl object is registered with the mover, but in such a case this wrapper gets the default parameters, which means, among other things, the resizing in any direction (certainly, if the two mentioned properties of a control allow to do it).

Depending on the type of the needed (organized) resizing, the covers at **figure 12.1** show the different number of nodes, but this is only on visualization. A cover for any SolitaryControl object always has eight small nodes along the frame. If you do not see part of these nodes next to some of the controls at the figure, then it means that the parameters of some of the nodes were changed so, as to make them invisible and not working, but the number of nodes did not change; this constant numbering makes the implementation of the MoveNode() method for the SolitaryControl class easier. These eight nodes are numbered from the top left corner and going clockwise, so the node in the top left corner has the number 0, in the middle of the upper side – 1, and so on. The **Form_SolitaryControls.cs** was designed especially for demonstration, so you can switch the visualization of covers ON / OFF, but in real applications I never show the covers, so the easiness of moving and resizing the controls depends on the design decision and the selection of several parameters.

The default size of those eight nodes (9 pixels) is bigger than the default width of the frame, which is equal six pixels. The enlarged size of the corner nodes makes them preferable places for resizing. The corner nodes are used not only for Resizing.Any, but for two other types of resizing (WE and NS); for these three types of resizing the cursors over the corner nodes are different and the resizing works according to the demand. The size of the nodes in the middle of the sides depends on the length of the appropriate side: the longer the side, the longer the node, so it would be easier to find it somewhere next to the big control.

I have explained that there is no parameter in the SolitaryControl constructor that determines the type of resizing; but the combination of sizes and two properties of the control determine it. Well, this is true in the majority of cases, but the SolitaryControl class has another constructor, which has the resizing parameter.

```
SolitaryControl (Control ctrl,          // control to become movable
                 Resizing resize,       // resize type
                 bool bMove,            // declares an object to be movable or not
                 int nodesize,          // size of the nodes in the corners
                 int frame)             // the frame's width
```

The trick is that this specified resizing cannot be wider than the resizing estimated in the standard way. If the sizes and two properties allow the full resizing of the control, you can limit the allowed resizing by passing as a parameter Resizing.NS, Resizing.WE, or Resizing.None. In such case the control gets the type of resizing that is passed as a parameter. But if the properties of control allow it to be resized only horizontally (Resizing.WE) and you try to change it with the Resizing.NS or Resizing.All parameter, this has no effect on the resizing of control.

The type of resizing for a control can be set not only at the moment of initialization of a SolitaryControl, but later with the help of the Resizing property. Why it can be needed? Consider a case of fully resizable control. While an application is running, you change the size of control to whatever you prefer and do not want to change its width accidentally after it. You can change its resizing status to Resizing.NS and there will be no change of the width, even if you press the corner node. The yellow button in the bottom right corner of **figure 12.1** can be used to demonstrate some possibilities. I purposely changed the background color of this button to distinguish it from others; this is the only control in the form, on which and around which the context menu can be called (**figure 12.2**).

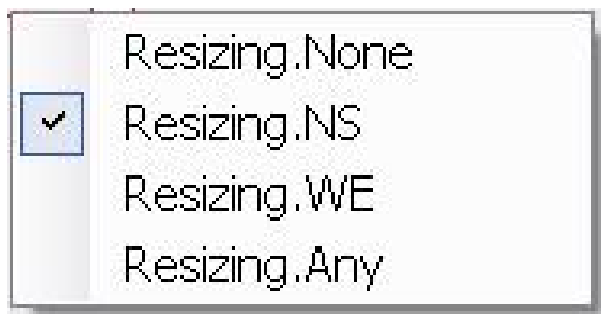

**Fig.12.2** Menu to change the control's resizing

The yellow button is designed to be fully resizable by setting the appropriate values for its two properties. Then I called the menu on this button and changed its resizing to Resizing.NS; you can see it by the view of its cover. With the help of the same menu I can change the resizing of this control to any other. This flexibility (all four choices) is possible only for the controls, which are originally fully resizable. If the comparison of the original sizes and its MinimumSize and MaximumSize properties would result in another level of original resizing, then some of the positions in menu would be disabled.

```
private void Opening_menuResizing (object sender, CancelEventArgs e)
{
    … …
    SolitaryControl solcon = mover [i] .Source as SolitaryControl;
    Resizing resizing = solcon .Resizing;
    ToolStripItemCollection items = menuResizing .Items;
    switch (solcon .ResizingFromMinMax)
    {
        case Resizing .WE:
            items ["miAny"] .Enabled = false;
            items ["miNS"] .Enabled = false;
            break;
        case Resizing .NS:
            items ["miAny"] .Enabled = false;
            items ["miWE"] .Enabled = false;
            break;
        case Resizing .None:
            items ["miAny"] .Enabled = false;
            items ["miWE"] .Enabled = false;
            items ["miNS"] .Enabled = false;
            break;
    }
    ((ToolStripMenuItem) items ["miAny"]) .Checked = resizing == Resizing .Any;
    ((ToolStripMenuItem) items ["miWE"]) .Checked = resizing == Resizing .WE;
    ((ToolStripMenuItem) items ["miNS"]) .Checked = resizing == Resizing .NS;
    ((ToolStripMenuItem) items ["miNone"]) .Checked = resizing == Resizing .None;
```

Two different properties of the SolitaryControl class allow to get two levels of resizing: ResizingFromMinMax property returns the original level; Resizing property returns the currently used level. The second property allows also to change the level.

```
private void Click_miNS (object sender, EventArgs e)
{
    SolitaryControl solcon = mover [iPressedObject] .Source as SolitaryControl;
    solcon .Resizing = Resizing .NS;
    Invalidate ();
}
```

The change of the control's resizing does not change anything in the view of control. The only reason for including the call for repainting into this method is the possibility of showing covers in the **Form_SolitaryControls.cs**; the view of the cover for the SolitaryControl class depends on the type of resizing.

There is one restriction on the size of the controls that I implemented with the SolitaryControl class: the controls cannot be reduced to less than 16 pixels. This was done to avoid the accidental disappearance of controls in applications. I do not think that users would appreciate such a disappearance; if any control has to be deleted, this must be done in a different way.

There is absolutely nothing special in the movement of the controls: press the mouse next to the border and relocate the control. There is nearly nothing special in resizing the majority of controls. For example, you will not find anything strange in resizing panels, buttons, and list views. But you may find something a bit strange in resizing, for example, the text boxes.

The problem is not in algorithm; which is used in exactly the same way with all the controls. The problem was placed at the basic level of those controls many years ago, when their developers implemented a strict link between two properties of the controls: font and size. These properties have to be absolutely independent, but somebody "too smart" at Microsoft decided to organize a link between them, and the change of a control's font often changes the size of control in addition. Specialists from different branches of engineering know very well that if you organize two independent sources of control and change for the same parameter, it is often the easiest way to organize a disaster. And the "genii" from Microsoft did it on purpose and with a sound mind! The best way to describe such a thing is to quote the former Russian prime-minister: "We wanted to do the best, but the result was as usual…" The worst results of that old decision become obvious, when you design some really complex forms with a lot of different controls inside. Yet, you have to work with what you get, and if you work with Visual Studio, you have to be aware of some of the traps.

For the majority of controls, there are no tricks with declaring minimum and maximum needed sizes, but for some of the controls you need to take into consideration the limitations, enforced by the system, in which you design your applications (Visual Studio). In all my real applications (not Demo versions), users can change all the parameters of visualization: this is one of the main features of user-driven applications (I will write about it further on). One of the things that users can change at any moment is the font for a set of controls or for any control on an individual basis. For a majority of controls, there are no conflicts between changing the font and declaring the ranges for the control's sizes, but for several types of controls there are some strange situations. Usually it happens with the controls with a single text line; some hidden parameters are deep inside their structure and are out of your control, when you use such controls in your design.

For example, take the NumericUpDown control. You can declare the minimum and maximum width of the control via the MinimumSize and MaximumSize properties, but the second number in both of them (the height) always has zero value and cannot be changed. Do not worry about this value: whenever you change the font of the control, its height is adjusted to the font, so you always see one line of text. I would not call the adjusting of the size to the size of a font a good decision; I did not ask for changing of the control's size and I do not like, when a system is doing anything by itself, whenever it was not required. But let us agree with this for a moment and decide that this is a normal and expected reaction. That is why a different behaviour from another, but very similar control, looks at least strange to me.

The standard TextBox control demonstrates this strange behaviour, but not always. Suppose that a TextBox control is registered with the Resizing.Any parameter, so it can be resized by the corners. If the Multiline property of this control is set to true, then the size can be changed in the range, declared by the two size properties MinimumSize and MaximumSize. As a developer, you set these two values in such a way that allow to change the height of this text box; after it users would decide, how many lines of text they want to see.

Suppose that you have another resizable TextBox control, in which a single line of text has to be shown, so its Multiline property is set to false. When users change the font for such a control, its height is not changed at all! It is not adjusted to the size of a font (and I think this is a correct decision), so users can change the height of such control as they do it for buttons or list boxes. For users to get a chance of changing the height of such text box, its two properties have to allow it by providing a wide enough range for the height of the control. So, in case of such `TextBox` control, the values in two properties have to be declared in different way, than in the mentioned `NumericUpDown` control. Though visually these two controls look alike: each one has a single line of text.

The mentioned things are only minor negative remarks to otherwise very important thing: all controls are easily turned into movable and resizable.

Few controls are used as stand alone movable elements. Much more often they are used in groups or in combinations with graphical objects. In the next chapter I am going to analyse the combination “control + textual comment”.

# Control + text

The combination of "control + painted text" is so widely used that I decided to take it out of the discussion of the more complex groups and organize a special chapter for this case.

## *Arbitrary positioning of comments.*

File: **Form_CommentedControls.cs**
Menu position: Individual *controls – Controls with arbitrary comments positioning*

Different controls are widely used for design of forms; a lot of these controls have some comments, which explain to the users the purpose of these controls. Different types of controls can be used with or without comments, but the probability of seeing some text at the side of the `TextBox` control is nearly 100 percent. These controls are often used for typing in some needed parameters; if there is more than one `TextBox` control in the form, each of them needs some explanation about the parameter that must be typed inside.

Textual comments next to the controls became the standard element of dialogues many years ago. There are two standard ways to organize such comments: either as `Label` controls or as painted texts. Visually these cases are undistinguishable, so the question is only in the easiness of using one way or another and your personal preferences as a programmer.

When you work in the realm of fixed design, the use of `Label` controls is easier: position all the needed controls with the help of Visual Studio and do not think about the `Paint` message.

With the design based on movable elements the situation is different. All the screen elements are movable. All the graphical objects are movable by any inner point; but the controls can be moved only by their borders. It is not the problem with the controls, which have visible borders (ordinary buttons, `ListView`, …), but the `Label` controls are often used without borders. If the text is organized as a `Label` control, then users have to remember that this object can be moved only by its invisible borders. It is not a huge problem, but it is obviously inconvenient. It would be much easier if any text can be moved by any point. For this reason I never use the `Label` controls now: all texts in my applications are painted. So, when I write about the *textual comment*, *comment*, or a *text*, it is always a painted object, but not a `Label`.

A pair "control + text" is a complex object. Moving / resizing of such objects can be described by several rules.

- The moving / resizing of the control of such pair are identical to the same operations with the solitary controls.
- On any moving / resizing of a control, the comment preserves its relative position to the control.
- The comment can be moved (and rotated) individually. There are no restrictions on positioning of the comment in relation to its partner – control except one: the comment cannot be totally covered by the control, because in such case it slips out from the user's control. To avoid this situation, whenever the comment is moved around and released while totally covered by its associated control, then it is forcedly moved slightly outside so that it becomes visible and accessible. I want to underline that this enforced relocation is used only when the comment is closed from view by its own associated control and not by any other.

**Figure 13.1** demonstrates the **Form_CommentedControls.cs** with several objects of the `CommentedControl` class. This class has a lot of different constructors, which allow to organize in different ways the initial positioning of comment in relation to the control and specify for this comment the font, color, and angle. A comment, used in the `CommentedControl` class, belongs to the `CommentToRect` class. As this class is derived from the `TextMR` class, rotation of the comment is done automatically without mentioning it anywhere in the code of a form.

A `CommentedControl` object is a complex object consisting of two parts: control and text. Control has exactly that type of cover, which was shown for the solitary controls at **figure 12.1**. And control in a pair can be declared with different types of resizing, as was explained in the previous chapter. As any complex object, this one cannot be registered with the mover by the simple `Add()` or `Insert()` methods, but instead the `CommentedControl.IntoMover()` method must be used.

The `CommentedControl` class looks simple; at the same time it has 50(!) constructors. Maybe it is a bit too much, but it gives a chance to select the one you really need in one or another case. Any constructor of this class has to describe (directly or with default values) two main things: a cover around the control and the position and view of the comment.

The cover around the control is of the same type that was demonstrated for the solitary controls in the previous chapter. There are not too many variants. You can specify the sizes of nodes and frame and you can declare the resizing, which gives not more choices than the one calculated from the sizes of a control and its two properties `MinimumSize` and `MaximumSize`. The majority of constructors are the variations on position and view of the comments.

First of all, a comment is a text, which has font, color, and angle. You can specify all these parameters or omit them in different combinations; the default values are the form's font, the form's `ForeColor`, and zero degree angle.

Positioning of comment provides a lot of different constructors. Comment belongs to the `CommentToRect` class, which is derived from the `TextMR` class, so any comment is located by its central point. Comment has two mechanisms of positioning: either by the coordinates or by coefficients that describe its relative position to the rectangular area of the dominant control. These two descriptions of comment's position exist all the time, but are used in different ways throughout different movements. If the comment is moved, then the coefficients are recalculated on the basis of the changing position; if the control is moved or resized, then the new comment's position is calculated on the basis of the unchanging coefficients. Constructors of the `CommentedControl` class can describe the position of comment in many different ways:

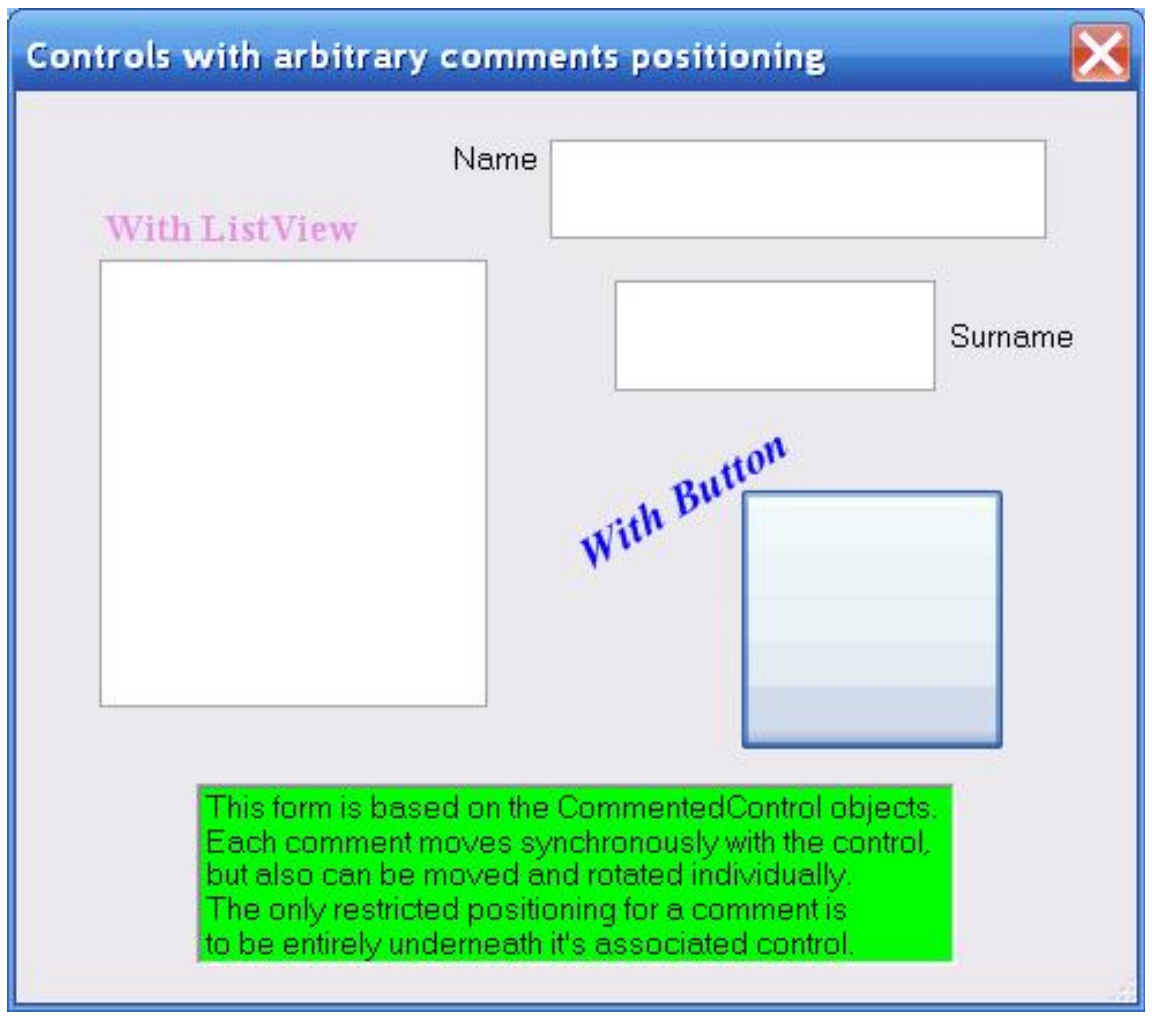

**Fig.13.1** `CommentedControl` objects

- Positioning coefficients in relation to the area of a control.
- Location point.
- Side of the control and additional space between the control and its comment.
- Side of the control and additional alignment (three variants) along this side.
- Side of the control and additional positioning coefficient along this side.

Each variant has the constructors with the full description of all the parameters and many combinations of partly omitted parameters, which are substituted by default values; the result is the total number of 50 constructors for the class. You can find the `CommentedControl` objects in many forms of this Demo application; the objects are used by themselves or as members of the group; overall there are only six different constructors in use. The **Form_CommentedControls.cs** with four objects of the `CommentedControl` class demonstrates four different constructors, but I need to mention that the view at **figure 13.1** is not the default view for all of them. Some movements were already made, so the lines below do not describe the view of these objects on the picture.

```
CommentedControl ccName = new CommentedControl (this, textboxName, Side.E, "Name");
```

An object in the top right corner uses nearly the minimum allowed set of parameters, as three of them must be present in any form of constructor. It is impossible to construct a `CommentedControl` object without specifying a `Form`, a control, and a text.

```
CommentedControl ccSurname = new CommentedControl (this, textboxSurname,
                                        Resizing .WE, Side .E, 4, "Surname");
```

The second from top object on the right side was designed around a control with identical sizes and parameters, as the first one, but its resizing is restricted to horizontal only. So the height of this control cannot be changed, while the first one was easily reduced.

```
CommentedControl ccList = new CommentedControl (this, listView1,
                              Side .N, SideAlignment .Left, 3, "With ListView",
                 new Font ("Times New Roman", 10, FontStyle .Bold), Color .Violet);
```

This example of constructor also set the font and the color of the comment.

```
CommentedControl ccBtn = new CommentedControl (this, btnResizeAny,
                              Side .E, SideAlignment .Center, 4, "With Button",
              new Font ("Times New Roman", 12, FontStyle .Bold | FontStyle .Italic),
                                                                   30, Color .Blue);
```

The comment for this big button on the right side of the form was definitely not at the place, where it is shown at the picture. Any comment can be easily moved to another location at any moment, so all those variations of constructors describe only the initial position of comment. This constructor does not add too much to the previous case; it includes an angle for initial positioning of the text, but any text can be rotated at any moment.

There are properties in the `CommentedControl` class, which allow to change the visibility parameters of both control and comment. For comment it is a color and a font; for control it is a background color and also a font. Usually the change is organized via the context menu, which is associated with `CommentedControl` objects in your program. I did not add any menus to the **Form_CommentedControls.cs**, but all the variants of tuning the `CommentedControl` objects are demonstrated in some examples further on, where a lot of such objects are used.

When you decide to change the font of a control, you use the `CommentedControl.ControlFont` property. It is possible to get the control itself through the `Control` property, so you can change the same font through `Control.Font` call, but do not do it! Though in many situations it can results in not more than not the best change of the whole object, in some situations it can cause a disaster: the comment can disappear from view without any chance to see it again. How and why? The cause is in the design of the `CommentedControl` class.

Suppose that you have a big control with a small comment, which is positioned next to the bottom of the control. Some of the controls change their size according to the size of a font; if you significantly increase the font of a control, then the size of a control also increases. If you change the font through the `ControlFont` property, the comment is informed about the new area of the control and changes its position so that it is seen in exactly the same way. If the change is done with the `Control.Font` call, then the size of a control is changed, but there is no warning for the comment to adjust its position. Control itself does not know that it is used as some part of the `CommentedControl` object, so there is no message to comment. Comment stays at the same place, but the increased control entirely closes it from view. After it there is no way to return the comment back into view. Any move of the control results in synchronous move of the associated comment. If you find the way to close the comment with its dominant control, there is no way to see this comment again. It is gone.

This is definitely not the thing that users of your programs would expect, so do not try to change the font of a control by the `Control.Font` call! The situation with the possibility of comment's disappearance is so special that I will write about it in the discussion of the rules of the user-driven applications in the second part of this book.

## *Limited positioning of comments.*

File: **Form_CommentedControlsLTP.cs**
Menu position: Individual *controls – Controls with limited comments positioning*

Objects in the **Form_CommentedControlsLTP.cs** (**figure 13.2**) look similar to the objects in the previous example. The only difference, if you pay attention to it, that none of the texts at this figure are rotated. All the texts are horizontal, but that is the normal view of nearly all the texts in the majority of applications, so this may not catch your attention at first. But you will immediately find something new at the very first moment when you try to move the objects in this form. At this moment you will find that by moving the text you move the whole pair "control + text". Objects in this form belong to the class `CommentedControlLTP`; abbreviation **LTP** stands for **L**imited **T**ext **P**ositioning. This means that a text cannot be placed arbitrary around the control, but only in some relatively predefined positions. There are exactly 12 predefined positions around any control; 3 positions on each side. These three positions include lining the text to one of the ends of the side, or putting the text in the center of a side.

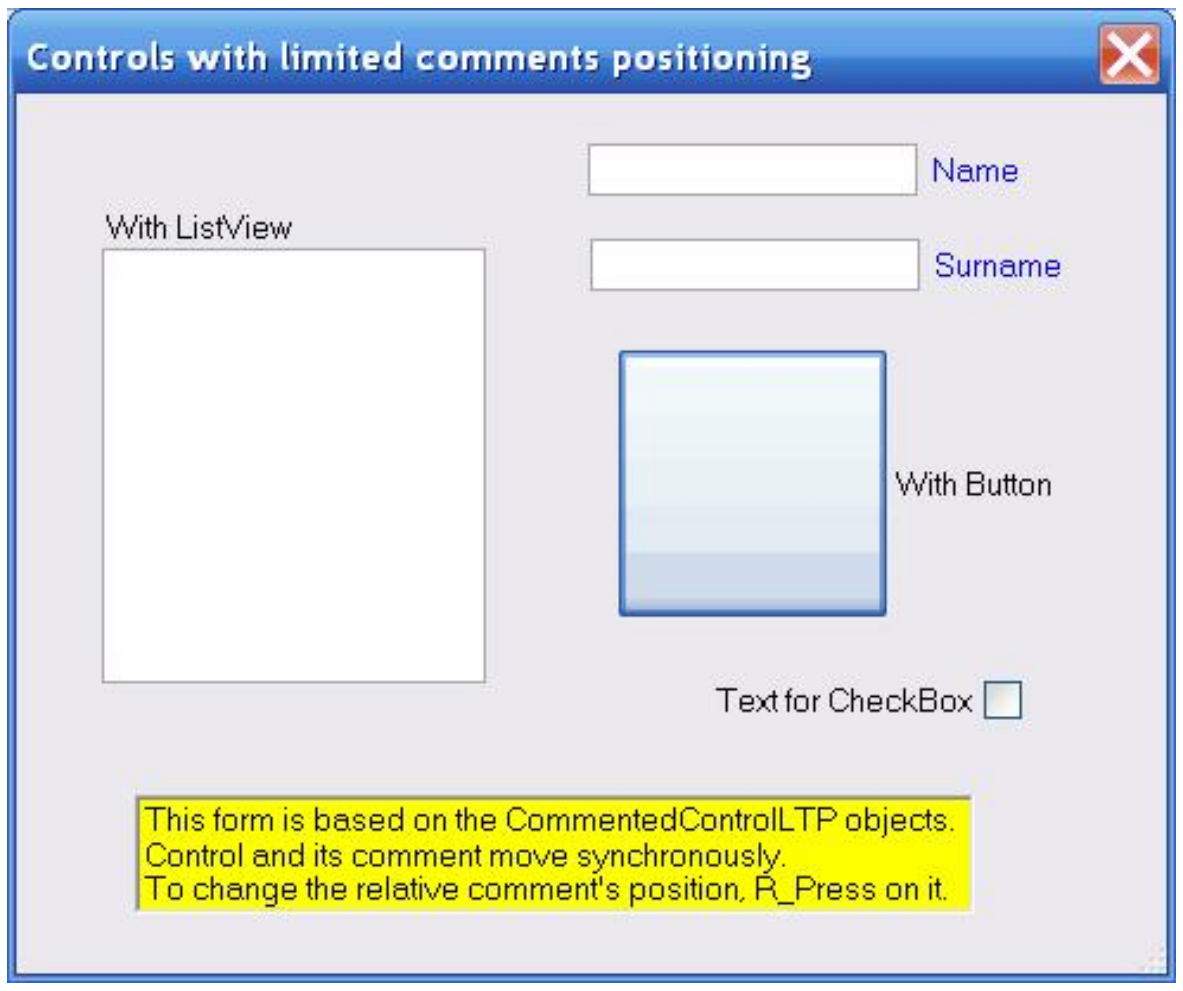

**Fig.13.2** Several `CommentedControlLTP` objects

Here are the rules for moving / resizing the `CommentedControlLTP` objects.

- The moving / resizing of the control of such pair are identical to the same operations with the solitary controls.
- On any moving / resizing of a control, the comment preserves its relative position to the control.
- Moving the comment causes the synchronous movement of the control.

The first two rules are the same as in the `CommentedControl` class; the only difference (and the important one!) is in the third rule. As the comment of such pair cannot be moved individually around the screen but starts the synchronous move of the associated control, then there must be another way of changing the position of comment in relation to the control. To do this, you click the text to be moved with the right button and select in the opened form* the new position for this text (**figure 13.3**). There are 12 rectangles in this form, which mark the possible positions of the comment in relation to the control; click the one you prefer. With a set of available positions being only around the control, the text of the `CommentedControlLTP` object is never blocked from view by its own control, so there is no need for enforced relocation in this class.

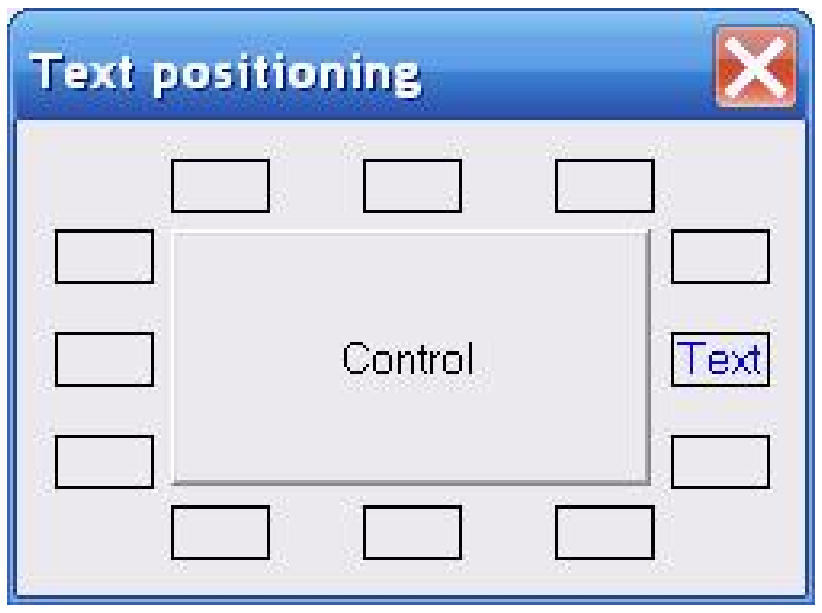

**Fig.13.3** Click any rectangle to choose the text's position in relation to the control

Obviously, the previously described `CommentedControl` class gives greater flexibility in positioning the text than the `CommentedControlLTP` class. Then why was the last class introduced? The answer is at **figure 13.2** in the case of a check box. This is not a standard `CheckBox`, but a combination of such an element without text at all plus the needed text, which is painted. Why do I prefer such a combination?

If you use the standard `CheckBox` object, it can be moved around only by the border of this control. The border is invisible, and though the changing of a cursor at the side of the text gives a clear signal that this is the right place to press the mouse, grab the check box, and move it, but it is still not the best decision. If you have a chance to move the same element by any point of its text, it is much easier and, from my point, much better.

Some people like to see the check mark on the left of the associated text, others – on the right; the proposed technique of changing the position of text to one or another of the available spots makes such change really easy. This form allows even to try the text being placed above or below the check mark. It looks strange, funny or… Maybe some people would like it, I am not sure.

The elements of the two classes – `ComemntedControl` and `CommentedControlLTP` – look similar, but they have the different rules for moving, so do not mix them in the same form; use either one or another. Which one better suits the goal of form's design depends on the type of controls that are going to be used. I never use these two classes together except for one form; I would prefer to have no exceptions, but I cannot find better solution for this case. I will write about it further on in the chapter *Applications for science and engineering*.

The `CommentedControlLTP` class is not the complex class, as I described it: there are no independently movable parts in these objects. Because of this, the objects of this class are registered with the mover by the simple methods `Add()` or `Insert()`.

```
mover .Add (new CommentedControlLTP (this, checkBox1, Side .W,
                                     "Text for CheckBox"));
mover .Add (new CommentedControlLTP (this, listView1, Side .N,
                                     SideAlignment .Left, "With ListView"));
mover .Add (new CommentedControlLTP (this, btnResizeAny, "With Button"));
```

* In the *Preface* of the book I have declared that ALL the objects in all the numerous examples are movable. Here is the whole **Form_TextLimitedPositioning.cs** in which nothing can be moved. There is no mover in this form! Did I break my promise? Well, first of all, this is not an example. You can say that this is a sophism. I wouldn't argue; I think that every form in this Demo application must be looked at as an example. But this is a special example with a single object inside. There are not 13 objects in this form but only one! It is like a human body. We have the main body plus hands, legs, and a head. If we have to draw a person, are you going to expect a mechanism of drawing extremities at any place or only where they supposed to be? In this form we have a body with 12 extremities and the places for all of them are predetermined. (The Creation is over; claims are not accepted.) If we have a form with a single object, it is possible to leave it unmovable. If you want, you can easily turn this single object into movable. Make a cover consisting of a single rectangular node and covering all the parts. The whole object will become movable, but the relative positions of its parts must stay unchanged.

# Groups of elements

Individual controls with or without text, which were discussed in the previous chapters, can be used for design of different forms, but more often the design is based on bigger blocks, which consist of several controls or combinations of controls and graphical objects. These blocks can vary from a simple pair of two elements to very complicated hierarchical groups of elements. The wide variety of elements from the simplest blocks to different groups is discussed in this chapter.

All the currently used applications are designed on the fixed elements. These can be the individual controls or the groups of controls united either by a `GroupBox` or a `Panel`. The set of controls is known for many years and includes around 25 – 30 different types. Of them not more than 10 are used very often; others you can see only occasionally in different applications. Only one or two new controls were introduced during the last five or six years; all other controls were developed 20 or more years ago. They were put into life together with the rules of their design and use; the good solid books were published about the rules of using those controls in applications; you would rarely see now an application, which risks to ignore those rules. This situation can be looked at as a big advantage, as users of any application know how to use any familiar element. On the other side, the system of strict rules in design prevents any progress and became the cause of stagnation in design of big complex applications.

The design of applications on the basis of movable / resizable objects blows up that sluggish system and opens an infinitive area of possibilities. The discussion of these new possibilities will be the subject of the second part of this book, but before starting that discussion I want to show the variants of new elements that can be used in new design. Some of the elements, which are demonstrated in this chapter, are widely used in design of user-driven applications. Others were designed somewhere on the way, but later replaced with the better versions. But *better* means only my point of view at the current moment. From time to time I return back to the ideas, which I used two or three years ago, but abandoned later. Those older ideas are not the dead ends in design; they can be used for development of new elements in the future, so I would like to mention them here in parallel with those elements that are at high demand now.

Before turning to absolutely new ideas, I want to demonstrate the application of movability to some of the well known elements; you will find them in the first two examples of this chapter.

## *Standard panel*

File: **Form_StandardPanel.cs**
Menu position: *Groups – Standard panels*

If you use the standard fixed design there are two ways to organize a set of controls into a group. The first of these two choices is to put the needed controls on a panel. The border of a panel, if shown, makes it obvious, which of the controls are included into the group.

**Form_StandardPanel.cs** demonstrates three `Panel` objects turned into movable; two of them are also resizable (**figure 14.1**). As any other control, a panel can be moved / resized only by its border. There is absolutely nothing special about the involvement of panels in the moving and resizing; the whole procedure is organized in the standard way for all other controls.

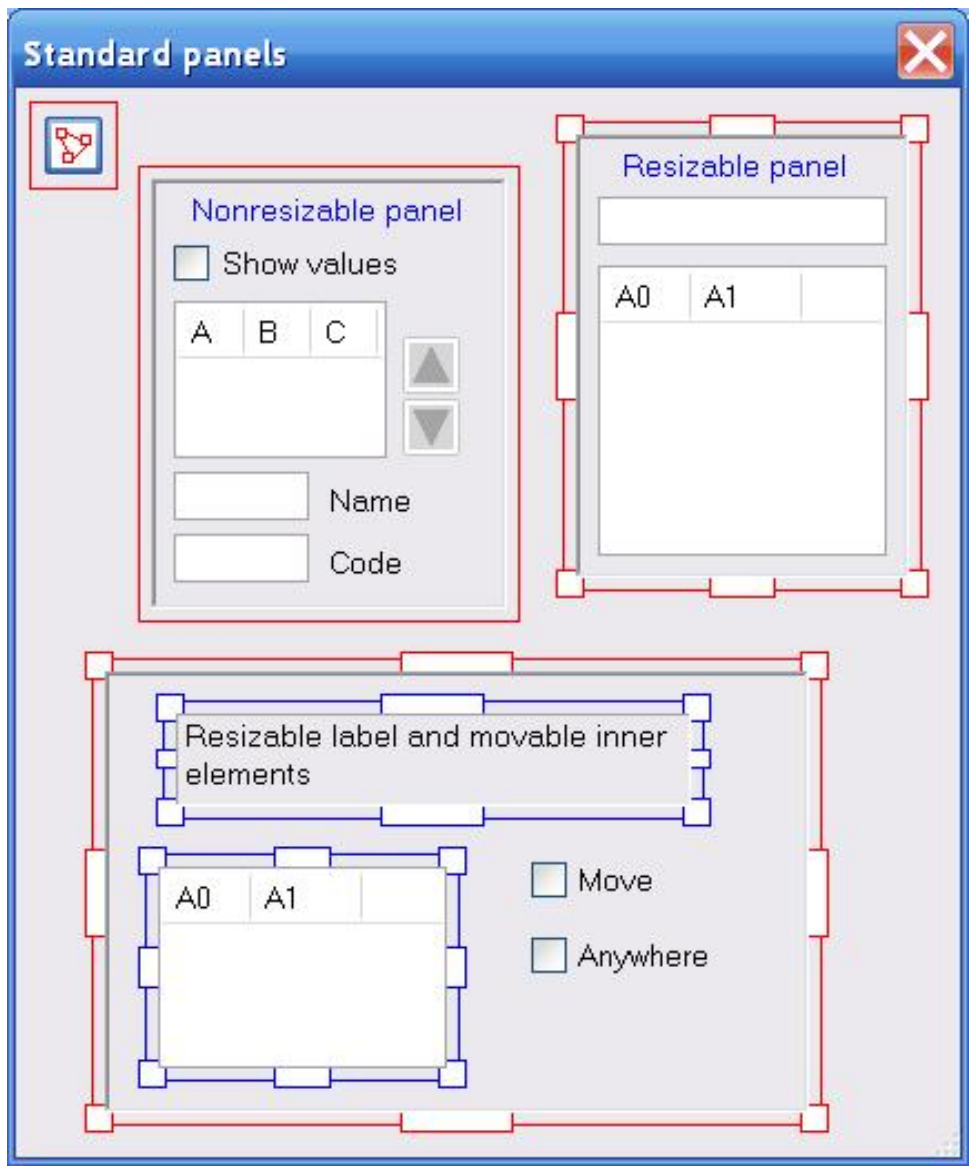

**Fig.14.1** Panels are moved and resized as solitary controls

Declare a mover and initiate it with the form, in which it works.

```
mover = new Mover (this);
```

Panel is a control, so it is registered as any other control.

```
mover .Add (panelResizeMoveInside);
mover .Add (panelResizable);
mover .Add (panelNonresizable);
```

Use the three standard mouse events and the simplest way of calling the needed mover's methods; they provide whatever is needed from the mover.

```
private void OnMouseDown (object sender, MouseEventArgs e)
{
    mover .Catch (e .Location, e .Button, true);
}
```

```
        private void OnMouseUp (object sender, MouseEventArgs e)
        {
            mover .Release ();
        }
        private void OnMouseMove (object sender, MouseEventArgs e)
        {
            if (mover .Move (e .Location))
            {
                if (mover .CaughtSource is SolitaryControl)
                {
                    Update ();
                }
                Invalidate ();
            }
        }
```

All three panels are movable; two of them are resizable, because the `MinimumSize` and `MaximumSize` properties of these panels get the appropriate values. With the non-resizable panel everything is simple: it can be moved around by the frame, but its inner elements never change their sizes or relative positions. Standard and expectable behaviour of the well known element. The inner elements on other two panels behave differently: on one of them the inner elements are fixed, where they were placed during the design in Visual Studio; elements on another panel have their own covers (**figure 14.1**), so they can be also moved and resized.

When you organize a resizable panel, you can make a decision about the possibility of moving / resizing for the inner elements of this group. The first option is to use the ideas of dynamic layout: use anchoring to regulate the sizes of the elements on a panel, when this panel is resized. This would be an automatic decision for everyone, who is currently using anchoring and docking in design of their applications. This is demonstrated with the panel in the top right corner of **figure 14.1**, but such decision is definitely against the rules of the user-driven applications and I would not recommend such a thing. There is much better way: use another mover to organize the moving / resizing for these elements on the panel. (The questions of using several movers in the same form are discussed in the *Appendix B*.)

Declare another mover, which works on this panel.

```
        moverInner = new Mover (panelResizeMoveInside);
```

Register with this mover those elements that are positioned on this panel.

```
    moverInner .Add (labelTitle);
    moverInner .Add (listView3);
    ccMove = new CommentedControlLTP (this, checkboxMove, Side .E, "Move");
    ccAnywhere = new CommentedControlLTP (this, checkboxAnywhere, Side .E,
                                                                     "Anywhere");
    moverInner .Add (ccMove);
    moverInner .Add (ccAnywhere);
```

To show that some of the objects are supervised by another mover, I changed the color of the covers for this mover. The questions of visualization the covers are explained in *Appendix A*.

```
    moverInner .Color = Color .Blue;
```

The same three mouse events, but applied to the panel, are used to organize the moving / resizing of the elements on this panel.

```
      private void MouseDown_panel (object sender, MouseEventArgs e)
      {
          moverInner .Catch (e .Location, e .Button, true);
      }
      private void MouseUp_panel (object sender, MouseEventArgs e)
      {
          moverInner .Release ();
      }
      private void MouseMove_panel (object sender, MouseEventArgs e)
      {
          if (moverInner .Move (e .Location))
          {
```

```
                (sender as Panel) .Invalidate ();
                if (moverInner .CaughtSource is SolitaryControl)
                {
                    (sender as Panel) .Update ();
                }
            }
        }
```

Any mover deals only with the objects, which are included into its queue, so two movers are working with the different groups of elements: one mover regulates the moving / resizing of three panels in the form; another mover supervises the moving / resizing of the elements on one of the panels.

One thing is important to understand in the case of panel: regardless of the number of elements on the panel and their behaviour, the panel itself is always looked at as an individual control. It looks like a group of controls and there can be a lot of elements on this panel, but for mover it is a solitary control, which is registered in the simplest way and which behaviour is quite simple.

For users it is also a very simple object, which can be moved and resized, if it is allowed, only by the parts of its border. Panels, as any other controls, cannot be moved by their inner points; this is the biggest flaw in otherwise simple design.

## *Standard GroupBox*

File: **Form_StandardGroupBox.cs**
Menu position: *Groups – Standard GroupBox*

`GroupBox` objects are as easy to include into moving / resizing process as already mentioned panels. It also has the same negative feature as a `Panel`: you cannot move a `GroupBox` by its inner points. But in addition the standard `GroupBox`, when turned into movable, has one more negative feature of its own. **Figure 14.2** demonstrates the view of the **Form_StandardGroupBox.cs**, in which two `GroupBox` objects are organized: one is non-resizable, another is resizable. The picture shows the covers of these two groups and makes the explanation of the problem easier. `GroupBox` is a standard control and has the same cover, as any other control. A `GroupBox` has its own frame, painted inside its area. This frame is supposed to mark the boundaries of the group. The main idea of this frame is to make it obvious, which controls are included into the group; for this purpose the frame works perfectly. It is a very rare situation that the background color of the `GroupBox` differs from the background color of the form, in which it is used. When these two colors are the same (and this happens nearly always), you cannot see the real border of the `GroupBox`. But in the standard design, based on fixed elements, you are not interested in the real borders of a `GroupBox`; what you are interested in is the set of controls that are included into this group; the frame makes it obvious, so the difference between the frame and the real border makes no problems under such design.

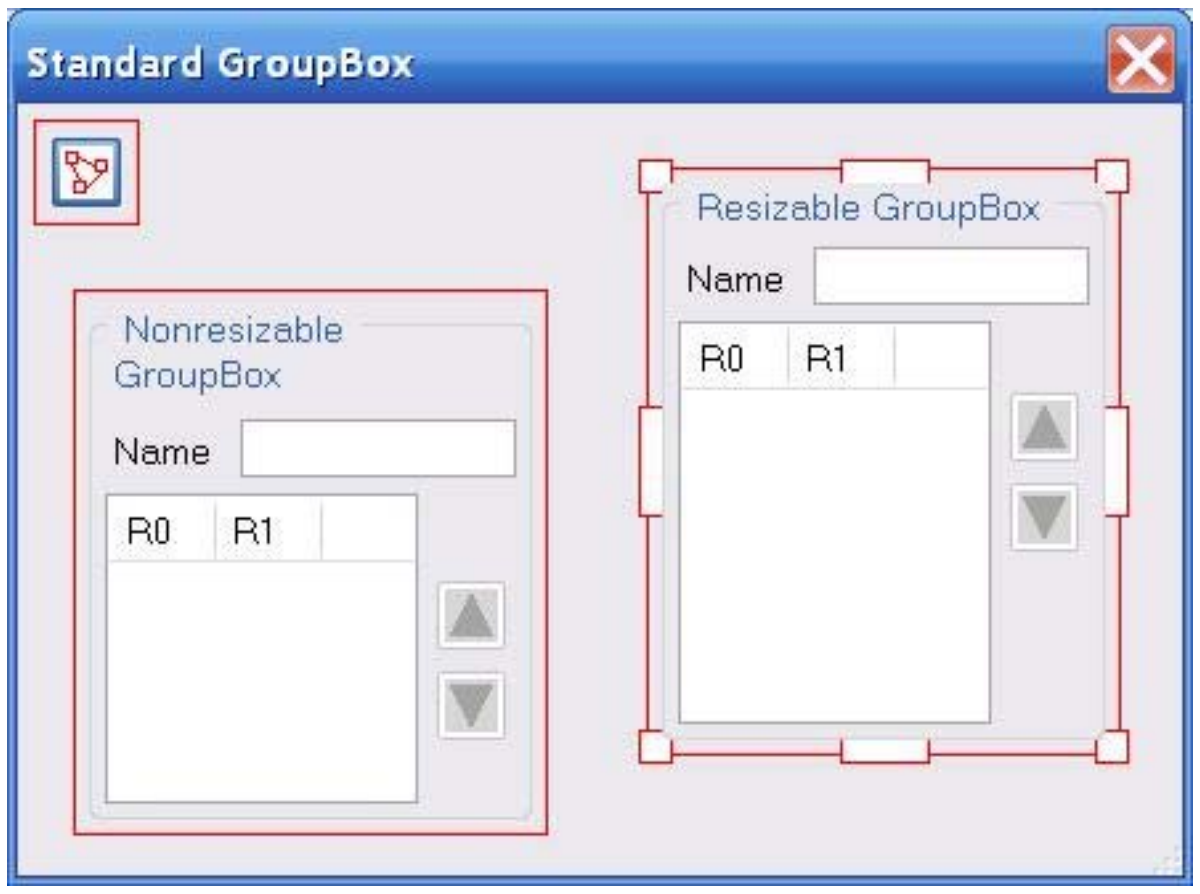

**Fig.14.2** Standard `GroupBox` objects are moved and resized as solitary controls

When you turn the `GroupBox` object into movable / resizable, it gets the cover next to its real borders. Usually the covers are not shown at all, so users try to grab a `GroupBox` for moving or resizing somewhere next to the visible frame. On three sides the cover is close to the visible frame, so on those sides anyone can grab this group "by the frame", but at the top the cover is far away from the frame (**figure 14.2**). If you try to grab the group with the invisible cover by its frame at the top, it would be impossible. You will find that you can grab such group, if the cursor is moved somewhere up, where there is no indication of such possibility (except for the changing of mouse cursor). This inability to grab the movable object at the place, which you think is absolutely right for it, becomes very annoying; especially, if there are several such objects in the form. That was the reason, why I quickly stopped using standard `GroupBox` objects among movable elements and started to design other types of groups.

## Non-resizable group

File: **Form_LinkedRectangles.cs**
Menu position: *Groups – Non-resizable groups*

From time to time it happens that you need to use a non-resizable group. Not too often but occasionally there is such a requirement. You see it rarely enough (or ever at all) in the really complex forms, but in the relatively simple previous examples I have used such elements several times. A non-resizable group of elements is represented in the **MoveGraphLibrary.dll** by the `LinkedRectangles` class. In several of the previous examples this class was used as an auxiliary element for better demonstration, when I needed a quick change of one or two parameters.

- The case of standard moving / resizing of rectangles (**Form_Rectangles_StandardCase.cs**, **figure 3.1**).
- The case of disappearing rectangles (**Form_Rectangles_StandardToDisappear.cs**).
- The discussion of movable texts (**Form_TextsIndividual.cs**, **figure 5.1**).
- The example of movements under restrictions (**Form_BallsInRectangles.cs**, **figure 11.4**).
- The case of solitary controls (**Form_SolitaryControls.cs**, **figure 12.1**).

In all these cases the `LinkedRectangles` class was used but never analysed; now it is the time to look into the details of its design.

Suppose that you have a set of elements, which need a synchronous move and nothing else. No resizing of elements and no change of relative positions. But at the same time their initial relative positions can be arbitrary: elements can stay next to each other, or they can overlap, or they can be placed far away from each other. Their positions do not matter at all; whenever any element of the set is moved, all others have to move synchronously. In a lot of situations the objects that require such a synchronous move are controls; often enough they are accompanied by some comments. The design of the `LinkedRectangles` class was caused by a request for synchronous move of several controls; later the graphical objects were added to the elements of this class, but it did not change the class at all.

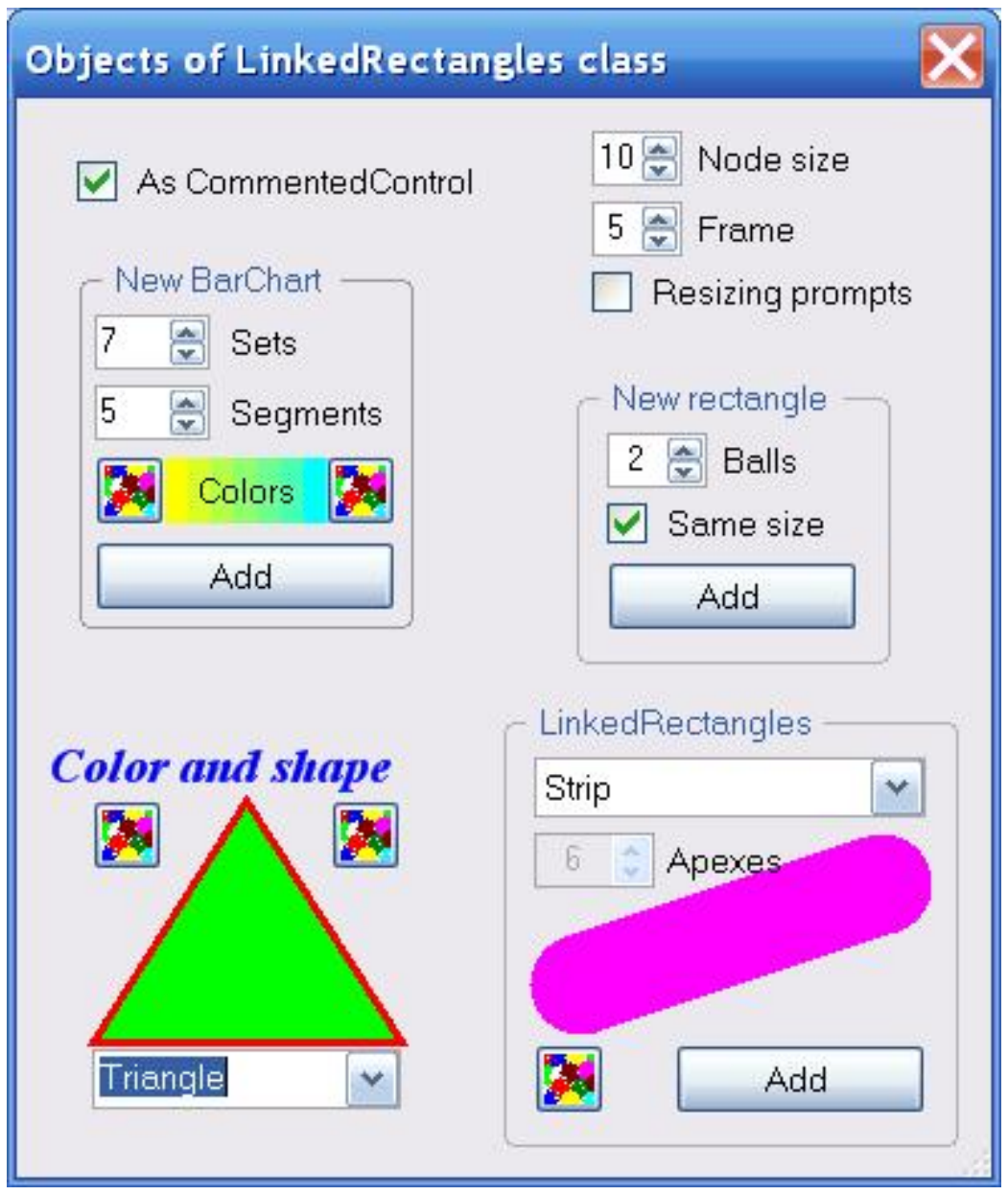

**Fig.14.3** A variety of `LinkedRectangles` objects

All the standard controls are the rectangular objects. If they have to be only movable, then each of them receives a rectangular cover, which is slightly wider than the control itself; such covers were demonstrated for solitary controls in the **Form_SolitaryControls.cs** (chapter *Individual controls*, **figure 12.1**). The group for synchronous movement has to unify the individual covers of its elements; that is where the name of the `LinkedRectangles` class came from. **Figure 14.3** shows the view of the **Form_LinkedRectangles.cs** with six different objects of this class. As you see, some of them include the graphical objects in addition to the controls. A frame can be also added to a `LinkedRectangles` object, but a frame is not a mandatory thing for such an element.

The `LinkedRectangles` class has several constructors. You can start with a single control, a single rectangle, or an array of controls or rectangles. After it you can add any number of controls or rectangles to this object, so there are a lot of possibilities. If a control is included into the `LinkedRectangles` object, it means an addition of the rectangular area around this control. So any `LinkedRectangles` object contains the lists of included controls and special rectangles. The only special parameter of these rectangles is the tag, by which they are identified and can be easily found.

```
public class LinkedRectangles : GraphicalObject
{
   List<Control> controls = new List<Control> ();
   List<TaggedRect> rects = new List<TaggedRect> ();
```

The cover design for such an object is easy enough. Each rectangle is covered by a single rectangular node; the cover consists of a set of these nodes.

```
        public override void DefineCover ()
        {
            CoverNode [] nodes = new CoverNode [rects .Count];
            for (int i = 0; i < nodes .Length; i++)
            {
                nodes [i] = new CoverNode (i, rects [i] .Rectangle);
            }
            cover = new Cover (nodes);
        }
```

Moving of the whole object must include the synchronous movement of all the rectangles and controls.

```
        public override void Move (int dx, int dy)
        {
            for (int i = 0; i < controls .Count; i++)
            {
                controls [i] .Left += dx;
                controls [i] .Top += dy;
            }
            for (int i = 0; i < rects .Count; i++)
            {
                rects [i] .Move (dx, dy);
            }
        }
```

The `MoveNode()` method for this class is extremely simple. It does not matter at all by which node of the cover the movement is started; in any case the `MoveNode()` method calls the `Move()` method.

```
public override bool MoveNode (int i, int dx, int dy, Point ptM, MouseButtons btn)
{
    if (btn == MouseButtons .Left)
    {
        Move (dx, dy);
    }
    return (false);
}
```

Though a `LinkedRectangles` object may include a lot of controls and rectangles and may look like a really complex object, from the point of mover it is not, because there is no independent move of the parts. Regardless of the number of inner elements, the cover of any object consists of a set of rectangular nodes. This set of nodes is fixed at the moment, when the `LinkedRectangles` object is registered in the mover's queue. For mover, any `LinkedRectangles` object is simple. The cover consists of a set of polygonal (rectangular) nodes, which never change, so this cover is registered by the `Mover.Add()` or `Mover.Insert()` methods.

Any control included into a `LinkedRectangles` object is not resizable. It is surrounded by a simple rectangular node which is slightly wider than the area of a control; the default width of the sensitive frame around is six pixels; the minimum width of the frame is two pixels. To make the moving of any `LinkedRectangles` object easier, it has to be moved by any inner point. To achieve this purpose, two things are often used:

- All texts inside are painted. No `Label` controls are used, if a painted text can be used instead. The textual part of such controls like `CheckBox` are deleted and substituted by a painted text.
- An additional rectangle can be added to close the whole area and avoid any insensitive gaps inside.

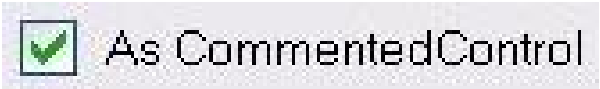

**Fig.14.4** One control and one rectangle

The simplest version of a `LinkedRectangles` object may consist of a single rectangle or a solitary control. This object (**figure 14.4**) looks like an ordinary `CheckBox` control, but it is not. According to the mentioned rule, the text of the `CheckBox` object was deleted, thus squeezing it to a tiny square box, on which a `LinkedRectangles` object was constructed. Then a rectangle was added to this object; the size of rectangle is determined by the text, which is painted there.

```
lrsAsCommentedControl = new LinkedRectangles (new Control [] { checkboxLabel });
lrsAsCommentedControl .Add (Auxi_Geometry .RectangleToRectangleSide (
                                  checkboxLabel .Bounds, Side .E,
```

```
                         Auxi_Geometry .MeasureString (this, strComment), 4), "Text");
```

The most often used case of a `LinkedRectangles` object is a set of controls, used to declare several related parameters (**figure 14.5**). Controls might be of the same class and size or they can be different. Controls may need some textual explanation, which, as a rule, need different areas for painting. Controls and comments can be placed in many different ways. If controls are lined along one side, then the opposite side of rectangles may have different coordinates. You can leave it as it is, and in this case the whole object can be moved by any point of any comment. You can add another rectangle, which is determined by the outside points of all the elements; in this case the sensitive area has the form of this ambient rectangle. For the case of three controls with comments shown at **figure 14.5**, the `LinkedRectangles` object is initially constructed of three controls, then three rectangles for comments are added, then one more rectangle is added to prevent any gaps inside. There is no need to calculate the big rectangle manually; the `LinkedRectangles` class can do it by using its own `AddUnionRectangle()` method.

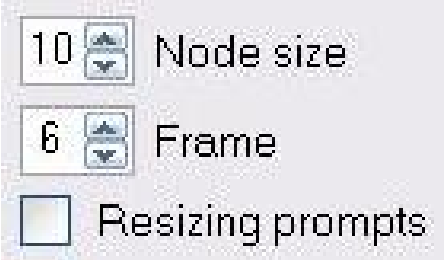

**Fig.14.5** Several controls with additional comments

```
lrsView = new LinkedRectangles (new Control [] { numericUD_NodeSize,
                                           numericUD_Frame, checkShowPrompts });
lrsView .Add (Auxi_Geometry .RectangleToRectangleSide (numericUD_NodeSize .Bounds,
                                           Side .E, sizefStrs [0], 4), "Node");
lrsView .Add (Auxi_Geometry .RectangleToRectangleSide (numericUD_Frame .Bounds,
                                           Side .E, sizefStrs [1], 4), "Frame");
lrsView .Add (Auxi_Geometry .RectangleToRectangleSide (checkShowPrompts .Bounds,
                                           Side .E, sizefStrs [2], 4), "Prompts");
lrsView .AddUnionRectangle ();
```

Some of the `LinkedRectangles` objects in the **Form_LinkedRectangles.cs** are shown with the frames. Frame in this class is not an option that can be added or not; frame is a part of the painting. As frame is painted inside an object and an object can be moved by any inner point (it is better to organize it in such a way), then there are no problems in moving such a group by any point of a frame.

In the case of a group from **figure 14.6,** it is enough to construct the `LinkedRectangles` object on a set of controls and add one rectangle slightly outside their borders; the frame is painted inside the rectangle.

```
    Control [] cntrls = new Control [] { comboUnicoloured,
                             numericUD_Vertices, btnUniColor, btnAddUnicoloured };
    lrsUnicoloured = new LinkedRectangles (cntrls);
    Rectangle rcFrame = Auxi_Geometry .FrameAroundControls (cntrls, spaces);
    rcFrame .Inflate (3, 3);
    lrsUnicoloured .Add (rcFrame, "Area");
```

Depending on the amount of needed painting in the `LinkedRectangles` object, it can be organized directly in the `OnPaint()` method of the form or a special method can be prepared and linked with the object.

The `LinkedRectangles` objects work perfectly, if you need a movable but non-resizable group of elements. The resizable groups are needed much more often, so I spent more time on developing such groups. At the beginning, they were still using the ideas of dynamic layout; next section demonstrates two classes of such groups.

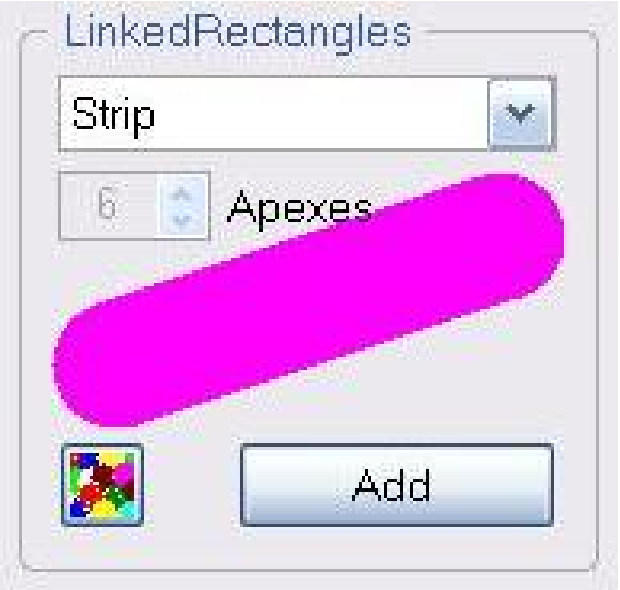

**Fig.14.6** Group with a frame

## *Resizable groups with dynamic layout*

File: **Form_GroupsWithDynamicLayout.cs**
Menu position: *Groups –Groups with dynamic layout*

Dynamic layout became very popular some years ago and there are a lot of programmers, who like to use the ideas of dynamic layout in design of applications. The **Form_GroupsWithDynamicLayout.cs** (**figure 14.7**) demonstrates two different classes of groups, which use these ideas for positioning of the inner elements, when the groups are resized. I tried to develop the groups, which:

- Can be moved by any inner point.
- Can be resized by the border.

- Provide different possibilities for resizing, but at the same time give some prompt on the type of resizing implemented for one or another group.

The first of these requirements was the most important, as the inability to move the standard `GroupBox` object by any inner point was the biggest flaw in using that standard class.

The first class, designed to fulfil the mentioned requirements, was the `GroupBoxMR` class; its representative is shown on the left at **figure 14.7**. It looks like an ordinary group, but has some additional curves in its frame. The curves indicate those places of the frame, by which the group can be resized. An object can be moved by all other stretches of the frame and by any inner point. The objects of this class can be used with four different variations of resizing, so there was a problem of visually distinguishing these four cases. To solve it, the curves of the frame depend on the type of resizing. First, the curves are shown in the middle of only those sides which can be moved (thus resizing the group along this axis). Second, the view of the curves in the corners depends on the type of resizing (Any, NS or WE). A group can be non-resizable, in which case there are no additional curves in the frame. If you switch ON the visualization of covers in the **Form_GroupsWithDynamicLayout.cs**, you will find out that this group has he same type of cover as an ordinary `GroupBox`. Plus this group can be moved by any inner point and that was the main idea of designing the `GroupBoxMR` class.

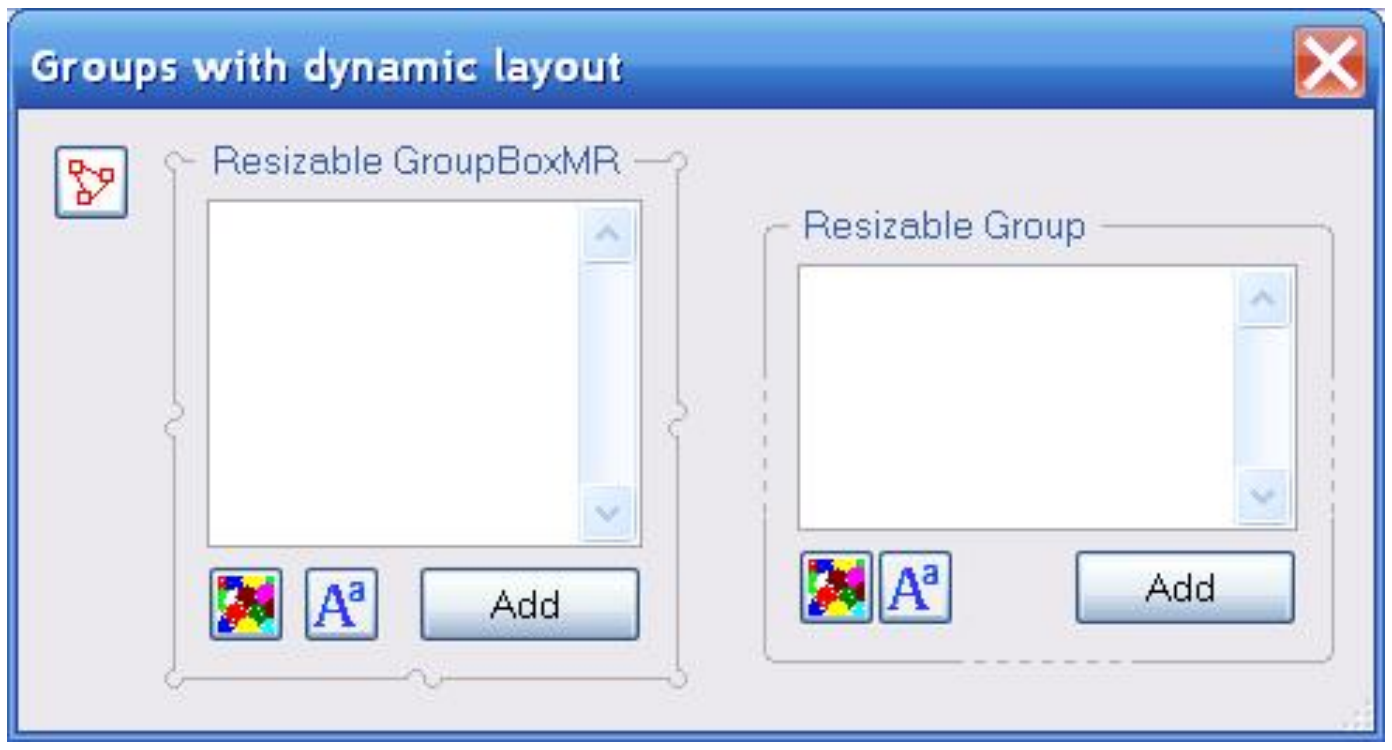

**Fig.14.7** Groups with dynamic layout

The type of resizing for any `GroupBoxMR` object is not declared as a parameter on initialization, but is determined by analysing the ranges for the group's size which are declared on initialization. Several parameters of the `GroupBoxMR` constructor determine some very important features of an object.

```
public GroupBoxMR (Form formSrc,
         Rectangle rcFrame,
         RectRange ranges,                     determines the range of the resizing
         bool bMoveByInner,                    determines if an object has to be moved by inner points
         Pen pen,
         bool bSpecial,                        determines if the curves of the frame have to be shown
         string str,
         StringAlignment align,
         int space,
         Font fnt,
         Color clrStr,
         Delegate_NoParams onMoveResize)       determines the moving / resizing of the inner elements
```

The `RectRange` object allows to set different variants of resizing, so the group can be resizable only horizontally, or only vertically, or both. Certainly, if you declare the range equal to the current size of the frame or omit this parameter, then such group is not resizable.

The dynamic layout for the elements of a group is organized not by setting the anchoring property for each control inside the group, but by writing your own method, which is passed as a parameter to the constructor. And anyway, nobody declared that the inner elements must be controls! They can be graphical objects of any kind, or there can be a mix of controls and graphical objects.

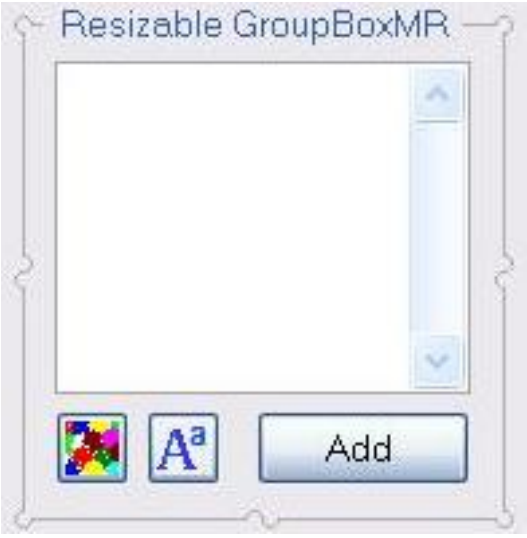

**Fig.14.8** `GroupBoxMR`

In the code of several examples you will see the use of the `spaces` variable; this is an object of the `Spaces` class, which is included into the **MoveGraphLibrary.dll**. Class provides some standard distances between the elements on the screen, which I consider to be the good ones for design of different applications. The majority of the values are set in pixels; several depend on the currently used font. From time to time I change some of these values, but at any moment they provide me with the same values for all examples in the applications, so all the forms would use the same default distances between the frames, the distances between the frames and inner elements, etc.

Here is some code to show how the resizing works with the `GroupBoxMR` object in the **Form_GroupsWithDynamicLayout.cs**. There are four controls in the group (**figure 14.8**). I made the decision about the minimum size of the group and also decided that the width of this group can be increased not more than three times and the height – not more than four times from the minimum sizes.

```
int minW = spaces .Left_inFrame + btnCommentClr .Width + btnCommentFont .Width +
                      2 * spaces .HorMin +  btnAdd .Width + spaces .Right_inFrame;
int minH = spaces .Top_inFrame + btnAdd .Height * 2 + spaces .VerMin +
                                                      spaces .Btm_inFrame;
Rectangle rcFrame = Auxi_Geometry .FrameAroundControls (new Control [] {
                 textNewComment, btnCommentClr, btnCommentFont, btnAdd }, spaces);
groupboxMR = new GroupBoxMR (this, rcFrame, new RectRange (minW, 3 * minW, minH,
                            4 * minH), "Resizable GroupBoxMR", MoveGroupMR);
```

Reorganizing of the inner elements on resizing the group is done in the `MoveGroupMR()` method which is passed as a parameter. In this case this method is based on simple ideas. Everything is calculated from the rectangle of the frame. Three buttons are always at the bottom; two of them are on the left with the fixed space between them; the third button is on the right. The `TextBox` control occupies the remaining part of the inner area.

```
void MoveGroupMR ()
{
    Rectangle rc = groupboxMR .RectAround;
    btnCommentClr .Location = new Point (rc .Left + spaces .Left_inFrame,
                      rc .Bottom - (spaces .Btm_inFrame + btnCommentClr .Height));
    btnCommentFont .Location = new Point (btnCommentClr .Right + spaces .HorMin,
                                          btnCommentClr .Top);
    btnAdd .Location = new Point (rc .Right - (spaces .Right_inFrame +
                                              btnAdd .Width), btnCommentClr .Top);
    textNewComment .Bounds = new Rectangle (btnCommentClr .Left, rc .Top +
      spaces.Top_inFrame, rc.Width - (spaces.Left_inFrame + spaces.Right_inFrame),
           btnCommentClr .Top - rc .Top - (spaces .Top_inFrame + spaces .VerMin));
}
```

After I have used the `GroupBoxMR` class for some time, I understood that there was no reason to use the border (frame) of such a group both for moving and resizing. The group can be moved by any inner point, so there is no need to do the same by some parts of the frame; the whole frame can be used for resizing only, and that would be according to the idea of moving and resizing for all the objects: move by inner points, resize by border points. The only question of such design was in visualization: I needed to show in some way the difference between the variants of resizing. Then I came to the idea of the dashed parts of the frame: the lines on those sides, which can stretch and squeeze, have the dashed parts in the middle. The only exception is the upper line: if a group has a title, then the dashed segment is not included into the upper line of a frame. Thus designed class is called `Group` (**figure 14.9**).

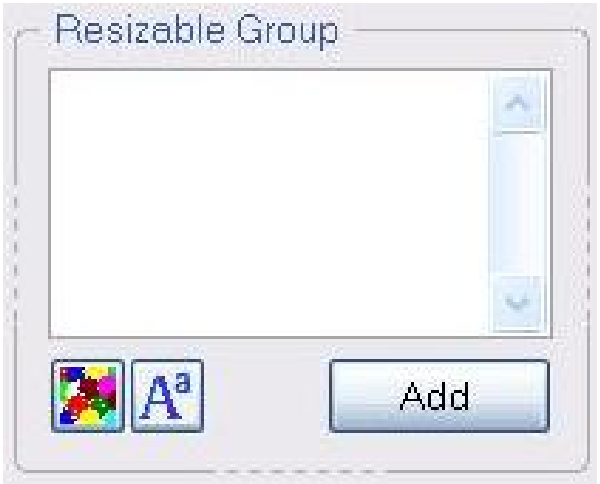

**Fig.14.9** `Group` object

I purposely included the similar `GroupBoxMR` and `Group` classes into the same form and filled these groups with the identical elements; this makes more obvious the similarities and discrepancies between two classes. Both groups can be moved by any inner point and resized by the borders. In both cases the resizing is done "by the frame", so there are no problems with it. The `GroupBoxMR` object can be resized by some special places of the frame; the `Group` object – by any point of the frame and this is definitely better.

In the **Form_GroupsWithDynamicLayout.cs** both classes - `GroupBoxMR` and `Group` - are represented by the objects with a few elements inside, but these classes can be used in much more complex cases. Slightly more than a year ago the most complex tuning forms for different plotting classes from the **MoveGraphLibrary.dll** library were designed on the `Group` class; I had no problems in using especially this class in the most complex cases. Though the group itself can include a lot of inner elements, but for the mover it is a simple object, because there are no individual movements of the inner elements. All the movements are described by the method, passed as a parameter on initialization, so registering of the groups of both classes is simple.

```
mover .Add (groupboxMR);
mover .Add (group);
```

After the `Group` class was designed, it became for some time the main class to develop the groups in all my applications. The `Group` class is still used in some of the tuning forms from the mentioned library, but the time of this class is over. Those several objects exist only until the moment, when I find the time to work on those several forms from the library. An amazing turn in using the class, which only a year ago looked like the best class for the resizable groups. The reason? Dynamic layout is not used in my applications any more.

The second part of the book is about the user-driven applications. In this type of applications, there is no place for dynamic layout. Whether you like it or not, but the dynamic layout means the work of application according to the designer's view on what is good and what is bad. But I think that only USERS have to decide what and how must be shown at the screen.

When a `Group` object is resized, its inner area is rearranged according to the developer's view and taste. All the changes are predetermined in that method, which was written by the developer and passed as a parameter on the initialization of the group. Users have no chances to make their own changes. If I decided that two small buttons had to stay in the bottom left corner of the group next to each other and another button had to stay in the bottom right corner (**figure 14.9**), this decision is final and cannot be changed by users. They can move and resize the group, but they cannot overrule my decision about the inner view of the group. In the user-driven applications this is absolutely wrong! Users must have the full control of the whole view and over any detail. The groups for user-driven applications must be based on different ideas. When I thought out, how to design such groups, it was the last step in design the real user-driven applications. Next several sections are about different kind of groups based on different ideas.

# Group with dominant control

File: **Form_DominantControls.cs**
Menu position: *Groups – Group with dominant control*

In the previous section I have shown the `Group` class, which satisfies two requirements for a good group design:

- Group can be moved by any inner point.
- Group can be resized by any border point.

Unfortunately, there is one feature of that class, which makes it inappropriate for user-driven applications: rearranging of a group as a result of its resizing is decided not by the users, but by the developer of a program. This cannot be allowed in the user-driven applications, so there must be some other decision.

The main goal of any group is not in resizing the frame, which somehow changes the inner view; the frame exists only as an auxiliary element of design to make the content of the group more obvious. The goal of rearranging a group is to position the inner elements in such a way, which is the best for the user. The user has to decide, how all those elements of the group must be resized and placed; after it a frame can be shown around all the elements regardless of their relative positions. In this way we are switching the main idea of a group upside down: the ruling is going not from the frame to the inner elements, but from those elements onto the frame. In this way the next several examples work. And to begin with, we will even get rid of the frame (only for some time!) and look at the group of controls, which influence each other in an interesting way. This is the `DominantControl` class.

Though there is a singular form of the word *control* in the name of this class, but this is a real group of controls. There is absolutely no sense in having less than two controls in a group; the upper limit for the number of elements does not exist. An object of this class represents a dominant control with an access to a set of subordinates; the subordinates belong to the `SubordinateControl` class. Any of the subordinate controls can be turned into the new dominant; one of further examples demonstrates it.

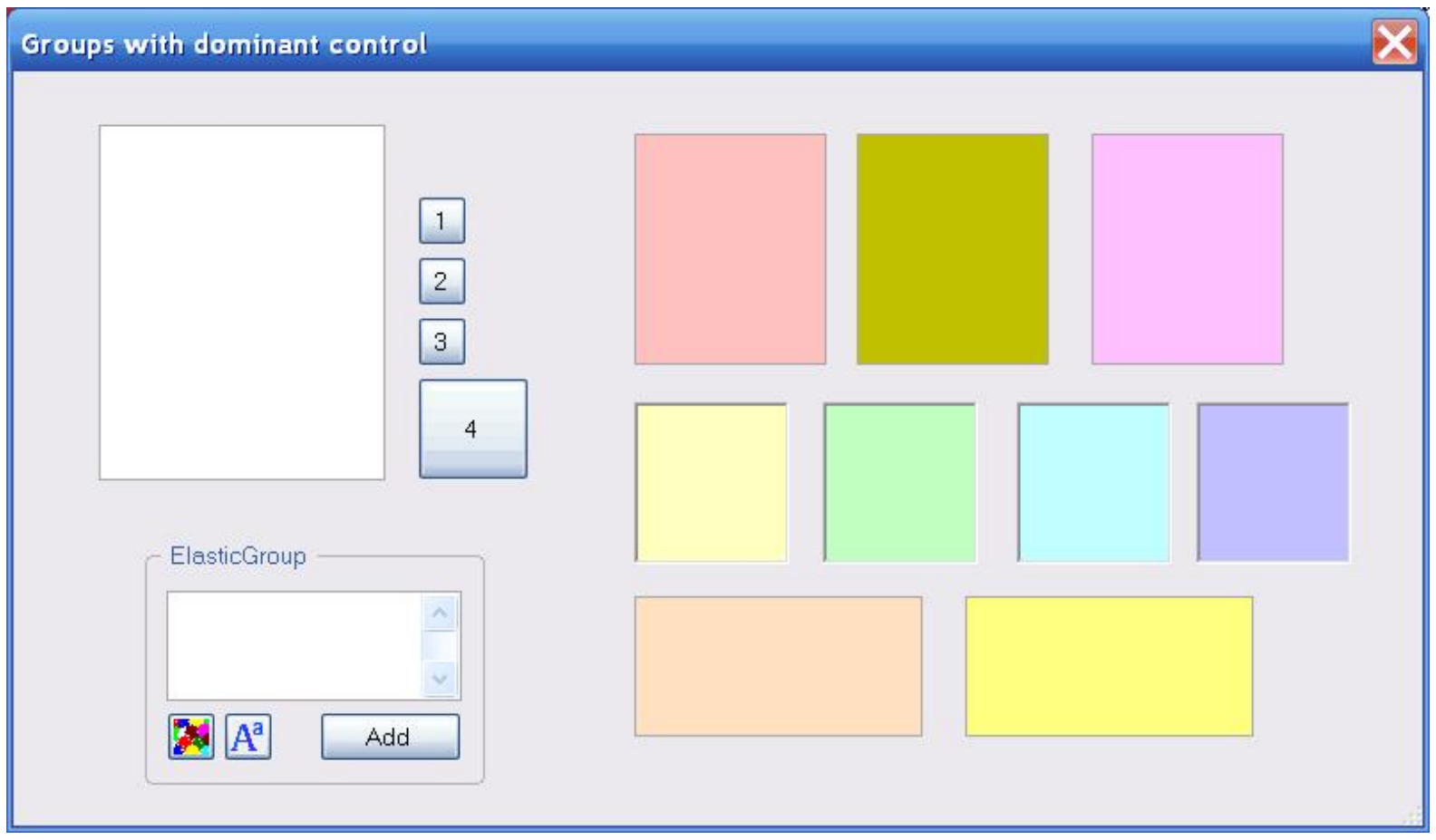

**Fig.14.10** **Form_DominantControls.cs** demonstrates three objects of the `DominantControl` class

```
public class DominantControl : GraphicalObject
{
    List<SubordinateControl> subordinates = new List<SubordinateControl> ();
```

There are three different objects of the DominantControl class in the **Form_DominantControls.cs** (**figure 14.10)**. In two of these three objects the dominant control of the group is obvious; two of three groups have no frames, but I think that even without them the sets of related controls make a clear-cut distinction between the groups. The most common case of using a DominantControl, though it is not a mandatory rule, is the situation with one big and several much smaller controls working as its satellites. This case is shown at **figure 14.11**; this is the best case to start the explanation.

The easiest way to construct a DominantControl object is to organize an array of controls, starting with the one, which is going to be dominant.

```
groupNums = new DominantControl (new Control [] { listviewDom, button1,
                                                  button2, button3, button4 });
```

The division on dominant and subordinate controls describes the difference in their individual and related movements.

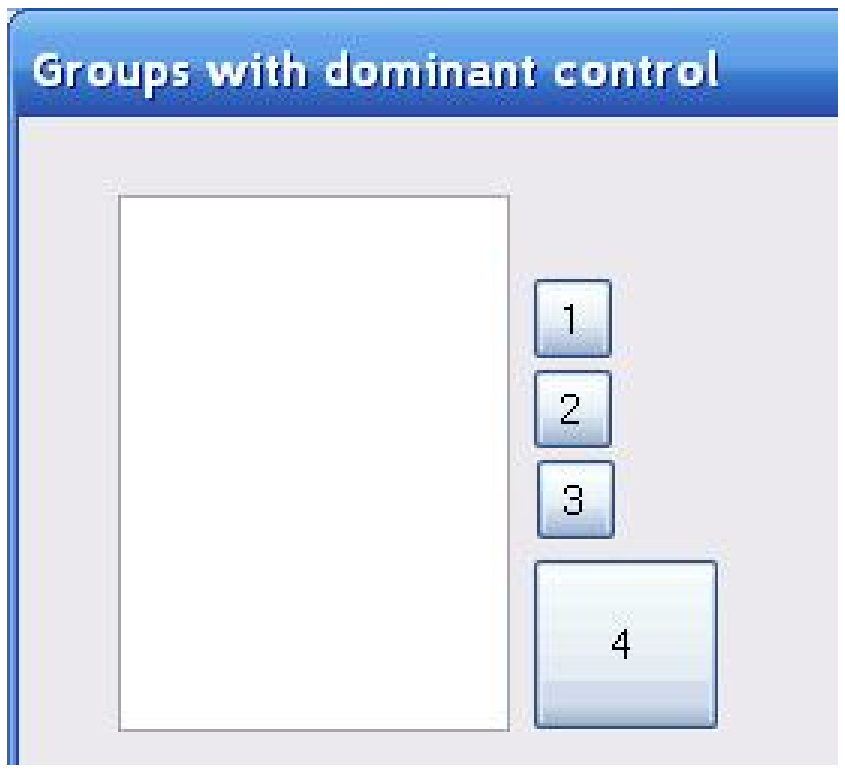

**Fig.14.11** Group with dominant control

- When the dominant control is moved, all the subordinates move synchronously.
- When the dominant control is resized, each of the subordinates tries to retain its relative position to the dominant control. An object of the SubordinateControl class has two coefficients, which describe the position of its left upper corner in relation to the rectangle, occupied by the dominant control.
- A subordinate control can be moved / resized individually and positioned anywhere in relation to the dominant control. But if a subordinate control is moved and released at such a place that it is fully inside the area of its dominant control (above or below – this depends on the Z-order of controls), then this subordinate control is forcedly relocated and placed next to the upper right corner of the dominant control. This rule is longer to read and understand, than to try; move any of the four buttons from the shown group to the area of the ListView control, release it there, and you will see the result.

The exception of the possible arbitrary positioning of a subordinate control happens only when the subordinate control finds itself in the area of <u>its own dominant control</u>. Pay attention to the underlined statement; if one of the subordinates is released over another or if it is released over any other control, to which it is not subordinate, then no enforced relocation is going to happen.

The same enforced relocation might happen in one more case. Consider the situation when the subordinate control is placed mostly inside the area of the dominant control with only tiny part of it looking outside the right or bottom side of the dominant control. Because a small part of subordinate control is visible, it can be placed in such a way without causing the enforced relocation. Then start increasing the dominant control. Because the positioning coefficients of the subordinate control are fixed and its size is also not changing, but the dominant control is increasing, then the subordinate control may find itself inside the increased area. As the result, if the subordinate control is inside the dominant control, when it is released, then the subordinate control will be relocated.

Both situations with the possibility of the enforced relocation are checked, when an object is released by the mover. When the subordinate control is released at the forbidden ground, then the SubordinateControl.CheckLocation() method ignites the relocation; when the dominant control is released, then the same relocation can be started by the DominantControl.CheckSubordinates() method.

```
private void OnMouseUp (object sender, MouseEventArgs e)
{
    if (mover .Release ())
    {
        if (e .Button == MouseButtons .Left)
        {
            GraphicalObject grobj = mover .WasCaughtSource;
            if (grobj is DominantControl)
            {
                (grobj as DominantControl) .CheckSubordinates ();
            }
            else if (grobj is SubordinateControl)
            {
```

```
                    (grobj as SubordinateControl) .CheckLocation ();
                }
                groupOfDominant .Update ();
                Invalidate ();
            }
        }
    }
```

By analysing the set of rules for the DominantControl class, you can see that its rules for enforced relocation are identical to the rules of the CommentedControl class. The difference between these two classes is in the number and types of the subordinates: the CommentedControl class has one painted text as a subordinate, the DominantControl class has an arbitrary number of controls.

Any control in the group can be made non-resizable or resizable in the same way, as was described for the SolitaryControl class. The mover determines the type of resizing for a particular control by analysing its sizes and the values of its MinimumSize and MaximumSize properties. If those properties are not changed from their default values (0, 0) or their values are the same as the size of a control, then this control is non-resizable. If there is a range for one direction only (width or height), then this control is resizable in this direction only; otherwise it is fully resizable.

In the group from **figure 14.11** the dominant control is obvious, but this happens not always. There can be situations, when you work with a group of controls and need to change the dominant control from time to time. Look again at the rules of resizing dominant control and subordinate controls. Suppose that there are several controls, which do not overlap but are positioned in some way not far from each other. When the dominant control is increased, then all the subordinates retain the space from the dominant control; all the subordinates are moved aside, so there will be no overlapping. When any subordinate control is increased, it is done on an individual basis without paying any attention to the positions of other elements in the group. The probability of overlapping is high enough.

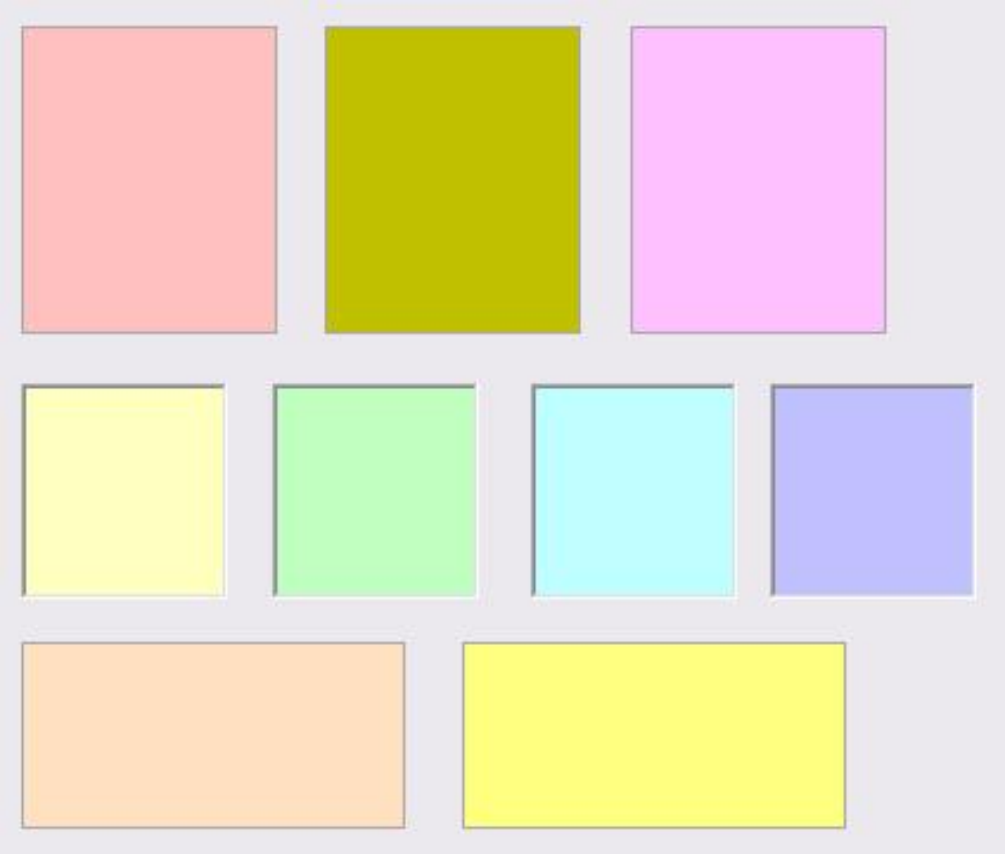

**Fig.14.12** Any member of this group can be declared a dominant control via the context menu.

Suppose that you have several ListView controls, buttons, and TextBox controls. During the work of an application, you need to increase one control, then another, but you do not want them to overlap, because they show the important information. The best solution would be to declare the control to be increased as the dominant control of such a group. In this case, you can change the sizes of any control without overlapping with others. This case is demonstrated with another DominantControl object (**figure 14.12**).

There are nine different controls in this group. The construction of the group is done in the same easy way by providing an array of controls. The first control of the array automatically becomes a dominant control of the group; in this case it is a ListView control in the top left corner.

```
groupClrs = new DominantControl (new Control [] { listView1, listView2,
                 listView3, panel1, panel2, panel3, panel4, textBox1, textBox2 });
```

At any moment any control of the group can be declared dominant by calling the context menu on it (**figure 14.13**); the pressed control becomes dominant.

```
    private void Click_miDeclareControlAsDominant (object sender, EventArgs e)
    {
        bool bShowPrompts = groupClrs .ShowPrompts;
        DominantControl groupChanged = groupClrs .SwitchDominant (ctrlPressed);
        if (groupChanged != null)
        {
            groupClrs = groupChanged;
            groupClrs .ShowPrompts = bShowPrompts;
            RenewMover ();
            Invalidate ();
        }
    }
```

Pay attention that the `DominantControl.SwitchDominant(ctrl)` method organizes the group with the new dominant control, only if one of the subordinates was pressed and passed as a parameter. If the pressed control is already dominant in the group or if a wrong control is passed as a parameter (the control, which does not belong to this group), then the same method will return `null`, so nothing must be done.

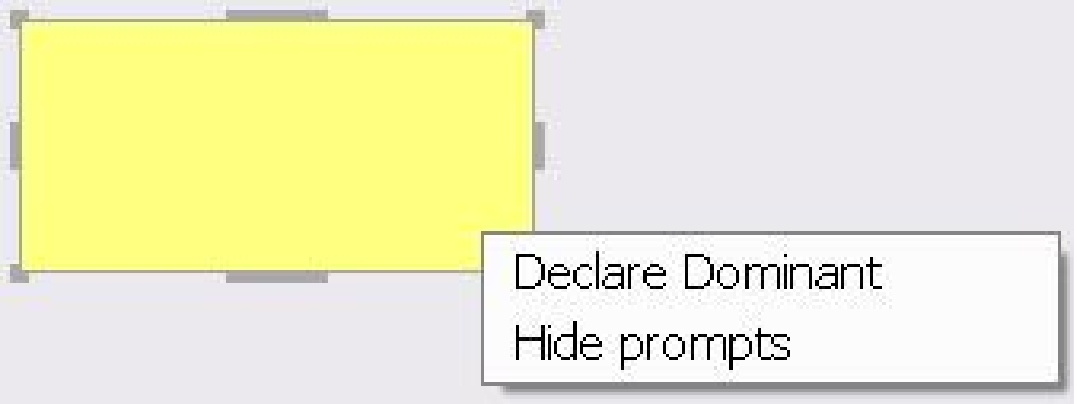

**Fig.14.13** This menu on controls allows to change the dominant control and to show / hide the prompts around the dominant control

**Figure 14.13** demonstrates that the same menu includes one more command – *Hide prompts*; the small darkened rectangles close to the border of the pressed control give a tip about this command. If you call the same menu on any control, when it is shown without such small additions as seen at **figure 14.12**, then the same menu line will declare *Show prompts*.

I mentioned in several places of this book that neither movability nor resizability of the screen elements in the real applications is ever shown in any way. It is enough to know that all the objects are movable by any inner point and resizable by any border point. I think that in such situation no visual prompts are needed. However, there can be special situations, when some other kind of visual prompt can be helpful. Consider the case of the group from **figure 14.12**, there is no visual difference between all those objects, but one of them is dominant and this means the different reaction on moving / resizing of this particular control and all others. Maybe some users would like to see an indication of the dominant control. I am not sure whether it is needed or not and I am demonstrating here some kind of visual prompt. It was not designed especially to mark dominant control, but it can do it by the way.

The `DominantControl` class includes the possibility of showing small prompts at the places, where the frame of the control can be caught for resizing. Though all the controls can be resized in similar way, regardless of whether it is dominant or subordinate, but the prompts are shown only around dominant control, so they can give a tip. Showing of those small darkened areas is regulated by the `DominantControl.ShowPrompts` property. On initializing the new `DominantControl` object, the `false` default value is imposed and the prompts are not shown. Especially for this reason, when the dominant control in the group is going to be changed, then the value of this property is saved from the existing group and applied to the new group after its initialization; see the code of the `Click_miDeclareControlAsDominant()` method on the previous page.

The demonstration of those small visual prompts inside the group was the only reason to include the drawing of this group into the `OnPaint()` method.

```
private void OnPaint (object sender, PaintEventArgs e)
{
    Graphics grfx = e .Graphics;
    … …
    groupClrs .Draw (grfx);
}
```

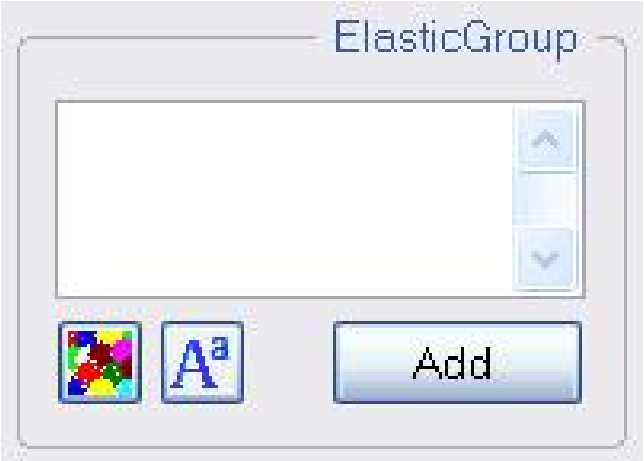

**Fig.14.14** The `DominantControl` is the basis of this group

`DominantControl` group consists exclusively of controls. There is nothing to draw in such a group, so the `DominantControl.Draw()` method is only about painting those small prompts, if they are needed to be shown. If you use the group of the same class without prompts, then you do not need any drawing; the `groupNums`, which is shown at **figure 14.11**, is not mentioned in the `OnPaint()` method.

Both representatives of the `DominantControl` class, which are shown in **figures 14.11** and **14.12**, have no frame. But at the beginning of this section I wrote about the switch of ruling direction (inner elements must set the position of the frame) and mentioned that I wanted only to start the discussion with the frameless groups, but not to abandon the frame entirely. There is the third group in the **Form_DominantControls.cs**, which has a frame (**figure 14.14**), but to be absolutely correct I have to admit that the content of this group is the `DominantControl` object, while the group itself (if a group is recognized as a frame with all its inner elements) belongs to the `ElasticGroup` class. The `DominantControl` class has no frame of its own, but it can get a frame, if it is used to organize a content of an `ElasticGroup` object.

```
groupOfDominant = new ElasticGroup (this, new ElasticGroupElement [] {
          new ElasticGroupElement (
              new DominantControl (new Control [] {textNewComment, btnCommentClr,
                                      btnCommentFont, btnAdd})) }, 12, "ElasticGroup");
```

The ElasticGroup class is the main theme of the next section; here are several more words about the last example of the DominantControl object.

The group at **figure 14.14** is purposely organized as being identical to the Group object from **figure 14.9**. The group with such content is used in several tuning forms from the **MoveGraphLibrary.dll**; the same group is used in the **Form_RegPolyWithComments.cs** (**figure 10.2**). The purpose of this group is to add the new comment to one or another screen object. You type the needed comment inside the TextBox control and then select the color and font for this comment. Button **Add** will add the new comment to the screen, when the group is used in the proper way. The main element in this group is definitely the control, in which the text is typed; three buttons are the auxiliary elements. The needed size of the TextBox control depends on the situation; in some cases the comments are short; in others they can be long enough and multilined. If the control is the main element of the group, then it makes sense to make this control the "ruler" of the whole group, so I made it dominant, while initializing this group. But the whole group itself is not going to look exactly in the way I gave it at the beginning. If users agree with the developer's ideas, they can leave the group as it is, but each user can easily change the overall view of the group, the relative positions of the elements, and their sizes.

The DominantControl class demonstrates an absolutely new type of organizing a group: such groups are determined only by users; the developer does not interfere in the process of changing the group but only provides such type of group which can be rearranged by a user in any possible way.

The best representative of such design is the ElasticGroup class, which is the theme of the next section.

## Elastic group

An elastic group has no predetermined area into which a set of elements has to be squeezed. On the contrary, the elements of such group can move and resize independently, but the area of a group always adjusts to all these changes and surrounds the united area of all those elements. If the frame of such a group is visualized, then you can watch how it automatically changes its position with all the moving / resizing of the inner elements. In such case a frame looks like an elastic string around a bunch of pencils, which keeps them all together. Such behaviour of a group gave the name to the ElasticGroup class. This class has been already used in several of the previous examples:

- **Form_NnodeCovers.cs** (**figure 7.1**)
- **Form_FillTheHoles.cs** (**figure 8.10**)
- **Form_SetOfObjects.cs** (**figure 9.4**)
- **Form_RegPolyWithComments.cs** (**figure 10.2**)

In all these cases the group consists of a set of individually movable controls. Though it is one of the most primitive ways of using the ElasticGroup class, let us have a quick look into such case before turning to much more interesting situations.

File: **Form_SetOfObjects.cs**
Menu position: *Graphical objects – Basic elements – Set of objects*

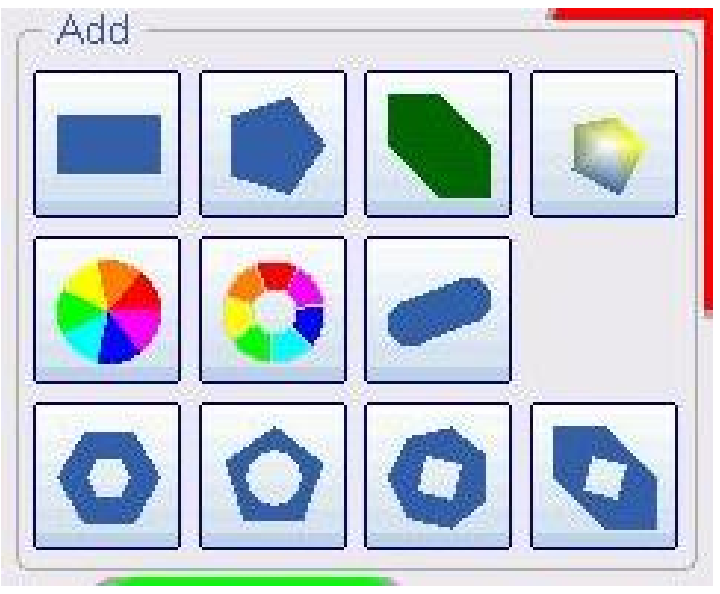

**Fig.14.15** A group of ordinary buttons, organized into an **ElasticGroup** object

Let us return once more to the **Form_SetOfObjects.cs.** In the chapter *Sliding partitions* this form was used to demonstrated a whole variety of different graphical elements (**figure 9.4**), but now we are interested not in them, but in another object of that form – the group of buttons, which allow to add all those different elements into the form. The buttons inside the group (**figure 14.15**) are individually movable, so you can place them in any way you want: put them vertically in one column, or horizontally in a row, change their order in any possible way, set the distances between the buttons according to your taste and requirements for good design. I hope you are aware that the reaction on the click of any button does not depend on the position of this button. For example, a click of the button with the colored circle on top will add a new circle into the form regardless of the location of this button. Thus the purpose of the form (to put into the form only the set of different elements, required by the user) and the composition of the form are entirely separated. The developer (sorry, it is me) is responsible for the correct work of the form according to its main purpose; you, as a user, are free to compose the group in any way you want. The unlimited freedom of composition is provided by the ElasticGroup class.

Suppose that you decide to organize such a powerful group in one of your applications. How much work will be needed for it? Let us check it with this group from the **Form_SetOfObjects.cs**; it is easy to do by checking all the places in the code, where the group is mentioned.

Step 1. Declare a group.

```
ElasticGroup groupAdd;
```

Step 2. Construct a group.

```
groupAdd = new ElasticGroup (this, new Control [] { btnRectangle,
                    btnRegularPoly, btnConvexPoly, btnChatoyantPoly, btnCircle,
                    btnRing, btnStrip, btnPerforatedPoly, btnRegPoly_CircularHole,
                    btnRegPoly_RegPolyHole, btnConvexPoly_RegPolyHole},
                                new int [] { 12, 6, 12, 12 }, "Add");
```

The constructor describes:

- A set of buttons, to be included into the group.
- An array of spaces between the inner elements and the frame of the group.
- The title of the group.

This is one of many possibilities to initialize such a group; all depends on the number of details, which you want to specify.

Step 3. Register a group with the mover, because we want all the elements to be movable.

```
groupAdd .IntoMover (mover, 0);
```

Step 4. Draw the group.

```
private void OnPaint (object sender, PaintEventArgs e)
{
    Graphics grfx = e .Graphics;
    … …
    groupAdd .Draw (grfx);
```

That is all! Not too much work for such a flexible group. Maybe this is because, as I mentioned, "a set of individually movable controls … is one of the most primitive ways of using the `ElasticGroup` class"? Next example is a real application that demonstrates different cases of using this class and a lot of details.

File: **Form_PersonalData.cs**
Menu position: *Applications – Personal data*

The `ElasticGroup` class can be used in much more interesting ways than in the previous example; that is why it became the main constructing element in all of my applications. The basic ideas of the `ElasticGroup` class sound simple enough:

- Any element of a group can be moved individually.
- The frame of a group is automatically adjusted to the positions and sizes of all the inner elements and always surrounds them.
- A group can be moved by any inner point.

What makes the `ElasticGroup` class extremely flexible and powerful, is the list of classes, objects of which can be used as inner elements of such group:

- `SolitaryControl` objects;
- `CommentedControl` objects;
- `CommentedControlLTP` objects;
- `DominantControl` objects;
- `ElasticGroup` objects.

This list of classes means that an element can be a control, a control with comment, a group of controls, of which one is dominant and others are subordinates, and an element can be a group of the same class. Especially this recursive definition of the group makes the `ElasticGroup` class the most valuable. I do not see any other types of relations between the controls that are not covered by these cases, so, from my point of view, any type of form can be organized with the `ElasticGroup` objects.*

**Figure 14.16** shows the view of the **Form_PersonalData.cs**. The form demonstrates the use of the `ElasticGroup` class with all possible types of inner elements except the `CommentedControlLTP`. At the same time, this is not an artificial example to show one or another discussed feature. This form deals with the collection of personal data; it is a situation, which is familiar to everyone throughout his life, so everyone can make his own decision about the usefulness of the proposed design on the basis of the movable / resizable elements. Certainly, each of us deals with the collection of personal data only occasionally, but there are people (from HR departments), who work with similar forms day after day; it would be very interesting to hear their opinion.

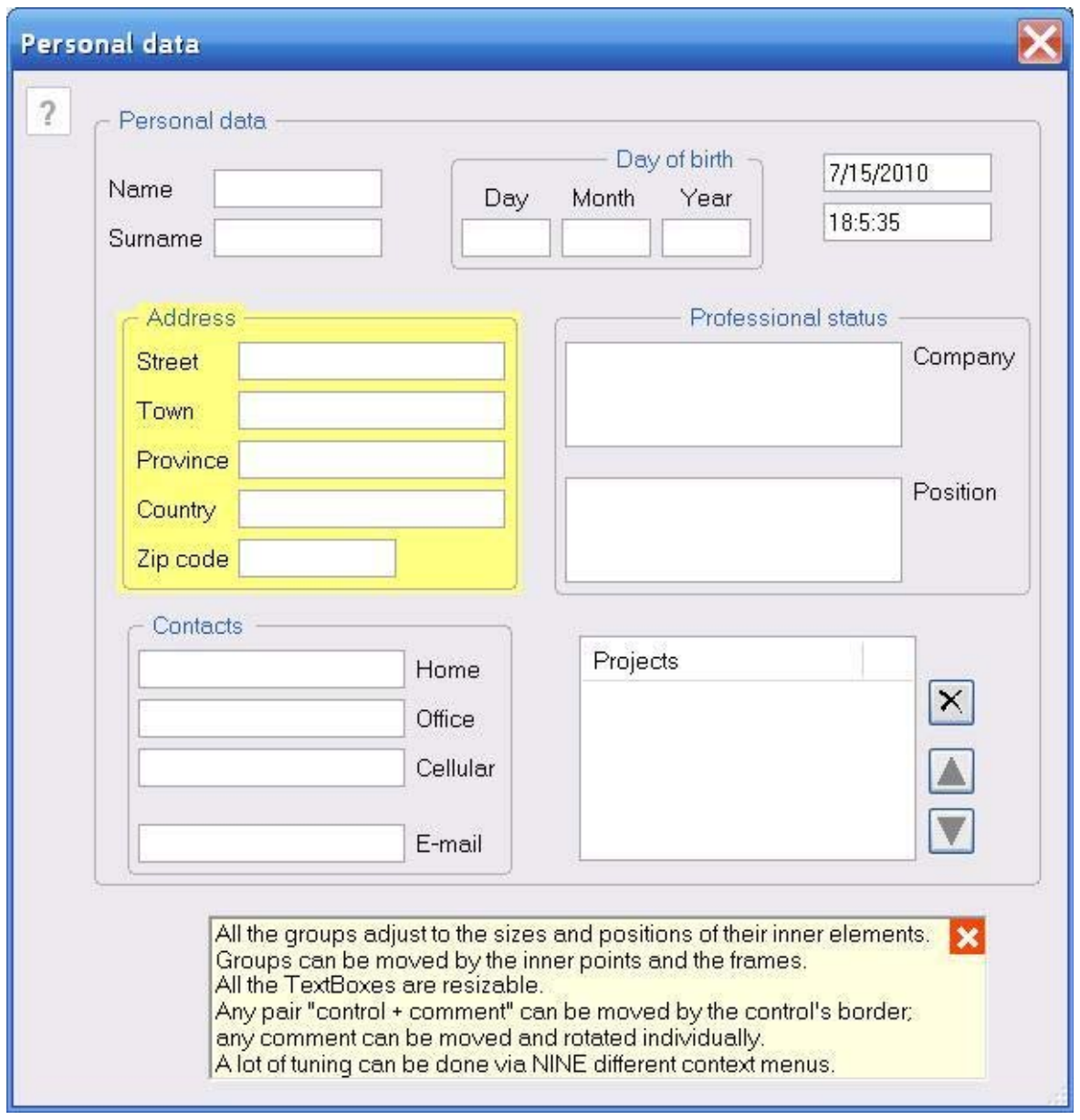

**Fig.14.16** **Form_PersonalData.cs** is based on objects of the `ElasticGroup` class

The **Form_PersonalData.cs**, which is obvious from its name, deals with the personal data. Depending on the case, it can include only a small piece of data or might require to include more than is shown here, but you are free to add any other needed groups of information (all the codes are available).

The `ElasticGroup` class has a significant number of constructors; close to 50 constructors in the current version of the library. Such a number of constructors is due to the vide variety of cases, in which the class can be used, and my attempt to make the initialization in many cases as easy as possible and avoid the request for a parameter, which can be set by default. When any object of the five classes, mentioned a bit earlier, is used as an element of the `ElasticGroup` object, it is first cast (directly or indirectly) into the `ElasticGroupElement` class. There are nearly 20 different constructors for this class, so together they provide such a variety, that you can find the easiest one to use for any type of the needed group whether it consists solely of the elements of one type or of some combination. In the shown form, the biggest (outer) group includes two `SolitaryControl` objects to show the date and time, two `CommentedControl` objects to show name and surname, and five inner groups.

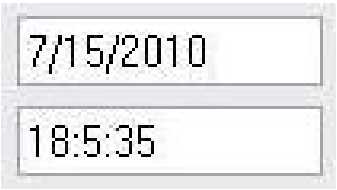

**Fig.14.17** Two elements represent the `SolitaryControl` case

Overall there are 23 controls in this form. I would not call it an extremely complex form (from time to time I have to design much more complicated forms), but at the same time I think that the majority of readers rarely have a chance to design a form with such a number of controls. There are many ways of using different constructors to design such a form. I tried to make the code as easy as possible for understanding, so, for the majority of elements, I use the constructors to develop the inner elements and then combine them into bigger groups. The only elements that avoided this preliminary construction are the two controls for date and time (**figure 14.17**); they are so simple that are organized during the construction of the outer group.

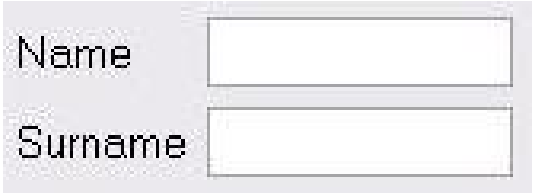

**Fig.14.18** Two elements represent the `CommentedControl` case

```
    new ElasticGroupElement (textDate, Resizing .WE),
    new ElasticGroupElement (textTime, Resizing .WE),
```

Name and surname are represented by the `CommentedControl` objects (**figure 14.18**).

```
CommentedControl ccName = new CommentedControl (this, textName, Resizing .WE,
                                                Side .W, "Name");
CommentedControl ccSurname = new CommentedControl (this, textSurname, Resizing .WE,
                                                   Side .W, "Surname");
```

* I want to emphasize that this is only for the forms which consist of controls only (and controls with comments). If, in addition to controls, there must be some graphical objects in the form, then this situation is not covered by the `ElasticGroup` objects and will be considered in the next section, where the `ArbitraryGroup` class is discussed.

Four of the inner groups – *Day of birth*, *Address*, *Contacts*, and *Professional status* - have similar design: each one consists of an array of `CommentedControl` objects. First I prepare the arrays of these objects; later these arrays are used during the construction of the outer group. Here is the code to prepare an array of elements for the *Contacts* group (**figure 14.19**).

```
CommentedControl [] ccsPhones = new CommentedControl [] {
     new CommentedControl (this, textHomePhone, Resizing .WE, Side .E, "Home"),
     new CommentedControl (this, textOfficePhone, Resizing.WE, Side.E, "Office"),
     new CommentedControl (this, textMobilePhone, Resizing.WE, Side.E, "Cellular"),
     new CommentedControl (this, textEMail, Resizing .WE, Side .E, "E-mail") };
```

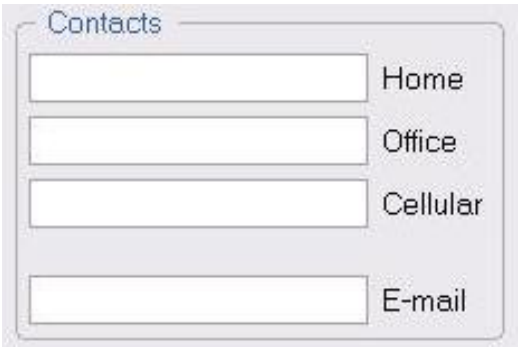

**Fig.14.19**

The elements of the last inner group – the *Projects* group – are united into the `DominantControl` object (**figure 14.20**). At the view of the whole form (**figure 14.16**) this group is shown without a frame, but the frame can be reinstated at any moment via the context menu. There is no question, which control in this group is dominant, and there is no sense in changing the dominant control in this group, so it was not implemented here. This group does not need any title, as the text in the header of the `ListView` control works as the title of the group.

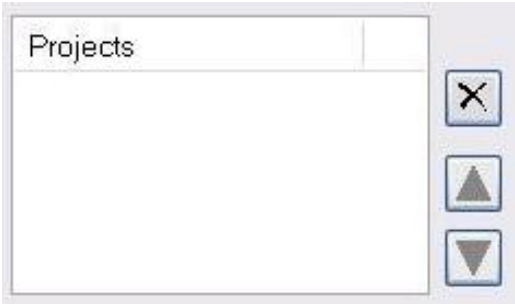

**Fig.14.20**

```
ElasticGroup groupProjects = new ElasticGroup (this, new DominantControl (
          new Control [] { listProjects, btnDelete, btnMoveUp, btnMoveDown }), "");
```

After all the inner elements are constructed, it is time to construct the big group. Regardless of the type of the inner element (`SolitaryControl`, `CommentedControl`, `DominantControl`, or `ElasticGroup`), each one of them is transformed into an `ElasticGroupElement` object; the group is constructed on the array of such objects.

```
groupData = new ElasticGroup (this, new ElasticGroupElement [] {
         new ElasticGroupElement (textDate, Resizing .WE),
         new ElasticGroupElement (textTime, Resizing .WE),
         new ElasticGroupElement (ccName),
         new ElasticGroupElement (ccSurname),
         new ElasticGroupElement (this, ccsDOB, "Day of birth"),
         new ElasticGroupElement (this, ccsPhones, "Contacts"),
         new ElasticGroupElement (this, ccsAddress, "Address"),
         new ElasticGroupElement (this, ccsProfessional, "Professional status"),
         new ElasticGroupElement (groupProjects) },
                              "Personal data");
```

One inner group at **figure 14.16** (the *Address* group) is shown in different color, but this is only to indicate the possibility of changing colors and other parameters of visualization. In the user-driven applications, and the **Form_PersonalData.cs** is definitely one of them, ALL the parameters of visualization are under users' control, so these parameters can be changed by users at any moment. There are two standard ways to change the parameters: either through the context menu or via some tuning form; the **Form_PersonalData.cs** demonstrates both of them.

Each class of objects has its own set of parameters, so, if the parameters are changed through the context menu, then each class requires its own context menu. It is an ordinary thing to have even not in a very complicated form around ten or more different context menus. There are nine context menus in the **Form_PersonalData.cs**; as usual, the menu selection is called from inside the `OnMouseUp()` method, when the right button is released.

```
private void OnMouseUp (object sender, MouseEventArgs e)
{
    ptMouse_Up = e .Location;
    double nDist = Auxi_Geometry .Distance (ptMouse_Down, ptMouse_Up);
    if (e .Button == MouseButtons .Left)
    {
        … …
    }
    else if (e .Button == MouseButtons .Right)
    {
        if (mover .Release ())
        {
            if (nDist <= 3)
            {
```

```
                        MenuSelection (mover .WasCaughtObject);
                    }
                }
                else
                {
                    if (nDist <= 3)
                    {
                        ContextMenuStrip = menuOnEmpty;
                    }
                }
            }
        }
```

The only menu that does not require any preliminary identification of the released object, is the menu, which is called at any empty place – `menuOnEmpty`; all others require such identification and more often than not the identification of the parent object for the pressed one. The `MenuSelection()` method has one parameter – the number of the pressed object in the mover's queue. This means that this type of menu selection is started only when the mover releases an object. For a lot of applications, even for the most complex, which I am going to demonstrate further on, this is the only way to organize a menu selection. It is the only way, when a form is filled with different graphical objects, which are turned into movable / resizable. For the applications with a lot of controls, and the **Form_PersonalData.cs** is one of them, the menu can be also called by direct click of a control. Such click of a control cannot call the same `MenuSelection()` method, as the control itself is never covered by a cover. The menu selection after the click of a control will de discussed a bit later.

The `MenuSelection()` method is based on the analysis of the class of the released object. For some of the objects the analysis is extremely easy, for example, because an object can be unique, like an object of the `ClosableInfo` class, which shows some explanation on this form. There is a single object of this class in the form, so checking of the object's class is enough.

```
        private void MenuSelection (int iInMover)
        {
            GraphicalObject grobj = mover [iInMover] .Source;
            … …
            else if (grobj is ClosableInfo)
            {
                ContextMenuStrip = menuOnInfo;
            }
```

Some of the objects do not require any identification beyond an object itself. Through the context menu of such object only some of the parameters can be changed, but all these changes do not affect any other objects and do not require any changes in the mover's queue. In such case the identification up to the object is enough, because all further changes are done via the methods of the class, to which this object belongs.

```
        private void MenuSelection (int iInMover)
        {
            GraphicalObject grobj = mover [iInMover] .Source;
            … …
            else if (grobj is CommentToRect)
            {
                cmntPressed = grobj as CommentToRect;
                ContextMenuStrip = menuOnComment;
            }
```

In this piece of code the `CommentToRect` object was identified as the pressed element and the `menuOnComment` is called. This menu contains only two lines. One of them allows to change the color of comment; this command definitely has no side effects so the reaction is simple. This is a classical case when only one property of the pressed element is going to be changed with no effect on anyone else.

```
        private void Click_miCommentColor (object sender, EventArgs e)
        {
            ColorDialog dlg = new ColorDialog ();
            dlg .Color = cmntPressed .Color;
            if (dlg .ShowDialog () == DialogResult .OK)
            {
```

```
            cmntPressed .Color = dlg .Color;
            Invalidate ();
        }
    }
```

The second command available throughout the `menuOnComment` allows to change the font of a comment. It looks as simple as the change of color, but there is one catch that makes the reaction on this command more complicated. There is one rare situation with the probability of its happening next to zero, but because it can happen, I have to deal with it.

I have already described the `CommentedControl` class, which consists of a pair "control + text". The text can be placed anywhere in relation to its parent control, but not entirely covered by this control. If you move the text and release it under the parent control, it is forcedly moved outside. Now consider a situation when a comment has a big font and placed so that it is mostly covered by its control with only a tiny part looking out. You press this visible "tail" of the comment with a mouse; the comment's menu is opened and you select a small font for this comment. The area of the comment will shrink and the comment will hide under the parent control without any chances to see it again. Nearly impossible situation, but with the probability greater than zero, so I have to prevent it.

The `CommentedControl` object of the pressed comment must be identified and the `CommentEnforcedRelocation()` method for this object must be called. This method compares the areas of control and its comment and accomplishes the enforced relocation only if it is needed.

```
    private void Click_miCommentFont (object sender, EventArgs e)
    {
        FontDialog dlg = new FontDialog ();
        dlg .Font = cmntPressed .Font;
        if (dlg .ShowDialog () == DialogResult .OK)
        {
            long idCmntCtrl = cmntPressed .ParentID;
            GraphicalObject grobj;
            for (int i = mover .Count - 1; i >= 0; i--)
            {
                grobj = mover [i] .Source;
                if (grobj is CommentedControl && grobj .ID == idCmntCtrl)
                {
                    CommentedControl cc = grobj as CommentedControl;
                    cc .CommentFont = dlg .Font;
                    cc .CommentEnforcedRelocation (mover);
                    groupData .Update ();
                    Invalidate ();
                    break;
                }
            }
        }
    }
```

The `CommentedControl` objects are used in several groups the **Form_PersonalData.cs**. Via the menu, the parameters of object can be changed in such a way that its "parent" group has to be updated. Also, an object can be hidden from view, which requires not only the updating of the group, but also the renewal of the mover's queue. For any of these actions, not only the `CommentedControl` object must be identified, but also the group, to which it belongs. The object itself can be identified in the standard way, but the identification of the group requires some search through the mover's queue. Two things are used for this search:

- All the members of the group precede in the mover's queue their group, so the additional search can be done only through the remaining part of the queue.
- All the elements of a group contain the **id** of their group; this is used for group's identification.

```
    private void MenuSelection (int iInMover)
    {
        GraphicalObject grobj = mover [iInMover] .Source;
        … …
        else if (grobj is CommentedControl)
        {
```

```
            ccPressed = grobj as CommentedControl;
            for (int i = iInMover + 1; i < mover .Count; i++)
            {
                if (mover [i] .Source is ElasticGroup &&
                    mover [i] .Source .ID == ccPressed .ParentID)
                {
                    groupParent = mover [i] .Source as ElasticGroup;
                    break;
                }
            }
            ContextMenuStrip = menuOnCommentedControl;
        }
```

Similar technique is used for the subordinate controls, only instead of the `ParentID` property, the dominant control can be obtained via the `Dominant` property and then its **id** is used.

As I have already mentioned, the same menus can be called not only on the covers of the controls, but on the controls themselves. In this case, there is no menu selection via the control's identification. Instead, the menu is mentioned in the `ContextMenuStrip` property of the control at the design stage. The process of control's identification is slightly different, as can be seen in the case of the `CommentedControl` objects. The mover's queue is searched through by comparison of the pressed control with the controls, associated with each `CommentedControl` object in the mover's queue. After the `CommentedControl` object with this control is found, the remainder of the identification is the same.

In the **Form_PersonalData.cs**, the `menuOnCommentedControl` is mentioned in the `ContextMenuStrip` property of every control, which is used to design a `CommentedControl` object, and the next method is marked as a reaction on **`MouseDown`** event for each of these controls.

```
private void MouseDown_commentedcontrol (object sender, MouseEventArgs e)
{
    Control ctrl = sender as Control;
    if (e .Button == MouseButtons .Right)
    {
        for (int i = 0; i < mover .Count; i++)
        {
            if (mover [i] .Source is CommentedControl)
            {
                ccPressed = mover [i] .Source as CommentedControl;
                if (ccPressed .Control == ctrl)
                {
                    for (int j = i + 1; j < mover .Count; j++)
                    {
                        if (mover [j] .Source is ElasticGroup &&
                            mover [j] .Source .ID == ccPressed .ParentID)
                        {
                            groupParent = mover [j] .Source as ElasticGroup;
                            return;
                        }
```

Looking through the menus of the **Form_PersonalData.cs**, you can find that they allow to do three main actions:

- Change the parameters of visualization.
- Hide / restore objects or the whole groups.
- Change the movability of the objects.

Changing of the visualization parameters (colors, fonts, transparency…) is an understandable and expected requirement. Users have different preferences, so this allows everyone to have an application with the best (or personally preferable) view. Changing of visibility and movability require a special discussion.

## Interesting aspects of visibility

Hiding and restoring of objects and groups allow to have at the screen and to deal with exactly that part of information which user wants to see at one or another moment and not to waste a very valuable screen space on the unneeded data. Any change in the visibility of objects, whether it is hiding from view or unveiling on the screen, changes the number of objects seen in the form. As any object in view is movable, then any such command requires the renewal of the mover's queue. This means that any action of such a type must include the call to the `RenewMover()` method. Here is the reaction on the demand to restore all the hidden objects of some group.

```
private void Click_miUnveilElements (object sender, EventArgs e)
{
    foreach (ElasticGroupElement elem in groupPressed .Elements)
    {
        if (!elem .Visible)
        {
            elem .Visible = true;
        }
    }
    groupData .Update ();
    RenewMover ();
    Invalidate ();
}
```

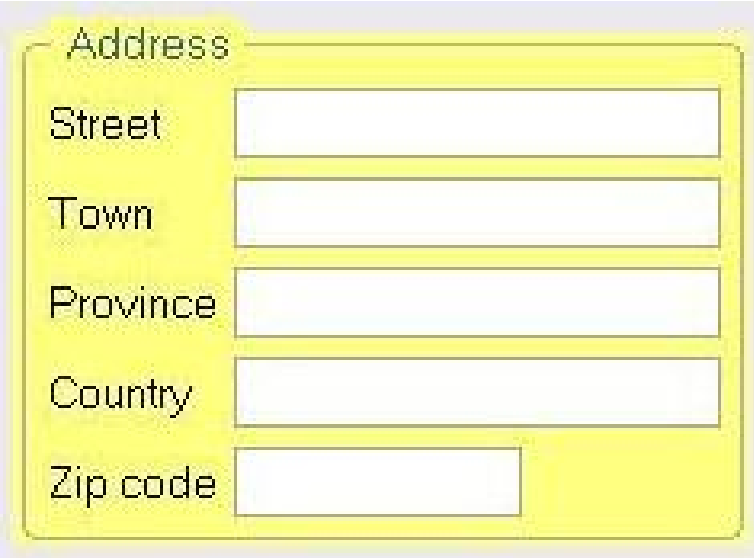

**Fig.14.21**

An object can be a stand alone element, or a member of some group or more complex object. Accordingly, there can be two different ways of changing the visibility of objects

- As a result of direct command for this particular object.
- As a consequence of changing the visibility of the surrounding group or parental object.

Because of this duality, it is not enough for an object to have a single visibility parameter, but there must be two different parameters to regulate the visibility of objects.

Consider a situation with the *Address* group, shown at **figure 14.21**. Suppose that you work with information on a group of people only from your country; in this case the name of the country is definitely excessive. Call the menu on the *Country* commented control (the menu is called anywhere around the control of a "control + text" pair) and hide this pair; this is the first command from menu at **figure 14.22**; the command is executed .by the `Click_miHideCommentedControl()` method.

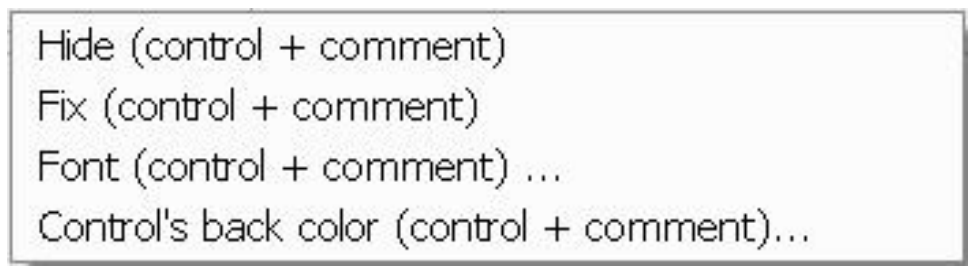

**Fig.14.22** Menu on `CommentedControl` objects

```
void Click_miHideCommentedControl (object sender, EventArgs e)
{
    ccPressed .Visible = false;
    groupData .Update ();
    RenewMover ();
    Invalidate ();
}
```

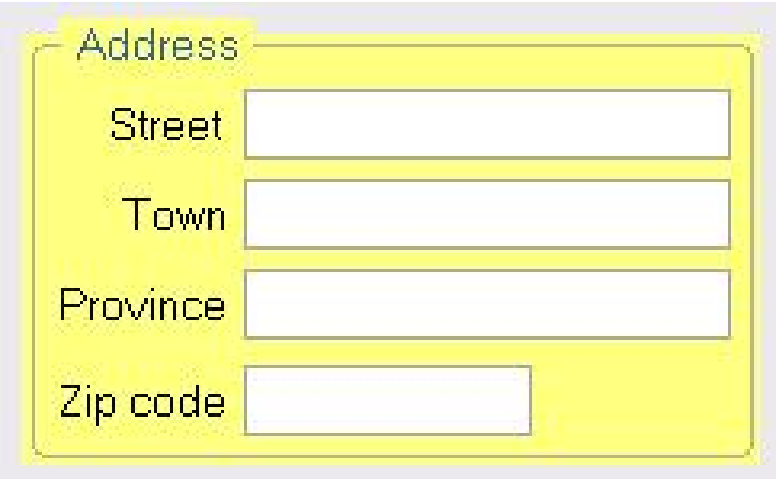

**Fig.14.23**

After hiding one `CommentedControl` object and moving another, you will get the new view of the group (**figure 14.23**). Now suppose that you do not need to keep the *Address* group on the screen, because mostly you communicate with these people by phone. Call the menu on this group by clicking at any empty spot inside the group and use the *Hide group* command of the opened menu (**figure 14.24**); this command calls the `Click_miHideGroup()` method.

```
private void Click_miHideGroup (object sender, EventArgs e)
{
    groupPressed .Visible = false;
    groupData .Update ();
    RenewMover ();
    Invalidate ();
}
```

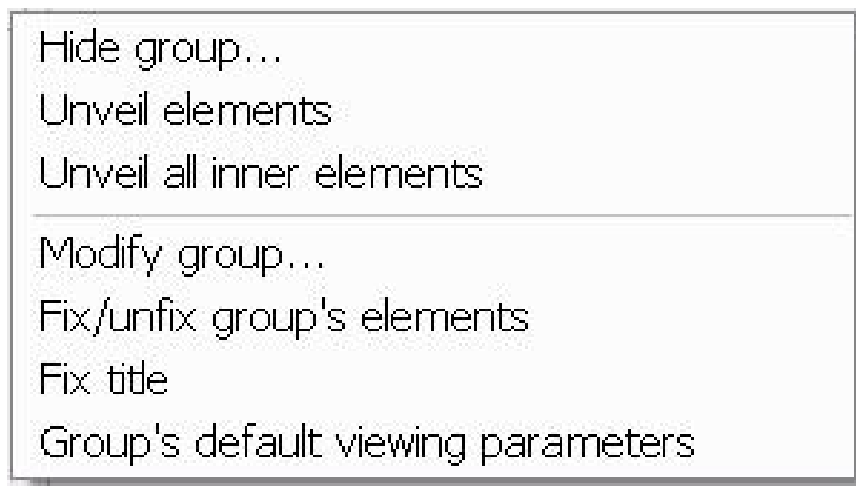

**Fig.14.24** Menu on group

At the end of the year you need an access to the address information, because it is the time for Christmas cards. You cannot call a menu on the invisible group, so you have to open the menu on the surrounding group – the *Personal data* group. In the opened menu you will see the list of the main elements with the checkmarks against those of them, which are currently visible (**figure 14.25**). The *Address* line is not marked; you have to click this line to return the *Address* group into view. Which view of the *Address* group you expect to see: with or without the *Country* information? I would prefer the view from **figure 14.23**, because it was the view of the group before it was hidden. But how can a program make such a decision, if all the `CommentedControl` objects inside the group were invisible at the moment, when you ordered to show the *Address* group back? How can a program decide to turn into visible only four of those five controls inside the group?

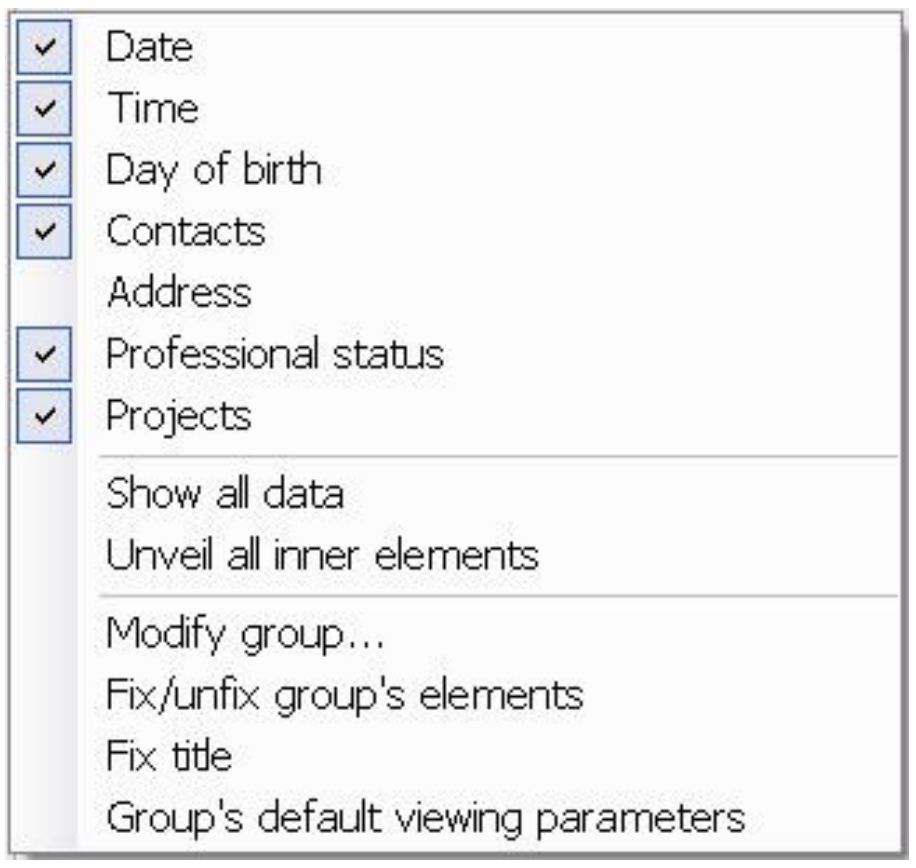

**Fig.14.25** Menu on *Personal data* group

The mechanism to make the correct decision is based not on one visibility parameter, but on two of them. Both parameters belong to the base `GraphicalObject` class and are inherited by any other class of movable objects, as all of them are derived (directly or through others) from that class. There are two properties to get / set these parameters.

- `Visible` property is to deal with the object's visibility parameter which is changed directly. This property of the `CommentedControl` object is set to `false`, when you hide the commented control through its own menu.
- `VisibleAsMember` property is to deal with the visibility which is changed indirectly, for example, by a surrounding group.

Any class inherits both properties from the base class, but for any class of complex objects you will have to write the new versions of them. For example, here is the `Visible` property for the `ElasticGroup` class.

```
new public bool Visible
{
    get { return (base .Visible); }
    set
    {
        base .Visible = value;
        bool bToMembers = base .Visible && base .VisibleAsMember;
        foreach (ElasticGroupElement elem in elements)
        {
            elem .VisibleAsMember = bToMembers;
        }
    }
}
```

Any object is shown only if both of its visibility properties return `true`. Taking this into consideration, it is now easy to understand how the *Address* group will be returned to view from **figure 14.23** in the above mentioned situation.

1. You set the `Visible` property of the pressed commented control (`ccPressed`) to `false`. The pressed commented control (*Country*) disappeared from view.
2. You set the `Visible` property of the *Address* group to `false`. Both visibility parameters of the group are used to determine the value, which has to be sent to all inner elements of the group; all of them get the `false` value into their `VisibleAsMember` properties.
3. You reinstall the visibility of the *Address* group to `true`. All the inner elements of the group receive the `true` value into their `VisibleAsMember` properties.
4. Four of the five inner elements have both of their visibility parameters set to `true`, so they appear in view. The *Country* member has the value of its `VisibleAsMember` property turned into `true`, but its `Visible` property is still set to `false`, so this commented control is not shown.

All the objects have two visibility parameters and two properties to deal with them, but not all the objects use both. If an object is used only as a stand alone, then its `VisibleAsMember` property is never used. It gets its `true` value on initialization of an object; but it is never changed, so it has no effect on visualization; the visualization of such an object is determined only by its `Visible` property.

The groups use both properties. It really does not matter, how many levels of nested groups you have; beginning from the second inner level, the command from the outer group to all the inner elements goes into the VisibleAsMember property, regardless of whether it was initially a direct or an indirect change. Here is the VisibleAsMember property for the ElasticGroup class. Compare this code with the previous piece and you will see that there is no difference beginning from the next level.

```
new public bool VisibleAsMember
{
    get { return (base .VisibleAsMember); }
    set
    {
        base .VisibleAsMember = value;
        bool bToMembers = base .Visible && base .VisibleAsMember;
        foreach (ElasticGroupElement elem in elements)
        {
            elem .VisibleAsMember = bToMembers;
        }
    }
}
```

Further on you will see that exactly the same mechanism of two visibility parameters works in the complex objects, used for different types of plotting.

Two more aspects of hiding and unveiling objects.

1. There is no visual indication in the groups that one or several of its inner objects are hidden. You might want to add such an indication and you can certainly do it, but I decided not to add such indication.
2. If you have a multilevel system of the nested groups and some of the objects on different levels were hidden, then the restoration of visibility for all of them will require several consecutive calls on different menus. To make this much faster, the ElasticGroup class has a VisibleAll property, which restores the visibility of all the elements at all the inner levels of a group.

## On movability

Changing the movability of the objects on a fly seems a bit strange after all the efforts to make everything movable, but the requirement for such option becomes more and more important with the increase of the complexity of the forms (applications). Return once more to **figure 14.16**. In the group *Professional status* you are not going to have any problems in trying to move the comments, the controls, or the whole group, because the elements stay clearly apart and there is enough empty space inside the group to grab it by a mouse.

However, look at the *Contacts* group (**figure 14.19**) or at the *Address* group (**figure 14.21**). All these groups contain the CommentedControl objects, so in all these cases you can call the same menus and move the same type of objects. But in two last cases the empty space inside the groups, where you can grab not an inner object but the group itself, is very limited. Chances are high that, while trying to move the group, you will accidentally grab and move some inner element. Usually, when you are moving the group, its inner elements are already placed in such a way, which you want, so you do not need to move any of them. To avoid the accidental move of the inner elements instead of the needed movement of a group, the inner elements can be fixed (via the menu).

Fixing / unfixing of the elements can be organized through two different menus (two different levels): it can be done individually via the menu of the particular object or it can be done for all objects of a group through the group's menu.

Turning even standard objects into movable can give some interesting and absolutely unexpected ideas.

The view of an ElasticGroup object is similar to the view of the standard GroupBox, which is known for many years: frame, title, and the familiar view of the inner elements. The standard GroupBox allows to show the title either on the left or on the right side. Several of my classes for groups (GroupBoxMR and Group are among them) used the same variants for positioning the title plus the possibility of the central positioning, so it was the parameter, which allowed three choices. But one day I asked myself a very simple question: "Why only three possibilities? What about those users, whose sense of perfection requires to put the title somewhere else?" I slightly changed the cover for the ElasticGroup class; now the title can be moved along the upper line of frame from the left corner to the right and placed anywhere inside this range. (Certainly, the title can be moved in the same range even when the frame itself is not shown.) The movability of title can be changed exactly in the same way, as the movability of any inner element: it can be fixed and unfixed at any moment. At the current moment, the movability of the title is changed via the group's menu; maybe it is a mistake and I will move it into the group's tuning form, where all the group's parameters of visualization can be changed. On the other hand, all the changes of movability and the switches to hide or show the elements are done via the context menus; the movability of title is of the same type of parameters.

There are two standard ways to change the parameters: either through the context menu or via some tuning form. When an object has only two or three changeable parameters, the menu is used. Textual comments, `CommentedControl`, `DominantControl`, `SubordinateControl`, and `SolitaryControl` objects belong to this group and each one of them has its own menu. But the `ElasticGroup` class has around 15 parameters to tune, so the tuning of such groups is organized via the tuning form. **Figure 14.26** demonstrates the default view of this **Form_ElasticGroupParams.cs**; the font of the form can be changed via the menu, which can be called outside of any group.

This tuning form contains three groups to deal with the different sorts of parameters: the left group to change the background color of the group and transparency, the middle group to set the colors and fonts of some parts and elements; the right group to set the spaces between the inner elements and the group's frame. Certainly, the name *left*, *middle* or *right* group can be applied only to their positions on this figure; the groups are movable and you can change the view of the form in any way you prefer.

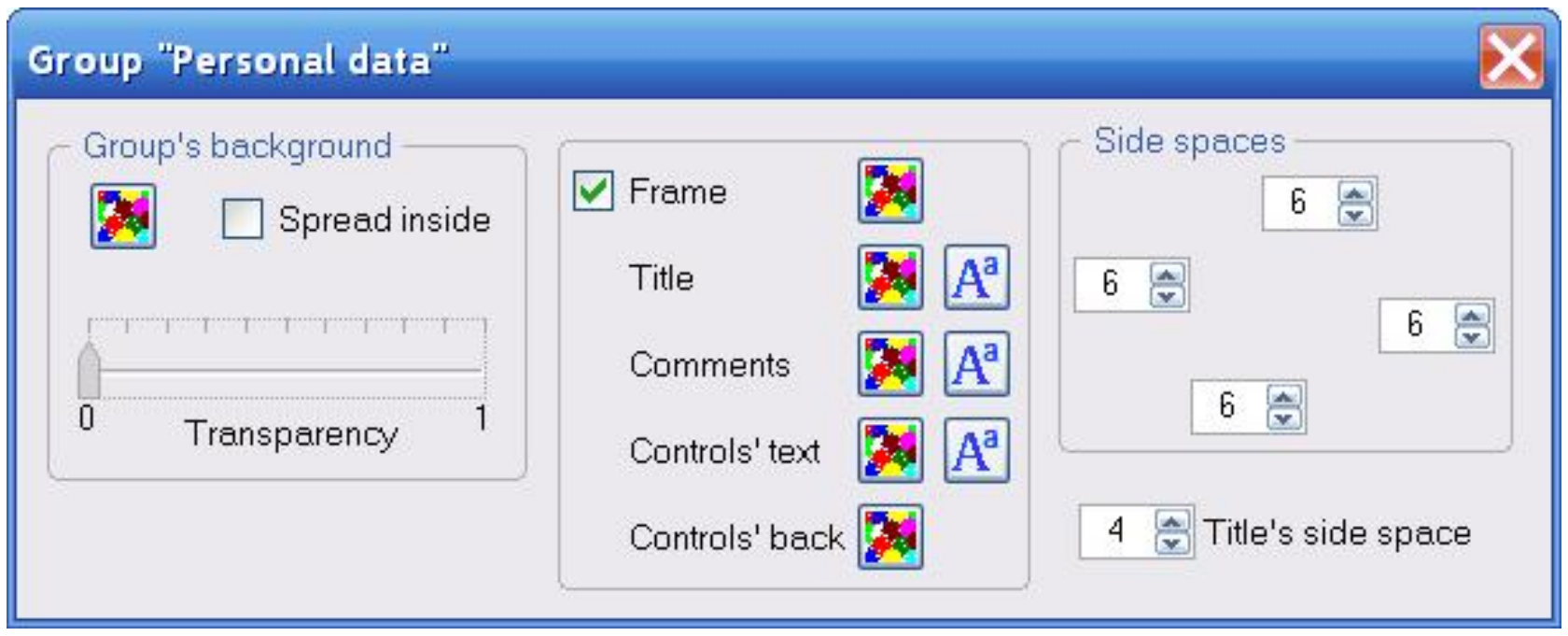

**Fig.14.26** Tuning form for the `ElasticGroup` objects

I have explained before that the frame of any `ElasticGroup` object automatically surrounds the whole set of inner elements regardless of their movements and the size change. A group can be always moved by any frame point or inner point; by increasing the distance between the inner elements and the frame you enlarge an area, where the group can be grabbed for moving. By changing the values in the four controls of the right group, you can change the distances between the elements and the frame of the tuned group.

The controls in the central group allow:

- To switch the drawing of the frame ON / OFF.
- To change the color and font of the title.
- To change the color and font simultaneously for all the comments of the tuned group.
- To change the background color, foreground color, and font for all the controls of the tuned group.

These two groups in the tuning form are the `ElasticGroup` objects with their `ElementsMovable` property set to `false`. Thus no elements inside these groups can be moved, but the groups can be.

Three elements in the left group at **figure 14.26** have such purposes:

- To change the background color of the tuned group.
- To spread (or not) this new background color on the inner groups; this checkbox is enabled only for the groups, which have inner groups.
- To change the transparency of the tuned group. This is an object of the `Trackbar` class, which was described previously (chapter *Complex objects*, section *Track bars*).

All three objects of this group are movable; the `Trackbar` object is resizable, so you can change the length of this track bar. The group can be moved around the form exactly as other two groups; on moving or resizing the inner elements, the frame of the group adjusts itself to the positions of the three inner elements. The group behaves exactly as any other `ElasticGroup`, but it cannot belong to this class. At the beginning of this section, while starting the discussion of the `ElasticGroup` class, I listed the classes of objects, which can be used as the inner elements of this class. All those choices are based on the controls; no arbitrary graphical object can be used as an element of the `ElasticGroup` class. Though the behaviour of the group is identical, it must be something else. This group is an object of the `ArbitraryGroup` class, which is discussed a bit further on. But before turning to another class of groups, I want to demonstrate the use of temporary groups of elements.

## The basis of total tuning

It is the vicious circle of explaining something really big, which has to include and consists of many details: if you begin with the declaration of the main thing, then it is not obvious that it has to be so and this misunderstanding causes the initial very strong denial of the whole thing; if you start from explaining all the needed details, then the main purpose of the long explanation is not obvious and the ongoing discussion of the numerous details causes the same negative effect.

The main theme of this book is the design of the user-driven applications, but in order to discuss the main rules of such applications I have to explain and demonstrate the design of many elements, without which those programs cannot be constructed. I have already mentioned the rules of user-driven applications. One of them is formulated in such a way.

**Rule 2**. All the visibility parameters must be easily controlled by the users.

I have demonstrated how this rule is implemented in several of the previous examples, but the explanation was mostly aimed at showing, how USERS can control all the visibility parameters and change the view of a working application. Certainly, any users' actions are transformed into the changes of the involved objects, so it must be based on the design of the associated classes. All the properties and methods of the involved classes are described in the **MoveGraphLibrary_Classes.doc** (see Programs and documents). I do not want to rewrite parts of that documents into this book, so I decided not to list in the text the properties and methods of the classes, which I demonstrate. They are mentioned in the code samples and you can easily find their descriptions in the mentioned document. The only exception, which I hope to make as short as possible, would be this subsection about the `ElasticGroup` class. This class became the most flexible and effective in design of absolutely different groups from the tiny to very complicated and consisting of several levels. Objects of this class are used in absolutely different applications; this is demonstrated in many examples of this book. I preferred first to demonstrate the use of many properties of this class and only after it to give some relatively short description of them. It will give some better understanding of the `ElasticGroup` class but also of other involved classes, because their design is based on exactly the same principles. If users are given the full control over the tuning of applications then it must be based on the same level of design for all the classes. After the demonstration of the **Form_PersonalData.cs,** it would be easier to understand, how all those commands of context menus and tuning form of the `ElasticGroup` class are transformed into the properties of this class.

The `ElasticGroup` objects consist of different combinations of controls; some of those controls are accompanied by their comments, so two groups of properties are aimed at the controls and comments.

| | | |
|---|---|---|
| `List<Control>` | `Controls` | Gets the list of controls, associated with the group. |
| `Color` | `ControlsBackColor` | Sets the back color for all the controls inside. |
| `Font` | `ControlsFont` | Sets the font for all the inner controls. |
| `Color` | `ControlsForeColor` | Sets the fore color for all the controls inside. |
| `int` | `ControlsNumber` | Gets the number of controls, used by the inner elements. |
| `Color` | `CommentsColor` | Sets the color of the comments on all inner levels. |
| `Font` | `CommentsFont` | Sets the font for all the inner comments. |
| `bool` | `CommentsMovable` | Sets the movability of the comments for the inner `CommentedControl` elements. |

Inner elements of a group can be of five different types, but they are all considered like elements (of the `ElasticGroupElement` class), so there are several properties to deal with these elements.

| | | |
|---|---|---|
| `List<ElasticGroupElement>` | `Elements` | Gets the List<> of inner elements. |
| `Font` | `ElementsFont` | Sets the font for the inner elements. |
| `bool` | `ElementsMovable` | Sets the movability of the inner elements only. If the elements are not movable, then their relative positions cannot be changed by moving any of them; the whole group is moved by any inner point. |
| `int` | `ElementsNumber` | Gets the number of the inner elements. |

The background color of the group, its transparency, and the spreading of the new background color on the inner groups are regulated by several properties.

| | | |
|---|---|---|
| `Color` | `BackColor` | Gets or sets the back color; spreading the same color on the inner groups depends on the flag. |
| `bool` | `BackColorSpreadInside` | Gets or sets the flag for spreading the back color of the group on the inner groups. If the new set value is `true`, then the inner groups get the same color. |
| `double` | `Transparency` | Gets or sets the group's transparency. |

Any group may have a frame; several properties deal with the parameters of the frame.

| | | |
|---|---|---|
| `Rectangle` | `FrameArea` | Gets the frame's area. |
| `Color` | `FrameColor` | Gets or sets the frame's color. |
| `int []` | `FrameSpaces` | Gets an array of frame spaces (left, top, right, bottom). |
| `Pen` | `Pen` | Gets or sets the pen for the frame. |
| `bool` | `ShowFrame` | Gets or sets the flag to show the frame line. |

Title is a separate element of the group, which is regulated by several properties. Title is not the part of the frame; title's appearance does not depend on the existence or non-existence of the frame.

| | | |
|---|---|---|
| `string` | `Title` | Gets or sets a title. |
| `double` | `TitleAlignmentCoef` | Gets or sets the positioning coefficient for the title ([0, 1]). |
| `Color` | `TitleColor` | Gets or sets the title's color. |
| `Font` | `TitleFont` | Gets or sets the title's font. |
| `bool` | `TitleMovable` | Gets or sets the movability of the title. |
| `int` | `TitleSideSpace` | Gets or sets the space between the title and the frame on its sides; the allowed range is [0, 12]. |

These are not all the properties of the `ElasticGroup` class. There are other properties to deal with the visibility and movability of the whole group and its elements. These properties are especially important and were discussed in two special subsections. There are also properties, which allow to deal simultaneously with the visibility parameters of the title and the inner elements.

I want to underline again that the above mentioned properties are not included here to show the full list of things that can be done with the `ElasticGroup` objects. This is only to give you the better understanding of the implementation of the basic rules of user-driven applications in the case of one of the classes, included into the **MoveGraphLibrary.dll**.

## *Temporary groups*

File: **Form_TemporaryGroup.cs**
Menu position: *Groups – Temporary group*

The previous example with the `ElasticGroup` class (**Form_PersonalData.cs**, **fig.14.16**) demonstrates the flexibility of the class and a lot of opportunities, which it provides in design of applications. But even with the variety of the groups in that example, the content of all of them was predetermined by the designer. Elements in each group can be hidden and returned back to view at any moment, but nothing can appear in any group from outside the list that was declared at the beginning. However, this can be not enough in many cases. There are situations, when there is a need to organize a group of any elements, which are currently on display. The **Form_TemporaryGroup.cs** demonstrates the use of the same `ElasticGroup` class for organizing such temporary groups.

**Figure 14.27** shows the initial view of the form. On opening the form, you see 20 buttons of the same size lined in two rows. The buttons are numbered to make the tracking of their moving and resizing easier. Each one of these buttons can be moved and resized individually like it was described before in the section *Moving solitary controls*. Buttons are not big but can be enlarged; the best way to do it is to grab any of them by the corner, though the places in the middle of any side can serve in similar way. Other parts of borders for all buttons are used for moving. The standard technique for moving controls individually.

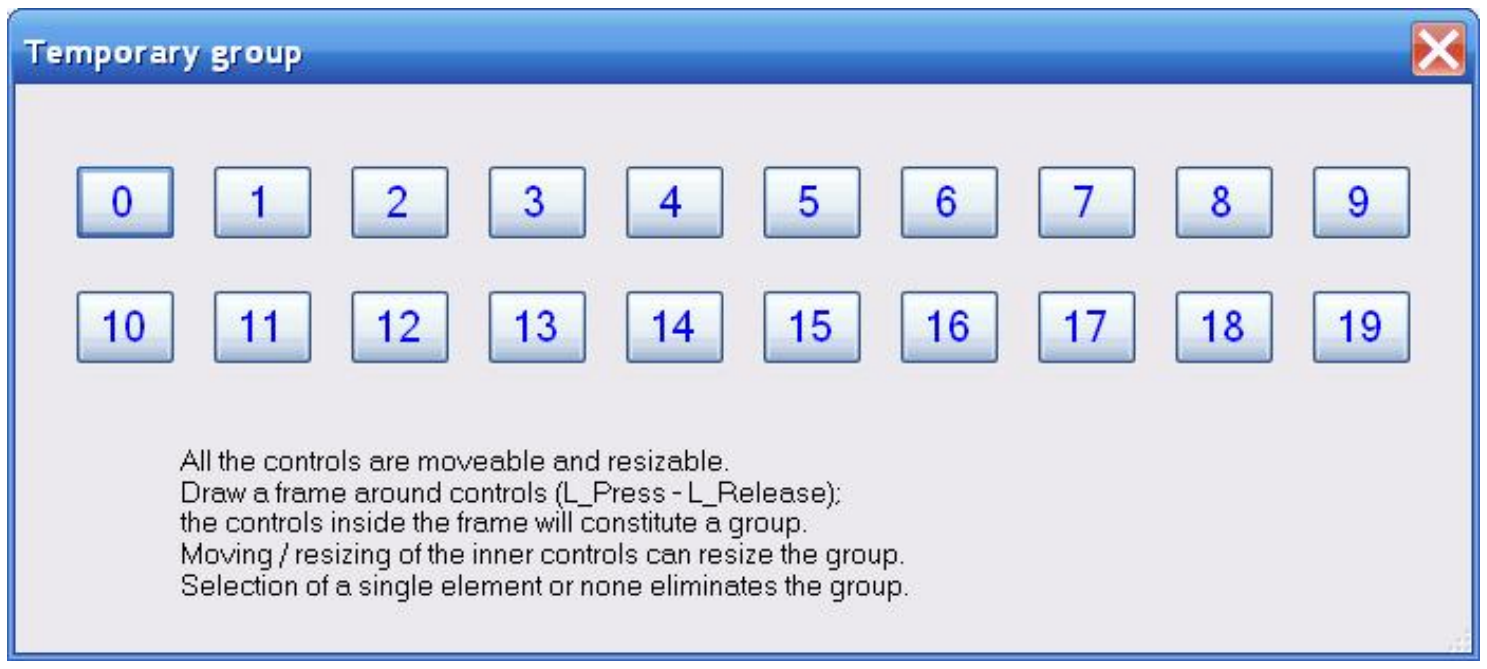

**Fig.14.27** The initial view of the **Form_TemporaryGroup.cs**

There are no groups in view, but you can move an arbitrary set of controls at any moment. A group is organized in a way, which is standard for such operations. Press the mouse button at one point, drag the mouse to another point, and release the button. The two points become the opposite corners of the rectangle. The rounded controls form an `ElasticGroup` object. As any control in the form can be moved and resized individually, then there is no sense in organizing a group with less than two buttons inside.

All the controls in the form are stored in some collection, to which there is an access via the `Controls` property. When the mouse is moved and released, I want to distribute all these controls between two lists. The first `List` will contain the controls, which are outside the painted rectangle; the second `List` contains only the controls, which happened to be inside.

```
List<Control> controlsSingle = new List<Control> ();
List<Control> controlsInFrame = new List<Control> ();
```

There is no rotation of any object in this form; there are also no context menus, so there is no difference on whether you would like to press left or right button to start rounding the controls.

```
private void OnMouseDown (object sender, MouseEventArgs e)
{
    ptMouse_Down = e .Location;
    if (!mover .Catch (e .Location, e .Button))
    {
        bRectInView = true;
    }
}
```

There are only 20 movable buttons and the movable text in this form. If none of these elements was caught, when the mouse button was pressed, then the flag `bRectInView` to draw the temporary frame is switched ON. The initially pressed point is one corner of the frame; the current point of the moving mouse is an opposite corner of the frame. If another group was organized before, it is still shown throughout the time of the new group designing.

```
private void OnPaint (object sender, PaintEventArgs e)
{
    Graphics grfx = e .Graphics;
    info .Draw (grfx);
    if (bRectInView)
    {
        grfx .DrawRectangle (Pens .Blue,
                             Math .Min (ptMouse_Down .X, ptMouse_Move .X),
                             Math .Min (ptMouse_Down .Y, ptMouse_Move .Y),
                             Math .Abs (ptMouse_Move .X - ptMouse_Down .X),
                             Math .Abs (ptMouse_Move .Y - ptMouse_Down .Y));
    }
    if (controlsInFrame .Count > 0)
    {
        group .Draw (grfx);
    }
}
```

The temporary frame is not the movable object! If the mouse is released without releasing any movable object, then it is the time to check the possibility of organizing the new group.

```
private void OnMouseUp (object sender, MouseEventArgs e)
{
    if (!mover .Release ())
    {
        if (bRectInView)
        {
            SetGroup (new Rectangle (Math .Min (ptMouse_Down .X, e .X),
                                     Math .Min (ptMouse_Down .Y, e .Y),
                                     Math .Abs (e .X - ptMouse_Down .X),
                                     Math .Abs (e .Y - ptMouse_Down .Y)));
            bRectInView = false;
        }
    }
    Invalidate ();
}
```

The two `Lists` of controls are cleared and all the controls of the form are checked against the area of the temporary frame. If only a single control was caught inside, then there is no sense in organizing such a group; this control is also moved to the `List` of individually movable buttons (`controlsSingle`). Everything else is going to happen in the

`RenewMover()` method but the crucial thing is whether there is anything in the `controlsInFrame` list or not, when this method is called.

```
private void SetGroup (Rectangle rc)
{
    controlsSingle .Clear ();
    controlsInFrame .Clear ();
    foreach (Control control in Controls)
    {
        if (rc .Contains (control .Bounds))
        {
            controlsInFrame .Add (control);
        }
        else
        {
            controlsSingle .Add (control);
        }
    }
    if (controlsInFrame .Count == 1)
    {
        controlsSingle .Add (controlsInFrame [0]);
        controlsInFrame .Clear ();
    }
    RenewMover ();
}
```

First, all the controls from the `controlsSingle` list are registered with the mover. If there are controls in the `controlsInFrame` list, then a group must be organized and include all these controls. The group (of the `ElasticGroup` class) has to be registered by its `IntoMover()` method; the group is registered ahead of all the controls, which are not included into the group. I also added a couple of code lines to highlight the area of the group and to make it slightly transparent; these two lines can be commented, if you do not like such an effect.

```
private void RenewMover ()
{
    mover .Clear ();
    foreach (Control ctrl in controlsSingle)
    {
        mover .Add (ctrl);
    }
    if (controlsInFrame .Count > 0)
    {
        group = new ElasticGroup (this, controlsInFrame);
        group .BackColor = Color .LightYellow;
        group .Transparency = 0.3;
        group .IntoMover (mover, 0);
    }
    mover .Add (info);
}
```

There are some programs, which allow to do the similar action on the controls. All these programs mark the selected elements with some additional signs and turn those programs into "Move" mode; after moving the elements, the signs are taken out, and the programs return to the normal mode of operation. I think it is a wrong decision. As you can see in my example, there is no other mode except the normal one. At any moment, whether there is a group or not, any control continues to work as it is supposed to do. If any methods are linked with the clicks of these controls, they can be started at any moment. Substitute these buttons with real controls from some real application; those controls can continue to fulfil their tasks regardless of their positions and sizes, of being inside some temporary group or outside. You receive exactly the same application, but users get a chance to rearrange its view.

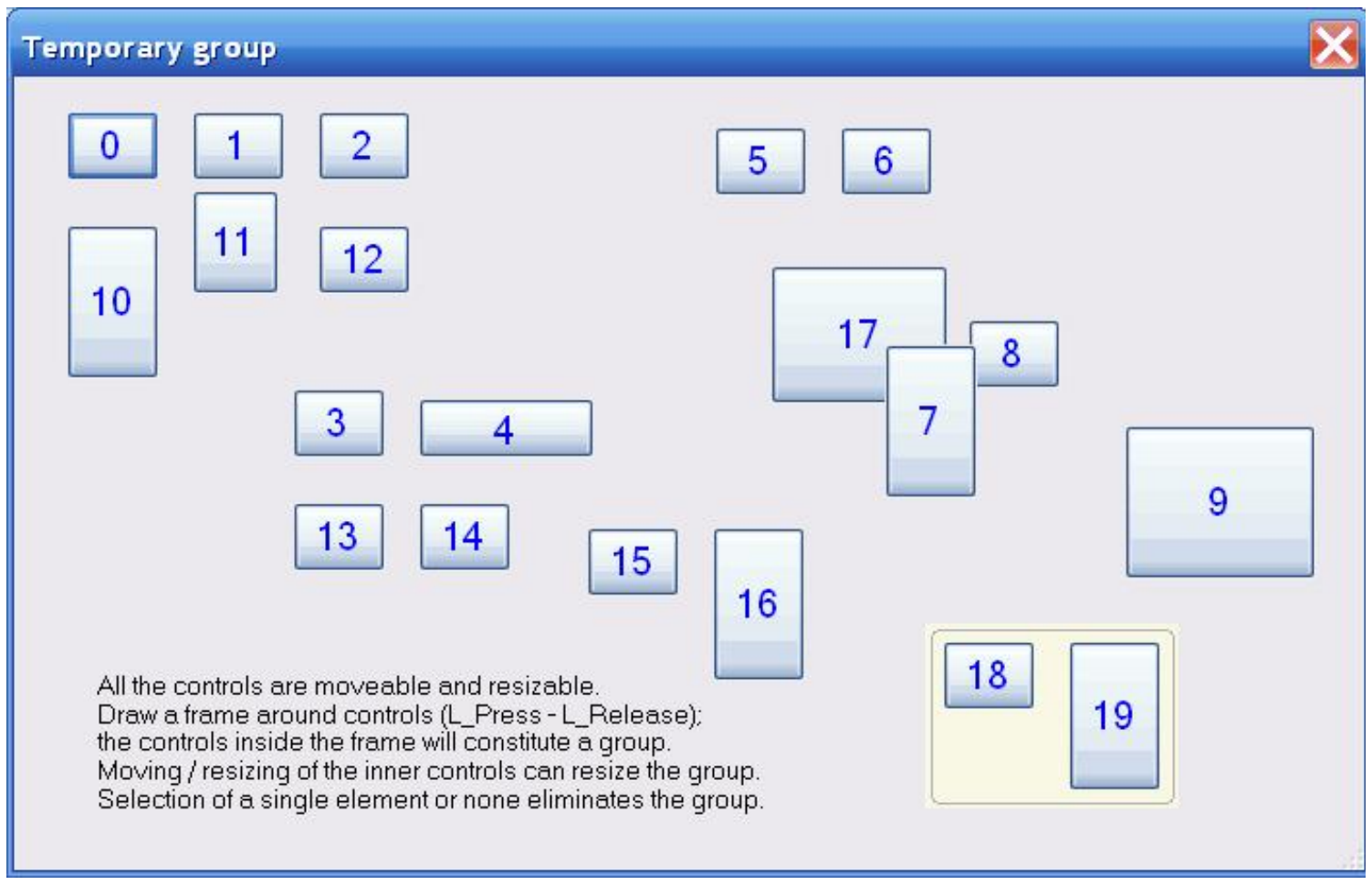

**Fig.14.28** The view of the form after several movements and resizing of some controls

The behaviour of the temporary group can be changed in different ways; I demonstrated it in some examples of the earlier programs, but decided not to include into this one in order to keep the code as simple, as possible. For example, this temporary group is organized at the moment, when the mouse is released; the content of the group is not changed until it is closed or another group is organized. The view of the group is easily changed by moving / resizing the inner controls, but not the list of controls in the group (**figure 14.28**). However, it is easy to add another checking of controls at the moment, when a group is moved somewhere and released or when any inner element is moved and released. At that moment the controls, which are still not in the group, can be checked against the area of the group and included into the group, if they are found to be inside the boundaries. In such a way the content of a group can be changed not only by drawing the new frame, but by annexation of other controls. You can find such an addition to the behaviour of a group in one of the further examples of this book (chapter *An exercise in painting*, **Form_Vilalge.cs**).

## Arbitrary groups

File: **Form_ArbitraryGroup.cs**
Menu position: *Groups – Arbitrary group*

The `ElasticGroup` class, discussed in the previous section, can be used for design of forms with any level of complexity, but there is one limitation on the inner elements of such groups: they must belong to one of the predetermined types.

- `SolitaryControl` object - a single control;
- `CommentedControl` object - a control with comment;
- `CommentedControlLTP` object - a control with comment, which has limitations on relative positioning;
- `DominantControl` object - a group of controls, one of which is dominant;
- `ElasticGroup` object - allows to organize a system of nested groups.

Such list of elements allows to organize any kind of groups, consisting of controls. The only possible inclusion of the graphical objects into such group can be done in the form of textual comments of the `CommentedControl` and `CommentedControlLTP` objects. The `ElasticGroup` class does not allow to include into a group an arbitrary graphical object. Though the `ElasticGroup` class is an excellent instrument for development of forms, this limitation becomes a real problem, when the design requires to unify into a group controls and some graphical objects. For these situations, the **MoveGraphLibrary.dll** contains the `ArbitraryGroup` class.

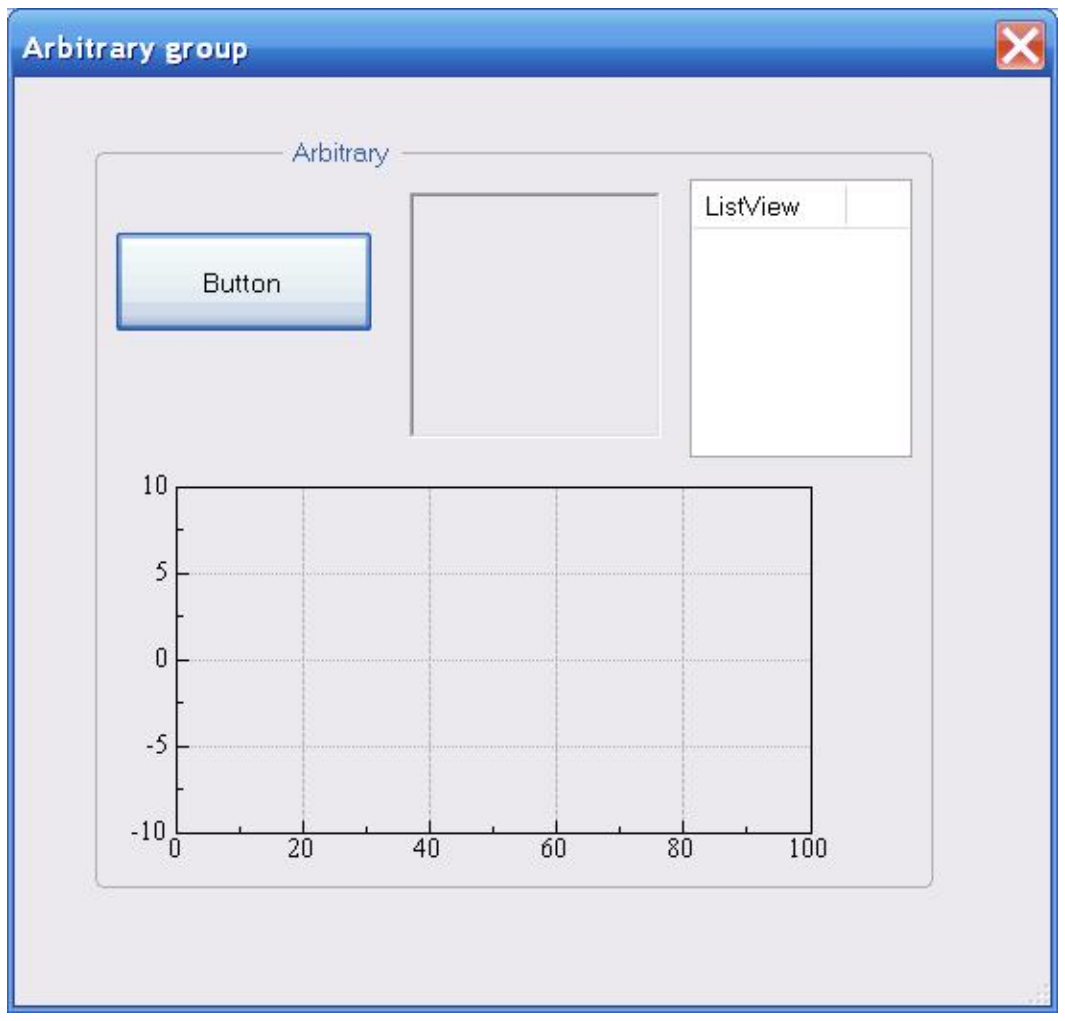

**Fig.14.29** An `ArbitraryGroup` object

I have already shown the use of this class in the tuning form of the `ElasticGroup` class (**figure 14.26**). The **Form_ArbitraryGroup.cs** allows to look into the details of using the `ArbitraryGroup` class (**figure 14.29**)

The main idea of the `ArbitraryGroup` class design is to get all the benefits of the `ElasticGroup` class, but to apply them to any type of inner elements. The visualization parameters of an arbitrary group include:

- Background color of a group.
- Spreading (or not) the background color into the inner groups (if there are such groups).
- Showing (or not) the frame around the group.
- Color of the frame, if it is shown.
- An array of spaces on four sides between the inner elements and the frame.
- Having (or not) a title.
- Title's font.

- Title's color.
- An additional space between the title and the line of the frame on its sides.
- Positioning coefficient for a title. A title can be moved between left and right sides with the coefficient taking a value from the [0, 1] range.
- Font and color for the inner elements.

These are all the same parameters that can be tuned for any ElasticGroup object. The only difference is in the technique of tuning these parameters. In the case of the ElasticGroup class, there is a special tuning form to do all the changes. In case of the ArbitraryGroup class, there is no such tuning form, at least, at this moment. All the parameters of visualization for an object of this class are saved in a GroupVisibleParameters field; an access to this field is through the property with the same name. By clicking the right button anywhere inside the group, but not on any element of the group, you open a context menu with a number of lines, allowing to change all the parameters. Each of these changes is going through the mentioned field, which keeps the set of parameters. For example, here is the change of the title's color.

```
private void Click_miTitleColor (object sender, EventArgs e)
{
    ColorDialog dlg = new ColorDialog ();
    dlg .Color = group .GroupVisibleParameters .TitleColor;
    if (dlg .ShowDialog () == DialogResult .OK)
    {
        group .GroupVisibleParameters .TitleColor = dlg .Color;
        Invalidate ();
    }
}
```

The rules of moving / resizing an ArbitraryGroup object are the same as for the ElasticGroup class.

- Any element of a group can be moved and resized individually. This action is regulated only by the rules of the class for each particular element.
- The frame of a group is automatically adjusted to the positions and sizes of all the inner elements and always surrounds them.
- A group can be moved by any inner point.

The rules are the same but the implementation must be different. The reason is obvious: for the fixed list of elements in the ElasticGroup class it was possible to define beforehand and put into the code the most crucial features of the group's behaviour:

- Synchronous movement of elements.
- Updating the group after any change of the inner elements.
- Drawing.
- Registering in the mover's queue.

With the open list of elements of the ArbitraryGroup class, none of these actions can be coded beforehand. Thus, the methods to do all these things must be provided at the moment of the group's initialization. Here is the constructor for this class.

```
public ArbitraryGroup (Form formSrc,                // form, in which the group is used
         Rectangle rcElems,                         // initial area of the inner elements
         GroupVisibleParameters vispars,            // group's parameters of visualization
         string strTitle,                           // group's title; it can be empty
         Delegate_Move onSynchroMove,               // method for synchronous movement of the elements
         Delegate_NoParams onUpdate,                // method to update the group
         Delegate_Draw onDrawElems,                 // method to draw the group
         Delegate_IntoMover onIntoMover)            // method to register the group with the mover
```

Let us see, how it all works in the case of group from **figure 14.29**. The group consists of three controls and a Plot object. It is better to have these controls in one array, as this array can be useful in several places.

```
            ctrls = new Control [] { button1, listView1, panel1 };
```

The **Form_ArbitraryGroup.cs** allows to save all the positions and parameters of visualization in the `Registry` to restore the same view on the next opening of the form, but let us begin with the situation, when nothing is stored yet. If nothing is restored from the `Registry`, then:

- Declare a plot (a `Plot` object).
- Calculate the united area of all the inner elements.
- Organize a `GroupVisibleParameters` object containing the group's parameters of visualization.
- Write four methods that are needed to initialize a group.
- Initialize a group.

```
public Form_ArbitraryGroup ()
{
    InitializeComponent ();
    mover = new Mover (this);
    ctrls = new Control [] { button1, listView1, panel1 };
    RestoreFromRegistry ();
    if (!bRestore)
    {
        plot = new MSPlot (this, new Rectangle (350, 300, 300, 200));
        Rectangle rcElems = Auxi_Geometry .RectAroundControls (ctrls);
        rcElems = Rectangle .Union (rcElems, plot .RectAround);
        GroupVisibleParameters visparams = new GroupVisibleParameters ("sample1",
            0, BackColor, true, true, true, Auxi_Colours .DefaultFrameColor (this),
            new int [] { 10, 6, 10, 10 }, true, Font, SystemColors .Highlight,
            0.25, 4, Font, ForeColor);
        group = new ArbitraryGroup (this, rcElems, visparams, "Arbitrary",
                                OnSynchroMove, GroupUpdate, OnDrawElems, OnIntoMover);
        visparams .GroupID = group .ID;
    }
    RenewMover ();
}
```

The `GroupVisibleParameters` object can be organized with the majority of parameters getting the default values; instead I decided to use the constructor with the direct setting of maximum of parameters.

The most interesting aspect of designing the `ArbitraryGroup` object is the preparation of those four functions, which describe the whole group's behaviour. The `GroupUpdate()` method is used from the very beginning to calculate the frame based on the positions of the inner elements and some parameters from the `GroupVisibleParameters` field. On three sides of the group's frame, it is enough to know the spaces between the inner elements and the frame; on the upper side the title and the title's font are also needed to calculate the position of the frame. If there is no title, then the upper space is the minimum distance from the inner elements to the frame; otherwise it is the minimum distance between the elements and the title.

```
private void GroupUpdate ()
{
    GroupVisibleParameters visparams = group .GroupVisibleParameters;
    int [] spaces = visparams .SideSpaces;
    Rectangle rcElems = Auxi_Geometry .RectAroundControls (ctrls);
    rcElems = Rectangle .Union (rcElems, plot .RectAround);
    int cxL = rcElems .Left - spaces [0];
    int cyT = rcElems .Top - (spaces [1] +  Auxi_Geometry.RoundMeasureString (this,
                                  group .Title, visparams .TitleFont) .Height / 2);
    int cxR = rcElems .Right + spaces [2];
    int cyB = rcElems .Bottom + spaces [3];
    group .FrameArea = new Rectangle (cxL, cyT, cxR - cxL, cyB - cyT);
}
```

It is easy to organize the synchronous movement of inner elements: for controls it is a simple change of location, while the `Plot` object has its own `Move()` method.

```
void OnSynchroMove (int dx, int dy)
{
    Size size = new Size (dx, dy);
    foreach (Control ctrl in ctrls)
    {
        ctrl .Location += size;
    }
    plot .Move (dx, dy);
}
```

The drawing method is the easiest for this group as no drawing is needed for the controls, while the `Plot` object has its own `Draw()` method.

```
private void OnDrawElems (Graphics grfx)
{
    plot .Draw (grfx);
}
```

Registering this group in the mover's queue is also simple, but it must obey several standard rules.

1. The controls must precede the graphical objects in the queue.
2. All the inner elements must precede the group itself.

I did not include anything else into the **Form_ArbitraryGroup.cs**, but in general there can be any number of other movable objects in the same form. Taking into consideration the whole set of movable objects in the form, you decide about their order in the mover's queue; let us say that out `ArbitraryGroup` object must be included on the `iPos` position in this queue. To place our group into the mover's queue in correct order, I start with the group itself and then insert before it all the elements that must precede it. Controls can be registered by the `Mover.Insert()` method; the `Plot` object, as any complex object, has to use its own `IntoMover()` method.

```
void OnIntoMover (Mover mv, int iPos)
{
    mover .Insert (iPos, group);
    plot .IntoMover (mover, iPos);
    foreach (Control ctrl in ctrls)
    {
        mover .Insert (iPos, ctrl);
    }
}
```

After the `ArbitraryGroup` object is initialized, it is registered in the mover's queue in the same way as any other complex object.

```
private void RenewMover ()
{
    mover .Clear ();
    group .IntoMover (mover, 0);
}
```

This `group.IntoMover()` method simply calls the `OnIntoMover()` method of the group, which was prepared beforehand. The same happens with the drawing, where the `group.Draw()` method calls the `OnDrawElems()` method of the group.

```
private void OnPaint (object sender, PaintEventArgs e)
{
    Graphics grfx = e .Graphics;
    group .Draw (grfx);
}
```

When the group is moved, the mover itself calls the `OnSynchroMove()` method, but with the moving / resizing of the inner elements the situation is different. In the `ElasticGroup` class, each inner element is cast into an ElasticGroupElement object; any moving or resizing of such object automatically causes the updating of the group, to which it belongs. In the ArbitraryGroup class, there is no casting into anything else. The inner elements in our arbitrary group belong to such classes that can be used in the group or as stand alone movable objects. There is no indication that any movement of these objects can affect some arbitrary group. Certainly, I could organize some checking, based on the identification numbers, but I decided to simplify the code and call the Update() method of the group whenever anything is moved by the mover. This update is done in an instant.

```
private void OnMouseMove (object sender, MouseEventArgs e)
{
    if (mover .Move (e .Location))
    {
        group .Update ();
        if (mover .CaughtSource is SolitaryControl  ||
            mover .CaughtSource is ArbitraryGroup)
        {
            Update ();
        }
        Invalidate ();
    }
}
```

There are several other things in the **Form_ArbitraryGroup.cs**, which are standard for all the user-driven applications. First, the tuning of the Plot object can be done; as usual, it is called by a double click on the main plotting area or on any scale. Second, as I have already mentioned, all the parameters of visualization are stored in the Registry to be used on the next opening of the form.

The ArbitraryGroup class has the same parameters of visualization as the ElasticGroup class but there is no standard tuning form for all of them. There is a context menu, which can be called on the group; commands of this menu allow to change a lot of visualization parameters. Some of the commands of this menu will call the standard dialogues for changing color or font; others will call a couple of small tuning forms which can remind you some parts of the tuning form for the ElasticGroup class (**figure 14.26**).

Short conclusion to this chapter.

Though I have included into this chapter some interesting and unusual examples, I do not want to make an impression that I have described all the possible variants of organizing groups. On the contrary, I think that the movability of the elements opens so many possibilities for design of groups that there will be no end of the fresh ideas. These ideas are going to be not only about some improvement of one or another variant but really new. Some of them might look a bit crazy at the beginning because you never saw and, certainly, never tried such a thing, but… Do not make any conclusions without trying. I have already received some reviews from the “big specialists” on the subject that “users do not need the movability of the elements inside the programs”. When the scientists from our department of mathematical modeling were reading these reviews, there were roaring with laughter, because these scientists now refuse to use any new program, in which something is not movable. They quickly understood the new opportunities for their research work and do not want to continue without them.

One “crazy” idea about the groups. It is not simply an idea, because I have already designed such groups and used them, but did not include any such example into this book.

It is not a rare situation when you have a whole group of parameters associated with some class (type, form) of objects. So it is an expected thing to organize a group, which has the name of these objects in the title and a bunch of controls inside the group to change the parameters. At the same time, these objects can be either shown or not. If they are shown, then all the inner elements of the group are enabled; if the objects are not shown, then the inner elements must be disabled. To switch objects ON / OFF, there is a CheckBox; usually this CheckBox has the same text as the group with the parameters. So you have a CheckBox, which switches ON / OFF the group with the same text. Why not to combine them by using the CheckBox as the title of the group?

You may think that this is a crazy idea, but I am using such groups and they do not look crazy at all. On the contrary, it makes the most natural and obvious link between the switch and the parameters, which are controlled by this switch.

# Programs and documents

Several applications are developed for demonstration of design and use of movable / resizable objects. In most cases these applications were developed as accompanying programs for some publications. There are also some other documents with a helpful information. All files are available at www.sourceforge.net in the project **MoveableGraphics** (names of projects are case sensitive there!). I renew the files there from time to time, when the new (better) version of the library or a new demonstration program is ready. Usually a newer version of DLL may appear once a month; the preparation of the big book "***World of Movable Objects***" delayed the new version for six months. I try to organize these files in such a way that all the programs with all their source codes use the same latest version of DLL, which is currently on display. As there are people who strongly oppose using the DOC format, I try to include all the documents both in DOC and PDF formats. Some of the documents may appear on other sites, but www.sourceforge.net is the only place, where I place the DLL and all the demonstration projects with all the available source files.

| | |
|---|---|
| **WorldOfMoveableObjects.zip** | File contains the book "***World of Movable Objects***" in DOC format (this is the current document) and the program to accompany this book (the whole project with all the codes in C#). All the examples in the book are from this project. If you want only to run this application, then two files are needed: **WorldOfMoveableObjects.exe** and **MoveGraphLibrary.dll.** |
| **Book_WorldOfMoveableObjects.zip** | File contains the book both in DOC and PDF formats. 368 pages |
| **TheoryOfUserDrivenApplications.zip** | File contains an article "***User-Driven Applications***" (in DOC and PDF formats; 34 pages) and an accompanying application (the whole project with all the codes in C#). To run an application, only two files are needed: **TheoryOfUserDrivenApplications.exe** and **MoveGraphLibrary.dll.** |
| **OnTheTheoryOfMoveableObjects.zip** | File contains an article "***On the Theory of Moveable Objects***" (in DOC and PDF formats; 31 pages) and an accompanying application (the whole project with all the codes in C#). To run an application, only two files are needed: **TheoryOfMoveableObjects.exe** and **MoveGraphLibrary.dll.** |
| **MoveGraphLibrary_Classes.zip** | File contains the description of classes, included into the **MoveGraphLibrary.dll**. Two versions are available (DOC and PDF, 149 pages). |
| **MoveGraphLibrary.dll** | The library. |
| **LiveCalculator.zip** | Calculator, which is included as one of the examples into the **WorldOfMoveableObjects.exe**, but presented here as a separate program. To use this application, it must be accompanied by the **MoveGraphLibrary.dll**, which is also inside the zip file. |
| **Test_MoveGraphLibrary.zip** | Before the book "***World of Movable Objects***" was written, it was the biggest collection of the examples with their explanations. Now it became redundant, but I will keep this file for some time more because of some of the included examples. File contains a big article "***Moveable and Resizable Objects***" (in DOC and PDF formats; 97 pages) and an accompanying application (the whole project with all the codes in C#). To run an application, only two files are needed: **Test_MoveGraphLibrary.exe** and **MoveGraphLibrary.dll.** |
| **TuneableGraphics.exe** | The older application, which demonstrates the movable / resizable objects from absolutely different areas. Some of the used examples were not included into later Demo applications. |
| **MoveGraphLibrary_OlderClasses.doc** | Description of the older classes, still included into **MoveGraphLibrary.dll**. |